\pdfoutput=1
\documentclass[a4paper,12pt,times,numbered,print,index,custommargin,custombib]{Classes/PhDThesisPSnPDF}

\input{Preamble/preamble}

\begin{document}

\graphicspath{{Figures/}}

\frontmatter

\includepdf[pages=-]{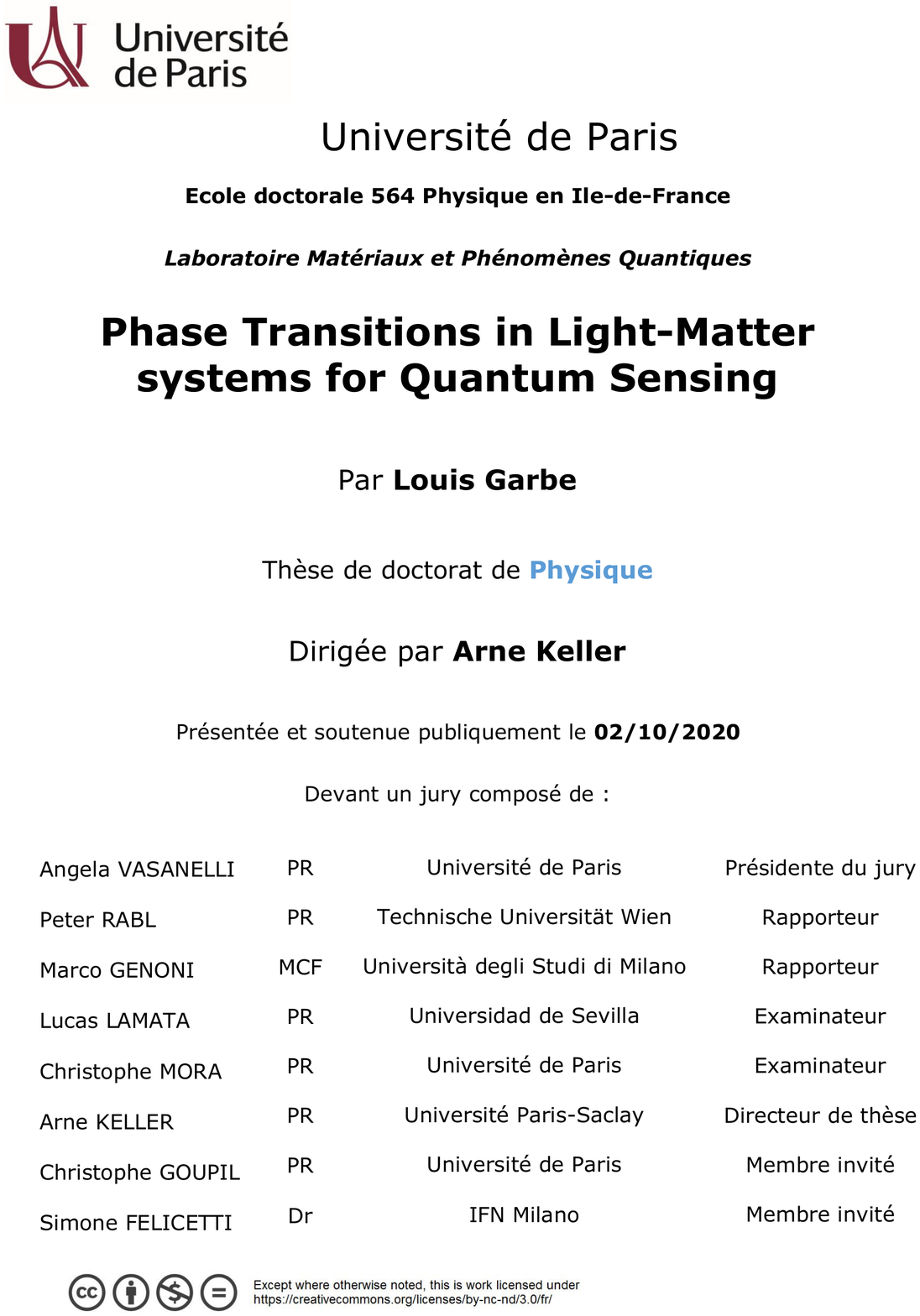}



\begin{dedication} 

\textit{\`A mes amis.} \\


\end{dedication}

\pagenumbering{gobble}
\epigraph{\textit{Dans quel monde souhaites-tu vivre? Un monde avec des pyramides, ou un monde sans pyramides?}}{Hayao Miyazaki}



\textbf{R\'esum\'e Court}

Lorsque lumière et matière sont faiblement couplées, elles peuvent être traitées comme des systèmes distincts, échangeant des quanta d’énergie. En revanche, lorsque le taux de couplage devient très élevé, les deux systèmes se mêlent pour former des excitations hybrides, qui ne peuvent être décrites isolément en termes de lumière ou de matière. Au long de cette dissertation, nous étudierons quelques-unes des propriétés exotiques qui surviennent dans ce régime. Nous accorderons notamment une grande attention à l’émergence de transitions de phase quantiques dans ces systèmes.
L’un des axes de recherche développés ici est l’étude d’un mécanisme de couplage à deux photons, par lequel des photons sont créés ou absorbés par paires. Ce mécanisme crée un diagramme de phase très riche, présentant à la fois une transition de phase et des instabilités.
Une autre étude porte sur l’utilisation de ces transitions de phase pour le développement de capteurs. En effet, à proximité du point critique, le système devient extrêmement sensible à des perturbations extérieures. Nous présenterons un protocole exploitant un unique système à deux niveaux, couplé à un champ bosonique. En dépit de sa simplicité, un tel système peut générer des transitions de phase. Près du point critique, la fréquence du champ et celle du système à deux niveaux peuvent toutes deux être mesurées avec une précision accrue. Ainsi, des transitions de phase dans des systèmes de taille finie pourraient être utilisées pour développer des capteurs de petite taille.
Enfin, en utilisant le formalisme des théories de ressources, nous étudierons comment la capacité d’un système physique à effectuer des tâches métrologiques pourrait être utilisée pour caractériser et quantifier la « non-classicalité » d’un tel système.\\

\hspace{20pt}

\textbf{Mots Clefs}: Interaction Lumière-matière, Couplage Ultra-fort,  Transitions de Phase Quantiques, Systèmes de Taille Finie, Métrologie Quantique, Capteurs Quantiques, Etats Gaussiens.


\textbf{Short Abstract}

When light and matter are weakly coupled, they can be described as two distinctive systems exchanging quanta of energy. By contrast, when their coupling strength becomes very large, the systems hybridize and form compounds that cannot be described in terms of light or matter only. In this Thesis, we will study some exotic properties which arise in this regime. In particular, we will be interested in the possibility to engineer quantum phase transitions in these systems. 
One direction we explore is the study of two-photon coupling, a mechanism in which photons are created or emitted in pairs. This mechanism creates a rich phase diagram containing both phase transitions and instabilities. Another point of interest is the possibility to use these transitions for sensing applications. Indeed, near the critical point, the system becomes extremely sensitive to external perturbations. We will present a protocol in which a single qubit coupled to a bosonic field. Despite its simplicity, this system displays a phase transition. Near the critical point, both the frequency of the qubit and the field can be measured with improved accuracy. Hence, finite-size transitions could be used to develop small-scale sensors.
As last topic, we study how the ability of a system to perform certain metrological tasks could be used to characterize and quantify nonclassicality, by using the formalism of resource theories.\\

\hspace{20pt}

\textbf{Keywords}: Light-matter Interaction, Ultrastrong Coupling Regime,  Quantum Phase Transitions, Finite-size Systems, Quantum Metrology, Quantum Sensing, Gaussian States.

     
\chapter*{Remerciements}
     
Me voilà donc arrivé à l’issue de ma thèse. Le manuscrit est écrit, la soutenance faite, les rapporteurs ont rapporté, les examinateurs ont examiné. Je tiens à remercier tous les membres du jury, notamment Peter Rabl et Marco Genoni, pour avoir pris le temps d'étudier mon travail en détails, et pour leurs commentaires et questions très constructives. La partie scientifique de la thèse étant achevée, il faut maintenant écrire la section de loin la plus importante, et prendre le temps de m'adresser à ceux qui m’ont accompagné au long de la route. \\

Les premiers remerciements, je les dois à ma famille. Pour mes parents d’abord, pour tous les efforts qu’ils ont consacré à notre éducation. Quatre enfants à élever, tous avec un goût prononcé pour les sujets bizarres et les études interminables, voilà un défi qui n’est récompensé par aucun prix Nobel, mais qui demande tout autant de patience, de travail, et de motivation. Et, je l’espère, une égale quantité de plaisir et de bonheur. Sans eux, sans l’accompagnement constant qu’ils m’ont procuré, je ne serais jamais arrivé là où j’en suis aujourd’hui. La formule est convenue, mais le fait est là. Je tiens donc encore à les remercier, eux avant tous les autres. Puis, il y a mes frères et sœurs. Un quatuor de fortes têtes, avec ses moments partagés, ses conflits parfois. Merci à eux pour leur soutien, à Clarisse notamment pour sa patience durant ces années de cohabitation.

Après ma famille, viennent mes maîtres. Ma propre expérience de l’enseignement, bien que brève, m’a fait réaliser à quel point former les autres est chose difficile. Ils n’en ont que plus de mérite. Je veux d'abord remercier Shéhérazade, pour avoir cru qu’on pourrait faire quelque chose du petit garçon bizarre que j’étais. Ensuite mes enseignants de lycée et prépa. Enfin, et surtout, tous mes professeurs de l’ENS de Lyon,  ceux qui m’ont pleinement initié à la physique moderne, dont cette mécanique quantique qui est finalement devenue mon domaine d’élection. Pour tout ce qu’ils m’ont fait découvrir, depuis la vie sexuelle des muons jusqu’aux opérateurs sandwichés, voire, en une mémorable occasion, au calcul quantique en présence d’une machine à remonter le temps dans des univers multiples (si si), et pour les renseignements qu’ils m’ont apporté sur la carrière de chercheurs, je tiens à les remercier tous. Et parmi eux, je suis particulièrement reconnaissant envers Tommaso Roscilde, Arnaud le Diffon, et Pascal Degiovanni. Enfin, je pense que l'endroit est adéquat pour remercier Olivier Revol, Florence Roger et Rosa de Vogüé, pour l'aide particulière et très précieuse qu'ils m'ont apportés.\\

Après ces années de formation, j'ai commencé ma thèse, cette situation particulière où l’on est à la fois chercheur, enseignant et étudiant. C'est à ce triple titre que je dois adresser mes remerciements. Tout d'abord, il y a tous ceux qui m'ont accompagnés dans mon travail de recherche ; et, en tout premier lieu, les personnels administratifs. On n’y pense pas toujours, ils n’ont pas leur nom sur les articles ou les thèses publiées, mais leur niveau de compétence a une influence considérable sur la vie du labo et le quotidien des chercheurs. Or, j’ai eu la chance d’avoir eu affaire à des personnes aussi compétentes que bienveillantes. Il y avait d’abord eu Jérôme Calvet et Fadela Djélloul, nos secrétaires du département de physique de l’ENS. Au laboratoire MPQ, ce furent Anne Servouze, Nathalie Merlet, Jocelyne Moreau et Sandrine DiConcetto qui prirent le relai. Quelles que soient les difficultés qu’ait pu poser ma thèse sur le plan scientifique, mes problèmes administratifs ont toujours été résolus avec efficacité grâce à elles. Il me faut également adresser une pensée à Ai Sato, qui a organisé de façon impeccable le voyage au Japon que mon travail m’a donné l’occasion d’effectuer. Je profite donc de ces lignes, les seules où j’en ai vraiment l’occasion, pour leur adresser à toutes mes remerciements et ma gratitude.

Un autre élément qui, pour un théoricien, se place juste derrière manger et juste avant dormir en termes d’importance, c’est le bon fonctionnement de son matériel informatique. Il me faut donc remercier nos informaticiens, Wilfrid Niobet et Loïc Noël, pour leur travail d’entretien du parc informatique, et l’aide qu’ils m’ont apportée lorsque ma bécane n’en faisait qu’à sa tête. 

Viennent ensuite tous les chercheurs avec lesquels j'ai eu l'occasion de travailler durant cette thèse. D’abord mon directeur de thèse Arne Keller, pour son aide, pour nos discussions stimulantes, et pour l’autonomie scientifique qu’il m’a laissé. Puis Simone Felicetti, pour son apport scientifique, ses nombreux conseils, et pour toutes les personnes que j’ai eu l’occasion de rencontrer grâce à lui. Nos discussions fréquentes et notre travail commun ont jouées un rôle déterminant dans l’élaboration et l’avancée de ma thèse. Ce travail m’a aussi donné la grande chance de pouvoir voyager, de découvrir d’autres groupes de recherche et d’autres pays. L’essentiel des résultats présentés dans ce manuscrit ont été obtenus grâce à ces collaborations ; je souhaite donc remercier chaleureusement tous ceux qui m’ont accueillis et au contact desquels j’ai eu l’occasion de me former. Il y a d’abord eu le voyage à Bilbao que j’ai effectué alors que je travaillais sur le tout premier projet de ma thèse : je tiens à remercier Enrique Solano pour son accueil, et I\~nigo Egusquiza pour son apport décisif à ce projet. Il y a ensuite eu mon séjour milanais, durant lequel j’ai eu le plaisir de travailler avec Matteo\&Matteo, Paris et Bina. Outre notre travail qui s’est avéré fructueux et stimulant, je garde un excellent souvenir de nos déjeuners à l’Union Club, dont j’espère avoir l’occasion prochaine de saluer de nouveau la patronne. A la fin de ce séjour, j’ai eu le plaisir de rencontrer Nathan Shammah, aussi bon physicien que cuisinier. Si bon, à vrai dire, que je suis allé jusqu'au Japon quelques mois plus tard pour le revoir. Je suis très reconnaissant à Franco Nori pour son invitation, qui m'a donné cette magnifique opportunité, et à Tarun pour m'avoir hébergé à la fin de mon séjour. Ce séjour fut la conclusion d’une collaboration menée avec Nathan et Fabrizio Minganti, l’une des personnes les plus sympathiques et des chercheurs les plus brillants que j’ai jamais eu le plaisir de rencontrer. Et parce que décidément, la nourriture compte, c’est le souvenir des dîners le soir à la cantine, et des repas au Gatten Sushi avec Fabrizio, Nathan et Ezio, que j’évoquerai ici. Enfin, je remercie Dominik \v{S}afr\'anek, que j'ai eu le plaisir de rencontrer au cours d'une conférence, et dont la thèse m'a été très utile dans mon propre travail.

Ces collaborations internationales ne me font pas non plus oublier les chercheurs du laboratoire MPQ. Je souhaite d'abord adresser ma reconnaissance envers Thomas Coudreau, en qui j'ai trouvé une personne de bon conseil, et toujours disponible pour répondre à mes doutes et à mes questions. Je veux aussi remercier Pérégrine Wade pour notre travail commun et nos discussions, et lui souhaiter le meilleur. Je remercie également Cristiano Ciuti pour notre collaboration au début de ma thèse, Pérola Milman pour l'encadrement qu'elle m'a fourni durant le stage précédant ma thèse, ainsi que tous les autres membres permanents de l'équipe QiTe: Sara Ducci, Florent Baboux, Maria Amanti, Lucas Guidoni et Jean-Pierre Likforman. Et il y a, enfin, toutes les autres personnes du laboratoire avec lesquelles j’ai eu le plaisir de discuter de la physique, des physiciens, et d’autres sujets connexes; en particulier Alexandre le Boîté, Massil Lakehal, Kevin Dalla Francesca, Alberto Biella et Pierre Allain.

La seconde activité d'un thésard est celle d'enseignant. La combinaison de l'enseignement et de la recherche constitue une richesse ; enseigner un sujet est un des meilleurs moyens pour le comprendre. Mais ces activités entrent aussi parfois en conflit, surtout lorsqu’il faut concilier une charge d’enseignement avec une invitation à une conférence ou une collaboration au bout du monde. Je tiens donc à remercier tous mes collègues, surtout François le Diberder, Frédérick Bernardot et Sara Ducci, pour leur aide et leur souplesse lorsqu’il a fallu échanger des créneaux, ainsi que tous mes étudiants pour leur patience. J’espère leur avoir été de quelque utilité au cours de ces séances de TD, et leur adresse mes vœux de réussite. 

Enfin, en quelques occasions, il m’a fallu me refaire moi-même étudiant et suivre des formations, dont certaines se sont révélées très enrichissantes. J'ai eu notamment l'opportunité d'animer un stand du Palais de la Découverte, une belle expérience pour laquelle je remercie le personnel du Palais. Je suis également reconnaissant à Olivier Darrigol, Jan Lacki et Vincent Jullien pour leur cours d'histoire des sciences. Enfin, je dois tout particulièrement remercier Laurence Viennot pour sa formation à la didactique de la physique, qui fut très stimulante intellectuellement. Comme elle a   elle-même l'habitude de le faire, je lui souhaite bon vent pour la suite. \\

Enfin, après la famille, les enseignants, les personnels administratifs et techniques, les collaborateurs et collègues, et se confondant souvent avec eux, viennent les amis. Ils sont toujours essentiels, mais peut-être tout particulièrement durant les années de thèse. Un chercheur en physique quantique, s’il a l’heur d’être expérimentateur, passe ses journées à manipuler des tubes et des petits miroirs dans l’espoir d’étudier des particules que personne ne peut voir. S’il a la malchance d’être théoricien, il passe ses journées à faire des calculs afin de comprendre comment il faut bouger les tubes. Certes, c’est un métier passionnant et riche. Mais il n’empêche, on se retrouve parfois (voire souvent) à se demander ce qu’on est en train de faire de sa vie. C’est pour cela que les amis sont si importants. Ceux qui sont aussi chercheurs, avec qui l’on peut partager les doutes, mais aussi retrouver le plaisir de faire de la physique ; et ceux qui ne le sont pas, avec lesquels on peut s’ouvrir, s’adonner à d’autres activités, envisager d’autres perspectives. Ce sont mes amis, avant tout, qui m’ont porté pendant mes années d’études et de thèse. C’est à eux que je dois mes meilleurs souvenirs. Et c’est à eux, à eux tous, que cette thèse est dédiée.

Il y a, en premier lieu, ceux de l’ENS. D’abord les trois inséparables, Thibaut, David et Paul. Je n’ai pas oublié ces jeudis après-midi, dans notre recoin de la bibliothèque, à préparer les TD de mécanique quantique du lendemain. On n’y comprenait peut-être pas grand-chose (je n’y comprends d’ailleurs pas beaucoup plus aujourd’hui) mais ces après-midi sont restés pour moi un des meilleurs moments de mes études. 
Il y a ensuite le Bouquet. Je trempe mon morceau de chocolat dans ma tasse de thé au litchi, et tout à coup le souvenir m’apparaît : celui de ces soirées musicales et surtout amicales, dont je veux saluer chacun de ses participants. Pierre, toi le gourou qui présidait aux destinées de cette sorte de secte. Emmanuel, toi l’Hêtre musical que je revoie encore composer une sonate lors d’une de ces soirées, imperturbable malgré le chaos ambiant. Flora, toi qui officiais au premier bouquet auquel j’ai assisté : je n’ai oublié ni Einaudi découvert à cette occasion, ni ton sourire.  Erwan, toi qui fus le sel de ma vie (parfois) et celui sur mes plaies (souvent). Et vous encore, Célia, Marion, Thibault, Guillaume, Corentin; le Cerisier vous salut et vous remercie. Et lorsqu’il fallut se séparer, lorsque le noyau se dispersa sous l’effet de ce dieu  jaloux qu’est le doctorat, ce fut pour revivre sous une autre forme, avec d’autres amis. \`Eve d'abord, que je remercie encore pour son accueil en Normandie et tous les bons moments passés ensemble. Puis Petrus, Deniz, Eva et Jeanne. Que ce soit pour débocher un matheux à Paris (opération finalement couronnée de succès!), pour aller étudier les organismes phosphorescents dans le Calvados, ou pour aller planer sur du Pink Floyd à Strasbourg tout en buvant du Gevrey-Chambertin 94, les mélomanes des deux cercles ont toujours répondu présent.
 
Il y a aussi le club des Miyazakistes (la deuxième secte). Se retrouver la nuit dans un amphithéâtre pour regarder des films d’animation en japonais, il faut être fou ou physicien pour y penser. Merci à Clément, à Méril, à Géraldine, à Thibaut encore, et à Lety\&Steve (bon courage à vous deux pour la suite de vos aventures au pays de l’oncle Sam !). 
Il y a aussi Loreena, Vincent, Carine, Ruben (ses gâteaux à l’acide formique, son système de surveillance intégré, ses capacités de cambrioleurs), et Pauline. Il en est certains que j’ai revu depuis, d’autres non, mais je leur adresse à tous mes remerciements et ma reconnaissance.

Enfin, et surtout, il y a toi Alex, mon coloc, mon compagnon de route, mon jumeau bénéfique, toi qui m’as fait découvrir le Bouquet et bien plus. Merci pour ton humour, ton soutien et ta patience face à mes emportements, et bien sûr pour ta modestie. Et merci, avant tout, pour ton amitié. Je me réjouis d'avance de vous revoir, \`Eve et toi, de vous accueillir à Vienne, et de toutes nos futures expériences!

Je me tourne maintenant vers les thésards et postdocs du laboratoire, à commencer par ceux de l’équipe QiTe. D’abord Gaël, Nicolas et Simone. A toi Gaël, merci pour toutes nos conversations, depuis le binning d’une mesure homodyne jusqu’à l’historique des vitraux de la cathédrale de Chartres, en passant par les mérites (ou leur absence) d’un salaire à vie ou quelques crêpages de chignon sur l’Union Européenne ; merci pour ton immense curiosité, et pour avoir autant élargi mes horizons. A toi Nicolas, merci pour ta franchise, tes connaissances, les papiers que tu m’as fait découvrir, pour toutes nos râleries communes et pour toute l’aide que tu m’as apporté. Et à toi Simone, merci non seulement pour ton aide scientifique que j’ai déjà eu plus haut l’occasion de mentionner, mais aussi pour ton calme et ton optimisme constant. 

Il y a ensuite nos meilleurs ennemis, les expérimentateurs, nos compagnons de bureau avec lesquels nous avons maintenu plus qu’une relation diplomatique pendant toutes ces années : Saverio, Félicien, le vélo de Félicien, Théo (rival de Nicolas, aussi bien pour la râlerie que pour les connaissances),  Arnau\underline{lt}, Jérém\underline{ie}, ainsi que Vincent et Iännis, les espions DON. Merci pour tous ces moments passés ensemble, pour tous ces échanges quotidiens, ces petits riens qui font tout.  
Je n’oublie pas non plus les anciens membres de l’équipe que j’ai eu le plaisir de côtoyer. D’abord Sergueï ; désormais, comme toi, je peux répondre lorsqu’on me demandera mon activité : I’m the doctor. Ensuite Ibrahim: je n’oublierai ni ta persévérance, ni ton courage, ni ta gentillesse.  Et puis Vincent, Tom, Sergueï, Adrien, Aurianne, Saulo, Andreas, Giorgio, Claire et Jonathan ; je vous souhaite à tous de vous épanouir dans la voie que vous avez choisis.

Et enfin, il y a tous les autres doctorants et postdocs de toutes les autres équipes, qui ont animé la vie du labo durant toutes ces années. Jacko, à laquelle je souhaite de ne jamais laisser s’éteindre le feu de sa passion. Chloé, en compagnie de laquelle j’ai pu tester une cavité Fabry-Pérot de stress, et qui m’a fait voir des échantillons fragiles sans que j’en casse aucun, un vrai miracle. Nicolas Auvray, dont j'envie les futurs élèves (et j'espère bien que tu leur apprendras comment améliorer le fonctionnement d'un troll grâce à des supraconducteurs haute Tc!). Et encore Pierre (et sa campagne), Jean-Côme, Massine, Cassia, Zakari, Nicola, Quentin, Ian, Corneliu, et tous les autres.

En-dehors du labo, il y a eu aussi l'Agape, cette retraite faite de nourritures terrestres et spirituelles. Merci encore au gourou Pierre pour l'organisation de cet événement, et merci à toutes les personnes passionnées et passionnantes que j'ai eu le plaisir d'y rencontrer ou d'y revoir; notamment Robin, Johannes, Titouan, Jérémy et Adrian.

Mais une thèse ne se résume heureusement pas seulement aux heures passées à faire de la physique. Ce sont désormais vers tous ceux que j’ai eu le bonheur de fréquenter durant mes activités annexe que je me tourne. D’abord ceux de la Fabryk, notre cher club de théâtre. Paul d’abord ; certains t’ont définis comme « bizarre mais pratique » ; j’ajouterais à cela excellent acteur et gars plus que sympa. Ensuite Daniele ; le contraste est impressionnant entre ta capacité à jouer les rôles de méchants et ton immense gentillesse. Je crois d’ailleurs que nous n’avons pas encore terminé de définir ce qu’est un chiffre, il faudra que nous y revenions un de ces jours. Et je pense également à Valentin, Jeanne, Fabrice, Khaleb, Zoé, Swan, ainsi bien sûr qu’à nos chers metteurs en scène, Marie-Line et Marc. Il y a aussi le voyage (encore un!) que j'ai eu le plaisir de faire au Vietnam, et dont je salue tous les participants, et surtout Ulli, pour son soutien lorsque je me débattais avec les subtilités de la langue allemande.
Enfin, je dois aussi remercier tous les membres du club d’éloquence de l’ENS (la fausse, celle de Paris) : d’abord Thibaut (toujours lui), qui m’a fait découvrir la confrérie ; Raphaëlle, qui, lorsqu’elle n’est pas occupée à soulever des voitures avec des dictionnaires, sait à merveille désarmer un adversaire de quelques phrases bien senties ; Noé, philosophe, physicien, rimeur, bretteur, musicien, à l’occasion voyageur aérien ; Xavier, notre arme secrète et futur président de la République (on ne sait pas encore de laquelle, mais il en trouvera une), Elise, Louis, Laetitia, et tous les autres, tous les anciens, tous les nouveaux, ces chers bizus qui nous rappellent, bien aimablement, que le temps passe.\\

A vous tous, pour m'avoir tant enrichi, pour m'avoir porté jusque-là, je vous dis encore merci et vous souhaite le meilleur.

 \chapter*{List of Publications}
	
	\begin{itemize}
		\item L. Garbe, I.L. Egusquiza, E. Solano, C. Ciuti, T. Coudreau, P. Milman, S. Felicetti,\\
		\textit{Superradiant phase transition in the ultrastrong-coupling regime of the two-photon
Dicke model},\\
 Physical Review A \textbf{95}, 053854 (2017).
 		\item L. Garbe , S. Felicetti, P. Milman, T. Coudreau, A. Keller,\\
 		 \textit{Metrological advantage at finite temperature for Gaussian phase estimation},\\
 		 Physical Review A \textbf{99}, 043815 (2019).
 		\item L. Garbe, M. Bina, A. Keller, M.G.A. Paris, S. Felicetti,\\
 		 \textit{Critical Quantum Metrology with a Finite-Component Quantum Phase Transition},\\
 		  Physical Review Letters \textbf{124}, 120504 (2020).
 		\item L. Garbe, P. Wade, F. Minganti, N. Shammah, S. Felicetti, F. Nori,\\
 		 \textit{Dissipation-induced bistability in the two-photon Dicke model},\\
 		  Scientific Reports \textbf{10}, 13408 (2020)
	\end{itemize}

 \chapter*{How to use this Thesis}
	
	In this document, I discuss the research I have conducted in the course of my Ph.D. thesis, on the topics of ultrastrong light-matter coupling, superradiant phase transitions, and quantum metrology. The first three chapters of the manuscript present the main concepts and key results of these fields of research. I have strived to write these chapters pedagogically; as such, they can be used by non-specialists or students as an introduction to these domains. Those wishing to go further will also find references to numerous review papers. The three other chapters present my own research contributions. Although most of those have already been published in the articles listed above, this manuscript also contains additional results, remarks, and perspectives, and should be considered as an improved version of the original papers.

 \tableofcontents

 \listoffigures

 \listoftables


\addcontentsline{toc}{chapter}{\nomname}


# LIST OF TABLES

# SYMBOLS AND PROPERTIES

**Bosonic operators**

$\hat{a}, \hat{a}^\dagger$     Creation and annihilation operators of a bosonic field

$[\hat{a}, \hat{a}^\dagger] = 1$

$\hat{x} = \frac{\hat{a}+\hat{a}^\dagger}{\sqrt{2}}, \hat{p} = \frac{i(\hat{a}^\dagger - \hat{a})}{\sqrt{2}}$    Quadratures of the bosonic field

$\hat{X} = \hat{a}^2 + \hat{a}^{\dagger 2}, \hat{Y} = i(\hat{a}^{\dagger 2} - \hat{a}^2)$

$|m\rangle$     Fock states $\hat{a}^\dagger \hat{a}|m\rangle = m|m\rangle$

$|\alpha\rangle$     Coherent states $e^{\alpha^* \hat{a} - \alpha \hat{a}^\dagger}|0\rangle$

$\hat{a}|\alpha\rangle = \alpha|\alpha\rangle$

$|\xi\rangle$     Squeezed vacuum states $e^{\frac{1}{2}(\xi^* \hat{a}^2 - \xi \hat{a}^{\dagger 2})}|0\rangle$

$|n, \alpha\rangle$    Displaced Fock state $e^{\alpha^* \hat{a} - \alpha \hat{a}^\dagger}|n\rangle = (\hat{a}^\dagger + \alpha^*)^n|\alpha\rangle$

$|n, \xi\rangle$    Squeezed Fock states $e^{\frac{1}{2}(\xi^* \hat{a}^2 - \xi \hat{a}^{\dagger 2})}|n\rangle$

$|n, \xi\rangle = \left(\cosh(\xi)\hat{a}^\dagger + \sinh(\xi)\hat{a}\right)^n|\xi\rangle$ if $\xi$ is real

$\langle\xi|\hat{x}^2|\xi\rangle = e^{-2\xi}$ and $\langle\xi|\hat{p}^2|\xi\rangle = e^{2\xi}$ if $\xi$ is real

**Spin operators**

$\hat{\sigma}^x = \begin{bmatrix} 0 & 1 \\ 1 & 0 \end{bmatrix}, \hat{\sigma}^y = \begin{bmatrix} 0 & -i \\ i & 0 \end{bmatrix}, \hat{\sigma}^z = \begin{bmatrix} 1 & 0 \\ 0 & -1 \end{bmatrix}$    Pauli matrices

$\hat{\sigma}^+ = \begin{bmatrix} 0 & 1 \\ 0 & 0 \end{bmatrix}, \hat{\sigma}^- = \begin{bmatrix} 0 & 0 \\ 1 & 0 \end{bmatrix}$    raising and lowering operators

$\hat{J}_x = \frac{1}{2}\sum_{j=1}^{N} \hat{\sigma}_j^x, \hat{J}_y = \frac{1}{2}\sum_{j=1}^{N} \hat{\sigma}_j^y, \hat{J}_z = \frac{1}{2}\sum_{j=1}^{N} \hat{\sigma}_j^z$    Collective spin operators

$[\hat{J}_a, \hat{J}_b] = i\varepsilon_{abc}\hat{J}_c$

$\hat{J}_\pm = \hat{J}_x \pm i\hat{J}_y$

$|j;m\rangle$   Angular momentum eigenbasis $\hat{J}^2|j;m\rangle = j(j+1)|j;m\rangle$ and $\hat{J}_z|j;m\rangle = m|j;m\rangle$

$|\uparrow\rangle, |\downarrow\rangle$   Eigenstates of $\sigma_z$

$|\leftarrow\rangle, |\rightarrow\rangle$   Eigenstates of $\sigma_x$

**Light-Matter coupling**

$g$   Individual coupling between qubit and bosonic field

$G = \sqrt{N}g$   Collective coupling

$\tilde{g} = Ng$   Two-photon coupling

$g_p, G_p, \tilde{g}_p$   Coupling value at the transition in the absence of dissipation

$g_p^D, G_p^D, \tilde{g}_p^D$   Coupling value at the transition in the presence of dissipation

$\lambda = \frac{g}{g_p} = \frac{G}{G_p} = \frac{\tilde{g}}{\tilde{g}_p}$   Renormalized coupling constant

$\omega$   Frequency of the bosonic field

$\Omega$   Frequency of the qubit

$\eta = \frac{\Omega}{\omega}$   Frequency ratio

$\tilde{\eta} = \frac{N\Omega}{\omega}$

$\alpha_0 = \frac{g}{\omega}$

$\Delta$   Frequency detuning

$\kappa$   Bosonic field dissipation rate

$\Gamma$   Qubit decoherence rate

$\beta, \zeta, \gamma$   Critical exponents

**Metrological quantities**

$\tau$   Duration of a metrological protocol

$\delta x$   Estimation error for the parameter $x$

$F_x$   Fisher Information associated with the estimation of the parameter $x$

$\mathcal{I}_x$   Quantum Fisher Information associated with the estimation of the parameter $x$

$\mathcal{Q}_x$   Signal-to-noise ratio associated with the estimation of the parameter $x$

| | |
|---|---|
| $\mathcal{F}$ | Resource monotone |
| $\mathcal{A}$ | Metrological advantage |

**Quantities in phase space**

| | |
|---|---|
| $\sigma$ | Covariance matrix of a Gaussian state |
| $\boldsymbol{d}$ | Displacement vector of a Gaussian state |
| $\mathcal{B}$ | Mode-mixing in phase space |
| $\mathcal{S}$ | Squeezing matrix in phase space |
| $\mathcal{R}$ | Phase-shift in phase space |

$$K = \begin{bmatrix} \mathbb{1}_q & 0 \\ 0 & -\mathbb{1}_q \end{bmatrix} \quad \text{Symplectic form}$$

**Physical constants**

| | |
|---|---|
| $e$ | Electron charge |
| $\varepsilon_0$ | Vacuum permittivity |
| $\mu_0$ | Vacuum permeability |
| $\hbar$ | Quantum of action (set to 1 throughout the manuscript) |
| $T$ | Temperature |
| $k_B$ | Boltzmann's constant |

**Others**

| | |
|---|---|
| $\hat{\rho}$ | Density matrix of the system |

$L[\hat{O}](\hat{\rho}) = \hat{O}\hat{\rho}\hat{O}^\dagger - \frac{1}{2}(\hat{O}^\dagger\hat{O}\hat{\rho} + \hat{\rho}\hat{O}^\dagger\hat{O})$  Dissipative term in Lindblad equation

| | |
|---|---|
| $\Lambda$ | Quantum channel (Completely positive trace-preserving map) |
| $a^*$ | Complex conjugate of $a$ |
| $\hat{U}^\dagger$ | Hermitian conjugate of $\hat{U}$ |

$\text{Var}(\hat{U}) = \sqrt{\langle \hat{U}^2 - \langle \hat{U} \rangle^2 \rangle}$  Variance of $\hat{U}$

| | |
|---|---|
| $\nabla$ | Nabla operator |

**Abbreviations**

| | |
|---|---|
| BEC | Bose-Einstein Condensate |
| BS | Beam-Splitter |
| CR | Cramér-Rao |
| DSC | Deep-Strong Coupling |
| FI | Fisher Information |
| FTQL | Finite-Temperature Quantum Limit |
| GHZ | Greenberger-Horne-Zeilinger |
| HOM | Hong-Ou Mandel |
| HP | Holstein-Primakoff |
| JC | Jaynes-Cummings |
| NV | Nitrogen-Vacancy |
| PLE | Passive Linear Elements (Beam-splitter and phase-shifters) |
| POVM | Positive Operator-Valued Measurment |
| QED | Quantum Electrodynamic |
| QFI | Quantum Fisher Information |
| QPT | Quantum Phase Transition |
| RWA | Rotating Wave Approximation |
| gRWA | generalized Rotating Wave Approximation |
| SC | Strong Coupling |
| SLD | Symmetric Logarithm Derivative |
| SNR | Signal-to-Noise Ratio |
| SQL | Standard Quantum Limit |
| SQUID | Superconducting Quantum Interference Device |
| SSB | Spontaneous Symmetry Breaking |
| SSR | Superselection Rules |
| TC | Tavis-Cummings |
| TOF | Time-Of-Flight |




\mainmatter


\chapter*{General Introduction}

\addcontentsline{toc}{chapter}{General Introduction}

Recent years have seen impressive progress in our ability to manipulate quantum states in many platforms. It is now possible to generate, manipulate, and fully characterize truly quantum states, which have permitted to test fundamental predictions of quantum mechanics such as superposition, entanglement, or violation of Bell's inequalities.
These fundamental phenomena have been witnessed with cold atoms, superconducting circuits, photons, trapped ions, nitrogen vacancies in diamond, just to cite a few platforms. The burgeoning field of quantum technologies now seeks to exploit these counter-intuitive effects for technological applications, in particular in the domains of computation, communication, and sensing. 
In parallel, there is a considerable research effort which strives to better understand and characterize these quantum correlations. The development of new tools to quantify entanglement, and the study of eluding effects such as nonlocality and contextuality, are but two examples.

The coupling between light and matter is at the heart of all these experiments. Sometimes, light is used as a tool to manipulate quantum information encoded in matter degrees of freedom. For instance, lasers are tools of choice to encode, manipulate, and read the electronic states of atoms or ions. Sometimes, light itself is in a non-classical state; the field of quantum optics has greatly contributed to these new advances. 

One important improvement concerns the interaction strength between light and matter. Quantum optics experiments have been traditionally limited to the weak coupling regime, in which light and matter retain their own identity and exchange excitations. For the last ten years, however, it has become possible to reach the so-called ultrastrong coupling regime, in which the interaction strength becomes comparable with the bare excitation energy of the photon field. In this regime, light and matter can no longer be described as separate entities, but must be considered through hybridized excitations. This regime comes with several exotic properties, which have been the subject of much theoretical work.

One of such properties is the superradiant phase transition. When the coupling is increased above a certain threshold, the bosonic field acquires a macroscopic population in its ground state. For a long time, this phenomenon was considered to be a collective effect that could only occur in the presence of multiple emitters. However, recent research has shown that this effect is actually a very general feature of ultrastrong coupling, and indeed can even occur with a single emitter coupled to a bosonic field.

The phenomenon of superradiance provides a connection between the fields of quantum optics and quantum phase transitions. Quantum phase transitions have been historically studied in many-body systems, typically in the context of condensed matter theory. The presence of many components allows for a very rich phenomenology. However, it also makes it much more challenging to access the truly quantum properties of the system: these are generally hidden in many-body correlations, not in macroscopic, directly observable quantities. To put it in an informal way, quantum mechanics is necessary to describe the behavior of graphene, but one does not test the Bell inequality by putting a piece of graphene in a superposition of two macroscopic states. There has been an increasing effort in recent years to apply tools from the quantum information community to study how these quantum properties could be extracted from these many-body systems.

In this Thesis, I will also be interested in this interplay between quantum information and quantum phase transition, but from the opposite perspective. I have considered simple, few-body quantum optics systems (or many-body systems which effectively behave as few-body), in which quantum information concepts can be naturally applied, and studied how these properties would be affected in the presence of quantum critical effects. I have put a specific focus on the use of these systems for quantum sensing and metrology.

In this context, I have conducted research in three different directions. First, I have considered interaction processes beyond the usual dipolar coupling. More specifically, I have studied two-photon processes, in which pairs of photons are simultaneously absorbed or emitted. This process leads to a dramatic effect, the spectral collapse, in which the energy levels coalesce as the system becomes unstable. I have investigated how this two-photon coupling could also lead to a superradiant phase transition, and the fate of both transition and spectral collapse when dissipation is considered.

Second, I have studied the superradiant phase transition from the perspective of sensing. In the vicinity of a critical point, a system becomes extremely sensitive to small perturbation. Using tools from the field of quantum metrology, I have characterized a sensing protocol exploiting critical light-matter interaction, and studied whether this could lead to enhanced sensitivity.

The last direction is more fundamental. It is established that non-classical correlations can be used to improve sensing protocols. Using the framework of resource theory, I have proposed that the ability to achieve a metrological advantage might in turn be used to define and quantify nonclassicality.

\begin{center} 

\textbf{Outline of the Thesis}

\end{center}

This Thesis is structured as follows. The first part is a review of important results and concepts which have been used in this work. In Chapter I, I introduce the concept of ultrastrong coupling between matter and light, from a theoretical and experimental perspective. I detail the definition of weak, strong, and ultrastrong coupling regime, and the phenomenology which is associated with this last regime. I also describe the various experimental platform in which ultrastrong coupling has been realized or approached. 
In Chapter II, I explain in detail what a superradiant transition is. I describe the models in which this transition has been observed or predicted, emphasizing the general connection with ultrastrong coupling. I put a specific emphasis on quantum simulation, which allows to realize the transition by sidestepping fundamental limitations.
Chapter III is an introduction to the conceptual framework of quantum metrology. I discuss the fundamental notions of Quantum Fisher Information and Standard Quantum Limit. I put a specific emphasis on the hypotheses which are used in deriving this limit, and I discuss how well-designed protocols can be used to circumvent it.

 The second part of the manuscript is composed of my own research contributions. In Chapter IV, I present my work on two-photon coupling. I first discuss the two-photon Dicke Hamiltonian, which describes the exchange of pairs of excitations between a collection of qubits and a bosonic field. I show that this model presents an interplay between a spectral collapse and a superradiant-like phase transition. Then, I study the fate of both transition and collapse in the presence of dissipation.
 Chapter V is focused on the use of superradiant transitions for quantum metrology. I study the behavior of the Rabi model near a critical point, and how it could be exploited for sensing tasks. I put a specific emphasis on the trade-off between better accuracy and longer protocol duration.
Finally, in Chapter VI, I discuss how the ability to perform certain metrological tasks could be used to characterize and quantify nonclassicality, by using the formalism of resource theories.

\chapter{Ultrastrong coupling between light and matter}


\epigraph{\textit{There is a crack in everything: that's how the light gets in.}}{Leonard Cohen}

\def\hbar{{\mathchar'26\mkern-9muh}}

   The coupling between light and matter is at the heart of many fields of research. In particular, the development of quantum technologies was ushered in by the achievement of the strong coupling regime, where the coupling strength becomes comparable with the losses. In this regime, experimental signatures arising from quantum coherence can be observed without the blurring effect of decoherence. However, even in these experiments, light and matter retained their own identity. In the last two decades, a new regime, the \textit{ultrastrong} coupling regime, has been considered theoretically and demonstrated experimentally, eventually opening a new field of research. In this regime, the coupling becomes comparable with the bare frequencies of the system; it is then no longer relevant to talk about light or matter independently. Instead, one must think in terms of \textit{hybrid} excitations. In this Chapter, we will present some concepts and achievements in that field. Several reviews \cite{kockum_ultrastrong_2019,forn-diaz_ultrastrong_2019,boite_theoretical_2020} have been published on this topic, and have been used in the writing of this Chapter.
   The interested reader may find a more complete presentation in those references.

  \section{Regimes of coupling between light and matter}

  \subsection{Introduction}

  Interaction between light and matter is ubiquitous in physics. The strength of this interaction has long been beyond experimental control. However, with the advent of cavity QED, it became possible to increase this strength by using mirrors or other photonic structures to confine the light around emitters (such as a cloud of atoms). Historically, experiments have been realized in the weak coupling regime, where the interaction strength $\gind$ is small compared with the dissipation rate $\Gamma$ of the atom and the loss rate $\kappa$ of the cavity. In this regime, a photon emitted by the atoms effectively escapes the cavity immediately, and the field acts only as an environment for the emitter. The presence of the cavity leads to an enhancement of the spontaneous emission rate, a phenomenon known as the Purcell effect.

Eventually, the development of cavity QED made it possible to reach the strong coupling (SC) regime, in which the interaction becomes larger than the decay rates. In this regime, the behavior of the system is dramatically modified. An atom will be periodically de-excited and re-excited, a behavior known as Rabi oscillation. Intuitively, this can be understood by the following argument: the coupling strength quantifies the typical rate at which an atom can emit or absorb a photon. Since this rate is large compared with the cavity losses, a photon emitted by the atom will remain in the cavity long enough to be absorbed again, re-exciting the atom.  Rabi oscillations then describe the coherent exchange of excitations between the matter and the light. Hence, in the SC regime, the field within the cavity can no longer be modeled by a dissipative bath, and the full quantum behavior of individual photons needs to be considered. The population of photons within the cavity can be described by bosonic creation (annihilation) operator $\adag$ ($\aop$). The atoms can be modeled by two-level systems (qubits), described by Pauli matrices $\hat{\sigma}_{x,y,z}$. In the simplest case of a single qubit interacting with the bosonic field, the behavior of the system can be described by the paradigmatic quantum Rabi model:

 \begin{equation}
 	\Hop=\Of\adag\aop+\frac{\Oq}{2}\sigz+\gind(\adag+\aop)\sigx,
 	\label{C1_QRM}
 \end{equation}
 where $\Of$ and $\Oq$ are the frequency of the field within the cavity and the qubit, respectively. 

 After the pioneering works realized with atoms in cavity, these ideas have been put to use in many other platforms, in particular superconducting circuits or solid-state systems. These experiments have allowed exploring entirely new regimes of coupling. The most prominent of those is the ultra-strong coupling (USC) regime, where the light-matter interaction becomes comparable to the bare frequencies of the system $\Of$ and $\Oq$ (a conventional criterion is $\gind\geq0.1\Of$). The description of the system in terms of individual qubit and field excitations, which efficiently explains the Rabi oscillations, ceases to be relevant. Instead, light and matter are so strongly coupled that they hybridize to form \textit{polaritons}. In this Chapter, we  will discuss the counter-intuitive properties which emerge in this regime, as well as some of the most recent experimental achievements.

 A remark should be made here: despite the name, USC does not mean SC with larger coupling. The two regimes compare the coupling with different quantities: the bare frequencies for the USC, and the dissipation rate for the SC. Therefore, the two regimes are not necessarily related. In practice, for most experiments considered here, we have $\Of\geq\kappa,\Gamma$, and achieving USC does imply achieving SC \cite{forn-diaz_ultrastrong_2019}. However, it is also possible to have USC without having SC; in particular, it was shown that the hybridization effects of USC can be preserved even in the regime of large dissipation \cite{liberato_virtual_2017}. In the following, we will assume $\gind\sim\Of,\Oq$, without making specific assumptions on the dissipation rates.

 \subsection{Jaynes-Cummings model}

 We will now discuss the properties of the Rabi model \eqref{C1_QRM} in the different coupling regimes. First, let us note that there exists an analytic procedure to derive exactly the spectrum of the Rabi model in all regime of parameters \cite{braak_integrability_2011}. However, the solutions are not found in a closed form, but as coefficients of an unbounded series, which in general are computed with numerical assistance. Accordingly, while it makes it possible to deduce some important properties of the eigenstates, it does not directly lead to an intuitive picture. 
Therefore, we will mostly focus here on other methods, which are better suited to develop physical intuition. 

 The interaction term in \eqref{C1_QRM} can be decomposed in two parts, the corotating term $\sigp\aop+\sigm\adag$ and the counter-rotating term $\sigm\aop+\sigp\adag$. The first conserves the total number of excitations $\adag\aop+\sigz$, but the second does not. For weak coupling value $\gind\ll\Of,\Oq$, the counter-rotating term can be neglected, yielding the Jaynes-Cummings (JC) model: 
 \begin{equation}
 	\Hop_{JC}=\Of\adag\aop+\frac{\Oq}{2}\sigz+\gind(\sigp\aop+\sigm\adag).
 	\label{C1_JCM}
 \end{equation}

 This so-called rotating wave approximation (RWA) can be understood in two equivalent ways. To simplify, we will set $\Of=\Oq$ at first. Starting from the Rabi Hamiltonian\eqref{C1_QRM}, we move to interaction picture by applying the unitary $\Uop=e^{i(\Of\adag\aop+\Oq\sigz)t}$. In this picture, the state evolves according to the following Hamiltonian:
 $$H_{\text{int}}=\Uop\Hop\Uop^{\dagger}-\Of\adag\aop+\Oq\sigz=\gind(\sigp\aop+\sigm\adag)+\gind(\sigm\aop e^{-2i\Of t}+\sigp\adag e^{2i\Of t})$$ Hence, while the corotating term remains constant, the counter-rotating term quickly oscillates at a frequency $2\Of\gg \gind$. Therefore, the effect of this term will cancel on average, and the term may be safely ignored \footnote{The state of the system evolves according to the Schrödinger equation \unexpanded{$\frac{d\ket{\psi}}{dt}=-i\tilde{H} \ket{\psi}$}. The coefficient in a given basis {$\ket{e_m}$} can be formally solved in time as: {$c_m(t)=\langle e_m|\psi(t)\rangle=c_m(t_0)-i\int_{t_0}^t dt^\prime\bra{e_m}\tilde{H}(t^\prime)\ket{\psi(t^\prime)}$}. 
 The coefficients $c_m$ will evolve at a typical rate $\gind$. The counter-rotating rotating term will act through integrals of the form {$\int_{t_0}^t dt^\prime c_p(t^\prime)e^{i2\Of t^\prime}$}; these integrals involve the product of a function $c_p$, which evolves slowly a rate $\gind$, with a term oscillating at a much higher rate $\Of$. These integrals will cancel for most evolution times, allowing us to neglect the effect of counter-rotating term.}. Alternatively, one may look at the energy level structure of the Hamiltonian. Without the interaction term, the eigenstates of the Hamiltonian are $\ket{\lspin n}$ and $\ket{\hspin n}$, with $\ket{\lspin}$ and $\ket{\hspin}$ the eigenstates of $\sigz$ and $\ket{n}$ the Fock states. The corotating term connect $\ket{\lspin n+1}$ and $\ket{\hspin n}$, which are degenerate; therefore, its action cannot be neglected, even if $\gind\ll\Of$. \footnote{Indeed, the correction to the eigenstates is small only if the ratio between the matrix element connecting two eigenstates and the energy difference between said eigenstates is small.} According to degenerate perturbation theory, the eigenstates of the system will be obtained by diagonalizing the corotating term within this degenerate spectrum. This yield the following eigenstates and eigenvalues: 
 \begin{align}
 	\ket{\pm_n}=\frac{\ket{\lspin n+1}\pm\ket{\hspin n}}{\sqrt{2}},\\
 	E_{n,\pm}=(n+1)\Of\pm \gind\sqrt{n+1}.
 	\label{C1_E_JC}
 \end{align}

 The ground state is the vacuum $\ket{\lspin0}$, with zero energy. By contrast, the counter-rotating term connects $\ket{\lspin n}$ and $\ket{\hspin n+1}$, which have an energy gap of $2\Of$. Therefore, as soon as $\gind\ll\Of$, this term will only have a perturbative effect. If the system is initialized in the state $\ket{0\uparrow}$, it will be in an equal superposition of the two eigenstates $\ket{+0}$ and $\ket{-0}$, leading to Rabi oscillations. The oscillation frequency will be given by the energy difference between the two states, that is, $2\gind$.\\

 The same reasoning can be made when the qubit and field are weakly detuned: when $\gind\ll\Of,\Oq$ but $\gind\gg\lvert\Of-\Oq\rvert$, the counter-rotating term have only a small perturbative effect, while the corotating term cannot be neglected. Let us define $\Delta_\pm=\Of\pm\Oq$. 
 At first order in $\frac{\Delta_-}{\gind}$, the eigenstates are given by (up to normalization factors): $$\left(1+\frac{\Delta_-}{4\gind\sqrt{n+1}}\right)\ket{\lspin n+1}+\left(1-\frac{\Delta_-}{4\gind\sqrt{n+1}}\right)\ket{\hspin n},$$  $$\left(1-\frac{\Delta_-}{4\gind\sqrt{n+1}}\right)\ket{\lspin n+1}-\left(1+\frac{\Delta_-}{4\gind\sqrt{n+1}}\right)\ket{\hspin n}$$ 
 with associated energies $n\Of+\frac{\Of+\Oq}{2}\pm\sqrt{\frac{(\Delta_-)^2}{4}+\gind^2(n+1)}$. Note that, as the detuning $\lvert\Delta_-\rvert$ increases, one of the eigenstates becomes increasingly similar to $\ket{\lspin n+1}$, and the other to $\ket{\hspin n}$.\\

 When the detuning becomes large, the corotating term too becomes perturbative. However, its effect remains larger than the one of the counter-rotating term. 
 In the dispersive limit $g\ll\lvert\Delta_-\rvert$, the JC model can be diagonalized by applying a unitary $\Uop=e^{\frac{\gind}{\Delta_-}(\adag\sigm-\aop\sigp)}$ to the Hamiltonian \eqref{C1_JCM}, which yields the ac shift Hamiltonian:

 \begin{equation}
  	\Uop^{\dagger}\Hop_{JC}\Uop\approx\Of\adag\aop+\frac{\Oq}{2}\sigz+\frac{\gind^2}{2\Delta_-}(2\adag\aop\sigz+\sigz)\approx\frac{1}{2}\left(\Oq+\frac{\gind^2}{\Delta_-}\right)\sigz + \left(\Of+\frac{\gind^2}{\Delta_-}\sigz\right)\adag\aop,
  \end{equation} 
plus terms of order $O\left(\frac{g^3}{\Delta_-^2}\right)$ and higher. This diagonalization is the first use we make of a technique called Schrieffer-Wolff transformation, which will be introduced in more details in Chapter 2.
  Hence, in the limit of large detuning, the effect of the coupling is a renormalization of the qubit and light frequency, called the dispersive shift. The field experiences a frequency shift depending on the state of the qubit, and vice-versa. \\

 To summarize, even with moderately large values of the coupling, the eigenstates of the system are given by dressed states, \textit{i.e.}, superposition of states containing both light and matter excitations. The hybridization depends on the detuning between light and matter: it is maximal in the resonant regime, where eigenstates are equal superpositions of light and matter. As the detuning increases, the light and matter become increasingly decoupled.  Note however that even in the resonant regime, the eigenstates involve only states which differ by one excitation; hence, the standard Fock basis $\ket{\lspin n}$ $\ket{\hspin n}$ is still a rather efficient description of the system. This is no longer the case in the USC regime, as we will see below.

 \subsection{Perturbative USC: the Bloch-Siegert Hamiltonian}

 When the coupling becomes comparable with the bare frequencies of the system, the counter-rotating terms can no longer be neglected. As we mentioned, the counter-rotating term does not conserve the total number of excitations. However, it does conserve the parity of the excitation number $(-1)^{\adag\aop+\sigz}$, which is associated with the symmetry transformation $e^{i\pi(\adag\aop+\frac{1}{2}\sigz)}$.

 Depending on the value of the coupling, some terms can be identified as perturbations. Based on which terms are dominant and which perturbation theory can be applied, a classification of the different coupling regimes has been proposed in \cite{rossatto_spectral_2017}. We will discuss some of these aspects in the following. For $\frac{\gind}{\Of}$ going from $0$ to around $0.3$, we are in the perturbative USC regime. For these coupling values, the counter-rotating terms may still be treated as a perturbation to the JC Hamiltonian. This is best-done through another Schrieffer-Wolff transformation. By applying the unitary $\Uop=e^{\frac{\gind}{\Delta_+}(\aop\sigm-\adag\sigp)}$ (where we recall $\Delta_+=\Oq+\Of$), we obtain: $\Hop=\Hop_{JC}+\frac{\gind^2}{\Delta_+}(\aop^2+\adagsq)\sigz$. An important fact can already be inferred from this Hamiltonian: the ground state is no longer the vacuum of photon. Instead, the presence of a square bosonic term suggests the presence of squeezing. The Hamiltonian can be diagonalized by applying a squeezing operator. Alternatively, we can simply apply the unitary $\Uop=e^{\frac{\gind}{\Delta_+}(\aop\sigm-\adag\sigp)+\frac{\gind^2}{2\Of\Delta_+}\sigz(\aop^2-\adagsq)}$ on the original Hamiltonian, and we obtain the Bloch-Siegert Hamiltonian:
 \begin{equation}
 	\Hop_{BS}=\Hop_{JC}+\Of_{BS}\sigz\adag\aop+\frac{\Of_{BS}}{2}\sigz-\frac{\Of_{BS}}{2},
 \end{equation}
 plus higher-order terms. Here we have defined $\Of_{BS}=\frac{\gind^2}{\Delta_+}$. The eigenvalues of this Hamiltonian are $E^{BS}_0=-\frac{\Oq}{2}-\Of_{BS}$ for the ground state and $E_{n,\pm}^{BS}=(n-\frac{1}{2})\Of-\Of_{BS}\pm\sqrt{\frac{(\Delta_-+2n\Of_{BS})^2}{4}+n\gind^2}$ for the excited states. Hence, the counter-rotating term has two consequences: the presence of squeezing in the ground state, and a shift of the energy levels, the so-called Bloch-Siegert shift.

 \subsection{Deep-strong coupling regime}
 In the limit of large $\gind$ (typically for $\frac{\gind}{\Of}$ larger than $1$), the system enters the deep-strong coupling regime. In this regime, the coupling term becomes dominant.
 This regime is efficiently described by another perturbative treatment \cite{irish_dynamics_2005,irish_generalized_2007,rossatto_spectral_2017}, whose principle can be summarized as follows. In the limit $\gind\rightarrow\infty$, the interaction term will tend to align the spin around the $x$-axis, and displace the field. \footnote{As discussed in a recent paper \cite{felicetti_universal_2020}, this behavior is quite general and can be obtained with other quantum optical systems: more on this in the next Chapter.} The free field term should be maintained to stabilize the Hamiltonian; by contrast, the free qubit term will be considered as a perturbation. The reason why this approximation is valid may not be immediately apparent, but is developed in details in Appendix A. Hence, we will consider the following Hamiltonian:

 \begin{equation}
 	\Hop_{DSC}=\Of\adag\aop+\gind(\adag+\aop)\sigx.
 \end{equation}
 It can readily be diagonalized by making a projection in the eigenstate of $\sigx$ and displacing the field. The eigenstates are:

 \begin{align}
 	\ket{\phi_{n,\pm}} & =\frac{\ket{\leftarrow}\ket{n,\alpha_0} \pm \ket{\rightarrow}\ket{n,-\alpha_0}}{\sqrt{2}},
 	\label{C1_gRWAeigenstates}
 \end{align}
 where $\ket{\leftarrow}$ and $\ket{\rightarrow}$ are the eigenstates of $\sigx$, $\ket{n,\alpha}$ are displaced Fock states, and $\alpha_0=\frac{\gind}{\Of}$. These states form degenerate doublets, with energies $E_{n,+} =E_{n,-}=n\Of-\alpha_0^2\Of$.

\begin{figure}
 	\includegraphics[angle=-90,width=\linewidth]{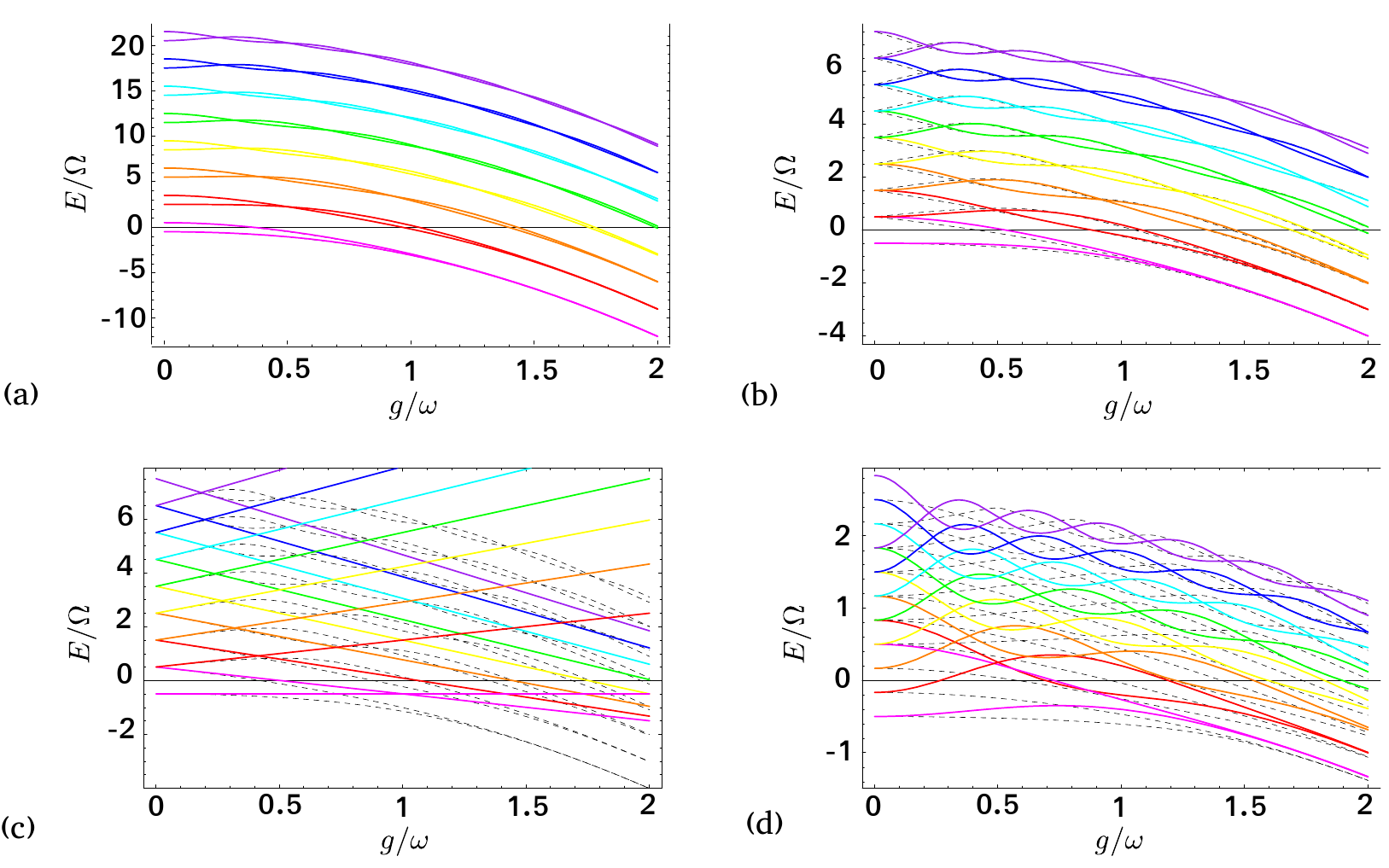}
 	\caption[Energy spectrum of the Rabi model given by various perturbation methods, versus exact numerical simulations.]{Energy spectrum, as given by various perturbation methods (solid line) and exact numerical simulation (dotted line), reproduced from \cite{irish_dynamics_2005}. (a)gRWA and exact spectrum for $\Oq/\Of=1/3$. The approximation accurately reproduces the exact spectrum, even for very small coupling values. (b) gRWA and exact spectrum for $\Oq/\Of=1$. The eigenstate structure shows a clear oscillatory pattern. The gRWA approximation becomes accurate for large $\gind$, and show some discrepancy with the exact spectrum for $\gind/\Of\leq1$. However, the agreement remains quite good even in this region. (c) Ordinary RWA and exact spectrum for $\Oq/\Of=1$. The agreement is good for $\gind/\Of$ up to $\sim0.3$ (more precisely up to the first crossing, also known as Juddian point), but clearly diverge from the true behavior for larger coupling: the RWA predicts that the energy levels should linearly increase or decrease with the coupling, instead of an oscillatory pattern. (d) gRWA and exact spectrum for $\Oq/\Of=3$. The method fails to reproduce the spectrum, except for very large coupling values.}
 	\label{C1_gRWAspectrum}	
 \end{figure}

 The qubit term $\frac{\Oq}{2}\sigz$ will then act as a perturbation. At first order, the sole effect of this perturbation is to lift the degeneracy between the states $\ket{\phi_{n,+}}$ and $\ket{\phi_{n,-}}$. The perturbed eigenenergies read:
 \begin{equation}
 	E_{n,\pm}=(n-\alpha_0)\Of\pm\frac{\Oq}{2}e^{-2\alpha_0^2}L_n(4\alpha_0^2),
 \end{equation}
 where $L_n$ are Laguerre polynomials. The presence of these polynomials leads to an oscillatory behavior in the spectrum of the Rabi model. In the limit $\gind\rightarrow\infty$, the qubit energy shift goes to zero, and the doublet becomes degenerate again. More generally, it is correct to treat the qubit term as a perturbation in the limit of large coupling $\gind\geq\Of$ or for small qubit frequency $\Oq\ll\Of$. However, it turns out that the method above can reproduce accurately the spectrum even for $\Oq=\Of$ and $\gind\leq\Of$, as shown in Fig.\ref{C1_gRWAspectrum}. With some refinements, this method, sometimes called the \textit{generalized Rotating Wave Approximation} or gRWA, can give good results even for $\Oq\leq\Of$ \cite{irish_generalized_2007}.\\

 In this regime, the behavior of the system is very different from the low-coupling regime. First, as is apparent from \eqref{C1_gRWAeigenstates}, in each eigenstate (including the ground state), light and matter are strongly entangled. Second, in the weak coupling regime, there is a simple link between the total number of photons and the energy: the more photons, the higher the energy. The ground state is just the vacuum of photons. This picture is lost in the DSC regime; in particular, the ground state has a nonzero number of photons, equal to $\lvert\alpha_0\rvert^2$, which increases with $\gind$. 

 Finally, the dynamical properties of this model also show distinctive features. In particular, many "typical" initial states will exhibit an oscillatory behavior with periodic collapse and revival. Let us take a simple example: we start from the vacuum of photons, with a spin aligned around the $x$ axis: $\ket{\leftarrow,0}$. This state can be decomposed in the DSC eigenbasis as: $\sum_n\frac{\alpha_0^n}{\sqrt{n!}}e^{-\frac{\alpha_0^2}{2}}\left(\frac{\ket{\phi_{n,+}}-\ket{\phi_{n,-}}}{\sqrt{2}}\right)$. It is then straightforward to study the time evolution of the state. In particular, we can compute the probability to find the system in the vacuum of bosons, and find, in the limit $\gind\rightarrow\infty$: $P(0,t)=e^{2\alpha_0^2(\cos{\Of t}-1)}$. This probability is almost always zero, except for $t=\frac{2p\pi}{\Of}$, with $p$ integer. Other initial spin polarizations yield the same result. Hence, a system prepared in the vacuum of photon initially will "spread" other the whole Fock state during the evolution, and, at periodic times, collapse again on the vacuum state. Similarly, a system prepared in a coherent or thermal state will experience oscillations with periodic collapse and revival \cite{irish_dynamics_2005}. 

 To summarize, the Rabi model exhibits completely different phenomenology in the weak-coupling and USC regimes. In the weak-coupling regime, the total number of excitation quanta is conserved. As the number of excitations increases, so does the energy; the ground state is simply the vacuum state. The system is best described in terms of the dressed state basis. The coupling leads to a lift of degeneracy in this basis, which can be witnessed by Rabi oscillations. In the USC regime, the counter-rotating terms come into play. The total number of excitations is no longer conserved; only the parity is. As a consequence, the ground state has a non-zero bosonic population, which increases with the coupling. For moderate values of $\gind/\Of$, the counter-rotating terms lead to squeezing of the ground state, and a shift in the energy levels captured by the Bloch-Siegert Hamiltonian. When $\gind$ is increased further, the energy levels display an oscillatory behavior, with several crossings. When we enter the DSC regime, the coupling term becomes dominant, and the qubit act only as a perturbation. The squeezing of the ground state is replaced by a Schrödinger-cat-like state, with strong entanglement between qubit and field. The dynamics of the bosonic field exhibits periodic collapse and revival features.

 \subsection{Multi-qubits models}

 So far we have discussed only the Rabi model, which describes a single emitter coupled with a bosonic field. In most realization, it is necessary to describe the behavior of multiple emitters. This task is highly non-trivial and has triggered considerable debate.
 One of the models which is most often discussed in this context is the Dicke model.
 This model describes the collective interaction of $N$ two-level systems with a single bosonic mode, according to the following Hamiltonian:

 \begin{equation}
 	\Hop=\Of\adag\aop+\frac{\Oq}{2}\sum_i^N\sigz_i + \frac{\gcoll}{\sqrt{N}}\sum_i^N\sigx_i(\adag+\aop)=\Of\adag\aop+\Oq\Jz + 2\frac{\gcoll}{\sqrt{N}}\Jx(\adag+\aop),
 	\label{C1_DM}
 \end{equation}
 where the index $i$ labels the different qubits. Here we have defined the collective spin operators $\hat{J}_a=\frac{1}{2}\sum_i^N\hat{\sigma}_i^a$ with $a={x,y,z}$, and the collective coupling $\gcoll=\gind\sqrt{N}$ (the reason for this definition will become clear in a moment). 
 This model is often studied at thermal equilibrium, with a finite or zero temperature (in the latter case, the model is in its ground state). For finite $N$, the photon population of the system increases smoothly with the coupling. In the limit of large $N$, however, a new behavior appears, which is characterized by the \textit{superradiant phase transition}. For $N\rightarrow\infty$ and $\gcoll\leq\gcoll_p=\sqrt{\frac{\Of\Oq}{4}}$, the system is in a so-called normal phase, where the number of photons does not grow with $N$. For $\gcoll\geq\gcoll_p$, the system enters the superradiant phase, in which the photon field acquire a macroscopic population, with an average photon number growing linearly with $N$. This behavior can be captured by the order parameter $\frac{\langle\adag\aop\rangle}{N}$; in the limit $N\rightarrow\infty$, this quantity is zero in the normal phase, and non-zero in the superradiant phase. At the critical point $\gcoll=\gcoll_p$, it has a non-analytical behavior, indicating a (second-order) phase transition. The properties of this transition will be the subject of the next Chapter.\\

Finally, let us mention one more property of the multi-qubit limit. The most salient property of a two-level system is the extreme anharmonicity of its spectrum. 
However, in the limit of many qubits, this feature tends to disappear. Indeed, the spectrum of an ensemble of qubits can be decomposed in several subspaces corresponding to the eigenvalues of the total angular momentum operator $\hat{J}^2=\Jx^2+\Jy^2+\Jz^2$. The eigenspace with maximal angular momentum $\hat{J}^2=\frac{N}{2}\left(\frac{N}{2}+1\right)$ is a ladder of $N+1$ states evenly spaced in energy. For $N$ large, if the qubits are only weakly excited \footnote{which is true far away from critical points and for small temperatures.}, they will only explore the bottom part of this ladder, which can then safely be considered as infinite. In this case, the qubits effectively behave as a bosonic field, losing their non-linear behavior. This is formalized through the Holstein-Primakoff (HP) transformation. For all $N$, in the $\hat{J}^2=\frac{N}{2}\left(\frac{N}{2}+1\right)$ sector, we have:

 \begin{align}
 	\Jz=-\frac{N}{2}+\bdag\bop,\\ \nonumber
 	\Jp=\bdag\sqrt{N-\bdag\bop},\\ \nonumber
 	\Jm=\sqrt{N-\bdag\bop}\hspace{5pt}\bop,
 \end{align}
 with $[\bop,\bdag]=1$. For small excitation number $\moy{\bdag\bop}\ll N$, we have $\Jp\sim\sqrt{N}\bdag$ and $\Jx=\sqrt{N}\frac{(\bdag+\bop)}{2}$. In this case, the Dicke model can be rewritten as: 
 \begin{equation}
 	\Hop=\Of\adag\aop+\Oq\bdag\bop+\gcoll(\bdag+\bop)(\adag+\aop)-\Oq\frac{N}{2}.
	\label{C1_Dickebosonic}
 \end{equation}

 That is, the system can be modeled as two linearly coupled bosonic fields. Note that if the original coupling was $\gind\Jx(\aop+\adag)$, we achieve an effective coupling $\sqrt{N}\gind$ between the two bosonic fields: the presence of many qubits effectively increases the coupling strength, which justifies the definition of the collective coupling $\gcoll=\sqrt{N}\gind$.  Since this Hamiltonian involves only quadratic products of bosonic operators, it can be diagonalized by a Bogoliubov transformation. We define new bosonic operators $\cop$ and $\dop$ which are linear combinations of the original $\aop$ and $\bop$ operators: $\cop=\alpha_c\aop+\beta_c\adag+\gamma_c\bop+\delta_c\bdag$, and similarly for $\dop$. An appropriate choice of coefficient allows diagonalizing the Hamiltonian: $\Hop=E_+\cdag\cop+E_-\opddag\dop$, with \cite{emary_chaos_2003}:

 \begin{equation}
 	E_\pm^2=\frac{1}{2}\left(\Of^2+\Oq^2\pm\sqrt{(\Of-\Oq)^2+16\gcoll^2\Of\Oq}\right).
 	\label{C1_polariton energy}
 \end{equation}

 Hence, the dynamics is best described in terms of hybrid light-matter excitations, called \textit{polaritons}. The hybridization is more or less important, depending on the parameters. For instance, when the two fields are highly detuned and the coupling is weak, the hybridization becomes negligible; for $\lvert\Of-\Oq\rvert\gg\gcoll$, one has $\cop\sim\aop$ and $\dop\sim\bop$. By contrast, when $\Of\sim\Oq$ and for higher couplings, each polariton is an almost equal mixture of light and matter. This is similar to what we obtained in the single-qubit case.

 From \eqref{C1_polariton energy}, it is clear that the energy difference between the two polaritons, $E_+-E_-$, increases with the detuning $\lvert\Of-\Oq\rvert$. This leads to a typical experimental signature of light-matter coupling, the anticrossing between the two polaritonic branches (see section I.2 below). \\

 Importantly, for $\gcoll=\gcoll_p=\sqrt{\frac{\Of\Oq}{4}}$, we have $E_-=0$: the gap of the system closes, and the behavior of the system is nonanalytic, which indicates that the superradiant transition is taking place. For $\gcoll\geq\gcoll_p$, the Hamiltonian \eqref{C1_Dickebosonic} becomes unstable. This is because the hypothesis $\moy{\bdag\bop}\ll N$ that we have made to derive this Hamiltonian breaks down. Indeed, for $\gcoll\geq\gcoll_p$, the system enters the superradiant phase, in which both light and matter acquire a macroscopic population.

 \subsection{$A^2$ and $P^2$ terms}
 The analysis above suggests that in general, a phase transition should occur when many emitters are coherently coupled to a bosonic field. The situation, however, is much more complex. As we will show below, among the experiments which achieved \textit{genuine} USC coupling so far, none have displayed a superradiant phase transition. 
 This indicates that the Dicke Hamiltonian is not a complete description of light-matter interaction, and that additional terms should be included. 

 The precise form of these terms is the subject of an ongoing controversy \cite{ciuti_quantum_2005,nataf_no-go_2010,viehmann_superradiant_2011,todorov_intersubband_2012,vukics_elimination_2014,vukics_fundamental_2015,todorov_dipolar_2015,bamba_superradiant_2016,bamba_circuit_2017,de_bernardis_breakdown_2018,de_bernardis_cavity_2018,stefano_resolution_2019,garziano_gauge_2020,rouse_avoiding_2020}. However, most studies have considered variants the so-called $A^2$ (or diamagnetic), and $P^2$ terms. To understand the origin and the general characteristics of these two terms, we will use an extremely simplified toy model. 
 Much of the debate mentioned above concerns the validity of approximations that are commonly used in the derivation of effective models. For the purpose of this presentation, we will first derive the $A^2$ and $P^2$ terms by making several uncontrolled assumptions, in order to give a flavor of the formalism used. We will later discuss some of the views that have been expressed concerning the validity of the assumptions.

  For the $A^2$ term, we will follow a derivation similar to the one found in \cite{nataf_no-go_2010}. We consider a single electron oscillating around a stationary nucleus. This electron interacts both with the nucleus and with the electromagnetic field. We will assume that the coupling takes place with only a single mode of the field. We will consider the Coulomb gauge, in which the electromagnetic field can be described by the vector potential $\bm{A}$ satisfying $\nabla.\bm{A}=0$. 
  We may write a minimal (classical) coupling Hamiltonian:

 \begin{equation}
 	H=H_{f}+\frac{(\bm{p}+e\bm{A}(\bm{r}))^2}{2m}+V(\bm{r}),
 \end{equation}
 where $H_f=\frac{1}{\epsilon_0}(\partial_t\bm{A})^2+\mu_0(\nabla\times\bm{A})^2$ describes the free Hamiltonian of the bosonic field, $m$ is the electron mass, $e$ its charge, $\bm{r}$ and $\bm{p}$ are the position and impulsion of the electron, respectively, $V(r)$ is the electrostatic potential created by the nucleus, and $\bm{A}$ is the vector potential of the electromagnetic field. If we assume that the electron remains localized on a distance much smaller than the wavelength of the field, one may set $\bm{A}(\bm{r})=\bm{A}(0)$. Going through a canonical quantification procedure, we replace the classical fields by operators: $\bm{A}\rightarrow\hat{\bm{A}}=\bm{A}(0)(\adag+\aop)$, $H_f\rightarrow\Hop_f=\Of\adag\aop$. We obtain the Hamiltonian $\Of\adag\aop+\frac{\hat{\bm{p}}^2}{2m}+V(\hat{\bm{r}})+\frac{e}{m}\bm{A}(0)\hat{\bm{p}}(\adag+\aop)+\frac{e^2}{2m}\bm{A}(0)^2(\adag+\aop)^2$. The term $\Hop_e=\frac{\hat{\bm{p}}^2}{2m}+V(\hat{\bm{r}})$ describes the electronic states $\ket{n}$ of the atom. If the potential $V$ is highly nonlinear, the spectrum of the atom will be highly anharmonic. As a consequence, we will assume that the bosonic field can only excite the transition between the two lowest eigenstates of $\Hop_e$, which we will call $\ket{0}$ and $\ket{1}$. Then the electronic motion can be described by a two-level system: $\Hop_e\sim E_0\ket{0}\bra{0}+E_1\ket{1}\bra{1}=\frac{\Oq}{2}\sigz+cst$, and $\hat{\bm{p}}=\bra{1}\hat{\bm{p}}\ket{0}\ket{1}\bra{0}+h.c.=\bra{1}\hat{\bm{p}}\ket{0}\sigx$. This is the so-called two-level approximation. We arrive at the following Hamiltonian:

 \begin{equation}
	\Hop=\Of\adag\aop+\frac{\Oq}{2}\sigz+\gind\sigx(\adag+\aop)+D_A(\adag+\aop)^2,
 	\label{C1_A2 term}
 \end{equation}
 with $\Oq=E_1-E_0$, $\gind=\frac{e}{m}\bm{A}(0).\bra{1}\hat{\bm{p}}\ket{0}$, and $D_A=\frac{e^2}{2m}\bm{A}(0)^2$
 This model is similar to the Rabi model, with an additional quadratic bosonic term. This term is known as the diamagnetic or $A^2$-term. 

 Let us define the parameter $r=\frac{\Oq D_A}{\gind^2}=\frac{m\Oq\bm{A}(0)^2}{2(\bm{A}(0).\bra{1}\hat{\bm{p}}\ket{0})^2}$. We will now use the Thomas-Reiche-Kuhn sum rule, which states $\sum_{n\neq j}\frac{\lvert\bra{n}\hat{p_\mu}\ket{j}\rvert^2}{E_n-E_j}=\frac{m}{2}$, for every component of the displacement $\mu={x,y,z}$, and for every electronic eigenstate $\ket{j}$ \footnote{This rule can be derived from the following argument. Since $\Hop_e=\frac{\hat{\bm{p}}^2}{2m}+V(\hat{\bm{r}})$, we have $\hat{p}_\mu=-i \hspace{2pt} m[\hat{x}_\mu,\Hop_e]$ for all $\mu$. This gives {$\bra{n}\hat{p}_\mu\ket{j}=-i \hspace{2pt} m(E_j-E_n)\bra{n}\hat{x}_\mu\ket{j}$}, and {$\sum_n\frac{\lvert\bra{n}\hat{p}_\mu\ket{j}\rvert^2}{E_n-E_j}=-\frac{i \hspace{2pt} m}{2}\sum_n\bra{j}\hat{x}_\mu\ket{n}\bra{n}\hat{p}_\mu\ket{j}-\bra{n}\hat{x}_\mu\ket{j}\bra{j}\hat{p}_\mu\ket{n}=-\frac{i \hspace{2pt} m}{2}\bra{j}[\hat{x}_\mu,\hat{p}_\mu]\ket{j}=\frac{m}{2}$} } From this sum, we can infer that $\lvert\bra{1}\hat{p}_\mu\ket{0}\rvert^2\leq m(E_1-E_0)=m\Oq$. We can then deduce that:

 \begin{equation}
 	r=\frac{\Oq D_A}{\gind^2}>1.
 \end{equation}

  This means that when the coupling strength $\gind$ increases, the diamagnetic term $\frac{r\gind^2}{\Oq}$ will increase at least like $\gind^2$, that is, \textit{faster} than the light-matter coupling term $\gind\sigx(\adag+\aop)$. This scaling has important consequences. While the $A^2$ term is negligible at small coupling, it actually becomes the dominant contribution for large $\gind$. This term favors small values of the bosonic field. In the DSC regime, this $A^2$ term will expel the field away from the emitters, effectively decoupling light and matter. As a consequence, for very large $\gind$, the emission rate of the emitters can actually decrease with $\gind$, a behavior opposite to the standard Purcell effect \cite{de_liberato_light-matter_2014}. This can be expressed formally as follows: the $A^2$ term can be absorbed by a Bogoliubov transformation of the field. We can then rewrite \eqref{C1_A2 term} as the standard Rabi model with \textit{renormalized} parameters:

 \begin{equation}
 	\Hop=\Of_{eff}\cdag\cop+\frac{\Oq}{2}\sigz+\gind_{eff}(\cdag+\cop)\sigx,
 	\label{C1_Rabirenorm}
 \end{equation}
  with $\Of_{eff}=\Of\sqrt{1+4\frac{D_A}{\Of}}=\Of\sqrt{1+4r\frac{\gind^2}{\Of\Oq}}$ and $\gind_{eff}=\frac{\gind}{(1+4r\frac{\gind^2}{\Of\Oq})^{1/4}}$. Hence, all the conclusions drawn from the Rabi model can be carried on directly in the presence of the diamagnetic term with a rescaling of the parameters (sometimes referred to as the depolarization shift in the literature). This rescaling, however, has several crucial consequences. For small $\gind$, we have $\gind_{eff}\sim\gind$ and $\Of_{eff}$ constant; raising the coupling does increase the effective coupling-frequency ratio. For larger $\gind$, however, the effect of the $A^2$ term kicks in, and the effective coupling only grows as $\sqrt{\gind}$, while $\Of_{eff}$ increases like $\gind$. In the limit $\gind\rightarrow\infty$, one has $\Oq\ll\gind_{eff}\ll\Of_{eff}$. Since $\gind\ll\Of_{eff}-\Oq$ and $\gind\ll\Of_{eff}+\Oq$, this effectively corresponds to a \textit{perturbative} USC regime. The properties of both light and matter do change with $\gind$; in particular, the light becomes increasingly squeezed. However, the dynamics of the light and matter part become gradually decoupled.\\

 We can also study the case of many qubits. We perform once more a HP transformation, and describe the system by coupled bosonic modes. This gives the following Hamiltonian (dropping constant terms):

 \begin{equation}
  	\Hop=\Of\adag\aop+\Oq\bdag\bop+\gcoll(\bdag+\bop)(\adag+\aop)+D_AN(\adag+\aop)^2,
  	\label{C1_Hopfield}
  \end{equation} 
 with $N$ the number of qubits, and $\gcoll=\sqrt{N}\gind$ the collective coupling. This Hamiltonian can be generalized to multiple modes $\aop_k$ and $\bop_k$, which gives the Hopfield Hamiltonian \cite{hopfield_theory_1958,ciuti_quantum_2005}, a model of great importance in solid-state physics. Here we will stick with the two-mode version, the treatment of the multi-mode case being very similar.
The presence of the diamagnetic term here has a crucial consequence. Let us consider again the renormalized parameters of \eqref{C1_Rabirenorm}. In the standard Dicke model, the superradiant phase transition occurs for $\gcoll\geq \gcoll_p=\sqrt{\Of\Oq/4}$. Including the renormalization, the system can enter a superradiant phase when $\frac{4\gcoll_{eff}^2}{\Of_{eff}\Oq}\geq1$. However, we have: 

 \begin{equation}
 	\frac{4\gcoll_{eff}^2}{\Of_{eff}\Oq}=\frac{4\gcoll^2}{\Of\Oq}\frac{1}{1+r\frac{4\gcoll^2}{\Of\Oq}}.
 \end{equation}
 It is straightforward to show that if $r>1$, this expression is always smaller than $1$. In other words, the system can not enter a superradiant phase; the diamagnetic term suppresses the phase transition altogether. This can also be seen in the energy of the two polaritons; when the $A^2$ term is included, we always have $E_->0$ as soon as $r>1$, meaning that the gap never closes. This result has been referred to as the no-go theorem for superradiant phase transition \cite{nataf_no-go_2010}.\\

 To derive the above results, we have worked in the Coulomb gauge, using the potential field $\bm{A}$ and its time-derivative as canonical conjugate variables for quantization. However, it is also possible to start from another gauge, by applying a gauge transformation before going through the quantization procedure. A common choice is the so-called dipolar gauge, in which the dynamics is described in terms of the polarization field $\bm{P}$ and the displacement field $\bm{D}=\epsilon_0\bm{E}+\bm{P}$. The quantization is performed with $\bm{D}$ and the magnetic field $\bm{B}$ as conjugate variables: $\bm{D}=\aop+\adag$ and $\bm{B}=i(\adag-\aop)$. When this procedure is applied for simplified atoms as we considered earlier, we obtain, after single-mode and two-level approximation, a Hamiltonian of the form \cite{de_bernardis_cavity_2018} :

\begin{equation}
 	\Hop=\Of\adag\aop+\Oq\Jz+2\gind(\adag+\aop)\Jx+4D_P(\Jx)^2,
 \end{equation}
 or, in a bosonized version:
 \begin{equation}
 	\Hop=\Of\adag\aop+\Oq\bdag\bop+\gcoll(\adag+\aop)(\bop+\bdag)+N D_P(\bop+\bdag)^2.
 	\label{C1_P2term}
 \end{equation}
 In this model, we have a quadratic term that involves the matter part of the Hamiltonian, instead of the field part. This term is referred to as the $P^2$ term in the literature. As in the case of the $A^2$ term, it can be argued on general grounds that $D_P\geq\frac{\gcoll^2}{\Of}$ \cite{de_bernardis_cavity_2018}. This scaling, once more, prevents the onset of phase transition; a no-go theorem also holds in this case. Intuitively, the $P^2$ term acts as an effective antiferromagnetic interaction, which prevents the appearance of a superradiant phase in which all qubits are aligned in the same direction. Note however that the terms $\aop$ and $\bop$ do not refer to the same quantities as earlier: $\aop$ now describe the displacement field, not the vector potential. Therefore, care needs to be taken when interpreting the results predicted in different gauges. \\

 In the last decade, the $A^2$ and $P^2$ terms, as well as the no-go theorem, have been the subject of considerable discussion \cite{ciuti_quantum_2005,nataf_no-go_2010,viehmann_superradiant_2011,todorov_intersubband_2012,vukics_elimination_2014,vukics_fundamental_2015,todorov_dipolar_2015,bamba_superradiant_2016,bamba_circuit_2017,de_bernardis_breakdown_2018,de_bernardis_cavity_2018,stefano_resolution_2019,garziano_gauge_2020,rouse_avoiding_2020}. Different derivations have been proposed for various systems, which sometimes yield very different predictions. In particular, some works \cite{nataf_no-go_2010,vukics_elimination_2014,vukics_fundamental_2015,bamba_superradiant_2016} have derived models in which the $A^2$ and $P^2$ terms are suppressed or weakened, allowing a phase transition to take place (albeit for higher coupling value $\gind$.)  Other works \cite{ciuti_quantum_2005,viehmann_superradiant_2011,todorov_intersubband_2012,todorov_dipolar_2015,de_bernardis_cavity_2018,garziano_gauge_2020}, instead, have concluded that a phase transition could not be observed. An intense discussion concerns the validity of the approximation used in the model \cite{viehmann_superradiant_2011,vukics_fundamental_2015,de_bernardis_breakdown_2018,stefano_resolution_2019}. Indeed, the "naive" derivations we have presented rely on several transformations, such as the gauge transformation from Coulomb to dipole gauge, or the projection into a two-level, single-mode subspace. These transformations do not commute with each other; therefore, depending on which transformations are applied and in which order, one may reach different conclusions. All of these choices are not equally valid; in particular, the use of two-level approximation has been shown to yield incorrect results in some gauges \cite{de_bernardis_breakdown_2018,stefano_resolution_2019}. Other factors need to be taken into account, such as the number of dipoles and the shape of their potential \cite{de_bernardis_breakdown_2018}, or the description of dipoles as point-like particles \cite{viehmann_superradiant_2011,vukics_fundamental_2015}.

 At the present time, no superradiant transition has been observed with \textit{genuine} light-matter coupling (it has, however, been achieved with quantum simulation, as we will see later on). The validity of the different models proposed is still a largely open, and platform-dependent, question. One of the reasons why the matter has not been settled yet is the difficulty to extract information from systems in the USC regime, as we will now discuss.\\

  \section{Probing the USC regime}
 In the previous section, we have reviewed some of the effects which have been predicted to arise in the USC regime. We will now discuss how these effects can be probed experimentally.

 \subsection{Photoemission in USC}

At first sight, one of the main features of USC is the fact that the ground state is no longer in a vacuum of photons. One may thus expect that entering the USC regime would lead to an emission of photons, a feature easily detectable experimentally. The situation, however, is much more complex. To understand the photoemission of a system in the USC regime, it is necessary to describe its coupling to the environment. When the environment can be modeled by a Markovian bath, and for weak system-bath coupling, this interaction can be captured by the Lindblad formalism. In this context, the state evolves according to an equation of the form:

 \begin{align}
 	\frac{\partial \rop}{\partial t}=-i[\Hop,\rop]+\sum_j \kappa_j L[\hat{A}_j](\rop), \\ \nonumber
 	L[\hat{A}](\rop)=\hat{A}\rop \hat{A}^\dagger - \frac{1}{2}(\hat{A}^\dagger\hat{A}\rop+\rop\hat{A}^\dagger\hat{A}),
 \end{align}
 where the $\hat{A}_j$ are referred to as jump operators. Obtaining the jump operators from microscopic interaction between system and bath is a non-trivial matter; standard derivations and discussions can be found in \cite{breuer_theory_2002,boite_theoretical_2020}. Here, we will discuss only one of their key properties. Let us consider the electromagnetic field inside of a cavity, with the field modes outside of the cavity acting as a bath. For illustrative purposes, let us assume a coupling of the form $\int dk \sum_k i(\bdag_k-\bop_k)(\adag+\aop)$, where $\bop_k$ are the various electromagnetic modes of the environment (a more realistic coupling should include a proper description of the density of modes, which we have omitted here for simplicity). Then the jump operators will be obtained by \textit{dressing} the quadrature $\adag+\aop$ coupled to the environment, with the eigenstates of the field within the cavity, and keeping only terms that induce jumps towards lower energy eigenspaces. More precisely, let us define $\ket{j}$ the eigenstates of the cavity. Then we can define terms of the form $\hat{X}^+(\Of_0)=\sum_{E_j=E_i+\Of_0}\bra{j}(\adag+\aop)\ket{i} \ket{j}\bra{i}$, with $\Of_0\geq0$. These terms induce jumps towards higher-energy levels, and hence increase the energy of the system. Similarly, the terms $\hat{X}^-(\Of_0)=\sum_{E_j=E_i-\Of_0}\bra{j}(\adag+\aop)\ket{i} \ket{j}\bra{i}$ decrease the energy of the system. Then for a bath at zero temperature, the jump operators will only involve the terms $\hat{X}^{-}(\Of_0)$; in other words, the interaction with the bath may only lower the energy of the system (for a finite-temperature bath, the system will continuously gain and lose excitations, at rates fixed by the temperature of the bath). In the example we have taken, the eigenstates of the system are the Fock states of the field within the cavity. Hence we have $\hat{X}^-(\Of_0)=\aop$ and $\hat{X}^+(\Of_0)=\adag$ if $\Of_0$ is equal to the frequency of the cavity, and $0$ otherwise. Hence, the jump operator will simply be the photon annihilation operator, and the system will evolve according to: $\frac{\partial\rop}{\partial t}=-i[\Of_{\text{cav}}\adag\aop,\rop]+\kappa L[\aop](\rop)$. The jump operators $\aop$ will progressively reduce the number of photons in the cavity, describing the emission of photons to the external world. Now, let us assume that we put an atom inside the cavity, ultrastrongly coupled to the field. The interaction between the field and the environment remains unchanged. However, the interacting quadrature $\adag+\aop$ now needs to be dressed by the eigenstates of the \textit{full} system. When qubit and field are only weakly coupled, the eigenstates are still given by Fock states; the jump operators remain of the form $\aop$. This is intuitive since, in the weak-coupling limit, it is still correct to describe the system in terms of atomic and field excitations. Reducing the number of field excitation does lead to lower energy for the system. By contrast, in the USC regime, the eigenstates are hybrid states with elements spanning the entire Fock space. During its evolution with the bath, the system will lose energy by losing hybrid excitations. This process is described by jump operators which strongly differ from the photon loss operator $\aop$. \footnote{Actually, a Lindblad equation involving $\aop$ as jump operator would describe a situation in which energy is being \textit{pumped into} the system.} The system will gradually lose energy through dressed jump operators, and reach its ground state \footnote{If the bath it at zero temperature. For a bath at finite temperature, the system will continuously lose and gain excitation and evolve towards a Gibbs state.}. Once the system has reached its ground state, the evolution will cease; no excitation will escape the system. Therefore, even though the analytical treatment predicts the presence of photons in the ground state, \textit{these photons can not escape the system when it is put into contact with an environment}. In the example of the atom-cavity system, a photodetector outside of the cavity would not click, even though the bosonic field of the cavity is predicted to be populated. More surprisingly, putting the detector \textit{inside} the cavity would not lead to detection events, either \cite{stefano_photodetection_2018}. The photons are tightly bound to the atom, and cannot be directly probed by usual techniques. Therefore, the photons that are predicted to arise in the USC regimes are generally referred to as \textit{virtual} particles \cite{ciuti_input-output_2006,liberato_virtual_2017,stefano_photodetection_2018,kockum_ultrastrong_2019}.

 \subsection{Spectral signatures}
 Once a system has reached its ground state, it ceases to leak information into the outside world. Therefore, one way to probe the system is to excite it. For instance, one may shine a laser on the cavity and look at its transmission properties. In general, transmission (or reflection) spectra are the most available and widely-studied signatures in USC experiments (see the next section).
 From a theoretical perspective, these processes can be studied by input-output theory, which describes how an input probe field will be modified by interacting with a given system \cite{boite_theoretical_2020}. Different signatures can then be studied, such as the intensity of the output field, or its noise properties. In particular, the transmission spectrum typically exhibits peaks corresponding to the different transitions excited by the probe. For instance, in the context of the Hopfield model, the transmission spectrum has two resonances which correspond to the two polaritonic modes excited by the probe \cite{ciuti_input-output_2006}. These resonances are located at the excitation energy $E_{\pm}$ of each polariton. For $\Of=\Oq$ and $\gind\rightarrow0$, there is a unique peak. As the coupling increases, the resonance is split into two peaks separated by a distance $E_+-E_-$, which increases with $\gind$. This phenomenon, known as Rabi splitting, is commonly used to evaluate the intensity of coupling achieved in a given experiment.

 Furthermore, the splitting also increases with the detuning $\lvert\Of-\Oq\rvert$. If the matter frequency $\Oq$ can be tuned from $\Oq\leq\Of$ to $\Oq\geq\Of$, one observes an avoided crossing as the splitting decreases, reaches a minimum at resonance $\Of=\Oq$, then increases again.

\begin{figure}
 	\includegraphics[angle=-90,width=\linewidth]{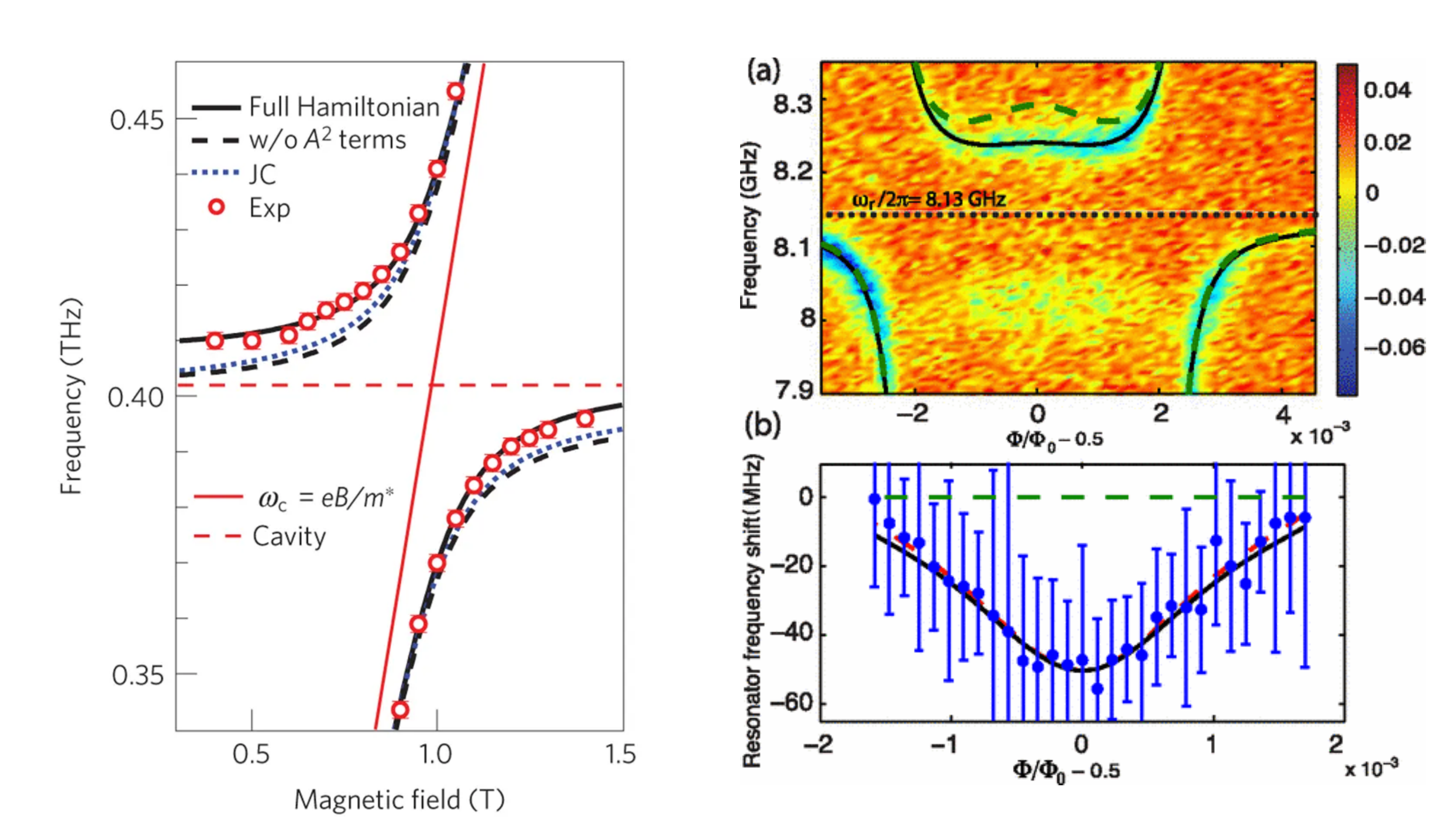}
 	\caption[Experimental spectral signatures of light-matter interaction.]{Spectral signatures of light-matter interaction. Left: inter-Landau level transitions. The transmittance spectrum is measured; the frequency of the transmittance peak gives direct access to the frequency of the polaritonic modes. The spectrum clearly exhibits two branches corresponding to the two polaritonic modes. When the applied magnetic field is increased, the matter frequency (here called $\Of_c$) changes, and the hybrid polaritonic branches exhibit an avoided-crossing behavior. In the absence of hybridization, one would have a pure light and a pure matter mode, with frequency describes by the dashed and solid red lines, respectively. Right: superconducting circuit. The magnetic flux $\Phi$ is used to tune the qubit frequency. The transmission spectrum exhibit also an avoided crossing. The green dotted line represents the predictions of the JC Hamiltonian; the observed spectrum shows a deviation corresponding to the Bloch-Siegert shift. Left figure reproduced from \cite{zhang_collective_2016}, right figure reproduced from \cite{forn-diaz_observation_2010}.}
 	\label{C1_spectral signatures}
 \end{figure}

 The Rabi splitting and the avoided crossing are also present both in the Rabi and the Hopfield models, and constitute a widely-studied signature of light-matter coupling (see Fig.\ref{C1_spectral signatures}). Note that these effects already occur in the JC model; however, in the USC regime, the splitting becomes comparable with the position of the initial resonance, which is sometimes called giant Rabi splitting in the literature.

 Finally, note that if the input is the vacuum, there will be no output, showing again that virtual excitations cannot be released in the absence of a driving field. Similarly, if the input field has no squeezing properties, so will the output \cite{ciuti_input-output_2006}. Therefore, the squeezing which is predicted to occur in USC systems also requires indirect probing methods.

 \subsection{Other methods}
 The above methods give information about the spectrum of the system, but little about the eigenstates. Hence, despite their great importance and experimental availability of spectral signatures, 
 much effort have been put into the design of alternative probing schemes.

 One widely-studied strategy is to periodically drive the system. For instance, it was suggested that a periodic modulation of the field frequency can turn the virtual excitations into real ones, which can then escape the system and be detected \cite{ciuti_quantum_2005,liberato_quantum_2007}.
  In general, a periodic driving of the system induces transitions towards excited states; when relaxing the system emits detectable excitations. 
   One feature of this process is that it is sensitive not just to the energy difference between eigenstates, but also to selection rules on the transition. Because of this, in the USC regime, the modification of eigenstates can lead to observable effects, such as exotic photon statistics for the output field \cite{ridolfo_photon_2012,ridolfo_nonclassical_2013,le_boite_fate_2016}\\

 Other works \cite{garziano_vacuum-induced_2014,lolli_ancillary_2015,felicetti_parity-dependent_2015} have proposed to coherently couple ancillary qubits to the system and use them as probes. When the system enters the USC regime, this leads to observable consequences such as the modification of the qubit frequency \cite{lolli_ancillary_2015}. 
 If the frequency of the ancilla is tuned with a specific two-level transition of the system, and a tomography of the ancilla is performed, it is possible to reconstruct the population and coherence of the two levels involved in the transition \cite{felicetti_parity-dependent_2015}. Repeating the procedure for all the system transitions allows us to fully reconstruct the state of the system.\\

 Finally, several studies \cite{stassi_spontaneous_2013,huang_photon_2014,falci_ultrastrong_2019} have suggested to use matter systems with an additional level $\ket{b}$ not coupled to the field (for instance, three-level systems with one transition $\ket{e}-\ket{g}$ resonant with the bosonic field, and a transition $\ket{g}-\ket{b}$ at a completely different frequency). Starting from the uncoupled level and no photon $\ket{b,0}$, driving fields are applied to drive the system from $\ket{b}$ to $\ket{g}$, and then back to $\ket{b}$. If the two levels $\ket{e}$ and $\ket{g}$ are ultrastrongly coupled to the bosonic field, this process can induce a Raman-like transition to the final state $\ket{b,n}$, with $n\neq0$; in other words, bosonic excitations are created. These excitations are real and can escape the system. By contrast, when the two levels $\ket{e}$ and $\ket{g}$ are only weakly coupled to the bosonic field, this transition is suppressed. This can be seen as another protocol to turn the virtual excitations into real ones.

  \section{Experimental achievements}
  Historically, the coupling between light and matter at a quantum level has been mostly studied with atoms in cavities (and indeed, so far we have used the terminology of atoms coupled to a light cavity mode). In this platform, it was possible to reach the strong-coupling regime, and to observe Rabi oscillations \cite{kaluzny_observation_1983}. These results have ushered in the development of quantum technologies, and the subsequent studies of quantum light-matter coupling. However, because the coupling between atoms and light is intrinsically small, cavity-QED platforms have never been able to approach the USC regime (the highest $\gind/\Of$ values reported are around $\sim10^{-6}$ \cite{brune_process_2008,tiecke_nanophotonic_2014}). Therefore, subsequent results in the USC regime have been achieved with other platforms. In this section, we will describe some of these platforms, and a few significant results that have been achieved. The experimental state-of-the art is discussed extensively in \cite{kockum_ultrastrong_2019,forn-diaz_ultrastrong_2019}.\\

 \subsection{Superconducting circuits}

  Superconducting circuits have emerged as a promising platform for quantum technologies. They can be used to engineer both qubits and harmonic oscillators.
  The basic components of electric circuits are inductors and capacitors. When a capacitor is charged (for instance by connecting it to a current generator), it accumulates electrical energy. This energy  can be quantified by the classical Hamiltonian $H_c=\frac{Q^2}{2C}$, where Q is the charge accumulating by the capacitor and $C$ its capacitance; or, alternatively, $H_c=\frac{CU^2}{2}$, with $U=\frac{Q}{C}$ the voltage drop across the capacitor. 

  By contrast, an inductor accumulates magnetic energy, quantified by the Hamiltonian $H_L=\frac{\phi_m^2}{2L}$, with $\phi_m$ the magnetic flux threading the inductor, and $L$ its induction. The magnetic flux is related to the electrical current flowing through the inductor: $\phi_m=LI$. When a voltage drop is applied at the ends of the inductor, the current progressively increases, as the inductor accumulates energy. 
 When inductor and capacitor are connected in a circuit, the total energy is expressed by:
 \begin{equation}
 	H=\frac{\phi_m^2}{2L}+\frac{Q^2}{2C}.
 \end{equation}
 Here, the charge and flux play the role of conjugate variables. Their evolution can be obtained through usual Hamiltonian evolution equation, giving: $\dot{Q}=\frac{\phi_m}{L}$ and $\dot{\phi_m}=\frac{Q}{C}$; or in terms of the more familiar variables $V$ and $I$, we have $\dot{V}=\frac{I}{C}$ and $\dot{I}=\frac{V}{L}$ (here and in the remainder of this section, the dot means time derivative). These equations of motion lead to an oscillatory behavior for both current and voltage, making the LC circuit an analog of a mechanical oscillator.

 Since $\phi_m$ and $Q$ are conjugate variables, the above equations can be readily quantized:

 \begin{equation}
 	\Hop=\frac{\hat{\phi}_m^2}{2L}+\frac{\hat{Q}^2}{2C},
 \end{equation}

 and we have $[\hat{\phi}_m,\hat{Q}]=i$. We can define annihilation operator $\aop=\frac{1}{\sqrt{\Of}}\left(\frac{\hat{\phi}_m}{\sqrt{2L}}-i\frac{\hat{Q}}{\sqrt{2C}}\right)$, with frequency $\Of=\frac{1}{\sqrt{LC}}$. This yields the familiar quantum harmonic oscillator Hamiltonian $\Hop=\Of(\adag\aop+\frac{1}{2})$. In general, the resistance of the circuit should be added, leading to the gradual loss of energy over time. The resistance, however, is suppressed in superconducting circuits. Therefore, LC superconducting circuits can be accurately modeled by quantum harmonic oscillators. This oscillator describes the exchange of energy between the different electromagnetic degrees of freedom.

 The LC circuit has a harmonic spectrum. However, the great interest of superconducting circuits is the possibility of engineering anharmonic spectrum as well, and in particular to create two-level systems. Several circuit designs can be used for this purpose; however, all of them use Josephson junctions as building blocks. We will present here a few key ideas, a more complete presentation of superconducting circuits may be found in \cite{gu_microwave_2017}. Josephson junctions are created by sandwiching a thin insulating layer between two superconductors. Cooper pair can tunnel across the junction, creating a current. The current and voltage across the barrier are related to $\delta$, the difference of superconducting phase across the barrier, by the Josephson equations:

   \begin{align}
   	I&=I_c \sin(\delta),\\ \nonumber
   	\dot{\delta}&=\frac{2e}{\hbar}V. 
   \end{align} 
   Crucially, the Josephson junction acts as a \textit{nonlinear} inductor, whose induction $L_J=\frac{\hbar}{2e\cos(\delta)}$ depends on $\delta$, and therefore on $I$. 
  The dynamics of the junction is described by the Hamiltonian $H=-E_J \cos(\delta)$, where $E_J=\frac{\hbar I_c}{2e}$ is the Josephson energy.

 When the Josephson junction is coupled to a capacity $C$ (which can be the capacity of the junction itself or an additional element), one obtains a nonlinear LC circuit. Because of the nonlinearity, the spectrum is non-harmonic, which makes it possible to isolate two eigenstates from the rest; this is the basic building principle of artificial qubits with superconducting circuits. The various designs are generally related to three archetypes: phase qubits, flux qubits, and charge qubits. In the phase qubit design, the phase difference across the junction and the number of Cooper pair tunneling through it act as continuous conjugate variables with oscillatory behavior. The two qubit states correspond to different oscillation amplitudes through the junction. In flux qubits, the phase difference can take two fixed values $\delta=\pm\delta_{f}$, which corresponds to the two qubit states (alternatively, this can be expressed in terms of the persistent current flowing through the junction having two possible values $I=\pm I_f$, or more commonly, in terms of the magnetic flux threading the circuit $\Phi=\pm\Phi_f$, hence the name flux qubit). These two designs are realized in the limit $E_J\ll\frac{e^2}{2C}$, that is, when the capacity is small and a large number of Cooper pairs can flow freely across the junction. Charge qubit are designed in the opposite limit of large capacity, where the tunneling probability is small. In that case, the two states of the qubit correspond to two different numbers of Cooper pairs in excess across the junction. The use of bias voltages, current, or magnetic flux, allows to tune the properties of these qubits, and to realize qubit gates. Many different designs extending and combining these basic principles have been studied.

 Artificial qubits can be coupled to a LC resonator; these systems are governed by equation exactly similar to the ones of cavity QED. As a consequence, this platform was named circuit QED. Thanks to the extreme confinement of the electromagnetic field, these systems are not bounded by the limitations of cavity QED, and can reach very high coupling values. Not only has the USC regime been achieved by numerous experiments, but the DSC regime has been reached as well, with a current record standing at $\gcoll/\Of=1.34$ \cite{yoshihara_superconducting_2017}. These systems have been mostly studied through spectral signatures, which are obtained for instance by looking at the transmission spectrum. A core observation is the position of energy levels as a function of the detuning, which exhibits an avoided crossing. The energy separation at the avoided crossing increases with the coupling, which allows to quantify it; the observation of a very large splitting indicates that the system operates in USC. Several spectral signatures arising from the counter-rotating terms were also clearly observed, such as the Bloch-Siegert shift \cite{niemczyk_circuit_2010,forn-diaz_ultrastrong_2019,bosman_multi-mode_2017}. Additional avoided crossings, created by the coupling of states with a different number of excitations, and not predicted by the JC model, have also been observed \cite{niemczyk_circuit_2010}. When the qubit is driven by a weak pump field, the transmission spectrum exhibits sidebands due to the ultrastrong-coupling \cite{chen_single-photon-driven_2017}.\\

 In very recent experiments \cite{yoshihara_superconducting_2017,yoshihara_inversion_2018}, the DSC regime has been achieved and studied through spectroscopic signatures. Crucially, these studies have demonstrated an effective ratio $\frac{4\gind_{eff}^2}{\Of_{eff}\Oq}$ larger than $1$, which shows that the diamagnetic term is absent or at least reduced in this setup. By contrast, these results are compatible with other models such as the one proposed in \cite{de_bernardis_cavity_2018}. This shows that the presence of the diamagnetic term is still an open (and platform-dependent) question.

 \subsection{Intersubband and Landau polaritons}

Solid-states systems are another major platform for the study of light-matter coupling. Those were the first to reach the USC regime, and hold the current record of coupling strength. Two setups are generally considered, which utilize intersubband and inter-Landau levels transitions, respectively.

\begin{figure}
 	\includegraphics[angle=-90,width=\linewidth]{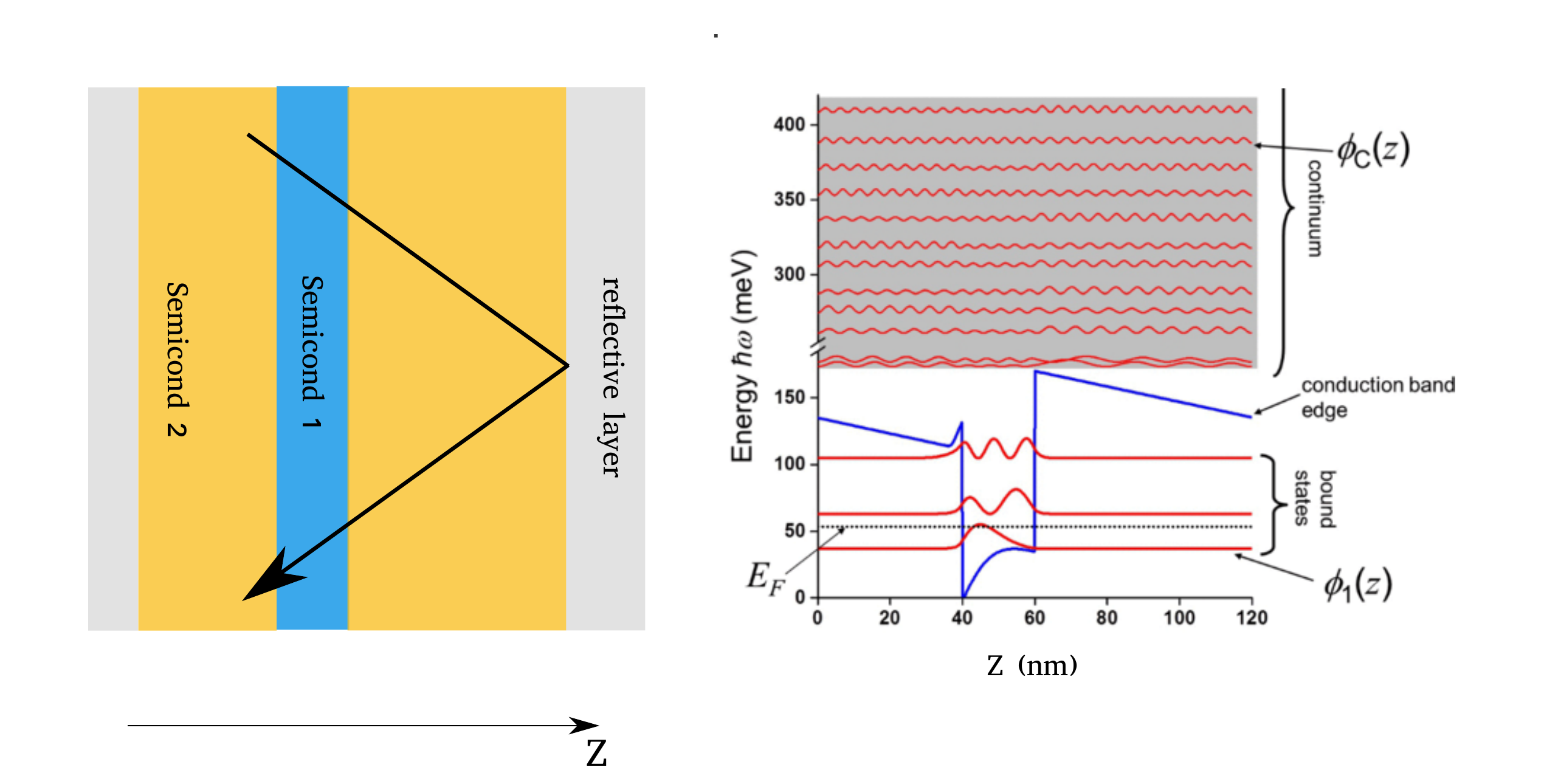}
 	\caption[Pictorial depiction of a quantum well.]{Pictorial depiction of a quantum well. Left: a layer of semiconductor is sandwiched between two layers with different properties. The ensemble is cladded between metallic layers acting as mirrors. An electromagnetic field propagates in the system, guided by the mirrors. $z$ represents the growth direction. Right: energy levels of the electrons in the system. The conduction band of the central (blue) layer has a bottom energy lower that of the surrounding (orange) layers, creating a quantum well. A discrete set of low-energy electronic modes are effectively confined inside the well; outside, they decay exponentially along $z$. Higher-energy modes are delocalized over the entire $z$ axis, and form a continuum. Along the $x$ and $y$ axis, the electrons are delocalized (not shown). The doping of the system is adjusted to put the Fermi energy of the system between the first and second confined levels. Right figure is reproduced from \cite{todorov_dipolar_2015}.}
	\label{C1_Qwells}
 \end{figure}

 Intersubband transitions can be realized by stacking multiple layers of different semiconductors on top of each other. Each semiconductor has a conduction and valence band with different energies. One of the layers is chosen such that the bottom of its conduction band has lower energy than the surrounding material. In this case, the energetic landscape forms a quantum well, trapping electrons within the layer (see Fig.\ref{C1_Qwells}). Within the confinement region, the continuous energy band splits into discrete subbands. In each subband, electrons are confined in the growth direction and delocalized in the perpendicular plane. The one-particle wavefunctions in the $n-th$ subband have the form $\psi_{n,\bm{k}}(x,y,z)\propto\phi_n(z)e^{i\bm{k}\bm{r}}$ with $z$ the growth direction, $\phi_n(z)$ a function which quickly decays out of the confinement zone, and $\bm{r}={x,y}$ are the coordinates in the perpendicular plane. This confined electronic state is generally called a 2-dimensional electron gas.

 In light-matter coupling experiments, the material is doped to populate the first subband. When light propagates through the material, electrons can hop from one subband to the other, in a so-called \textit{intersubband transition}. By confining light in the vicinity of the quantum wells (for instance by covering the structure with a metallic layer acting as a mirror), it is possible to achieve an extremely strong light-matter coupling. 
 Note that light does not excite one electron at a time. Each subband is populated with many electrons with different in-plane wavevectors $\bm{k}$. Light excites collectively many electrons at once, while changing their wavevectors. These collective excitations are best described by collective creation operators of the form \cite{ciuti_quantum_2005}:
 \begin{equation}
	\bdag_q\propto\sum_{\lvert k\rvert<k_F}c^\dagger_{2,k+q}c_{1,k},
 \end{equation}
 with $k_F$ the Fermi wavevector, and $q$ the momentum of the light field. Here, the fermionic operator $c_{1,k}$ annihilates an electron in the \textit{lower} subband with wavevector $k$, and $c^\dagger_{2,k+q}$ creates an electron in the \textit{upper} subband with wavevector $k+q$. In the limit where few electrons are excited, these operators obey bosonic commutation relations; this is very analogous to the HP transformation, which allows us to model a large number of weakly-excited two-level systems by a bosonic mode. Note that here we have considered a single mode for simplicity; however, for a full treatment, it is necessary to describe multiple light and electronic modes, and gives more intricate expressions for the bosonic modes \cite{todorov_dipolar_2015}.
   
 To sum up, in these systems, the light interacts with \textit{collective} bosonic matter modes. This coupling is normally described by a Hopfield-like model, of the form \eqref{C1_A2 term} or \eqref{C1_P2term}. The light and electronic modes hybridize to form \textit{intersubband polaritons}. This platform has reached the USC regime \cite{vasanelli_ultra-strong_2016}, with the current record for collective coupling at $\frac{\gcoll}{\Of}=0.45$ \cite{askenazi_midinfrared_2017}.

\begin{figure}
 	\includegraphics[angle=-90,width=\linewidth]{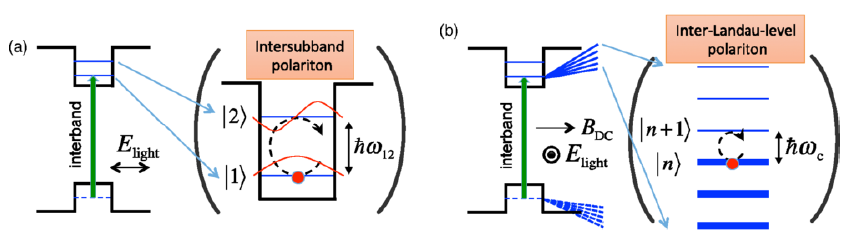}
 	\caption[Schematic depiction of interbband, intersubband and Landau level transitions.]{Schematic depiction of the transitions within quantum wells. a) The cavity field induces jumps between different subbands. b) An applied magnetic field split the subbands into Landau levels. The cavity field induces jumps between the highest occupied Landau level and the lowest unoccupied level. Instead of \textit{intra}band transition, it is also possible to consider \textit{inter}band transitions (green arrow). Although these transitions have a lower $\gind/\Of$ ratio, they have also reached USC. Reproduced from \cite{forn-diaz_ultrastrong_2019}.}
	\label{C1_Solidsyst_transition}
 \end{figure}

 Even higher couplings can be reached in very similar setups, using the so-called cyclotron resonance. A magnetic field is applied to the structure to create multiple Landau levels within a subband. The field induces transition between the highest occupied Landau level and the lowest unexcited level (see Fig.\ref{C1_Solidsyst_transition}). As earlier, these systems are described in terms of collective electronic excitations, which hybridize with light to form \textit{inter-Landau-level polaritons}. By controlling the magnetic field, it is possible to tune the matter field frequency. These polaritons have reached the largest coupling-to-frequency ratios so far \footnote{This was true during the writing of this dissertation. However, just a few days before the completion of the manuscript, a new result was published \cite{mueller_deep_2020}, reporting an even higher value of $\gind/\Of=1.83$ in a setup utilizing crystals of nanoparticles. In this setup, the electrons form plasmons, \textit{i.e.}, collective surface excitations. These excitations hybridize with light propagating through the crystal to form plasmon polaritons. This stands as a testimony of the fast evolution of this field of research.}, with a record standing at $\gind/\Of=1.43$ \cite{bayer_terahertz_2017}.
  As for circuit QED, the main experimental signature is the transmission spectrum of the system. This spectrum typically shows the anticrossing between the two polaritonic branches. The effect of the counter-rotating term has been quantified and isolated from the effect of the diamagnetic term \cite{zhang_collective_2016,li_vacuum_2018}. In the DSC regime, expulsion of the light field outside of the quantum wells have been reported \cite{bayer_terahertz_2017}, similar to predictions of \cite{de_liberato_light-matter_2014}.

 The results in the above works are well-fitted by the Hopfield model with $A^2$ term. However, some works have shown some deviation with respect to this standard behavior. In particular, in \cite{keller_landau_2020}, an effective lowering of the $A^2$ term was observed. Thus, it is still unclear whether the $A^2$ or $P^2$ terms must always be present in light-matter systems, and whether a superradiant transition could be achieved in these platforms.

 \subsection{Other platforms}
 Circuit QED, intersubband and Landau polaritons are the most advanced systems for the study of USC today\footnote{Or were, until the results of \cite{mueller_deep_2020}.}. However, recently, several other platforms have also reached this regime. For instance, USC coupling has been reported for molecules in microcavities, with $\gcoll/\Of$ up to $0.27$ \cite{george_ultra-strong_2015}.
 In solid-state systems, it is possible to study \textit{interband} transitions, instead of intersubband (see Fig.\ref{C1_Solidsyst_transition}). Electron jump from one band to the other, creating electron-hole bound states known as excitons; these hybridize with light, forming \textit{exciton polaritons}. Although the coupling strength is much lower than with intersubband or Landau polaritons, the USC has been recently achieved in organic semiconductors, with a record at $\gcoll/\Of=0.27$ \cite{gambino_exploring_2014}. 
 Optomechanical systems have also been considered; instead of electronic transitions, a \textit{mechanical} mode is coupled with light. Using a single molecule oscillating in a picocavity, a ratio $\gcoll/\Of=0.3$ has been reported \cite{benz_single-molecule_2016}. 
 Finally, some works have studied the coupling between light and spin waves in magnetic materials. The strong-coupling regime has been realized in these systems, with a ratio $\gcoll/\Of$ of a few $\%$, approaching the USC regime \cite{zhang_strongly_2014,tabuchi_hybridizing_2014}.

 \section{Simulating the USC}
 In the previous, section, we have presented experiments that achieve \textit{genuine} light-matter coupling. However, it is also possible to use quantum simulation to engineer \textit{effective} couplings \cite{georgescu_quantum_2014,mezzacapo_digital_2014,pedernales_quantum_2015,felicetti_spectral_2015,fedortchenko_quantum_2017,lamata_digital-analog_2017,aedo_analog_2018,lv_quantum_2018,jaako_quantum_2020}. There are two broad kinds of quantum simulations schemes: digital and analog. In digital schemes, the effect of a target unitary evolution is reproduced by applying a discrete sequence of simple quantum gates on a set of qubits. These simulators are interesting because of their universality; indeed, any quantum evolution can in principle be simulated by these systems. By contrast, in analog simulators \cite{pedernales_quantum_2015,felicetti_spectral_2015,fedortchenko_quantum_2017,aedo_analog_2018,lv_quantum_2018}, a specific system is used to mimic the properties of another system. The continuous dynamics of the simulator reproduces the Hamiltonian one wants to simulate. In this case, the physical properties of the original systems are typically conserved by the simulation. For instance, we can use trapped ions to simulate cavity-QED systems \cite{pedernales_quantum_2015,aedo_analog_2018,lv_quantum_2018}, in which the role of the atom is played by the internal state of the ion, and the role of the cavity field is played by the vibrational degrees of freedom of the ion. In this case, entanglement in the system to simulate will be represented by actual entanglement in the simulator, which can be detected at the device output (and even exploited as a resource in quantum protocols). The preservation of physical properties is an important difference with digital simulation, in which quantities of interest are encoded in the degrees of freedom of an ensemble of qubits, which require proper decoding. Finally, it was recently suggested that hybrid schemes, called digital-analog, could be used to combine the advantages of digital and analog protocols \cite{mezzacapo_digital_2014,lamata_digital-analog_2017,langford_experimentally_2017,parra-rodriguez_digital-analog_2020}.\\

Analog and digital-analog schemes have so far attracted more attention for the simulation of USC physics, notably because of their ability to preserve physical quantities. In the  remainder of this section, we will focus on these protocols. 
 The rationale of analog simulation for USC physics can be understood with the following model, which was studied in \cite{fedortchenko_quantum_2017}. Let us consider a three-wave mixing between three bosonic fields, with a coupling constant $\chi$. One of the fields is very intense and can be modeled by a classical pump with an amplitude $p$, while the other two are described by quantum operators $\aop$ and $\bop$. The Hamiltonian reads:

 \begin{equation}
 	\Hop=\Of\adag\aop+\Oq\bdag\bop+\chi(p(t)+p^*(t))(\adag+\aop)(\bdag+\bop).
 	\label{C1_3wavemixing}
 \end{equation}
 The pump will excite the transitions with which it is resonant. By appropriately tuning the frequency of the pump, it is possible to excite only certain processes, and to engineer the dynamic. For instance, if the pump is at frequency $\Of-\Oq$, the absorption of one photon from the pump will create one excitation in the $\aop$ field and destroy one excitation of the $\bop$ field. Therefore, the pump will excite the process described by the coupling term $\adag\bop$. Similarly, for a pump frequency $\Of+\Oq$, the absorption of one photon from the pump will simultaneously create one excitation in both fields, a process described by the terms $\adag\bdag$. In this case, the pump excites counter-rotating terms. Crucially, the above argument can be valid \textit{even} if the three-wave mixing amplitude $\chi$ is weak, or if $\aop$ and $\bop$ are widely detuned. Hence, by applying two pumps at frequencies close to $\Of+\Oq$ and $\Of-\Oq$, one can engineer an effective dynamic in which both resonant and counter-rotating terms play a role. 

 Let us make this reasoning more formal. We will consider two pumps at frequencies $\Of_R=(\Of-\Oq)+\Delta_R$ and $\Of_B=(\Of+\Oq)+\Delta_B$, which will be referred to as the red-detuned and blue-detuned pump, respectively. We will assume that both pumps are in phase; the full pump amplitude can then be expressed as $p(t)=p_R e^{-i\Of_Rt}+p_B e^{-i\Of_B t}$. We inject this expression in \eqref{C1_3wavemixing}, and move to the interaction picture. We obtain:

 \begin{align}
 	\Hop_{IP}= & \chi p_R\left[\adag\bop e^{-i\Delta_Rt}+\aop\bdag e^{-i(2(\Of-\Oq)+\Delta_R)t}+\adag\bdag e^{i(2\Oq-\Delta_R)t} + \aop\bop e^{-i(2\Of+\Delta_R)t}\right] \\ \nonumber
 	& + \chi p_B\left[\adag\bdag e^{-i\Delta_Bt} + \aop\bop e^{-i(2(\Of+\Oq)+\Delta_B)t}) + \adag\bop e^{-i(2\Oq + \Delta_B)t} + \aop\bdag e^{-i(2\Of+\Delta_B)t}\right] + h.c..
 \end{align}

 If the detuning $\Delta_R$ and $\Delta_B$ are small, all but two of these terms oscillate quickly. Retaining only the slow-oscillating terms, we obtain: $\Hop_{IP}=G_R\adag\bop e^{-i\Delta_Rt}+G_B\adag\bdag e^{-i\Delta_Bt}+h.c.$, where we defined $G_{R,B}=\chi p_{R,B}$. The time-dependence of this Hamiltonian may be suppressed by one last transformation $\Uop=e^{i(\frac{\Delta_B+\Delta_R}{2})\adag\aop t + \frac{\Delta_B-\Delta_R}{2})\bdag\bop t}$. We finally arrive at the following effective Hamiltonian:

 \begin{equation}
 	\Hop_{eff}=\Of_{eff}\adag\aop + \Oq_{eff}\bdag\bop + G_R(\adag\bop+\bdag\aop) + G_B (\adag\bdag+\aop\bop),
 	\label{C1_effectiveRabi}
 \end{equation}
 with $\Of_{eff}=-\frac{(\Delta_R+\Delta_B)}{2}$ and $\Oq_{eff}=\frac{\Delta_R-\Delta_B}{2}$. Hence, in this rotating frame, the dynamics of the system is given by a Rabi-like model with \textit{effective} frequencies.
 We now make a key observation: the effective frequencies may be controlled by changing the detuning of the pumps. In particular, they can in principle be brought arbitrarily close to zero. This means that a very large ratio $G/\Of_{eff}$ can be attained, \textit{even} if the underlying physical coupling $\chi$ is very small. This is the operating principle of many simulation protocols of USC: take a system with \textit{weak} light-matter coupling, and drive it with time-dependent pumps (for instance lasers), in order to excite selectively certain processes. Then the system will be governed by an effective dynamic, whose parameter can be tuned by changing the amplitudes and frequencies of the drive. This makes it possible to reach regimes of parameters largely unconstrained by the strength of the underlying light-matter coupling. 

 Moreover, techniques of the sort can not only be used to implement different regimes of parameters, but also different \textit{models}. This is already visible in \eqref{C1_effectiveRabi}: if we set $p_R=p_B$, we have $G_R=G_B$, and the effective dynamics is described by the Rabi model. If, however, we choose pumps with different amplitudes, we will have $G_R\neq G_B$: the effective dynamics will be given by an \textit{anisotropic} Rabi model. In general, thanks to quantum simulation, it is possible to engineer the dynamics to create models that are not easily produced with genuine systems. For instance, genuine systems may or may not exhibit a $A^2$ term depending on the specifics of the experiment; predicting which precise model describes best a given setup is a challenging task. With a simulated system, however, it is in principle possible to \textit{choose} to include an $A^2$ term or not. This allows us to test the predictions associated with various models in a controlled way; in particular, as we will discuss in the next Chapter, quantum simulation has allowed to confirm that in the absence of the diamagnetic term, the Dicke model does exhibit a superradiant phase transition.

 Another essential feature of engineered systems is their interaction with the environment. Because the system is driven, and the underlying real couplings can be small, the arguments related to the photoemissions of systems in USC do not hold anymore. For engineered systems, the photons appearing in the USC regime can be \textit{real} excitations. A system in the engineered ground state can emit excitations into the environment; this does not contradict energy conservation, since the drive continuously pump energy into the system. Contrary to genuine systems which will be driven towards a Gibbs equilibrium state when interacting with a thermal bath, engineered systems are fundamentally out-of-equilibrium. This makes them a good platform to observe effects that would be inaccessible with equilibrium systems. In particular, with analog simulation scheme, even ground-state properties such as entanglement can be physically present and unambiguously observed in the simulator.

 Finally, because the USC regime is associated with real excitations, engineered systems are generally more straightforward to probe than their genuine counterparts. Observables are not limited to spectral signatures, but can also include information about the state such as photon statistics. All of these features make quantum simulation a good way to test predictions and observe new phenomena.\\

 These concepts have been put to use in multiple experiments. One of the first simulations of the Rabi model was realized in superconducting circuits \cite{braumuller_analog_2017}, following a proposal laid down in \cite{ballester_quantum_2012}. The setup is composed of a single artificial qubit coupled to a LC resonator. The physical coupling is not ultrastrong, and the system can be described by the JC Hamiltonian. Periodic drives are applied to the qubit, which allows us to engineer strong counter-rotating terms. The system can simulate the Rabi model with an effective coupling tunable from weak to deep-strong coupling regime. Both the qubit and bosonic populations can be measured; this has made it possible to observe the collapse and revival predicted to take place in the DSC regime.\\

 Still in the context of circuit QED, effective USC between two bosonic modes have also been realized \cite{markovic_demonstration_2018}, following the proposal of \cite{fedortchenko_quantum_2017}. Josephson junctions are used to realize a three-wave mixer, which is feed with two quantum field and a two-tones strong pumping field. The behavior of the physical system is described by the Hamiltonian \eqref{C1_3wavemixing}; following the reasoning we have outlined above, one arrives at an effective dynamics described by \eqref{C1_effectiveRabi}. The effective USC coupling generates both one- and two-mode squeezing, which have been experimentally observed.\\

 The two previous setups involve the coupling of light with an artificial atom, or of two light modes. However, the properties of the Rabi model remain the same when bosonic fields other than light are considered. In particular, the Rabi model can also be engineered with two-level systems coupled to their vibrational degree of freedom; one can then achieve ultrastrong \textit{phonon}-matter interaction. Atomic systems are excellent platforms for this, thanks to their great controllability, and the use of quantum engineering allowing to compensate for the naturally low couplings. In particular, the Rabi model has been realized by coupling the internal and vibrational degrees of freedom of trapped ions \cite{lv_quantum_2018}, following a proposal from \cite{pedernales_quantum_2015}. The rotating and counter-rotating parts of the dynamics are engineered by two pump lasers at frequency $\Oq\pm f$, with $\Oq$ and $f$ the frequencies of the internal transition and the vibrational mode, respectively. Multiple predictions of USC and DSC have been observed in this experiment, such as spectral signatures, collapse and revival dynamics, or entanglement between the qubit and the boson.\\

 A similar experiment has been realized, using atoms instead of ions \cite{schneeweiss_cold-atom-based_2018,dareau_observation_2018}. The setup consists of an optical fiber in which two counterpropagating fields are sent. The evanescent field around the fiber allows to trap an atom in its vicinity. The field also induces a position-dependent shift of the atom frequency, which creates an effective coupling between the internal and vibrational degrees of freedom. In this setup, it is possible to tune both the qubit frequency and the effective coupling in-situ. Experimentally, the effective USC regime has been achieved, and has been studied through spectral signatures. This system could be further enhanced to reach the DSC or dispersive USC regime (\textit{i.e.}, USC with $\Oq\ll\Of$).\\

 The setups we have discussed so far are all analog simulations: they implement a continuous dynamics which mimics the Hamiltonian one wishes to simulate. 
Digital-analog simulation of the Rabi model was also proposed \cite{mezzacapo_digital_2014,lamata_digital-analog_2017} and realized \cite{langford_experimentally_2017}. A system freely evolves according to the JC dynamics for a time $T$, which corresponds to a unitary operation $\Uop_{JC}=e^{-i(\adag\sigm+\aop\sigp)T}$ (in interaction picture). Then a gate $e^{-i\frac{\pi}{2}\sigx}$ is applied, followed by a free evolution for a time $T$, and another gate $e^{i\frac{\pi}{2}\sigx}$. These three steps combined corresponds to an evolution given by $\Uop_{AJC}=e^{-i(\adag\sigp+\aop\sigm)T}$. This mimics a free evolution under anti-JC dynamics, with only the counter-rotating terms. By alternating JC and anti-JC sequences, it is possible to simulate the behavior of the Rabi model. This technique has been demonstrated experimentally using superconducting circuits \cite{langford_experimentally_2017}. In this experiment, DSC regime features well beyond spectral signatures have been observed, including large photon population and qubit-photon entanglement. Wigner tomography of the field have been realized, conditioned or not on the qubit state; this allowed to visualize the Schrödinger-cat-like dynamics of the eigenstates \eqref{C1_gRWAeigenstates}.\\

 Finally, other simulation schemes have been put forward. In \cite{felicetti_quantum_2017}, an all-motional scheme was proposed, involving cold atoms trapped in a potential which is the sum of a harmonic component and a periodic component. The state of motion can be expressed in Bloch basis, where it is decomposed into a discrete and continuous part. This describes different energy bands, each of which containing an infinite number of quasimomentum states. If the dynamics can be confined within the first two bands, this system simulates Rabi-like dynamics, the presence in one or the other band playing the role of the qubit, and the continuous quasimomentum being the bosonic field.  \\

 Although the effort has been mostly focused on the Rabi and Dicke Hamiltonians, other models have also been considered. For instance, in \cite{felicetti_spectral_2015}, a scheme to implement a \textit{two-photon} Dicke model was proposed. This model is obtained when the standard one-photon coupling of the Dicke model is replaced by a two-photon term $\sum_i\sigx_i(\adagsq+\aop^2)$. The proposal has been made for trapped-ion platforms and is very similar to the one used to simulate the usual Rabi model \cite{lv_quantum_2018}. The main difference is that the driving fields have a frequency $\Oq\pm2f$, which means that two phonon excitations are associated with each qubit flip. This proposal could also be extended to consider multi-photon couplings. The properties of the two-photon Dicke model will be the subject of Chapter 4.

  \chapter{Superradiant phase transition}


\epigraph{\textit{Car lumineux est ce qui joint en clarté deux sources de lumière.}}{Suger}

 One of the most striking predictions of light-matter models such as the Dicke model is the superradiant phase transition, during which the bosonic field acquires a macroscopic population. Despite extensive research on the subject, however, this effect is still elusive, and its experimental implementation is still challenging. Only recently has the superradiant transition been observed, using quantum simulation. Furthermore, several similar but non-equivalent definitions of the word \textit{superradiance} can be found in the literature. In this Chapter, we will try to clarify these issues and discuss the various properties of superradiant phase transitions. First, we will introduce the notion of superradiance from a historical point of view. We will present the perspective that will be adopted in this work. Next, in Sections \ref{Sec2.2} and \ref{Sec2.3}, we will discuss the properties of \textit{equilibrium} superradiant transition. We will defend that this phenomenon should be considered less a many-body phenomenon than a generic property of light-matter interaction. To illustrate this, we will first discuss how a superradiant transition can be achieved with a single emitter, before treating the many-emitter limit. Sec.\ref{Sec2.4} will be devoted to the out-of-equilibrium superradiant transition. In Sec.\ref{Sec2.5}, we will discuss a generalized version of the Dicke model, which has an enlarged $U(1)$ symmetry. Finally, Sec.\ref{Sec2.6} will be devoted to the experimental realizations of the superradiant transition.
 This Chapter, like the precedent, has made use of several review papers on the subject; \cite{kirton_introduction_2019}, in particular, has been particularly useful.

 \section{Introduction to superradiance}
 \label{Sec2.1}
 We will start by defining the term \textit{superradiance}. This word is often used in two different, although related, contexts. The use of the term is generally traced back to the 1954 paper by R.H. Dicke \cite{dicke_coherence_1954}. In this work, Dicke studied the spontaneous radiation rate of a collection of two-level systems. When the emitters are coupled independently to the light, qubits emit photons in random, uncorrelated events. The number of excited qubits and the intensity of emitted light decrease gradually, following an exponential decay law $\Gamma e^{-\Gamma t}$, with $\Gamma$ the single-atom decay rate. The instantaneous decay rate is maximum at the beginning of the emission, then gradually decrease following the exponential law. 

 By contrast, if the qubits are confined in a length smaller than the light wavelength, all emitters feel the same field, and interact coherently through a coupling of the form $\hat{J}_-\adag+\hat{J}_+\aop$, where $\hat{J}_\pm=\sum_i^N\hat{\sigma}^i_\pm$ are collective spin operators, and $\adag$ ($\aop$) creation (annihilation) operators of the electromagnetic field (note that in general, all the modes of the field need to be considered). In this case, the decay rate is given by $\Gamma \moy{\hat{J}_+\hat{J}_-}$, with the average taken over the state of the qubits. This state can be expressed in the angular momentum basis $\ket{j ; m}$, with $j(j+1)$ and $m$ the eigenvalues of $\hat{J}^2$ and $\hat{J}_z$, respectively. Initially, the qubits are in state $\bigotimes\ket{\hspin}_i$, which corresponds to $j=\frac{N}{2}(\frac{N}{2}+1)$ and $m=\frac{N}{2}$. At the end of the evolution, one has $m=-\frac{N}{2}$, which corresponds to the state $\bigotimes\ket{\lspin}_i$. Both states are separable, with all spins  aligned around one direction. The instantaneous decay rate $\Gamma\moy{\hat{J}_+\hat{J}_-}$ is equal to $\Gamma N$ for $m=\frac{N}{2}$, and $0$ for $m=-\frac{N}{2}$. 
 However, since the qubits evolve collectively, the total angular momentum is conserved during the decay process, while its component along $z$ gradually decreases: when $p$ photons have been emitted, the system is in a state $\ket{j ; m=\frac{N}{2}-p}$. When $m\sim0$, the system is in a highly collaborative state, which increases drastically its emission properties. For instance, for $m=0$, one has a decay rate $\Gamma\moy{\hat{J}_+\hat{J}_-}=\Gamma (\frac{N}{2}+1)\frac{N}{2}\sim\Gamma \frac{N^2}{4}$, a decay rate \textit{higher}  than the starting rate $\Gamma N$. Hence, in the early state of the emission process, the system will evolve towards more and more radiative states, and the emission process will become faster and faster. The fastest rate is achieved for $m=0$; after that, the emission rate starts to decrease, and finally reaches $0$ when all spins are deexcited and $m=-\frac{N}{2}$. Compared with a standard decoherence processe, in which the atoms are deexcited at a constant rate $\Gamma$, this process shows two important differences: first, the decay rate is not constant, first increasing to a maximum then being reduced as time goes on. Second, the average decay rate depends on $N$; it is much higher than the single-atom rate $\Gamma$, and increases with the number of atom. This means that the atoms decay much faster when they are coupled collectively to the environment. If we look at the intensity emitted when atoms are decaying individually, it will be maximal at the beginning and then gradually decrease. By contrast, with collective coupling, the intensity will quickly increase then decrease, in a much shorter burst.
 The name \textit{superradiance} was used by Dicke to describe this phenomenon of enhanced emission. In this context, the word refers to a \textit{transient} effect, which arises as a consequence of the \textit{collective} behavior of several spins, \textit{weakly} coupled to light. An in-depth discussion of this perspective can be found in \cite{gross_superradiance_1982}.\\

 Several years after Dicke, several authors have put forward a notion of \textit{steady-state} superradiance \cite{hepp_superradiant_1973,wang_phase_1973}. These authors considered a large collection of atoms interacting with a bosonic field inside of a cavity, with a coupling $\Of\adag\aop+\Oq\Jz+\gind(\Jm\adag+\Jp\aop)$, with $\gind$ the individual coupling of each qubit with the field. It was theoretically predicted that, by increasing the coupling strength beyond a threshold value, the field would spontaneously build up inside the cavity and acquire a macroscopic value, \textit{i.e.}, proportional to the number of atoms $N$. The threshold value was found to be $\gind=\sqrt{\frac{\Of\Oq}{4N}}$, indicating that the transition occurs for smaller coupling when the number of atoms increases. This is an effect of the collective enhancement of the coupling which we have mentioned in the previous Chapter. Defining the \textit{collective} coupling $\gcoll=\sqrt{N}\gind$, we obtain the so-called Tavis-Cummings (TC) model:

 \begin{equation}
 	\Hop=\Of\adag\aop+\Oq\Jz+\frac{2\gcoll}{\sqrt{N}}(\adag\Jm+\aop\Jp).
 	\label{TC}
 \end{equation}

 The transition in the Tavis-Cummings model was quickly shown \cite{hioe_phase_1973} to take place also when counter-rotating terms are present, in the Dicke (or Hepp-Lieb-Dicke) model:

 \begin{equation}
 	\Hop=\Of\adag\aop+\Oq\Jz+\frac{2\gcoll}{\sqrt{N}}(\adag+\aop)\Jx.
 	\label{Dicke}
 \end{equation}

Later on, the increasing amount of work devoted to ultrastrong coupling physics has shed new light on this transition. Indeed, while superradiance could initially be considered as a collective effect between many emitters, the study of USC and DSC regime has shown that \textit{even a single qubit}, when ultrastrongly coupled to light, would lead to the appearance of (potentially virtual) photons. Crucially, several recent studies have shown that these single-qubit systems \cite{ashhab_superradiance_2013,hwang_quantum_2015,hwang_dissipative_2018}, or other models such as two bosonic fields with small nonlinearity \cite{felicetti_universal_2020}, could exhibit precisely the same behavior as the Dicke model. Under appropriate parameter conditions, a notion of macroscopicity can be recovered even in these \textit{finite-size} systems, which leads to a phase transition. Similar critical effects can also be observed in a system composed of a single Kerr resonator (\textit{i.e}, a nonlinear optical cavity) with appropriate drive and dissipation \cite{bartolo_exact_2016}. From this point of view, collective effects do not play an essential role in the onset of superradiant transition; instead, the essential feature of these models is the very strong coupling between light and a nonlinear system.\\

  With these works, the focus has been shifted in a subtle but important manner: the light-matter system is no longer considered from the perspective of collective emission and interference, but from the point of view of \textit{Quantum Phase Transitions} or QPT. In this context, these systems offer a different, interesting, perspective. QPT are often studied in the context of many-body systems, in which components typically interact through short-range interactions (the Ising model being a paradigmatic example). The analysis of these models must explicitly take into account their spatial structure, for instance through renormalization group analysis \cite{sachdev_quantum_2011}. By contrast, in the Dicke model, the qubits have an effectively infinite-range interaction and obey a permutation symmetry. As a consequence, they behave collectively as a single body; the system is effectively zero-dimensional. This is even more apparent in the Rabi model \cite{hwang_quantum_2015}, where there is actually one single qubit involved, or in systems involving only two coupled bosonic fields \cite{felicetti_universal_2020}. Therefore, these optical systems offer a platform where critical effects can be observed in the absence of spatial structure. We immediately mention that there is some debate about whether these critical effects are "true" QPT \cite{larson_remarks_2017}. Although this question is to some extent a matter of semantic, we will discuss in this Chapter how at least some features of the QPT can be properly defined for these systems. 
 Furthermore, and precisely because of their simplicity and absence of spatial structure, these quantum optics systems are extremely controllable and can be fully characterized. In particular, they provide an ideal platform for quantum information experiments. Lately, the application of quantum information concepts to many-body systems has attracted considerable attention \cite{vidal_entanglement_2003,eckert_quantum_2008,eisert_colloquium_2010,islam_measuring_2015}. In this context, studying the behavior of quantum correlations, such as many-body entanglement, in the vicinity of phase transitions is a particularly interesting question. However, these correlations can be challenging to probe.

 Here, we follow the opposite direction: starting from quantum optical, few-body systems in which quantum correlations are routinely produced and characterized, we want to study how these systems could also exhibit critical effects, and how this will affect their quantum properties. This in turn could provide new insights to study their many-body counterparts.

 To summarize, in this Thesis we will be interested in putting together ideas from quantum information and quantum phase transitions, in the context of simple, few-body models. We will also study models involving many components, such as the Dicke model; however, in most of our analysis, the presence of many qubits will not be thought of as an essential element of the dynamic, but rather a mean to achieve a larger collective coupling, which can also be obtained by other ways. Instead, we will focus on the behavior of a bosonic field, in the presence of nonlinearity, and on the hybridization of light and matter (see Table \ref{Tableperspective}). In particular, we will study superradiant phase transitions as a generic property of quantum optical systems, rather than a many-body effect specific to the Dicke model. 
 To illustrate this point of view, we will purposely delay the study of the Dicke model, and start instead by discussing the emergence of a phase transition in the Rabi model.

 \begin{table}
 \begin{center}
 \begin{tabular}{|c|c|}
 	\hline \textbf{Original perspective} & \textbf{Perspective of this Thesis}\\
 	\hline
 	\hline Weak or strong coupling & Ultrastrong coupling, driven systems \\
 	\hline Collective effect requiring many atoms & Generic property of light-matter systems\\
 	\hline Transient dynamics & Steady-state property\\
 	\hline
 \end{tabular}
 \end{center}
 	\caption[Comparison between the original perspective on superradiance and the one developed in this Thesis.]{Comparison between the original perspective on superradiance and the one which will be developed here.}
 	\label{Tableperspective}
 \end{table}

  \section{Phase transition in the Rabi model}
  \label{Sec2.2}
 
  The behavior of the Dicke model \eqref{Dicke} is very dependent on two parameters: the number of qubits $N$ and the frequency ratio $\Oq/\Of$. As we will see, a phase transition can occur when either of these parameters is allowed to go to infinity. First, we will study the case $N=1$, \textit{i.e.}, the Rabi model, with the condition $\Oq\gg\Of$. Several works \cite{ashhab_superradiance_2013,hwang_quantum_2015,hwang_dissipative_2018} have studied the emergence of a phase transition in this regime, which we will now discuss in some length.

  \subsection{Quantum Phase Transitions}

  Before we discuss the properties of the Rabi Model, let us start by giving some words of context about QPT. The study (and the very definition) of a QPT is a field of research in itself \cite{sachdev_quantum_2011}, which we will not enter in details here. However, we will remind here a few key properties of a QPT, which can be observed in the systems we are interested in.

  First, a QPT is characterized by a \textit{non-analytic behavior for a given parameter value}. What we mean here is that when the parameters which describe the system are moved across a certain, definite, threshold value, some properties of the system (typically, the average value of certain operators in the ground or steady-state) will change in a non-analytical way.  Here, we are not considering situations in which the model we used to describe the system becomes unstable at a certain point, which indicates that the model itself is no longer valid. Instead, we want to study how a single model (typically a Hamiltonian) can describe the transition from one stable ground state to another, still stable but very different, one. We are not considering either smooth crossovers in which the system continuously evolves from one state to the other, but abrupt change that arise for a definite parameter value.

  Second, a QPT is driven by two or more non-commutative terms within the Hamiltonian. These terms will compete with each other, and favor different ground states. The evolution of the parameters will increase one term compared to the other, activating the transition. This is different from classical phase transitions, which are driven by a competition between energy minimization and entropy maximization. Indeed, these QPT can even occur at zero temperature.

  Third, many QPT (at least, the second-order ones) exhibit $\textit{spontaneous symmetry breaking}$ (or SSB). This is a situation in which a symmetry is obeyed by the equation describing the dynamics of the system, but not by the solution of the dynamics (for instance, when the Hamiltonian satisfies a certain symmetry, but not its ground state).

  Finally, the non-analytical change near the critical point is accompanied by diverging fluctuations, which are characterized by \textit{critical exponents}. No physical system, however, can be truly divergent. What we have instead are large fluctuations, which become infinite in a certain \textit{thermodynamic limit}, which, in many-body systems, corresponds to the limit of infinite system size.\\

  \subsection{Symmetries of the Rabi model}

 Let us now see how these concepts can be applied to the Rabi model:
 
 \begin{equation}
 	\Hop=\Of\adag\aop+\frac{\Oq}{2}\sigz+\gind(\adag+\aop) \sigx.
 	\label{Rabi}
 \end{equation}
 We start by describing the symmetry of the model. The Rabi Hamiltonian is invariant by the transformation $\Uop=e^{i\pi(\adag\aop+\sigz)}$, which has the effect $\aop\rightarrow-\aop$, $\sigx\rightarrow-\sigx$, and leave $\adag\aop$ and $\sigz$ invariant. This transformation applied twice gives the identity, and thus this describes a $\mathbb{Z}_2$ symmetry group. When $\gind\sim0$, the Rabi model can be accurately approximated by the Jaynes-Cummings model, and it has a unique ground state $\ket{\lspin0}$, which obeys the $\mathbb{Z}_2$ symmetry. By contrast, for large coupling, the system can be described by gRWA (see the previous chapter). When $\gind\rightarrow\infty$, the spin term acts only as a perturbation, and the Rabi model can be approximated by $\Of \adag\aop + \gind(\adag+\aop)\sigx$. This model obeys a larger symmetry group, since it is also invariant by the transformation $\Uop=e^{i\theta\sigx}$ in addition to the $\mathbf{Z}_2$ symmetry. This Hamiltonian has two degenerate ground states $\ket{\leftarrow}\ket{\alpha_0}$ and $\ket{\rightarrow}\ket{-\alpha_0}$, where the field is in a coherent state with $\alpha_0=\frac{\gind}{\Of}$. These two ground states do not obey the $\mathbf{Z}_2$ symmetry.  
 \footnote{The degeneracy of the Hamiltonian is a generic requirement to achieve SSB. Indeed, SSB can be defined as a situation in which the ground state of the Hamiltonian does not obey a symmetry of the Hamiltonian. Let us call {$\Uop$} the operator describing the effect of this symmetry, and {$\ket{\psi_E}$} an eigenstate of {$\Hop$} with energy {$E$}. Since {$[\Uop,\Hop]=0$}, we have {$\Hop\Uop\ket{\psi_E}=\Uop\Hop\ket{\psi_E}=E\Uop\ket{\psi_E}$}; hence {$\Uop\ket{\psi_E}$} is also an eigenstate of {$\Hop$} with energy {$E$}. This leaves only two possibilities: either the eigenspace of energy {$E$} is not degenerate, in which case we must have {$\Uop\ket{\psi_E}=e^{i\phi}\ket{\psi_E}=$}: the state {$\ket{\psi_E}$} obey the symmetry, and therefore we do not have SSB. Or the eigenspace is degenerate: in which case {$\Uop\ket{\psi_E}$} may be different (even orthogonal) to {$\ket{\psi_E}$}. Only in this last situation can we have an eigenstate that does not respect a symmetry obeyed by the full Hamiltonian.\\

 This can also be linked with the presence of additional symmetries {$\Vop$}, which are incompatible (\textit{i.e.}, do not commute) with {$\Uop$}. In this case, it is not possible for a given eigenstate to be invariant under all symmetries simultaneously, meaning that one of them at least must be broken. For instance, in the case of the Hamiltonian {$\Of \adag\aop + \gind(\adag+\aop)\sigx$}, both {$\Uop=e^{i\pi\sigz\adag\aop}$} and {$\Vop=e^{i\theta\sigx}$} are symmetries. The ground states {$\ket{\leftarrow}\ket{\alpha_0}$} and {$\ket{\rightarrow}\ket{-\alpha_0}$} obey the {$\Vop$} symmetry, but break the {$\Uop$} symmetry. By contrast, the states {$\ket{\leftarrow}\ket{\alpha_0}\pm\ket{\rightarrow}\ket{-\alpha_0}$} are also degenerate ground states of the Hamiltonian; they \textit{do} obey the {$\Uop$} symmetry, but break the {$\Vop$} symmetry.} 
This suggests that a SSB could take place when the coupling constant of the Rabi model is increased. However, in the regime where the gRWA is usually considered, $\Oq\leq\Of$, the symmetry breaking can take place only for $\gind\rightarrow\infty$;  when $\gind$ increases, the system only experiences a smooth, continuous crossover. By contrast, in the regime $\Oq\gg\Of$,  an abrupt SSB can take place for finite $\gind$, as we will now see.

  \subsection{Mean-field approximation}

We now consider the regime $\Oq\gg\Of$.
  The large difference of frequencies indicates that the back-action of the bosonic field on the spin will be small. As a first guess, we may assume that the fluctuations of the field will play only a negligible role in the dynamics, and consider only its mean-field value. We will see later on that this claim is indeed justified. 
  We will replace the field $\aop$ by its classical, mean-field value $\alpha$. This amounts to project the field into a coherent state $\ket{\alpha}$, to obtain an effective qubit Hamiltonian:

  \begin{equation}
  	\Hop_{\text{eff}}(\alpha)=\bra{\alpha}\Hop\ket{\alpha}=\Of\lvert\alpha\rvert^2+\frac{\Oq}{2}\sigz+\gind(\alpha+\alpha^*)\sigx.
  	\label{EffspinRH}
  \end{equation}

  This Hamiltonian can be readily diagonalized, and its ground-state energy is found to be $E_G(\alpha)=\Of\lvert\alpha\rvert^2-\sqrt{\frac{\Oq^2}{4}+\gind^2(\alpha+\alpha^*)^2}$. The next step is to minimize the energy $E_G(\alpha)$ with respect to $\alpha$. We find that, for $\gind\leq\gind_p=\frac{\sqrt{\Of\Oq}}{2}$, the minimum always corresponds to $\alpha=0$. However, for $\gind\geq\gind_p$, two minima are found, on 

  \begin{equation}
  	\alpha=\pm\alpha_g=\pm\sqrt{\frac{\gind^2}{\Of^2}\left(1-\left(\frac{\gind_p}{\gind}\right)^4\right)}=\pm\sqrt{\frac{\Oq}{\Of}}\sqrt{\frac{1}{4\lambda^2}(\lambda^4-1)},
  \end{equation}
  where we have defined $\lambda=\gind/\gind_p$. This shows that, for $\lambda\geq1$, the field will acquire a non-zero value. Importantly, the displacement of the field is proportional to $\sqrt{\frac{\Oq}{\Of}}$, which is very large. In turn, this large value will induce a spin rotation: the Hamiltonian \eqref{EffspinRH} can be diagonalized by defining the rotated spin operators 
  \begin{align}
  	\hat{\tau}_z^\pm & =\frac{1}{\sqrt{\frac{\Oq^2}{4}+4\gind^2\alpha_g^2}}\left(\frac{\Oq}{2}\sigz\pm2\gind\alpha_g\sigx\right)=\frac{2}{\Oq\lambda^2}\left(\frac{\Oq}{2}\sigz\pm2\gind\alpha_g\sigx\right) \label{Rotatedspin},\\ \nonumber
  	\hat{\tau}_x^\pm & =\frac{2}{\Oq\lambda^2}\left(\frac{\Oq}{2}\sigx\mp2\gind\alpha_g\sigz\right),\\ \nonumber
  	\hat{\tau}_y^\pm & =\sigy.
  \end{align}

 For a given $\gind$, the spin state will be given by $\ket{\lspin^{\pm}}=\pm\sqrt{\frac{\lambda^2+1}{2\lambda^2}}\ket{\hspin}+\sqrt{\frac{\lambda^2-1}{2\lambda^2}}\ket{\lspin}$, the lowest eigenstate of  $\hat{\tau}_z^\pm$. 

 This analysis allows us to develop a general picture of the phase transition. For $\gind\leq\gind_p$, the system is in a state similar to $\ket{\lspin}\ket{0}$, the symmetry is unbroken. For $\gind\geq\gind_p$, the behavior of the system can be described by two states $\ket{\lspin^+}\ket{\alpha_g}$ and $\ket{\lspin^-}\ket{-\alpha_g}$. Immediately after the critical point, these two states are not orthogonal. The (complete) Hamiltonian $\Hop$ is non-degenerate, and we can expect its unique ground state to be a Schrodinger-cat-like superposition of the two states above. However, when $\gind$ increases, $\alpha_g$ will very quickly acquire a macroscopic value, and the two states $\ket{\lspin^+}\ket{\alpha_g}$ and $\ket{\lspin^-}\ket{-\alpha_g}$ will become orthogonal. At this point, both states are degenerate eigenstates of $\Hop$. Since these states do not obey the $\mathbb{Z}_2$ symmetry, we have a SSB, during which the bosonic field acquires a large value. For $\lambda\leq1$, the ground-state energy is $E_G^{(N)}=-\frac{\Oq}{2}$; for $\lambda\geq1$, it is equal to $E_G^{(S)}=-\frac{\Oq}{4}\frac{\lambda^4-1}{\lambda^2}$. Both $E_G$ and $\partial E_G/\partial\lambda$ are continuous on $\lambda=1$. However, $\partial^2 E_G/\partial\lambda^2$ is discontinuous at the critical point. Similarly, the field displacement $\alpha$ and the orientation of the spins are continuous at the transition, but their derivative is not. This indicates that a second-order phase transition is taking place, even in this finite-size system. 

 \subsection{Field fluctuations}
 To make the general picture above more precise, we will now study the quantum fluctuations of the field $\aop$. This will also allow us to compute the critical exponents of the transition.

  Let us first consider the normal phase $\gind\leq\gind_p$. The very large frequency difference between the spin and the field suggests that the behavior of the field can be studied by adiabatically eliminating the spin. This can be done in several manners, here we will use a Schrieffer-Wolff (SW) transformation \cite{boite_theoretical_2020}, which we have already briefly met in the previous Chapter. The general philosophy of the SW technique goes as follows: because of the very different energy scales, the eigenspaces of the fast degree of freedom (here the spin) are well-separated in energy, and only weakly coupled through the interaction with the slow degree of freedom. The idea then is to completely decouple the spin subspaces, then make a projection within the lowest spin subspace. This allows us to eliminate the spin and obtain an effective dynamic for the field only. In practice, this is achieved by applying to the Hamiltonian \eqref{Rabi} a unitary transformation $e^{i\Sop}$, with $\Sop$ an \textit{infinitesimal} hermitian operator such that $\Hop\sp{\prime}=e^{i\Sop} \Hop e^{-i\Sop}$ commutes with $\hat{\sigma}_z$, up to a certain perturbation order. A general approach to SW, which allows to derive the appropriate transformation for several models, can be found in Appendix A. For the Rabi model, the relevant operator is $\Sop=\sqrt{\frac{\Of}{\Oq}}\frac{\lambda}{2}\sigy(\adag+\aop)$. We obtain:

 \begin{equation}
 	\Uop \Hop \Uop^{\dagger}=\frac{\Oq}{2}\sigz+\Of\frac{\lambda^2}{4}\sigz(\aop+\adag)^2+\Of\adag\aop,
 \end{equation}
 plus some corrections of order $O(\Of\sqrt{\Of/\Oq})$. The Hamiltonian now commutes with $\sigz$. We can eliminate the spin by projection in the lowest spin subspace $\sigz\rightarrow -1$, and we obtain the effective field Hamiltonian:

 \begin{equation}
 	\Hop_f^{(2)}=\Of\adag\aop-\Of\frac{\lambda^2}{4}(\adag+\aop)^2=\Of\left(\frac{\pop^2}{4}+(1-\lambda^2)\frac{\xop^2}{4}\right),
 	\label{Hrabiquad}
 \end{equation}
 where we have defined the field quadratures $\xop=\adag+\aop$ and $\pop=i(\adag-\aop)$. This Hamiltonian describes the oscillations of a harmonic oscillator with a quadratic potential, and can be easily diagonalized through a Bogoliubov transformation. We define $\xop_c=(1-\lambda^2)^{1/4}\xop=\cop+\cdag$ and $\pop_c=\frac{1}{(1-\lambda^2)^{1/4}}\pop=i(\cdag-\cop)$, and we obtain $\Hop=\Of\sqrt{1-\lambda^2}\cdag\cop$. The ground state of the system is a vacuum squeezed state obeying:
 \begin{align}
 	\moy{\xop^2}& =(1-\lambda^2)^{-1/2},\\ \nonumber
 	\moy{\pop^2}&=(1-\lambda^2)^{1/2},\\ \nonumber
 	\moy{\xop}&=\moy{\pop}=0.
 \end{align}
  Excited states are squeezed Fock states with the same squeezing parameter. The gap of the system is given by $\Of\sqrt{1-\lambda^2}$. As we come closer to the critical point, the fluctuations of the field increase, while the gap closes.\\

 A similar treatment can be made in the second (superradiant) phase, for $\gind>\gind_p$. This time, the system has two possible configurations, and we want to study the fluctuations of the field around its two possible average values. This is done by first displacing the field with respect to its mean-field value $\aop\rightarrow\pm\alpha_g+\aop$, and rewriting the Hamiltonian in terms of the rotated spins defined in \eqref{Rotatedspin}. This gives \cite{hwang_dissipative_2018} : 

 \begin{equation}
 	\Hop_\pm=\Of\adag\aop \pm \sqrt{\Of\Oq}\sqrt{\frac{\lambda^4-1}{4\lambda^2}}(\adag+\aop)(1+\hat{\tau}_z^\pm) + \frac{\Oq\lambda^2}{2}\hat{\tau}_z^\pm + \frac{\sqrt{\Of\Oq}}{2\lambda}(\adag+\aop)\hat{\tau}_x^\pm+\Of\lvert\alpha\rvert^2.
 \end{equation}

 We can now perform a SW transformation using the unitary $\Uop=e^{i\sqrt{\frac{\Of}{\Oq}}\frac{1}{2\lambda^3}(\adag+\aop)\hat{\tau}_y^\pm-i\frac{\Of}{\Oq\lambda^5}\sqrt{\frac{\lambda^4-1}{4\lambda^2}}(\adag+\aop)\hat{\tau}_y^\pm}$, and obtain: $\Uop\Hop\Uop^{\dagger}=\Of\adag\aop+\frac{\Of}{4\lambda^4}(\adag+\aop)^2\hat{\tau}_z^\pm+\Of\lvert\alpha\rvert^2$ (plus higher-order corrections). After a projection in the $\hat{\tau}_z^\pm$ lowest eigenspace, we obtain:

 \begin{equation}
 	\Hop_{f, \text{sup}}^{(2)}=\Of\left[\frac{\pop^2}{4}+\left(1-\frac{1}{\lambda^4}\right)\frac{\xop^2}{4}\right],
 	\label{potsuperrad}
 \end{equation}
 and we obtain squeezing behavior similar to what we obtained in the normal phase, with increasing squeezing and closing gap near the critical point. The system is described by two ladders of displaced squeezed Fock states, centered around each of the two minima $\moy{\xop}=\pm\alpha_g$.\\

 We have studied the behavior of the system in each phase, away from the critical point. To complete the description, let us now consider what happens in its vicinity. As the fluctuations of the field become large near the critical point, higher-order terms need to be considered. These can be obtained by including more terms in the SW transformation. From now on, we will define the frequency ratio $\rat=\Oq/\Of$. By applying the operator $\Uop=e^{i\frac{\lambda}{2\sqrt{\rat}}\sigy(\aop+\adag)-\frac{\lambda^3}{6\eta\sqrt{\eta}}\sigy(\aop+\adag)^3}$, we obtain:

 \begin{equation}
  	\Uop \Hop \Uop^{\dagger}=\frac{\Oq}{2}\sigz+\Of\frac{\lambda^2}{4}\sigz(\aop+\adag)^2+\Of\adag\aop -\Of\frac{\lambda^4}{16\rat}\sigz(\aop+\adag)^4,
  	\label{order4Rabi}
 \end{equation}
 plus higher-order terms. Once again, we can project the Hamiltonian into the lowest spin subspace, and we obtain an effective field Hamiltonian with a quartic potential:

 \begin{equation}
 	\Hop_f^{(4)}=\Of\left(\frac{\pop^2}{4}+(1-\lambda^2)\frac{\xop^2}{4}+\frac{\lambda^4}{16\rat}\xop^4\right).
 	\label{potquartic}
 \end{equation}
 Near the critical point, the quartic term will stabilize the field, and prevent the divergence of fluctuations. 

 We can estimate when this term will start to play a role: far away from the critical point, the quartic term is negligible, and the state, as we said, is described by a squeezed state with $\moy{\xop^2}=\frac{1}{\sqrt{1-\lambda^2}}$. By Wick's theorem, we have $\moy{\xop^4}\propto\moy{\xop^2}^2\sim \frac{1}{(1-\lambda^2)}$. This gives us an estimate of the relative importance of the quadratic and quartic terms: $$\frac{(1-\lambda^2)\moy{\xop^2}}{\lambda^4\moy{\xop^4}/\rat}\sim\frac{(1-\lambda^2)^{3/2}\eta}{\lambda^4}\sim(1-\lambda)^{3/2}\eta,$$
 where we have used $(1-\lambda^2)=(1-\lambda)(1+\lambda)$ and $\lambda\sim1$ near the critical point (in the following we will systematically drop constant prefactors and focus on scaling laws). Hence, we can deduce that the quartic term will play a negligible role for $1-\lambda\gg\eta^{-2/3}$; in this region, the field becomes more and more squeezed as the coupling increases. When $1-\lambda$ becomes comparable to $\eta^{-2/3}$, the fluctuations of the field and the gap are given by, respectively:
 \begin{align}
 	\moy{\xop^2}&\sim\eta^{1/3} \label{mingap},\\  \nonumber
 	\Delta E_m=E_{1}-E_0&\sim \Of\eta^{-1/3}.
 \end{align}
  At this point, the quartic term starts to play a role, stabilizing the field and preventing its fluctuations to increase further. In this region, the Hamiltonian can not be exactly diagonalized. However, the system is still described by an oscillator with a symmetric confining potential. Therefore, we can expect that the field state will not be too different from a Gaussian centered on $0$. To study the fluctuations of the field, we make a squeezed state ansatz \cite{hwang_quantum_2015} by setting $\ket{\psi}=\ket{\xi_a}$ and by minimizing $\bra{\xi_a}\Hop\ket{\xi_a}=\frac{\omega}{4}(e^{-2\xi_a}+(1-\lambda^2)\frac{e^{2\xi_a}}{4}+\frac{\lambda^4}{4\eta}3e^{4\xi_a})$ with respect to $\xi_a$. Writing $y=e^{2\xi_a}=\bra{\xi_a}\xop^2\ket{\xi_a}$, we find: 
  $\partial_y\left(\bra{\xi_a}\Hop\ket{\xi_a})=\frac{\omega}{4}(-\frac{1}{y^2}+\frac{(1-\lambda^2)}{4}+\frac{3\lambda^4}{2\eta}y\right)$. The problem of finding the fluctuations of the bosonic field thus reduces to finding the roots of a cubic polynomial. A quick numerical analysis confirms that, over the interval $1-\eta^{-2/3}\leq\lambda^2\leq1$, the (real) root of the polynomial is given by $y=\eta^{2/3}f(\lambda)$, where $f$ varies slowly over the entire interval. Hence, this ansatz indicates that, as soon as $1-\lambda^2$ will become comparable to $\eta^{-2/3}$, the fluctuations (as well as the gap) will stabilize around the value $\langle\xop^2\rangle\sim\eta^{1/3}$ (respectively $\Delta E_m\sim\omega\eta^{-1/3}$), a value which will remain approximately constant all the way to the critical point.

    When the critical point is crossed, $\lambda\geq1$, the system is described by a double-well potential. The maximal height of the barrier is $V_0\sim\Of\rat(\lambda-1)^2$, and the two minima are separated by a distance $D\sim\sqrt{\rat(\lambda-1)}$.  Very close to the critical point, the double well has a very low depth, and the two minima are very close to each other. As a consequence, the system can freely hop between the two minima: the eigenstates are delocalized states, and the Hamiltonian is non-degenerate. The ground state and first eigenstate are symmetric and antisymmetric superposition of the two well positions, and are separated by the energy gap $\Delta E_m$, as given by \eqref{mingap}. As $\lambda$ increases, both $V_0$ and $D$ increase; tunneling is quickly suppressed, and the system becomes localized in either one of the two potential minima. The first two eigenstates become degenerate, and the symmetry is spontaneously broken. As the coupling increase, more and more states become localized in each well, and the spectrum corresponds to two ladders of squeezed displaced Fock states, described by \eqref{potsuperrad}.

 The point at which the Hamiltonian becomes degenerate and the symmetry is broken can be estimated with the following reasoning. Let us consider a massive particle in a two-well potential $V(x)$. Around each minimum, the particle oscillates at a frequency $\omega_0$. Let us consider a vibrational state in each well, at the same energy. Tunneling across the barrier hybridize the two states and opens a gap $\omega_0 e^{-\phi}$, with $\phi=\int \sqrt{2m(V(x)-\Delta E_m)}$, where the integral is taken over the barrier \cite{griffiths_introduction_2018}. Although the precise expression depends on the value of $\Delta E_m$, we can give an estimate: for a barrier of width $D$ and height $V_0$, we have $\phi\sim\frac{D^2}{x_f^2}\sim\frac{\Delta E_m}{V_0}$,
with $x_f=\frac{1}{\sqrt{m\omega_0}}$ the zero-point motion of the particle in each well. Hence, in our case, we expect that the tunneling rate will become small when the height of the barrier will become comparable with the excitation energy of the system. In our case, when the double-well structure appears at $\lambda=1$, we have a gap $\Delta E_m\sim\Of\rat^{-1/3}$. Since the barrier height increases like $V_0\sim\Of\rat(\lambda-1)^2$, we expect that tunneling will be suppressed for $V_0\sim \Delta E_m$, \textit{i.e.}, for $\lambda-1\sim\eta^{-2/3}$. Alternatively, we may also approach the critical point starting from the superradiant phase. In this case, Eq.\eqref{potsuperrad} predicts fluctuations $\moy{(\xop^2-\moy{\xop}^2)}\sim(1-\frac{1}{\lambda^4})^{-1/2}$. The distance between the two wells is $D=2\alpha_g\sim\sqrt{\rat(\lambda-1)}$. The gap will open when the fluctuations $\moy{(\xop^2-\moy{\xop}^2)}$ become comparable with $D^2$, which yields again $\lambda-1\sim \rat^{-2/3}$. Importantly, for $\lambda>1+\eta^{-2/3}$, we will have very quickly $\moy{(\xop^2-\moy{\xop}^2)}\ll\alpha_\gind$, which means that in the superradiant phase, the fluctuations of the field will be negligible with respect to its average value. Furthermore, the field fluctuations are also very small with respect to the frequency ratio $\rat$, and therefore they can only have a perturbative back-action on the spin. Outside the region $\lvert\lambda-1\rvert<\eta^{-2/3}$, only the average value of the field plays a role, which means that the mean-field treatment we made at the beginning is indeed justified in this region.\\

 To summarize, for $\lambda<1-\rat^{-2/3}$, the field is effectively described by a quadratic potential which softens as the coupling increases, reducing the energy gap and increasing fluctuations like $\moy{\xop^2}\sim(1-\lambda)^{-1/2}$. For $1-\rat^{-2/3}<\lambda<1+\rat^{-2/3}$, the quartic term plays a non-negligible role, and the potential assume a shallow double-well structure. The field fluctuations of the system are of order $\moy{\xop^2}\sim\rat^{1/3}$. Note that higher-order terms could be obtained by including even more terms in the SW analysis. We could expect that, as we get closer to the critical point, more and more terms need to be taken into account to properly describe the system. However, it turns out not to be the case: taking only the quadratic and quartic term into account is sufficient to describe the system all the way to the critical point. Indeed, because the quartic term prevents the fluctuations to diverge near the transition, higher-order terms will remain small even at the critical point. Formally, this can be checked in a self-consistent manner: higher-order terms are of the form $\Of\rat^{1-p}\xop^{2p}$. If we assume that $\moy{\xop^2}\sim\rat^{1/3}$, then we can estimate the ratio between the quartic and higher-order terms: $$\frac{\Of\rat^{1-p}\moy{\xop^{2p}}}{\Of\rat^{-1}\moy{\xop^4}}\sim\rat^{2-p}\moy{\xop^2}^{p-2}\sim\rat^{\frac{2(2-p)}{3}},$$
 which is very small for $p>2$. In other words, even in the region of maximum fluctuation, higher-order terms will always be much smaller than the quartic term and will act only perturbatively.

 Finally, for $\lambda>1+\rat^{-2/3}$, the double-well potential becomes deep enough to suppress tunneling, the model becomes degenerate and the symmetry is spontaneously broken. The field acquires a macroscopic population, with $\moy{\xop}=\alpha_g\propto\sqrt{\rat(\lambda-1)}$, and $\adag\aop\propto\rat$.  \\

 This analysis makes it possible to compute the critical exponents of the model. Three exponents are generally considered, \textbf{\textbf{\textbeta}}, \textgamma and \textbf{\textzeta} (we follow the notation convention of \cite{kirton_introduction_2019}), which describe the scaling of the field displacement and fluctuations with respect to the coupling and system size. They are defined respectively by: $\moy{\xop}\sim(\lambda-1)^\text{\textbf{\textbf{\textbeta}}}$ in the superradiant phase, $\moy{\xop^2-\moy{\xop}^2}\sim\lvert\lambda-1\rvert^{-\text{\textgamma}}$ away from the critical point, and at the critical point $\moy{\xop^2}_{\lambda=1}\sim\rat^{\text{\textbf{\textzeta}}}$. From the above considerations, we have \textbf{\textbf{\textbeta}}$=1/2$, \textgamma$=1/2$, and \textbf{\textzeta}$=1/3$. The scaling exponents are gathered in Table \ref{Tableexponent}. 

 To sum up, the quartic potential \eqref{potquartic} fully describes the critical behavior of the Rabi model; the predicted scaling exponents were shown to be in excellent agreement with exact numerical simulations \cite{hwang_quantum_2015}. This potential can also be obtained by a phenomenological argument \textit{à la} Landau; here, however, it was derived by a controlled approximation.

  \begin{figure}
  	\includegraphics[angle=-90,width=\linewidth]{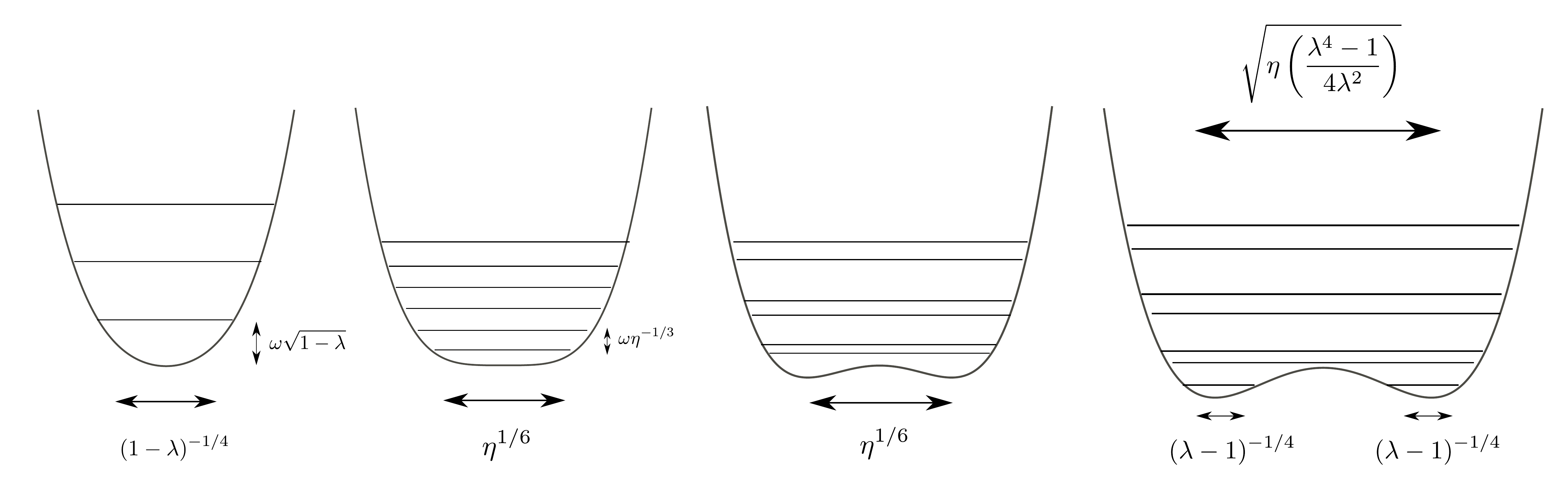}
  	\caption[Effective field potential in the Rabi model, showing the appearance of a double-well structure.]{Effective field potential for various coupling values. From left to right: $\lambda<1-\rat^{-2/3}$, $1-\rat^{-2/3}<\lambda<1$, $1<\lambda<1+\rat^{-2/3}$ and $\lambda>1+\rat^{-2/3}$. For $\lambda<1-\rat^{-2/3}$, the quadratic potential softens when $\lambda$ increases: the field fluctuations increases, while the gap decreases. In the region $\lvert\lambda-1\rvert<\rat^{-2/3}$, the dynamics is dominated by the quartic term. The potential becomes a shallow double-well potential. The eigenstates are grouped into doublets, and are all delocalized over the two wells. For $\lambda<1-\rat^{-2/3}$ , the separation between the two wells becomes larger than the ground-state fluctuations within each well (or, alternatively, the energy barrier becomes larger than the initial separation between the ground and first excited states). Tunneling is suppressed, and the system now admits two degenerate ground states, localized in the two wells.}
  \end{figure}

 \begin{table}
 \begin{center}
 \begin{tabular}{|c c | c c c|}
 	\hline \textbf{Exponent} & \textbf{Definition} & \textbf{QPT} & \textbf{CPT} & \textbf{NESS}\\
 	\hline
 	\textbf{\textbf{\textbeta}} & $\moy{\xop}\sim(\lambda-1)^\text{\textbf{\textbf{\textbeta}}}$ & $1/2$ & $1/2$ & $1/2$  \\
 	\hline \textgamma & $\moy{\xop^2-\moy{\xop}^2}\sim\lvert\lambda-1\rvert^{-\text{\textgamma}}$ & $1/2$ & $1$ & $1$  \\
 	\hline \textbf{\textzeta} & $\moy{\xop^2}_{\lambda=1}\sim\rat^{\text{\textbf{\textzeta}}}$ & $1/3$ & $1/2$ & $1/2$\\
 	\hline \hline$W$ & $\frac{(1-\lambda)\moy{\xop^2}}{\moy{\xop^4}/\rat} \lvert_{\lambda=1-W}\sim1$ & $\rat^{-2/3}$ & $\sqrt{\frac{k_B T}{\omega}}\rat^{-1/2}$ & \\
 	\hline
 \end{tabular}
 \end{center}
 	\caption[Critical exponents of the Rabi, Dicke, and Lipkin-Meshkov-Glick models.]{Critical exponents of the Rabi, Dicke, and Lipkin-Meshkov-Glick models, with a transition point at $\lambda=1$, for several regimes of temperature. $W$ is the width of the "critical zone", \textit{i.e.}, the range of coupling values over which the quartic term plays a significant role.
 	QPT: quantum phase transition, $k_BT\ll\Of\eta^{-1/3}$. CPT: classical phase transition, $k_BT\gg\Of\eta^{-1/3}$. NESS: non-equilibrium steady-state. Partially adapted from \cite{kirton_introduction_2019}.}
 	\label{Tableexponent}
 \end{table}

 \subsection{Finite temperature}

 So far we have focused on the zero-temperature case, which is dominated by the ground-state behavior of the system. Let us now discuss what happens at finite temperature. We assume that the system is in thermal equilibrium at temperature $T$, \textit{i.e.}, the system is in a Gibbs state

 \begin{equation}
 	\rop=\frac{1}{Z}e^{-\frac{\Hop}{k_B T}},
 \end{equation}
 with $Z=Tr[e^{-\frac{\Hop}{k_B T}}]$ the partition function of the system, and $k_B$ the Boltzmann constant. As previously, we will start with a mean-field treatment. We project the field into a coherent state, and obtain the mean-field partition function:

 \begin{align}
 	Z(\alpha)=Tr[e^{-\frac{\Hop_{eff}(\alpha)}{k_BT}}]=2e^{-\frac{\Of}{k_B T}\lvert\alpha\rvert^2}\cosh\left(\frac{\Oq}{2k_B T}\sqrt{1+\frac{4\gind^2}{\Oq^2}(\alpha+\alpha^*)^2}\right),
 \end{align}
where $\Hop_{\text{eff}}$ has been defined in Eq.\eqref{EffspinRH}.
 We can then define the free energy $F(\alpha)=-k_B T \ln(Z(\alpha))$, and minimize it with respect to $\alpha$.   For $\gind\leq\gind_p(T)=\sqrt{\frac{\Of\Oq}{4}\coth(\frac{\Oq}{2k_BT})}$, we find again a single minimum located in $\alpha=0$. For $\gind\geq\gind_p(T)$, the free energy has two minima, centered around two values $\pm\alpha_g(T)\neq0$. Hence, the mean-field analysis predicts that a phase transition will take place for the value $\gind_p(T)$. Note that for $T=0$, we find again $\gind_p=\sqrt{\frac{\Of\Oq}{4}}$. Near the critical point, $\alpha_g(T)$ evolves like $(\gind-\gind_p(T))^{1/2}$ \cite{kirton_introduction_2019}. Hence, at finite temperature, the scaling coefficient \textbf{\textbf{\textbeta}} is still equal to $1/2$. \\

 To study the fluctuations of the field, we take again the quadratic potential \eqref{Hrabiquad}, but this time with a thermal occupation. The eigenstates are squeezed thermal states, and we have:

 \begin{equation}
 	\moy{\xop^2}=\frac{1}{\sqrt{1-\lambda}}\coth\left(\frac{\Of\sqrt{1-\lambda}}{k_BT}\right),
 	\label{fluctuationT}
 \end{equation}
 and we have redefined $\lambda=\gind/\gind_p(T)$. In the limit $T\rightarrow0$, we find again $\moy{\xop^2}\sim(1-\lambda)^{1/2}$. By contrast, when $k_BT\gg\Of\sqrt{1-\lambda}$ (\textit{i.e.}, when the thermal energy is much larger than the gap), we obtain: $$\moy{\xop^2}\sim\frac{k_BT}{\Of(1-\lambda)},$$ which corresponds to a scaling coefficient \textgamma$=1$.\\

 To understand the behavior of the system at the critical point, we can once again study when the quartic term starts to play a role in the potential. Using \eqref{fluctuationT}, we find $\frac{\moy{\xop^4}}{\rat\moy{\xop^2}(1-\lambda)}\sim\frac{k_BT}{\Of\rat(1-\lambda)^2}$. Hence, the quartic term becomes comparable with the quadratic one when $1-\lambda \sim \sqrt{\frac{k_BT}{\Of\rat}}$. When this occurs, the fluctuations of the field are $\moy{\xop^2}\sim\sqrt{\frac{k_BT}{\Of}}\rat^{1/2}$. Assuming again that the fluctuations remain approximately constant as soon as the quartic term dominates, we infer that \textbf{\textzeta}$=1/2$. Alternatively, this scaling can also be obtained using the equipartition theorem. \\

 Hence, the case of zero and finite temperature correspond to very different behaviors, with different critical exponents. More precisely, when the thermal energy $k_B T$ is smaller than the gap, then the system is effectively at zero temperature and its behavior is dominated by its ground-state properties. The phase transition in this case is generally referred to as a Quantum Phase Transition or QPT. By contrast, when $k_B T$ is larger than the gap, the thermal occupation of excited states become dominant. In this case, the term Classical Phase Transition, or CPT, is often used in the literature. 

 Note that if $\rat$ is truly infinite, the gap exactly closes at the critical point, meaning that the system will always be thermally excited in its vicinity. However, no system can exhibit a truly infinite frequency ratio (or a truly infinite size). In practice, we can hope to observe critical and symmetry-breaking effects by implementing systems with large but finite $\rat$. In that case, two situations need to be distinguished:

 \begin{itemize}
 	\item If $k_BT\geq\Of\rat^{-1/3}$, the thermal energy is larger than the minimal gap predicted by the zero-temperature treatment. As a consequence, the system will be thermally excited near the transition.
 	\item If $k_BT\leq\Of\rat^{-1/3}$, the thermal energy is always smaller than the gap, even at the critical point. The behavior near the critical point is dominated by quantum fluctuations. 
 \end{itemize}

 \begin{figure}
 	\includegraphics[angle=-90,width=\linewidth]{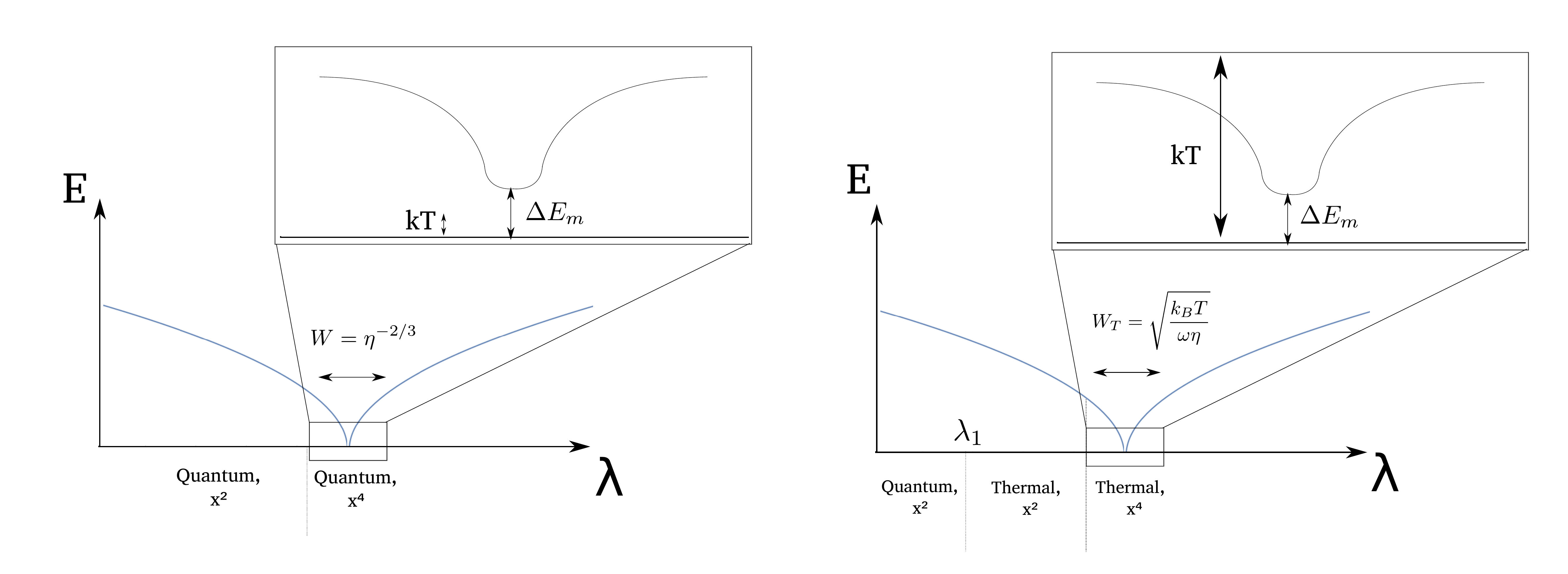}
 	\caption[Schematic behavior of the excitation energy in the Rabi model, showing the interplay between quantum and thermal excitations.]{Schematic behavior of the excitation energy versus coupling (here we focus on the energy gap between the ground and \textit{second} excited state). Left: $k_BT\leq \Delta E_m=\Of\rat^{-1/3}$. For all coupling value, the thermal energy remains small compared to the gap. The behavior of the system is split into two regions, dominated by the quadratic and the quartic terms respectively. Right: $\Of\geq k_BT\geq \Of\rat^{-1/3}$. For small couplings, the gap remains large with respect to thermal energy. For $\lambda>\lambda_1=1-\left(\frac{k_BT}{\Of}\right)^2$ , the system becomes thermally excited. The behavior is dominated by the quadratic term for $\left(\frac{k_BT}{\Of}\right)^2\leq1-\lambda\leq\sqrt{\frac{k_BT}{\Of\rat}}$, and by the quartic term for $1-\lambda\geq\sqrt{\frac{k_BT}{\Of\rat}}$. For $k_BT\geq\Of$, the system is thermally excited for all $\lambda$ (not shown).}
 \end{figure}

 Hence, very different behavior can be expected, depending on the ratio $\frac{k_BT}{\Of\eta^{-1/3}}$. In the following, the term CPT will more specifically refer to the former case, and the term QPT to the latter case.

 \subsection{Quantum fluctuations and nature of the transition}

To summarize, we have shown that the Rabi model can experience a transition which possesses many properties associated with a QPT. When the coupling constant is increased, the system experiences a transition towards a phase in which the $\mathbb{Z}_2$ symmetry of the Hamiltonian is spontaneously broken. In the limit of large frequency ratio $\Oq/\Of$, this evolution becomes a sharp transition, which occurs for a definite coupling value $\gind_p$. At this point, the order parameters (here, the field displacement and the qubit rotation) remain continuous, but their derivative is not, indicating that the transition is second-order. Even in this finite-size system, we can still define a thermodynamic limit, in which the large frequency ratio plays the role of large system size. The mechanism driving this transition is a competition between the bare qubit and field energy terms, $\adag\aop$ and $\Jz$, and the interaction term $(\adag+\aop)\Jx$, both terms favoring different ground states. This mechanism involves only the Hamiltonian and can take place even at zero temperature. Finally, near the critical point, the Rabi model shows large fluctuations, to which critical exponents are associated.\\

 However, some remarks are in order concerning the quantum nature of the transition. As we have shown, the fluctuations of the field can be very large, especially near the critical point. However, in the superradiant phase, the field fluctuations are negligible with respect to the average displacement $\alpha_g$. Hence, the behavior of the system is dominated by its mean-field properties, which justifies that the mean-field treatment that we have used yields the correct critical exponent \textbf{\textbf{\textbeta}}$=1/2$. The transition is said to be of the \textit{mean-field} type. Only in the region $\lvert\lambda-1\rvert<\eta^{-2/3}$ are the quantum fluctuations dominant, and this region vanishes for very large $\rat$. By contrast, in spin models with short-range interaction, like the Ising model, fluctuations are generally large for every system size, and the critical exponents can not be computed with mean-field analysis. As a consequence, there is some debate about whether the transition of the Rabi (and Dicke) model should be considered a "true" quantum phase transition \cite{larson_remarks_2017}.

 As we argued, it is important in this case to keep in mind the order of magnitude of the temperature and frequencies of the system. For large but finite $\rat$ and very small $T$, the system can still show (approximate) symmetry breaking, and does exhibit vacuum fluctuations and nonclassical field statistics, which dominate the dynamics near the critical point. It is in this restricted sense that we will use the terms \textit{Quantum Phase Transition} in the remainder of this Thesis.

 \section{Phase transition in the Dicke and Lipkin-Meshkov-Glick models}
 \label{Sec2.3}

 The study of the Rabi model has shown how a SSB and a large population of the bosonic field could take place at zero temperature, even in a finite-size system. We will now discuss how similar phenomena can take place in other models, including, at last, the Dicke model.

 \subsection{Dicke model}
 We start from Hamiltonian \eqref{Dicke}. This time, we will let the qubit number $N$ go to infinity, while the frequency ratio $\Oq/\Of$ will be kept finite. We perform HP transformation (see the previous Chapter) and express the qubit excitations in terms of a bosonic field $\bop$. We have a series expansion $\Jp=\bdag\sqrt{N-\bdag\bop}=\sqrt{N}\bdag(1-\frac{\bdag\bop}{2N}+...)$. We keep only terms of order $1$ and $2$ in $\bop$, which yields a Hamiltonian $\Hop=\Of\adag\aop+\Oq\bdag\bop+\gcoll(\bdag+\bop)(\adag+\aop)-\Oq\frac{N}{2}$. Again, the behavior of the superradiant phase can be captured by a mean-field treatment. The only difference with the Rabi model is that the qubit and field excitations now play a similar role. We project the $\aop$ and $\bop$ fields in coherent states $\ket{\alpha}$ and $\ket{\beta}$, respectively: $\bra{\alpha,\beta}\Hop\ket{\alpha,\beta}=\Of\lvert\alpha\rvert^2+\Oq\lvert\beta\rvert^2+\gcoll(\alpha+\alpha^*)(\beta+\beta^*)$. Minimizing this classical potential with respect to both $\alpha$ and $\beta$, we find:
 \begin{align}
 	\beta&=\pm\beta_g=\sqrt{\frac{N}{2}\left(1-\frac{1}{\lambda^2}\right)},\\ \nonumber
 	\alpha&=\beta \frac{2\gcoll}{\Of}\sqrt{1+\frac{1}{\lambda^2}}
 \end{align}
 for $\lambda>1$, and $\alpha=\beta=0$ for $\lambda<1$, where we still have: $$\lambda=\gcoll/\gcoll_p=\frac{2\gcoll}{\sqrt{\Of\Oq}}.$$ Hence, the critical coupling is the same in the Dicke and Rabi model. In the normal phase, both fields are centered around $0$; in the superradiant phase, they both acquire macroscopic and correlated values.\\

 As of the fluctuations, they can be obtained by considering hybrid polaritonic excitations. Following \cite{emary_chaos_2003}, we define:

 \begin{align}
 	\qop_1=\xop_a\cos(\theta)-\xop_b\sin(\theta)&, \hspace{5pt} \qop_2=\xop_b\cos(\theta)+\xop_a\sin(\theta),\\ \nonumber
 	\tan(2\theta)&=\frac{4\gcoll\sqrt{\Of\Oq}}{\Oq^2-\Of^2},
 \end{align}
 with $\xop_a=\aop+\adag$ and $\xop_b=\bop+\bdag$. We define the associated quadratures $\pop_1$ and $\pop_2$ in the same way.
 This yields the following quadratic Hamiltonian:

 \begin{align}
 	\Hop_{\text{Dicke}}^{(2)}=\frac{1}{4}\left[{E_-^{(N)}}^2\qop_1^2+\pop_1^2+E_+^{(N)}\qop_2^2+\pop_2^2\right]
 	\label{Dickequadratic}
 \end{align}
 (up to additional constant terms), with:

 \begin{equation}
 	{E_\pm^{(N)}}^2=\frac{1}{2}\left[\Of^2+\Oq^2\pm\sqrt{(\Oq^2-\Of^2)^2+16\gcoll^2\Of\Oq}\right]=\frac{\Of^2+\Oq^2}{2}\left[1\pm\sqrt{1+\frac{(2\Of\Oq)^2}{(\Of^2+\Oq^2)^2}(\lambda-1)}\right].
 	\label{gapDicke}
 \end{equation}

 The Hamiltonian \eqref{Dickequadratic} can be diagonalized once again by a Bogoliubov transformation. The ground state is a squeezed vacuum state for both fields $\qop_1$ and $\qop_2$, and satisfies:

 \begin{align}
 	\moy{\qop_1^2}=\frac{1}{E_-^{(N)}}, \hspace{5pt} \moy{\qop_2^2}=\frac{1}{E_+^{(N)}}.
 	\label{fluctDicke}
 \end{align}
 The gap is given by $E_-^{(N)}$; from \eqref{gapDicke} we can see that it goes to zero as $\gcoll$ goes to $\gcoll_p$. Near the critical point, we have $E_-^{(N)}\sim\sqrt{\frac{(\Of\Oq)^2}{(\Of^2+\Oq^2)}}\sqrt{(1-\lambda)}$, and $\moy{\qop_1^2}\sim(1-\lambda)^{-1/2}$. From this one can infer that both $\moy{\xop_a^2}$ and $\moy{\xop_b^2}$ will scale like $(1-\lambda)^{-1/2}$, which corresponds to a critical exponent \textgamma$=1/2$, identical to what we had in the Rabi model.

 The superradiant phase can be studied by first displacing the two fields by their average mean-field value: $\aop\rightarrow\alpha+\aop$ and $\bop\rightarrow\beta+\bop$. Then a similar treatment yield again field fluctuations scaling like $(\lambda-1)^{-1/2}$, and a gap $E_-^{(S)}$ scaling like $\lambda-1$ (the complete expressions can be found in \cite{emary_chaos_2003}).\\

 Close to the critical point, the proper treatment of fluctuations requires including more terms in the HP series expansion. The first term in the expansion will be of the form $\frac{(\adag+\aop)(\bop^{\dagger2}\bop+\bdag\bop^2)}{N}$. When expressed in terms of the transformed operators $\qop_{1,2}$, this yields a non-trivial combination of quartic and quadratic terms: $\frac{\qop_1^4}{N}$, $\frac{\qop_2^4}{N}$, $\frac{\qop_1^2\qop_2^2}{N}$ and so on. However, this term can be simplified by the following argument: near the critical point, the fluctuations of $\qop_1$ are large, while those of $\qop_2$ remain finite, according to \eqref{fluctDicke}. Therefore, the dominant term in the correction will be the term quartic in $\qop_1$. In general, since the fluctuations of the field $\qop_1$ are vastly superior to the one of $\qop_2$, the fluctuations of the system can be obtained by discarding the latter field and focusing on the former. With this reasoning, we arrive at the following Hamiltonian: 

 \begin{equation}
  	\Hop_{\text{Dicke}}^{(4)}\sim \frac{1}{4}\left[\pop_1^2+E_-^{(N)}\qop_1^2+a\frac{\qop_1^4}{N}\right],
  \end{equation} 
  with $a$ a positive term of order $1$, whose precise expression is irrelevant for this discussion. Hence, we have obtained an expression very similar to the quartic potential of the Rabi Hamiltonian \eqref{potquartic}, where the large parameter is the number $N$ instead of the frequency ratio. We can thus make the same reasoning: the quartic term will start playing a role for $1-\lambda\sim N^{-2/3}$. In this region, we will have a gap $\propto N^{-1/3}$, and fluctuations $\moy{\qop_1^2}\sim N^{1/3}$. By contrast, the fluctuations of $\qop_2$ remain finite. The fluctuations of the physical fields $\xop_a$ and $\xop_b$ are a combination of $\moy{\qop_1^2}$ and $\moy{\qop_2^2}$; however, since $\moy{\qop_1^2}\ll\moy{\qop_2^2}$, and for $\Oq\sim\Of$, we will have $\moy{\xop_a^2}\sim\moy{\xop_b^2}\sim\moy{\qop_1^2}$, hence $\moy{\xop_a^2}\sim\moy{\xop_b^2}\sim N^{1/3}$. Note that here, \textit{both} fields acquire large fluctuations in the critical region. This corresponds to a critical exponent \textbf{\textzeta}$=1/3$, again identical to what we had in the Rabi model. Furthermore, in this case as well, higher-order terms will remain small compared to the quartic one, and act only as perturbations. Finally, note that in the superradiant phase, the fluctuations of both spin and field will be small compared with their average values $\alpha$ and $\beta$, which again justifies that mean-field will yield correct results in this regime.

 Hence, the universal properties of the superradiant transition are the same in the Rabi and Dicke models. There are, however, a few noteworthy differences. In the Rabi model, there is a strong asymmetry between the qubit and the field. Even near the critical point, both spin and field retain their individual identities. While the fluctuations of the field are very large, those of the spin remain small, because of the vast frequency difference.  By contrast, in the Dicke model, both components play a similar role; near the critical point, the two systems become highly hybridized. This can be seen by studying the qubit-field entanglement. In the Rabi model, the entanglement is always finite, due to the finite size of the qubit and its small fluctuations. By contrast, in the Dicke model, the entanglement diverges near the critical point \cite{bakemeier_quantum_2012}.

 \subsection{Lipkin-Meshkov-Glick model}

 So far we have studied the behavior of the Dicke model in the limits $\Oq/\Of\rightarrow\infty, N=1$, and $N\rightarrow\infty, \Oq/\Of\sim1$. There is one last regime of interest from the point of view of phase transition, the limit $N\rightarrow\infty$ and $\Oq/\Of\rightarrow0$. This regime can be seen as the counterpart of the Rabi model. A large number of qubits means that they will act collectively as a bosonic field. By contrast, since the field frequency is very large with respect to the spin frequency, only the first two states of the field $\aop$ will play a role. Therefore, in this regime, the qubits and the field will effectively switch their behavior.   
 We can express this by adapting the SW transformation we used to study the Rabi model. We apply the unitary operation $\Uop=e^{\sqrt{\frac{\Oq}{\Of}}(\adag-\aop)\Jx}$, and obtain a Hamiltonian:
 \begin{equation}
 	\Uop\Hop\Uop^\dagger=\Of\adag\aop-\Oq\frac{\lambda^2}{4}\Jx^2+\Oq\Jz.
 \end{equation}
 The spin part of this Hamiltonian is known as the Lipkin-Meshkov-Glick model. It describes a collection of spins with infinite-range interaction. As the coupling $\lambda$ increases, the ensemble of spin becomes squeezed. For a finite number $N$, squeezing remains finite. However, when $N$ goes to infinity, the spin fluctuations can diverge, and the system experiences a phase transition. These fluctuations can be treated with a HP transformation, which yields again the same critical exponents as the Rabi and Dicke models. We will encounter again this limit of large qubit number and $\Of/\Oq\ll1$ in Chapter 4.

 \subsection{General features}

 As the previous discussion has illustrated, the superradiant transition is a rather general feature of systems involving coupled qubits and bosonic fields. In all cases, the transition could be understood as the consequence of a quartic potential involving one or two bosonic fields. The nonlinearity of the qubits was only needed to introduce a quartic field term; higher-order terms acted only perturbatively. As a consequence, it can be expected that any model involving two bosonic fields with a small nonlinear term could achieve the same results. This was highlighted in \cite{felicetti_universal_2020}; in this work, Hamiltonian of the form 
 \begin{equation}
 	\Hop=\omega_1\adag\aop+\omega_2\bdag\bop+\gcoll(\aop+\adag)(\bop+\bdag)+\epsilon_1\adag\adag\aop\aop+\epsilon_2\bdag\bdag\bop\bop,
 \end{equation}
 were $\epsilon_{1,2}$ are very small parameters, were considered. It was shown that these models could generally exhibit a phase transition during which one or both of the fields acquire a large value. Dicke and Rabi transitions are specific cases of this general transition. Dicke-like transitions correspond to the case where both fields acquire a macroscopic value. This corresponds to Hamiltonian of the form:

 \begin{equation}
 	\Hop=\omega_1\adag\aop+\omega_2\bdag\bop+\gcoll(\aop+\adag)(\bop+\bdag)+\frac{\epsilon_1}{N}\adag\adag\aop\aop+\frac{\epsilon_2}{N}\bdag\bdag\bop\bop,
 \end{equation}
 with $N$ some large constant. By contrast, Rabi-like transition corresponds to one field acquiring a macroscopic value, but not the other. This transition can be obtained with Hamiltonian of the form:

  \begin{equation}
 	\Hop=\omega_1\adag\aop+N\omega_2\bdag\bop+\sqrt{N}\gind(\aop+\adag)(\bop+\bdag)+\frac{\epsilon_1}{N}\adag\adag\aop\aop+\frac{\epsilon_2}{N}\bdag\bdag\bop\bop,
 \end{equation}
 with $\omega_1\sim\omega_2\sim\gind$. This corresponds indeed to the situation of the Rabi model, in which one of the field has a very large frequency with respect to the other, the coupling $\gind\sqrt{N}$ being intermediate between these two frequencies.\\

 With the insight provided by all the previous discussion, we can now propose the following idea: superradiant transition is a generic property of systems involving bosonic fields. The general recipe for such a transition to take place involves three basic elements. The first one is the presence of (at least one) bosonic field, which becomes populated at the phase transition. A bosonic field is required to achieve an unbounded spectrum, which is necessary to define a proper thermodynamic limit in which the population of the field becomes macroscopic. The second ingredient is a non-harmonic spectrum. To clarify this point, let us consider a Hamiltonian composed only on linear and quadratic field terms, for which the spectrum is harmonic. Then we can express the dynamics in terms of a potential quadratic in both quadratures $\xop$ and $\pop$: $\Hop= a_x\xop + a_p\pop + b_x\xop^2+ b_p\pop^2$ (plus possible cross-terms $\xop\pop$, which can be removed by a Bogoliubov transformation). Then only two cases are possible: either the quadratic terms $b_{x,p}$ are positive, in which case the ground state can only evolve smoothly; or one of the terms $b_{x,p}$ becomes negative, in which case the system becomes unstable. In neither case, we obtain a phase transition. For this purpose, it is necessary to have higher-order terms, which will stabilize the dynamics. In all the examples we have seen so far, the transition was caused by the interplay between the quadratic and quartic terms $\xop^2$ and $\xop^4$. Hence, higher-order moments (and therefore, some level of non-harmonicity in the spectrum) is a requirement to achieve a phase transition. Interestingly, the discussion in \cite{felicetti_universal_2020} has shown that even an infinitesimally small amount of non-harmonicity is sufficient to achieve the desired effect.
 Finally, the third ingredient is the presence of competing terms, of comparable strength, which drive the transition. In the example we considered, the competing terms are the interaction and the bare energy terms. The requirement of comparable strength in this competing term corresponds precisely to the USC regime. In dissipative systems \cite{bartolo_exact_2016}, this competition can be provided by drive and dissipation instead.

 \section{Non-equilibrium transition}
 \label{Sec2.4}

 \subsection{Driven-dissipative quantum dynamics}
 All the discussion so far took place in the frame of \textit{equilibrium} transition, when the system is in thermal equilibrium with its environment and can be cast in a Gibbs state (at zero or finite temperature). This is well suited to describe genuine, time-independent quantum systems, which naturally relax to their ground state when coupled to a large environment. However, as discuss in the last section of this Chapter, most implementations of the Dicke model rely on simulating, driven systems. This fact has profound consequences on the dissipation, and completely change the picture described in the previous Chapter. 

 To illustrate this, let us consider again the three-mode mixing process studied in \cite{fedortchenko_quantum_2017}. Two bosonic fields $\aop$ and $\bop$ are mixed by a weak pump of intensity $p$: 
 $\Hop=\Of\adag\aop+\Oq\bdag\bop+\chi(p+p^*)(\adag+\aop)(\bdag+\bop)$, with $\chi p\ll\Of,\Oq$. Since the mixing is very small, the ground state is just the vacuum for both $\aop$ and $\bop$. Hence, interaction with an environment at zero temperature will lead to a loss of excitations, according to a Lindblad equation $\frac{d\rop}{dt}=-i[\Hop,\rop]+L[\aop](\rop)+L[\bop](\rop)$.

 Now let us consider what happens when the pump $p$ is modulated in time like $p=p_R e^{i\omega_R t}+p_B e^{i\omega_B t}$, with still $\chi\hspace{4pt} p_{R,B}\ll\Of,\Oq$. Then the system evolves according to a Lindblad equation with a time-dependent Hamiltonian:

 \begin{equation}
 	\frac{d\rop}{dt}=-i[\Hop(t),\rop]+\kappa_aL[\aop](\rop)+\kappa_bL[\bop](\rop).
 \end{equation}

 As we discussed in the last Chapter, the time-dependence of $\Hop$ can be removed by considering a well-chosen rotating frame. This involves applying an operation $\Uop=e^{i(A\adag\aop+B\bdag\bop)t}$, and we have $\Uop\Hop\Uop^\dagger=\Hop_{eff}=\Of_{eff}\adag\aop + \Oq_{eff}\bdag\bop + G_R(\adag\bop+\bdag\aop) + G_B (\adag\bdag+\aop\bop)$. Importantly, the jump terms are not modified when the transformation is applied: $L[\Uop\aop\Uop^\dagger]=L[\aop]$ and $L[\Uop\bop\Uop^\dagger]=L[\bop]$. Hence, in the rotating frame, the system evolves according to the following, time-independent equation:

 \begin{equation}
 	\frac{d\rop}{dt}=-i[\Hop_{eff},\rop]+\kappa_aL[\aop](\rop)+\kappa_bL[\bop](\rop).
 \end{equation}
 However, since the counter-rotating terms play an important role in $\Hop_{eff}$, the Lindblad equation does $\textit{not}$ drive the system towards the ground-state of $\Hop_{eff}$. Actually, the terms $L[\aop(\bop)]$ describe energy being pumped into the system. Hence, we can have two different perspectives: in the frame of the laboratory, we have a system driven in time, which irreversibly emits excitations into the environment. But in a frame rotating with the drive, we have a time-independent system, which can both gain and lose energy while interacting with its environment. This does not contradict energy conservation because physically, any energy gained in this rotating frame can be attributed to the time-dependent drive.\footnote{This can be obtained more rigorously by re-deriving the interaction between system and field. One generally model the bath by a collection of harmonic oscillators with \textit{positive} frequencies. Moving to the rotating frame effectively displace the entire frequency distribution of the oscillators. As a consequence, components with a \textit{negative} frequency appear, which describe the pumping of energy into the system.}
 In general, engineered, effective dynamics rely on driven systems. When interacting with the environment, these systems follow a non-equilibrium dynamics, which does not bring the engineered Hamiltonian to its ground state. This dynamic can often be expressed in terms of undressed jump operators. In the case of the Rabi and Dicke model, three dissipation channels are often considered: photon loss, spin decay, and spin dephasing. This leads to the following Lindblad equation:

 \begin{equation}
 	\frac{d\rop}{dt}=-i[\Hop,\rop]+\sum_i \Gamd^i L[\sigm_i](\rop)+\sum_i \Gamp^i L[\sigz_i](\rop)+\kappa L[\aop](\rop).
 	\label{LindbladDicke}
 \end{equation}

 In the remainder of this section, we will study the impact of these different dissipation channels.

 \subsection{Case without spin dissipation, $\Gamd=\Gamp=0$}

 First, we will focus on the behavior of the system when only photon loss is present. We discuss here results for the Rabi model, the Dicke model can be treated in a very similar way.
 In general, solving the full Lindblad equation, even with no spin dissipation, can be a challenging task. However, much information can already be gathered by studying the evolution of a few observables. Following \cite{hwang_dissipative_2018}, let us study the evolution of the first-order moments $\moy{\aop}$, $\sigz$, and $\moy{\sigp}$. We obtain:

 \begin{align}
 	\frac{d\moy{\aop}}{dt}=(-i\Of-\kappa)\moy{\aop}-i\gind(\moy{\sigp}+\moy{\sigm}),\\\nonumber
 	\frac{d\moy{\sigp}}{dt}=i\Oq\moy{\sigp}-i\gind\moy{(\aop+\adag)\sigz},\\\nonumber
 	\frac{d\moy{\sigz}}{dt}=-2i\gind\moy{(\aop+\adag)(\sigp-\sigm)}.
 \end{align}

 This system of equations is not closed: indeed, it involves both first-order moments and second-order ones, like $\moy{\aop\sigp}$. In general, the dynamics of all moments will involve an infinite hierarchy of coupled equations. To truncate this hierarchy, we can once more resort to mean-field approximation: if we have $\aop=\moy{\aop}+\delta\aop$ and $\sigp=\moy{\sigp}+\delta\sigp$, then $\moy{\aop\sigp}=\moy{\aop}\moy{\sigp}+\moy{\delta\aop\delta\sigp}$. If we neglect the fluctuation correlations, we arrive at the following, closed, system of equations:

 \begin{align}
 	\frac{d\moy{\aop}}{dt}=(-i\Of-\kappa)\moy{\aop}-i\gind(\moy{\sigp}+\moy{\sigm}),\\\nonumber
 	\frac{d\moy{\sigp}}{dt}=i\Oq\moy{\sigp}-i\gind(\moy{\aop}+\moy{\adag})\moy{\sigz},\\\nonumber
 	\frac{d\moy{\sigz}}{dt}=-2i\gind(\moy{\aop}+\moy{\adag})(\moy{\sigp}-\moy{\sigm}).
 \end{align}

 It is now straightforward to find the steady-state of this equation. We find three possible solutions: $\moy{\aop}=0$ and $\moy{\aop}=\pm\alpha^D_g=\pm\sqrt{\rat}\frac{\gind/\sqrt{\Oq\Of}}{1-i(\kappa/\Of)}\sqrt{1-(\frac{\gind^D_t}{\gind})^4}$ and $\moy{\sigp}=\pm s_+=\mp\frac{1}{2}\sqrt{1-(\frac{\gind^D_t}{\gind})^4}$, with:

  $$\gind^D_t=\sqrt{\frac{\Of\Oq}{4}}\sqrt{1+\frac{\kappa^2}{\Of^2}}.$$

 Hence, the dissipative  Rabi (and Dicke) model has two possible steady-states in which the field acquires a mean-field value proportional to $\sqrt{\rat}$. The cases $\moy{\aop}=0$ and $\moy{\aop}=\pm\alpha^D_g$ are the driven-dissipative versions of the normal and superradiant phases.\\

  To study the onset of phase transition, we also need to consider the stability of these two phases. This is done by displacing each quantity around its steady-state value by a small amount: $\moy{\aop}=\pm\alpha^D_g+\delta\moy{\aop}$ or $\moy{\aop}=0+\delta\moy{\aop}$. The system of equations can then be cast in a linear form:

 $$(\delta\moy{\aop},\delta\moy{\adag},\delta\moy{\sigp},\delta\moy{\sigm},\delta\moy{\sigz})^T=M(\delta\moy{\aop},\delta\moy{\adag},\delta\moy{\sigp},\delta\moy{\sigm},\delta\moy{\sigz})^T,$$

 with $M$ a $5\times5$ matrix. A phase is stable against small perturbation if and only if the corresponding $M$ has only negative eigenvalues.

 Although these eigenvalues can easily be obtained numerically, it is difficult to obtain analytical expressions. Alternatively, if $\kappa\sim\Of$ and $\rat\rightarrow\infty$, the problem can be considerably simplified by applying first a SW transformation $\Uop=e^{i\frac{\gind}{\Oq}\sigy(\adag+\aop)}=e^{i\sqrt{\frac{\Of}{\Oq}}\frac{\lambda}{2}\sigy(\adag+\aop)}$ to the Lindblad equation. The Hamiltonian part will now be described by \eqref{Dickequadratic}. The transformation will also act on the jump operators, in a non-trivial way; however, if $\kappa\sim\Of$, this will only lead to higher-order corrections. Hence, we obtain a dynamics described by the following Lindblad equation: $$\frac{d\rop}{dt}=-i[\Of\adag\aop+\frac{\gind^2}{\Oq}(\adag+\aop)^2,\rop]+\kappa L[\aop](\rop).$$
 Starting from this equation, we can reduce the problem to the study of two variables only, $\moy{\aop}$ and $\moy{\adag}$, and obtain analytical, closed-form expressions for the stability matrix. In this case, it was found \cite{hwang_dissipative_2018} that, for $\gind\leq\gind^D_t$, the normal phase is stable and the two superradiant phases are unstable. For $\gind\geq\gind^D_t$, the normal phase becomes unstable and the superradiant phases stable. Hence, the system will experience a dissipative phase transition for $\gind=\gind^D_t$. Since, at $\gind=\gind^D_t$, we have $\alpha^D_g=0$, the field will not experience a discontinuous jump during the transition, which indicates a second-order transition. Very close to the transition, we have $\moy{\aop}\propto\sqrt{\gind-\gind^D_t}$, and we can deduce the first critical exponent \textbf{\textbf{\textbeta}}$=1/2$. 

 Similarly, closed set of equations can be obtained for the second-order moments of the field, $\moy{\adag\aop}$, $\moy{\aop^{\dag2}}$ and $\moy{\aop^2}$. This gives us access to the fluctuations of the field in the steady-state. Near the critical point, these fluctuations diverge like $\gind-\gind^D_t$, which gives a critical exponent \textgamma$=1$.\\

 The previous treatment becomes exact in the limit of large frequency ratio $\rat\rightarrow\infty$. For finite $\rat$, the fluctuations of the system can be studied with techniques such as Keldysh formalism \cite{torre_keldysh_2013,hwang_dissipative_2018,kirton_introduction_2019}. This allows to show that the fluctuations of $\moy{\xop^2}$, at $\gind=\gind^D_t$, scales like $\sqrt{\rat}$, which corresponds to a critical exponent \textbf{\textzeta}$=1/2$. Note that the three critical exponents \textbf{\textzeta}, \textgamma and \textbf{\textbf{\textbeta}} are identical in the non-equilibrium scenario, and in the equilibrium scenario with a finite temperature. This suggests that, close to the steady-state, the non-equilibrium system behaves like an equilibrium system at a finite, effective temperature. This property is discussed in more details in \cite{torre_keldysh_2013,kirton_introduction_2019}. \\

 \subsection{Spin dissipation}
 Finally, let us consider what happens when the Dicke model is considered, with dissipation acting on each qubit individually. We will  assume that the dissipation rate is the same for each qubit: in \eqref{LindbladDicke}, $\Gamd^i=\Gamd$ and $\Gamp^i=\Gamp$ for each $i$. Again, we can apply study the average spin and field values thanks to a mean-field treatment. A stability analysis can still be performed to witness the emergence of a phase transition.

  When it comes to fluctuations, however, a difficulty arises: because the dissipation acts on each spin individually, we can no longer consider only collective spin excitations. In other words, we cannot simplify the problem as earlier, by applying HP transformation and turning the spins into a collective bosonic mode. Several analytical and numerical techniques have been developed to tackle this issue \cite{kirton_introduction_2019}. 
 From the analytical sides, these systems have been studied using \textit{cumulant expansion} \cite{kirton_suppressing_2017}, a mean-field method in which the average value of higher-order moments are decomposed into products of average values of smaller-order moments. For instance, we may set $\langle\adag\aop\sigz\rangle\sim\langle\adag\aop\rangle\langle\sigz\rangle$, or $\langle\aop\sigy\sigz\rangle\sim\langle\aop\sigy\rangle\langle\sigz\rangle$. The many-body dynamics can then be broken down in a closed system of equations which are analytically tractable.

 From the numerical side, if the dissipation is the same for each spin, the Lindblad equation is still symmetric with respect to the permutation of two spins. This permutation symmetry allows for some simplification, and, in particular, makes it possible to implement efficient numerical simulations. Several works have studied the Dicke model using these techniques \cite{kirton_suppressing_2017,shammah_superradiance_2017,shammah_open_2018,gegg_superradiant_2018,kirton_superradiant_2018}). Under appropriate conditions on the dissipation, a superradiant transition can still take place. In particular, it was shown that the interplay between spin decay and dephasing could be used to suppress and restore the superradiant transition \cite{kirton_suppressing_2017}. More specifically, the superradiant transition disappears when only dephasing is present, but it is restored when both dephasing and decay are present. 

 Furthermore, in the absence of dissipation, the transition is a second-order, continuous one. By contrast, in the presence of dissipation, the system may exhibit a bistable, discontinuous behavior \cite{bowden_first-_1979,gelhausen_dissipative_2018}.
 Finally, it was also shown that these models could also exhibit transition towards lasing states \cite{zhiqiang_nonequilibrium_2017,kirton_superradiant_2018,kirton_introduction_2019}, in which the system is described by a limit cycle instead of a single steady-state. 

 \section{Coupling with both quadratures}
 \label{Sec2.5}

 In the Dicke model, the qubits and field are coupled only via a single field quadrature. However, it is also possible to engineer systems in which the coupling involves both quadratures. For instance, let us consider the following Hamiltonian \cite{baksic_controlling_2014,fan_hidden_2014}:

 \begin{equation}
 	\Hop=\Of\adag\aop+\Oq\Jz+\frac{\gcoll_E}{\sqrt{N}}(\adag+\aop)\Jx+\frac{\gcoll_M}{\sqrt{N}}i(\adag-\aop)\Jy,
 \end{equation}
 or, after a HP transformation (dropping constant terms):

 \begin{equation}
 	\Hop=\Of\adag\aop+\Oq\bdag\bop+\gcoll_E(\adag+\aop)(\bdag+\bop) + \gcoll_M(\adag-\aop)(\bop-\bdag).
 \end{equation}

 The behavior of this model is governed by the interplay between the two couplings $\gcoll_E$ and $\gcoll_M$. For $\gcoll_E=0$ or $\gcoll_M=0$, we find again the ordinary Dicke model, with a $\mathbb{Z}_2$ symmetry. However, for $\gcoll_E=\gcoll_M$, the Hamiltonian corresponds exactly to the TC model \eqref{TC}. This model has a larger $U(1)$ symmetry, which corresponds to the conservation of the total excitation number $\adag\aop+\Jz$. Thus, an equal coupling with both quadratures allows us to realize the Tavis-Cummings model, even in the region where the RWA can no longer be applied.

 To determine the phase diagram of the system, we can project both $\aop$ and $\bop$ in coherent states, and minimize the resulting energy $\bra{\alpha,\beta}\Hop\ket{\alpha,\beta}$. First, minimization with respect to $\alpha$ yields $\alpha=\left(\frac{\gcoll_E}{\Of}(\beta+\beta^*)+\frac{\gcoll_M}{\Of}(\beta-\beta^*)\right)\sqrt{1-\frac{\lvert\beta\rvert^2}{N}}$, and gives an effective potential $E_G(\beta)$. When the minimization with respect to $\beta$ is performed, a rich behavior emerges. For $\gcoll_E$ and $\gcoll_M$ smaller than $\frac{\sqrt{\Of\Oq}}{2}$, the energy has a single minimum on $\beta=0$, corresponding to the normal phase. For $\gcoll_E>\gcoll_M$ and $\gcoll_E\geq\frac{\sqrt{\Of\Oq}}{2}$, the potential has two minima, located on two real values $\pm\beta_E$. These two solutions are analogous to the two domains of the ordinary Dicke model, and coincide exactly with them for $\gcoll_M=0$. By contrast, when $\gcoll_M>\gcoll_E$ and $\gcoll_M\geq\frac{\sqrt{\Of\Oq}}{2}$, the potential has two minima, corresponding to two \textit{imaginary} values $\pm\beta_M$. When the system enters either regime of parameters, the $\mathbb{Z}_2$ symmetry is spontaneously broken. The symmetry breaking, however, is different in both cases. The $\mathbb{Z}_2$ symmetry can be decomposed in two other symmetries, $\mathcal{T}_E$ and $\mathcal{T}_M$, which act respectively as follows: $$\mathcal{T}_E:\left(\adag+\aop,i(\adag-\aop),\Jx,\Jy\right)\rightarrow\left(-(\adag+\aop),i(\adag-\aop),-\Jx,\Jy\right),$$
 $$\mathcal{T}_M:\left(\adag+\aop,i(\adag-\aop),\Jx,\Jy\right)\rightarrow\left((\adag+\aop),-i(\adag-\aop),\Jx,-\Jy\right).$$

 These symmetries can be broken separately. In the region $\gcoll_E>\gcoll_M$, the $\mathcal{T}_E$ is broken, but the $\mathcal{T}_M$ symmetry is preserved. In the region $\gcoll_M>\gcoll_E$, $\mathcal{T}_M$ is broken, but $\mathcal{T}_E$ is preserved.

 For $\gcoll_E=\gcoll_M\geq\frac{\sqrt{\Of\Oq}}{2}$, the effective potential $E_G(\beta)$ depends only on $\lvert\beta\rvert$. In the $(Re(\beta),Im(\beta))$ plane, this potential assumes a sombrero-shape potential, with an infinite number of minima, organized around a circle $\lvert\beta\rvert=\beta_0$. Hence, in this region, the fields can acquire infinitely many average values, which corresponds to the spontaneous breaking of the continuous $U(1)$ symmetry. 

\begin{figure}
 	\includegraphics[angle=-90, width=\linewidth]{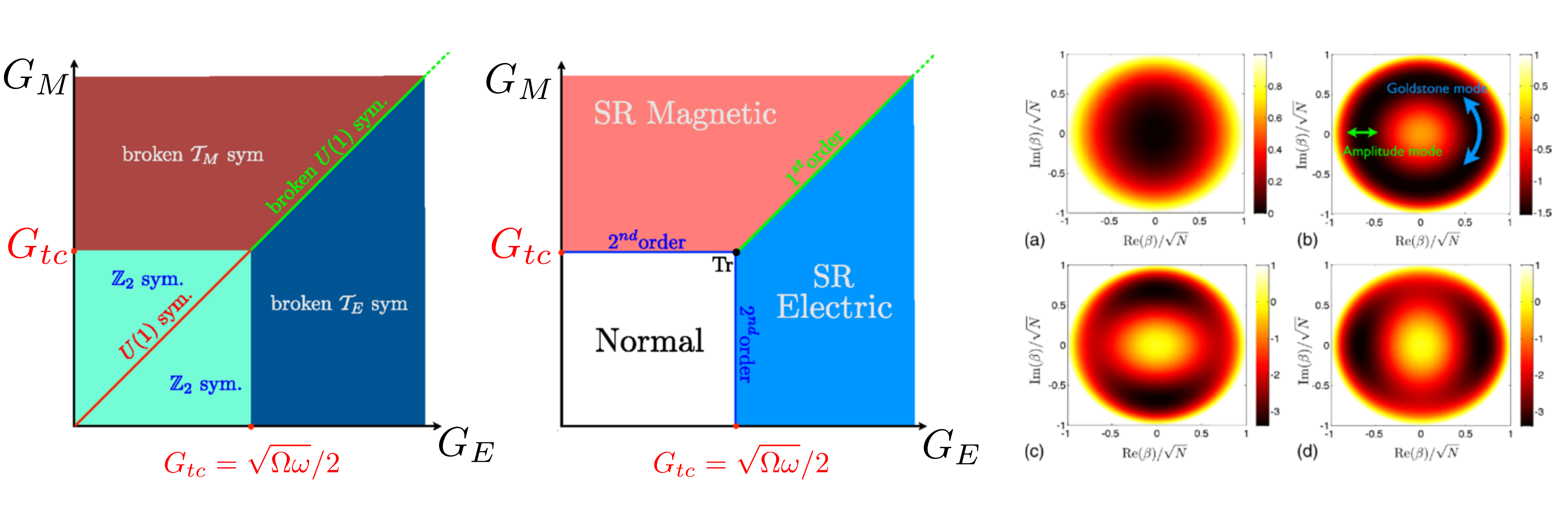}
 	\caption[Symmetries and phases of the Dicke model with coupling to both quadratures.]{Left and center: symmetries and phases of the model. For $\gcoll_E$ and $\gcoll_M$ smaller than $\gcoll_{tc}=\sqrt{\Of\Oq}/2$, the model is in a normal phase, and obey a $\mathbb{Z}_2$ symmetry, which is promoted to a $U(1)$ symmetry on the $\gcoll_E=\gcoll_M$ line. When $\gcoll_E$ ($\gcoll_M$) is increased beyond $\gcoll_{tc}$, the system experiences a second-order transition in which the $\mathcal{T}_E$ ($\mathcal{T}_M$) symmetry is spontaneously broken. If one goes from one domain to the other by crossing the $\gcoll_E=\gcoll_T$ line, the system experiences a first-order transition. Finally, by moving along the $\gcoll_E=\gcoll_T$ line, the $U(1)$ symmetry is broken. Right: effective potential in the $(Re(\beta),Im(\beta))$ plane. a) Normal phase, the potential is minimum on the origin. b) Broken $U(1)$ phase, the potential has a sombrero-like shape. c) Broken $\mathcal{T}_M$ phase, the potential admits two minima corresponding to imaginary values for $\beta$. d) Broken $\mathcal{T}_E$ phase, the potential admits two minima corresponding to real values for $\beta$.}
 	\label{FigTC}
 \end{figure}

 Hence, depending on the regime of parameters considered, this model can spontaneously break a discrete or a continuous symmetry. All these transitions are second-order; however, when both $\gcoll_E$ and $\gcoll_M$ are larger than $\frac{\sqrt{\Of\Oq}}{2}$, and if one goes from the broken $\mathcal{T}_E$ region to the broken $\mathcal{T}_M$ region by crossing the $\gcoll_E=\gcoll_M$ line, the system experiences a first-order transition (see Fig.\ref{FigTC}).

 Finally, it is also possible to study the fluctuations of the system around its mean-field value. In the broken $\mathcal{T}_E$ and the broken $\mathcal{T}_M$ regions, the fluctuations are similar to what we obtained in the ordinary Dicke model. In the region of broken $U(1)$ symmetry, the excitations can be decomposed in two polaritonic modes: a Higgs mode with a finite gap, and a Goldstone mode with zero excitation energy.

 Finally, note that the transition of the TC model is fragile against dissipation. In the Dicke and Rabi model, the counter-rotating terms compete with the loss of photons, activating the driven-dissipative transition. By contrast, in the TC model, the dynamics is unprotected against dissipation: in the presence of photon losses, the system will simply evolve towards the vacuum of photons. There is no steady-state transition for the dissipative TC model.

 \section{Experimental realizations}
 \label{Sec2.6}

 It is unclear whether the superradiant transition can be achieved with a genuine quantum system, due to the possible presence of additional $A^2$ or $P^2$ terms. However, as we discussed in the previous chapter, these controversies can be sidestepped by using quantum simulation protocols. In these protocols, the coupling $\gind$ is not a natural quantity, but the combination of different parameters involved in the simulator. A large effective coupling can thus be achieved even if the true physical interaction is weak, which implies that 1) the $A^2$ and $P^2$ terms can generally be omitted, and 2) the interaction with the environment involves undressed jump operators.  
 In this section, we will review some of these platforms in which the superradiant transition has been proposed or experimentally realized.

 \subsection{Raman transitions for atoms in cavity}
\begin{figure}
 	\includegraphics[angle=-90,width=\linewidth]{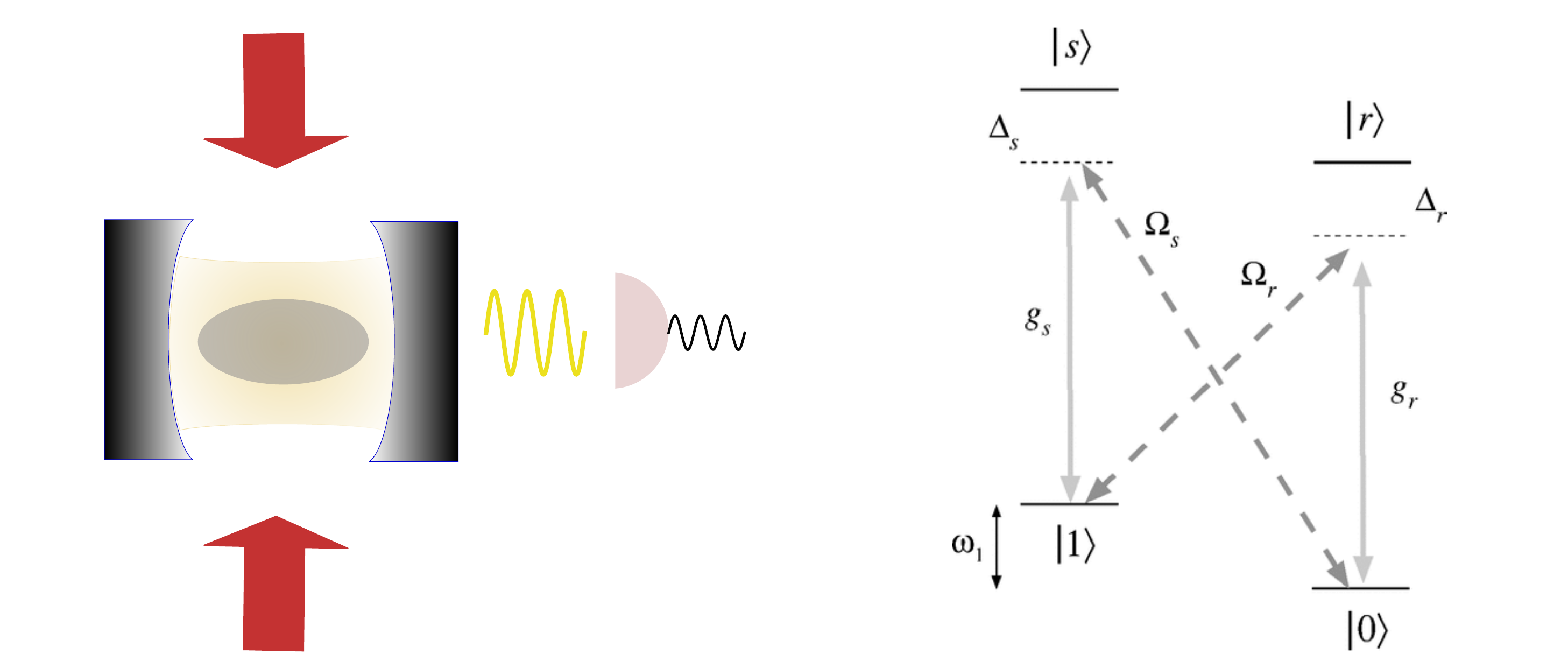}
 	\caption[Use of Raman transitions to engineer the Dicke model.]{Use of Raman transitions to engineer the Dicke model. Left: schematic depiction of the experiments. Two lasers $\Omega_r$ and $\Omega_s$ are used to drive a cloud of atoms inside a cavity. Right : atomic level structure. The drives excite rotating and counter-rotating interaction between the atom and the cavity field, respectively. When the system is driven across the transition, the cavity field builds up; photons escape the cavity and are detected.}
 	\label{Raman}
 \end{figure}

 The first setup we will consider utilizes atoms trapped in a cavity. Although the coupling between the atomic state and the cavity field is naturally low, it can be enhanced by drive-induced Raman transitions \cite{dimer_proposed_2007}. The operating principle is sketched in Fig.\ref{Raman}. The atoms have a four-level structure, the lowest two $\ket{0}$ and $\ket{1}$ being used to define the qubit. Two drives are applied, with Rabi frequencies $\Omega_s$ and $\Omega_r$. The drive excites the system towards the higher-excited states $\ket{s}$ and $\ket{r}$. When excited, the system relaxes coherently towards to $\ket{0},\ket{1}$ subspace by emitting a photon inside the cavity, at a rate $\gind_{s,r}$. All processes are coherent and can occur in both directions. If we consider the process driven by $\Omega_r$, this will effectively bring the qubit from $\ket{1}$ to $\ket{0}$ while emitting a photon, or vice-versa: this process corresponds to an interaction of the form $\adag\sigm+\aop\sigp$. By contrast, the drive $\Omega_s$ brings the qubit from $\ket{0}$ to $\ket{1}$ while emitting a photon, and thus creates a counter-rotating interaction $\adag\sigp+\aop\sigm$. The combination of both processes allows to engineer the Dicke model, and to achieve an effective ultrastrong coupling. When the system is driven across the transition, the electromagnetic field spontaneously builds up inside the cavity. These photons are real and can escape the cavity. The output field is the main experimental signature of this setup.

 Note that here we considered two distinct drives; however, such a process can also be realized with a single drive, as long as the two excitations processes remain distinct. More complex level structures can also be considered. Furthermore, one can choose to use a single drive instead of two: hence, this setup allows to implement the Tavis-Cummings model in addition to the Dicke model.

 These ideas were developed theoretically in \cite{dimer_proposed_2007}, and implemented in several experiments \cite{baden_realization_2014,zhiqiang_nonequilibrium_2017}. Both experiments have observed photon emission from the cavity when the effective coupling strength is tuned across a threshold value (note however that the emission could not be sustained indefinitely, and was typically lost after a few tens of milliseconds). Furthermore, in \cite{zhiqiang_nonequilibrium_2017}, an additional, lasing-like state was observed, in which the system, instead of reaching a steady-state, adopts an oscillatory behavior (more information on this state can be found in \cite{kirton_superradiant_2018}).

 \subsection{BEC localization}

 In the previous example, we considered that all atoms were coupled with the same field mode, for instance, if the atoms are localized on a distance much smaller than the cavity field wavelength. However, in practice, the size of the atomic cloud is often larger than the wavelength. Atoms then feel a different field depending on their position. This can lead to effects such as atomic self-organization, which actually provides an excellent platform for the implementation of the Dicke model. The operating principle is the following : let us consider a collection of atoms, modeled by two-level systems, in a cavity. The atomic transition is driven by a strongly red-detuned pump field. The qubits will scatter light in one of the cavity modes, which has a spatial dependence (say) $\cos(kx)$. Atoms located at different positions will emit photons with different phases; in particular, atoms located on the even and odd antinodes of the cavity will emit field with a phase difference $\pi$. If the atomic distribution is homogeneous inside the cavity, all the emitted fields interfere destructively, and the resulting field is zero. However, random fluctuations can bring an excess of atoms in one of the antinodes of the cavity mode, resulting in a nonzero total field. When the field starts to build up, the atoms experience an effective potential which drives them towards the already occupied antinodes. This triggers a self-amplified process which finally makes the atoms localized in either the even or the odd antinodes, while a large field is built inside the cavity. This effect was first predicted in \cite{domokos_collective_2002} and observed experimentally soon afterward \cite{black_observation_2003}. 

 Later on, it was shown that using a Bose-Einstein Condensate (BEC), this idea could be used to implement the Dicke model \cite{nagy_dicke-model_2010}. Instead of a point-like particle, a BEC is described by a delocalized wavefunction $\psi(x)$. The position of the BEC and the cavity field are coupled via the process we just described. For small coupling, the BEC is delocalized over the entire cavity, $\psi(x)=cst$. When the effective light-position coupling (which is controlled by the pump) is increased above a threshold value, the BEC localizes on either even or odd antinode sites. The localization on even (odd) antinode corresponds to a wavefunction $\psi(x)\propto1+\cos(kx)$ ($\psi(x)\propto1-\cos(kx)$), where again $k$ is the spatial frequency of the cavity field. This localization is associated with a buildup of the cavity field.
 We can then define an effective two-level basis, in which $\psi(x)=cst$ and $\psi(x)\propto\cos(kx)$ are the eigenstates of $\sigz$, and $\psi(x)\propto1\pm\cos(kx)$ are the eigenstates of $\sigx$. Starting from the lower eigenstate of $\sigz$ (that is, the completely delocalized state) and zero cavity field, the BEC localization brings us to an eigenstate of $\sigx$, correlated with a displacement of the cavity field, which is precisely a superradiant phase transition. 

 A somewhat more compact description of this phenomenon was proposed in \cite{kirton_introduction_2019}: we label each antinode site as even or odd, and introduce the effective variable $\sigx_j$ which indicates whether the atom $j$ is on an even ($\sigx_j=1/2$) or odd ($\sigx_j=-1/2$) site. Atoms on the even (odd) sites scatter light with a phase of $0$ ($\pi$). Hence, the scattering of light by the atoms can be described by a coupling $\adag(N_{\text{even}}-N_{\text{odd}})+h.c.=2\adag\sigx+h.c.$, where $N_{\text{even}}$ and $N_{\text{odd}}$ quantify the number of atoms in even and odd sites, respectively. The atoms can also tunnel between the sites, which is encoded by the operator $\sigz_j$. Combining these terms leads to the Dicke model.

 One of the great advantages of this setup is the possibility to measure both the field escaping the cavity, and the atomic density distribution via time-of-flight (TOF) measurements. Hence, it is possible to access the state of the system, instead of only spectral signatures.

  \begin{figure}
  	\includegraphics[angle=-90,width=\linewidth]{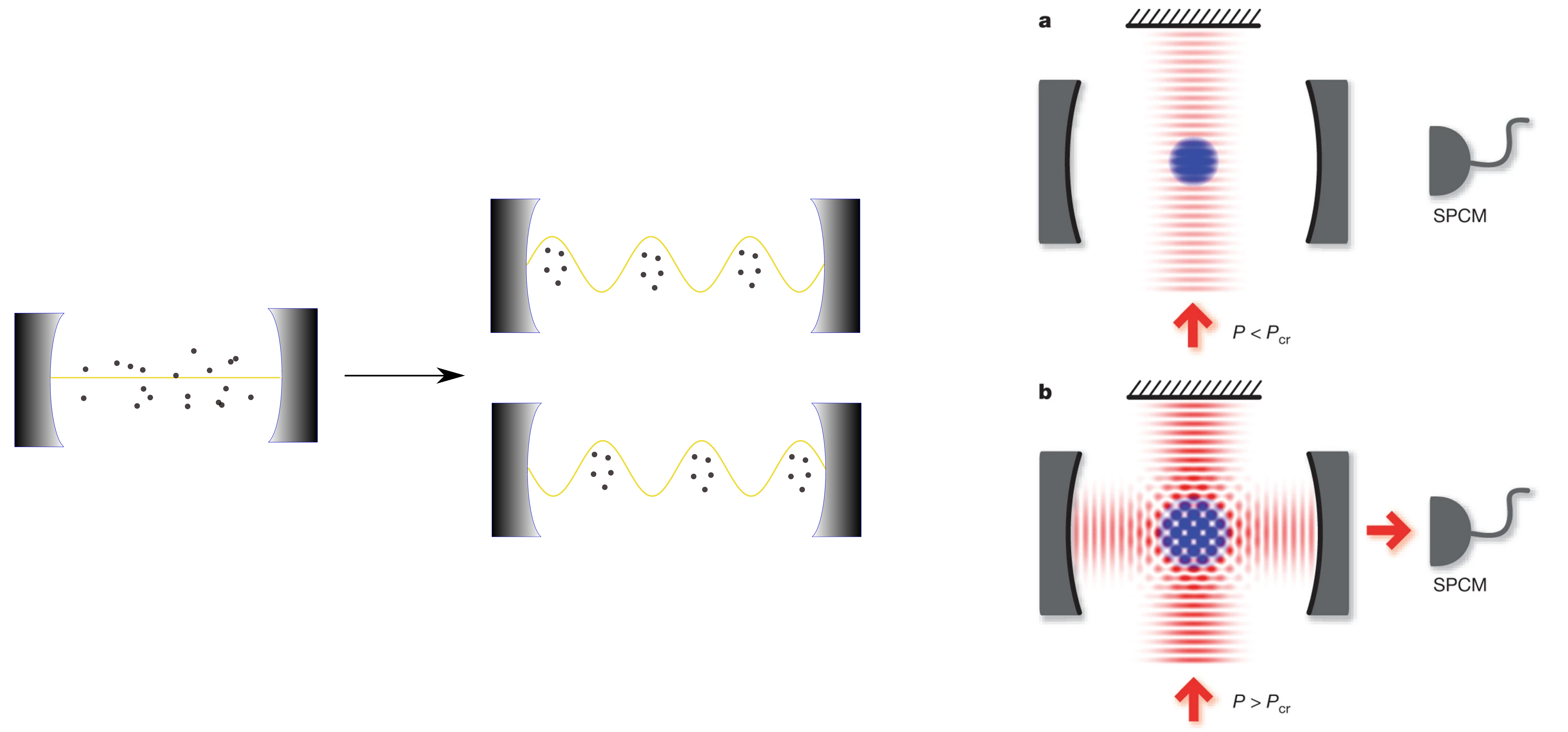}
  	\caption[Use of atom localization in a cavity to simulate the Dicke model.]{Left: pictorial representation of atom localization in a cavity. For small atom-cavity interaction, the atoms are scattered in the cavity, and the field amplitude (thick yellow line) is zero. For larger interaction, the atoms localize on the even or odd antinodes of the cavity, and the field acquires a nonzero amplitude, with a phase difference of $\pi$ between the two cases. Right: Experimental implementation with BEC. A transverse standing-wave pump is sent perpendicular to the cavity axis. a) For small pump power, the BEC is delocalized over the empty cavity. b) When the pump power is increased over a threshold, the cavity field is populated, while the BEC organizes in a checkerboard pattern. Right figure reproduced from \cite{baumann_dicke_2010}.}
  \end{figure}

 These ideas have been implemented experimentally several times. In \cite{baumann_dicke_2010}, a BEC in a cavity was driven by a transverse pump, which was itself a periodic stationary wave with a spatial dependence $\cos(kz)$. The spatial modulation of both the cavity and the pump field created a checkerboard potential. When increasing the pump power, the localization of the BEC on either the even or odd lattice sites, and the buildup of the cavity field with a corresponding $0$ or $\pi$ phase, was observed. Later on, the fluctuations of the field and of the atomic density were studied \cite{brennecke_real-time_2013}, and the critical exponents of the transition were computed. The findings were consistent with the predictions for the dissipative Dicke model. Finally, the out-of-equilibrium dynamics of the model was also studied \cite{klinder_dynamical_2015}. When sweeping the system across the transition, a hysteresis behavior obeying a power-law scaling was observed.

 \subsection{Spin-motion coupling}

 In the last two implementations, the bosonic field was a cavity electromagnetic field, while the qubit was either the internal state or the position of the atoms. However, it is also possible to implement the Dicke model with spin-phonon coupling. In particular, trapped-ions simulation of the Dicke model has been proposed \cite{genway_generalized_2014} and realized \cite{safavi-naini_verification_2018}. The qubit is defined by an internal transition of the ion, which can be tuned by an externally applied magnetic field. The system can be brought towards a superradiant phase by tuning the magnetic field across a critical value. The internal state of the ions can be monitored by observing the ion fluorescence. The time-evolution of the internal state also gives access to the spin-phonon correlation. We will discuss a similar scheme in Chapter 4.\\

 A similar experiment has been conducted, using BEC with spin-orbit coupling \cite{hamner_dicke-type_2014}. Note that contrary to localization experiments, here the vibrational degree of freedom was fully exploited, and was not confined to a two-state subspace. Through TOF measurement combined with a Stern-Gerlach apparatus, it is possible to monitor both the qubit (internal) state, and the bosonic (quasi-momentum) state. A transition from a spin-polarized phase with non-zero quasi-momentum, to a spin-balanced phase with zero quasi-momentum, was observed, similar to the transition from superradiant to normal phase.
  \begin{table}
 	\hspace{-30pt} \begin{tabular}{|c | c | c | c |}
  	\hline & Qubit & Boson & Observables\\
  \hline Raman transition for atoms in cavity \cite{baden_realization_2014,zhiqiang_nonequilibrium_2017} & Internal atomic state & Cavity field & Output field \\
  \hline BEC localization\cite{baumann_dicke_2010,brennecke_real-time_2013,klinder_dynamical_2015,leonard_supersolid_2017,leonard_monitoring_2017} & Atomic position & Cavity field & TOF, output field\\
  \hline BEC with spin-orbit coupling \cite{hamner_dicke-type_2014} & Internal atomic state & Atomic position & TOF, Stern-Gerlach\\
  \hline Trapped ions \cite{safavi-naini_verification_2018} & Internal ion state & Ion position & Ion fluorescence \\ \hline
  \end{tabular} \hfill
  \caption[Overview of the platforms used for simulating the superradiant transition.]{Overview of the platforms used for simulating the superradiant transition.}
  \end{table}

 \subsection{U(1) symmetry breaking.}

\begin{figure}
 	\centering
 	\includegraphics{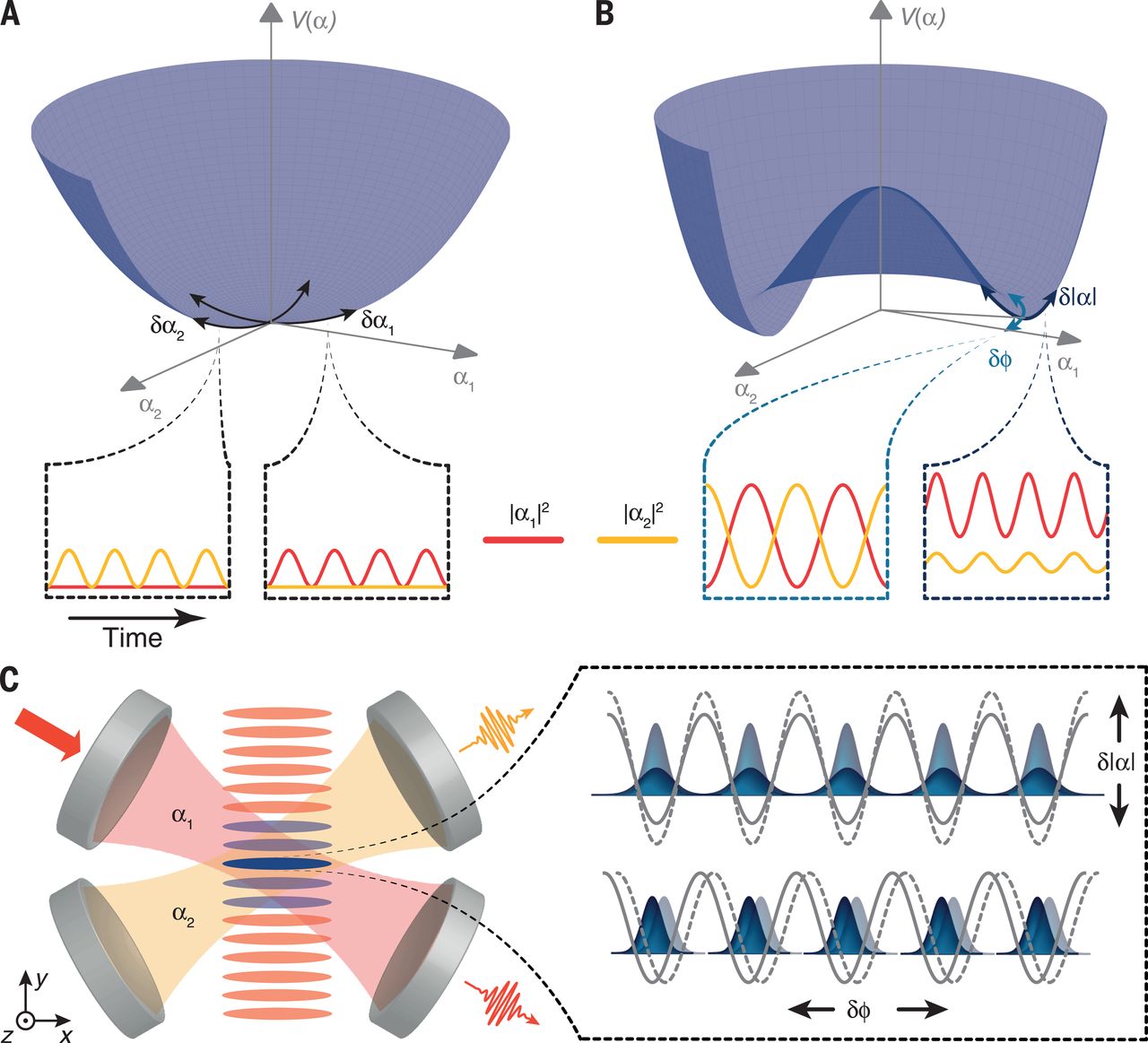}
 	\caption[Experimental observation of U(1) symmetry breaking, using a BEC.]{Observation of U(1) symmetry breaking with a BEC. The upper figure represents the effective field potential, the lower panel is a sketch of the experimental apparatus. $\alpha_{1,2}$ represents the field amplitude in both cavities. A) Normal phase: both fields are localized on $0$. B) Symmetry-broken phase: the potential assumes a sombrero-like shape. The field amplitudes $\alpha_1$ and $\alpha_2$ adopt a given value, in a manifold of equivalent possibilities $\lvert\alpha_1\rvert^2+\lvert\alpha_2\rvert^2=const$. For the field, the Goldstone modes is associated with rotation around the origin, in which the two fields exchange their amplitudes. The Higgs mode is associated with oscillation in the transverse direction. C) Sketch of the experiment. The BEC is organized in tubes modulated in the $y$ direction. Each position for the field in the sombrero potential corresponds to a position of the BEC along the $y$ directions. Inset: for the BEC, the Higgs and Goldstone modes are associated with fluctuations of the strength and position of the density modulation, respectively. Reproduced from \cite{leonard_monitoring_2017}.}
 	\label{FigSombrero}
 \end{figure}

 BEC localization has also permitted to study the phase diagram of the TC model \cite{leonard_supersolid_2017,leonard_monitoring_2017}. In these experiments, the BEC was simultaneously coupled to \textit{two} cavities, described by operators $\aop_1$ and $\aop_2$. When the coupling to both cavities is identical, the system exhibits a U(1) symmetry corresponding to the exchange of photons between the two cavities (which can be expressed by $e^{i\theta(\adag_1\adag_2+\adag_2\aop_1)}$). This is very similar to the continuous symmetry of the TC model, only replacing the two field quadratures by two different fields.
 By increasing the coupling of the BEC to a single cavity, the by now usual $\mathbb{Z}_2$ symmetry breaking is observed. When the coupling to both cavities is increased simultaneously, the U(1) symmetry is spontaneously broken. Physically, both cavity fields get populated;  the population of each cavity is random and differ from one experimental realization to the other. However, the two populations are correlated according to $\adag_1\aop_1+\adag_2\aop_2\sim const$. This shows the breaking of the symmetry: the system randomly "chooses" a position among the range of equivalent values for $\adag_1\aop_1$ and $\adag_2\aop_2$. Furthermore, the excitation energy of the system was measured; two modes with zero and nonzero excitation energy, respectively, where observed. These modes correspond to Goldstone and Higgs excitations, as described in Fig.\ref{FigSombrero}.

 \subsection{Other proposals}
 Other platforms have also been proposed to simulate the Dicke model, including arrays of NV centers \cite{zou_implementation_2014}. Trapped-ions platforms could also be used to engineer generalizations of the Dicke model, for instance by including two-photon coupling \cite{felicetti_spectral_2015} or effective spin-spin interaction \cite{jaako_quantum_2020}. Finally, digital and digital-analog simulations have also been considered \cite{mezzacapo_digital_2014,lamata_digital-analog_2017}.

 However, no experience so far has managed to observe the superradiant transition in the Rabi model. Many experiments have implemented this model with genuine light-matter interaction. However, in addition to being susceptible to additional $A^2$ terms, these platforms generally operate near resonance, far away from the $\rat\rightarrow\infty$ regime needed to observe the transition. Hence, here as well, simulators are the most promising candidates, in particular cold-atoms \cite{dareau_observation_2018} and trapped-ions \cite{lv_quantum_2018} platforms.

\chapter{Introduction to quantum metrology}

\def\hbar{{\mathchar'26\mkern-9muh}}

\epigraph{\textit{When you can measure what you are speaking about and express it in numbers you know something about it; but when you cannot measure it, when you cannot express it in numbers, your knowledge is of a meager and unsatisfactory kind.}}{Lord Kelvin}

Quantum mechanics places fundamental constraints on the precision of sensing protocols. However, the appropriate use of quantum correlations also permits to sidestep these limitations and achieve improved sensing performances. The field of quantum metrology studies how quantum correlations can be exploited to estimate a parameter encoded in a quantum probe with maximal precision. In this Chapter, we will review some key concepts and results of this field. It is organized in six sections: first, we give some introduction and motivation to quantum metrology problems. Second, we introduce the general framework in which quantum parameter estimation problems are formulated, and the central notion of Quantum Fisher Information. A few key properties of this quantity are reviewed in the next section. In the fourth section, we introduce the problem of quantum metrology with finite resources. We discuss the notion of Standard Quantum Limit, and the various strategies which have been proposed to overcome this limit. The fifth section is devoted to the specific case of metrology with quantum states of light. Finally, in the sixth section, we review some of the experimental realizations in this domain.

For the reader wishing to find more details on the topics, many excellent reviews have been published on the subject. Among those, Ref.\cite{paris_quantum_2011,demkowicz-dobrzanski_chapter_2015,safranek_gaussian_2016,pezze_quantum_2018,braun_quantum-enhanced_2018} have been particularly useful in the writing of this Chapter.

\section{Introduction}
Quantum mechanics imposes fundamental bounds to the accuracy with which measurement tasks can be performed. 
In textbook quantum mechanics, this problem is generally expressed in terms of an operator $\hat{O}$ describing an observable one wants to evaluate. When this observable is measured on a system of density matrix $\rop$, a random outcome is obtained. The statistics of these outcomes is given by the diagonal matrix elements of the state $\rop$ in the eigenbasis of $\hat{O}$. These outcomes are centered around the mean value $\moy{\hat{O}}=Tr[\rop\hat{O}]$, and have a square variance $Var^2(\hat{O})=Tr[\rop(\hat{O}-\moy{\hat{O}})^2]$. This variance then quantifies the uncertainty associated with the measure of the observable; in physical terms, it describes how measurement results will vary from one experimental round to the other. A generalized version of this approach is the positive-operator valued measure (POVM) formalism \cite{peres_quantum_2006}, which allows us to describe incomplete measurements that do not fully resolve the state of the system. Although of high fundamental relevance, this formalism cannot always be used to describe actual sensing experiments, for several reasons.\\

First, the POVM formalism only describes the measurement of average quantities which are \textit{linear} in the system state $\rop$. However, many quantities of interest, such as purity or entanglement entropy, are non-linear functions of $\rop$, and cannot be directly described by an observable or a POVM \cite{paris_quantum_2011}. Time is another quantity for which an observable cannot be created (at least not in a straightforward way, see \cite{holevo_probabilistic_2011,maccone_quantum_2020} for discussions on this topic). Thus, the POVM formalism is not sufficient on its own to understand how quantum mechanics limits us from measuring such quantities. \\

Second, even if the quantity to study (which will be referred to as \textit{the signal} in this paragraph) can be described by an observable, there is not always an experimentally accessible process that implements the corresponding projective measurement. 
More precisely, all sensors are made of physical systems, which couple to the signal one wishes to measure. Instead of implementing directly a POVM, what we are doing in practice is coupling the signal to a probe, then reading the probe. Then of course, how do we read the probe? This requires to couple it to a third physical system (such as the light escaping the system and entering the experimentalist's eye), which itself needs to be coupled to another one, a process that could be extended all the way to the brain of the experimentalist. Fully describing this process amounts to solving the measurement problem, which is still a much-debated topic today.

Fortunately, for many practical measurements, it is possible to adopt a pragmatic perspective and to abstract out this chain of coupled systems as a single, well-defined POVM. For instance, the arrival of atoms on a plate can be reasonably described as a projective measurement of the atom's momentum. In this case, the measurement uncertainty comes from the fluctuations of the atoms' momentum, \textit{i.e.}, of the signal itself. By contrast, the effect of the probe, here the plate, is negligible. For other protocols, however, the effect of the probe is non-trivial and needs to be taken into account. For instance, in a Ramsey interferometric experiment, a magnetic field $B$ is measured by coupling it to an ensemble of spins (for instance, a cloud of atoms). The precession rate of the spins is evaluated by measuring the final polarization, and is used to reconstruct $B$. Since the spins themselves have fluctuations (both classical and quantum), and are imperfectly coupled to the magnetic field, this will add further uncertainty on top of the intrinsic quantum fluctuations of $B$. In many relevant scenarios, the fluctuations related to the probe actually become dominant. In this case, it is justified to model the magnetic field as a classical, noiseless signal, and the polarization measurement as a POVM. Between the two, we have the atomic probe, which contains all the quantum fluctuations.

In more colloquial terms, a measurement process involves a chain of quantum systems coupled together. Quantum mechanics tells us that all of these systems have intrinsic fluctuations, which limits the precision of sensing protocols. The question then is to identify the dominant source of \textquote{quantumness} in this chain of systems, and where to put the boundary between quantum and classical systems. In the usual POVM description, one considers the fluctuations of the measured signal itself, the rest of the chain being modeled by a classical apparatus with no intrinsic fluctuations. By contrast, in the case described above, the focus is on the fluctuations of a probe, which is coupled to a classical signal and looked at by classical measurement.\\

\section{Quantum parameter estimation}

With these ideas in mind, we will now introduce the general estimation framework in which quantum metrology problems are usually formulated. We want to estimate some classical parameter $x$. This quantity need not be associated with an observable: for instance, it can be a duration, or the amount of entanglement in a two-qubit system. 
In the course of this Thesis, we will be interested in estimating a single parameter, but a similar formalism can be derived for multi-parameter estimation \cite{paris_quantum_2011}.

\begin{figure}
 	\centering
 	\includegraphics[angle=-90,width=\linewidth]{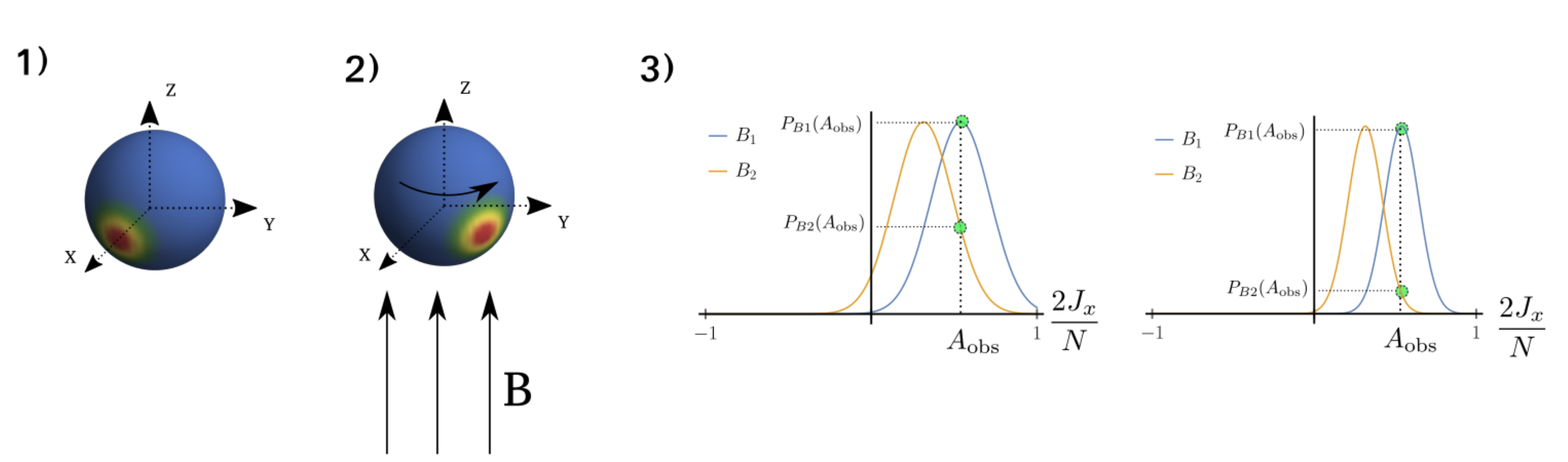}
 	\caption[Schematic depiction of an estimation protocol using atom interferometry.]{Schematic depiction of an estimation protocol using atom interferometry. 1) A cloud of atoms is prepared in a spin-coherent state, polarized along the $x$-axis. 2) A magnetic field $B$ is applied, inducing a rotation of the spins around the $z$-axis (depending on the protocol, the field could be static or oscillating, and applied directly or through a three-pulse Ramsey sequence. Here we show only the net rotation effect for simplicity). 3) Left: the average spin population $A=\frac{2\Jx}{N}$ is measured. Even when classical noise has been removed, the population is a random variable, whose probability distribution depends on the applied field (in a magnetometry experiment, the population depends on the amplitude of the field; in an atomic clock, it depends on its oscillation frequency). When the measurement is made, a value $A_\text{obs}$ is observed; both scenarios $B_1$ and $B_2$ have a nonzero probability to yield this outcome. Therefore, a single measurement does not allow to reconstruct the applied field with certainty. Instead, an estimate $\bar{B}$ is made, such that $P_{\bar{B}}(A_\text{obs}$) is maximal. Right: when the number of atoms is increased, the probability distributions become easier to discriminate, increasing the precision.}
	\label{SketchRamsey}
 \end{figure}

The estimation protocol is modeled by a four-stage procedure:
\begin{itemize}
	\item Preparation: a physical system, which will act as a probe, is prepared in some state $\rop_0$.
	\item Evolution: the probe is left to evolve under some physical process which depends on the parameter $x$ we want to measure. The evolution can be expressed by some map $\Lambda_x(t)$ : at any time $t$, the state of the system is now $\rop_x(t)=\Lambda_x(t)[\rop_0]$.
	\item Measurement: after a time $T$, the system is measured, according to a POVM.
	\item Estimation: an estimator for the parameter $x$ is constructed using the measurement results.
\end{itemize}
This procedure is very general, and can be used to describe both classical and quantum sensing schemes. Let us consider a specific example: in an atomic clock \cite{pezze_quantum_2018}, the parameter we want to measure is the frequency of a local oscillator field $\Omega_{LO}$. The probe is a cloud of $N$ atoms, described by its total magnetic polarization $\hat{J}_a$, with $a=x,y,z$. The system is initialized in the spin-polarized state $\rop=\ket{\downarrow}\bra{\downarrow}^{\otimes N}$. The evolution step is a Ramsey interferometry protocol, during which the local oscillator field is applied to the system. During this procedure, each qubit evolves according to an effective unitary dynamics (in rotating frame): $\rop_\Delta(\dur)=e^{-i\Delta \dur\sigy}\rop_0 e^{i\Delta \dur\sigy}$ with $\dur$ the duration of the Ramsey pulse, and $\Delta=\Omega_{LO}-\Omega$ the detuning between the local oscillator field and the qubit frequency. If the atoms are uncorrelated initially, each atom has a probability $\cos^2(\frac{\Delta \dur}{2})$ of being excited at the end of the evolution. During the measurement phase, the fraction of excited atoms (\textit{i.e.}, the polarization of the spins) is measured; finally, the measurement results are used to infer $\Delta$. This process can then be used to lock the oscillator field on the atomic frequency, allowing to realize the best frequency standards available today. The various steps are sketched in Fig.\ref{SketchRamsey}, for a similar but simplified protocol.

However, this entire process is probabilistic in nature. The average polarization $A=\frac{2J_x}{N}$ at the end of the experiment is not a deterministic quantity. It obeys a probability distribution $p_\Delta(A)$, centered around the value $\cos^2(\frac{\Delta \dur}{2})$, and whose variance is given by the central limit theorem. Estimating $\Delta$ with perfect accuracy can only be done by reconstructing completely the probability distribution of $A$.
However, \textit{no finite number of measurements allows to reconstruct completely a probability distribution}. Therefore, the estimation of $\Delta$ will necessarily come with some error, even if all the technical imperfections of the system were to be removed. The error can be reduced by increasing the number of atoms or iterating the entire process several times, but with a finite amount of resources, we cannot do better than a guess. How good can this guess be? \\

To answer this question, we will now see how each step of the protocol may be optimized, starting from the last one. As we said, if the \textquote{true} parameter value is $\Delta$, the measurement outcome will be distributed according to the distribution $p_\Delta(A)$. Now, we make the measurement and find an outcome $A_{\text{obs}}$; from this piece of data, we make an estimate of $\Delta$, which we call $\bar{\Delta}(A_{\text{obs}})$. If we redo the same measurement several times, we will obtain a slightly different $A_{\text{obs}}$ each time. Therefore, $\bar{\Delta}$ is itself a random variable. We will consider unbiased estimator, that is, estimators centered around the \textquote{true} value $\Delta$: $\int_A p_\Delta(A)\bar{\Delta}(A)=\Delta$. The estimator, however, has a non-zero variance, which leads us to define the \textit{mean square error}:

\begin{equation}
	(\delta\Delta)^2=\int_A p_\Delta(A) (\bar{\Delta}(A)-\Delta)^2.
\end{equation}
This quantity describes how the estimated value will fluctuate around the true value from one experimental round to the other. The goal then is to define the best possible estimator for $\Delta$, that is, the estimator that will lead to the smallest mean square error. 
According to a central result of classical estimation theory \cite{demkowicz-dobrzanski_chapter_2015}, the mean square error associated with this reconstruction process can be bounded by the Cram\'er-Rao (CR) bound:

\begin{equation}
 	\delta \Delta\geq\frac{1}{\sqrt{\nbrep F_\Delta}}\hspace{20pt} F_\Delta=\sum_{A} \frac{1}{p_\Delta(A)}\left[\frac{d p_\Delta(A)}{d\Delta}\right]^2,
 \end{equation} 
 where $F_\Delta$ is the \textit{Fisher Information} or FI, and $\nbrep$ is the number of independent iterations of the process. Intuitively, the more peaked $p_\Delta (A)$, the higher the FI. 
  If an estimator saturates the CR bound, it is guaranteed to be optimal. 
 Furthermore, we can construct explicitly at least one optimal estimate, the \textit{Maximum Likelihood Estimator} $\bar{\Delta}(A_{\text{obs}})$, which is defined by: $p_{\bar{\Delta}}(A_{\text{obs}})=\underset{\Delta}{\text{Max}} \hspace{5pt} p_\Delta(A_{\text{obs}})$. 
 \footnote{More precisely, the Maximum Likelihood Estimator is only \textit{asymptotically} optimal, \textit{i.e.}, it only saturates the CR bound in the limit of many repetitions. For each new iteration, both the error and the CR bound decrease, and the difference between the two tends to zero in the limit $\nbrep\rightarrow\infty$.}\\

In the course of this thesis, we adopt a \textit{frequentist} perspective, in which the parameter we want to estimate has a single, well-defined value. In this viewpoint, probability distributions describe the statistics of experimental results from one round to the other. There is another important way to interpret this probability distribution, the \textit{Bayesian perspective}. In this viewpoint, probability distributions describe primarily our \textit{knowledge} of the world. We do not consider the single, \textquote{true} value for the parameter $\Delta$; rather, we define a prior probability $p(\Delta)$ which describes the state of our knowledge about the parameter $\Delta$ before making the measurement. The more peaked the probability, the most confident we are that the parameter has a given value. The goal of the measurement protocol is to use the result data $A_\text{obs}$ to update our knowledge into a posterior probability, which can be computed using Baye's formula:

\begin{equation}
	p(\Delta|A_\text{obs})=\frac{p(A_\text{obs}|\Delta)p(\Delta)}{p(A_\text{obs})}=\frac{p(A_\text{obs}|\Delta)p(\Delta)}{\int_\Delta p(A_\text{obs}|\Delta)p(\Delta)}.
	\label{posterior}
\end{equation}

After the measurement, we can guess the value of $\Delta$, for instance by taking simply the average of the posterior distribution: $\bar{\Delta}(A_\text{obs})=\int_\Delta \Delta p(\Delta|A_\text{obs})$. The variance of the posterior distribution gives us the uncertainty on the parameter after the protocol.
\footnote{Note that since the posterior distribution depends on the value $A_\text{obs}$ we obtained, it will change from one experimental round to the other; the posterior distribution is itself a random quantity, and so is its average value {$\bar{\Delta}$}. Once more, we can define a mean-square error which describes the fluctuations of {$\bar{\Delta}$} from one round to the other: $(\delta\Delta)^2=\int_\Delta\int_A p(A|\Delta)p(\Delta)(\bar{\Delta}(A)-\Delta)$, where this time we need to take the average with respect to both $A$ and $\Delta$. We can prove that if {$\bar{\Delta}$} is just the mean of the posterior distribution, then $\delta\Delta$ is simply given by the variance of the posterior distribution \cite{demkowicz-dobrzanski_chapter_2015}. In other words, the posterior distribution also gives us the fluctuations of the estimator from one experimental round to the other. We can also show that choosing the mean of the posterior distribution as an estimator is optimal, in the sense that it minimizes the mean-square error. Note however that cost functions other than the mean-square error can be used; in this case, another estimator should be chosen to optimize the cost function considered.} \footnote{This estimator can often be linked with the maximum-likelihood estimator; indeed, according to \eqref{posterior}, the maximum-likelihood estimator also maximizes {$p(\Delta|A_\text{obs})$}; in other words, this estimator corresponds to the peak of the posterior distribution. For typical, symmetric posterior distribution (for instance, Gaussian), this peak value coincides precisely with the average value.} We will not discuss this perspective in details here, but further information can be found in \cite{demkowicz-dobrzanski_chapter_2015}.\\

Returning to our example, the probability distribution $p_\Delta(A_\text{obs})$ is associated with the measurement of the polarization. The measurement stage, however, could be performed in several different ways: we could, for instance, measure an observable other than the polarization, have imperfect detectors, perform weak measurement, etc. Each measurement strategy can be described by a different POVM $\hat{\Pi}$, and is associated with a FI $F_\Delta(\hat{\Pi})$. As long as the probe and detector are noisy and can be described as a classical system, the precision is only limited by the technical specifics of the protocol. We can improve the measurement strategy by choosing a different POVM, which just means choosing a detector or a probe with less technical imperfections. We cannot derive any fundamental, general limits on the FI; everything is dominated by the properties of the system at hand. 

When technical noise is further and further reduced, however, the quantum nature of the system will eventually come into play. At this point, we will find that the system will be affected by a residual error, which will be present even when perfect, projective measurements are implemented. This residual error, which quantifies the uncertainty after the measurement step has been fully optimized, defines the Quantum CR bound:

\begin{equation}
 	\delta \Delta\geq\frac{1}{\sqrt{\nbrep \pazocal{I}_\Delta}}\hspace{20pt} \pazocal{I}_\Delta=\underset{\{\hat{\Pi}\}}{\text{Max}} F_\Delta(\hat{\Pi}),
 	\label{QCR}
 \end{equation} 

 with $\pazocal{I}_\Delta$ the \textit{Quantum Fisher Information} or QFI.

The same reasoning can be made for any parameter estimation. The QFI depends only on the probe state $\rop_0$ and the encoding procedure $\Lambda_x$. It quantifies the noise of purely quantum origin, which remain when \textit{both} the measurement and estimation steps have been optimized. A central task in quantum metrology is to study how quantum-correlated probes, or well-chosen encoding processes, can be used to reduce this quantum noise. Interestingly, since the QFI depends only on the probe state and the encoding protocol, it is much less dependent on the technical details of the system. Therefore, it is possible to find very general properties  of the QFI, as well as recipes to achieve the quantum CR bound. We will give some examples in the next sections.\\

The quantum noise we have focused on arises from the fundamentally stochastic nature of quantum measurement. Other types of quantum noise, however, can be present in some experiments, which add to the one already cited and reduce further the achievable precision. In particular, measuring the probe system inevitably modifies its quantum state. It is not an issue if the probe is measured once and discarded (or reinitialized) afterward, which is a reasonable hypothesis in atomic interferometry experiments. The situation in interferometric experiments with light like LIGO-Virgo is more complicated. In these setups, the observable is the position of a movable mirror, coupled with the light acting as a probe. The continuous monitoring of the light perturbs its state, which in turn affects the state of the mirror. This is a second effect of purely quantum origin, known as backaction noise, which plays a central part in quantum optomechanics \cite{clerk_introduction_2010}. In the following, we will focus on situations when the backaction noise can be neglected (for instance, measure-and-discard protocols, or light interferometric experiments with immovable mirrors).

\section{Properties of the QFI}

\subsection{Symmetric Logarithm Derivative}

In this section, we will give some general properties of the QFI, before detailing specific cases in the rest of the Chapter. The initial state $\rop_0$ evolves according to the map $\Lambda_x$ for a time $\dur$. At the end of the evolution, it reaches the state $\rop_x(\tau)$. In this section, we will drop the explicit time dependence. Hence, the possible values of the parameters can be mapped to an ensemble (more precisely, a manifold) of states $\rop_x$. When the parameter $x$ evolves across the parameter space, so does $\rop_x$. This evolution can be unambiguously described by the \textit{Symmetric Logarithm Derivative} (or SLD) $\hat{\mathcal{L}}[\rop_x]$, which is defined as:

\begin{equation}
	\frac{\partial\rop_x}{\partial x}=\frac{\hat{\mathcal{L}}[\rop_x]\rop_x+\rop_x\hat{\mathcal{L}}[\rop_x]}{2},
\end{equation}

A standard result in quantum estimation theory \cite{braunstein_statistical_1994} shows:

\begin{equation}
 	\underset{\{\hat{\Pi}\}}{\text{Max}} F_x(\hat{\Pi})=\pazocal{I}_x=Tr\left[\rop_x\hat{\mathcal{L}}[\rop_x]^2\right].
 	\label{QFISLD}
 \end{equation}

This result can be interpreted as follows: the SLD quantifies how quickly the state evolves when the parameter $x$ is modified. Intuitively, if a small change of $x$ leads to a large variation of $\rop_x$, the probe is very sensitive, and $x$ can be evaluated with precision. Hence, Eq. \eqref{QFISLD} express a fundamental link between estimation theory and the geometry of the set of quantum states. 

Furthermore, it was also proven \cite{braunstein_statistical_1994} that a projective measurement in the eigenbasis of $\hat{\mathcal{L}}[\rop_x]$ is optimal, in the sense that it allows us to saturate the quantum CR bound. Hence, we can not only derive the optimal precision achievable with any measurement, but also an optimal measurement strategy.

 Crucially, the SLD, in general, depends on $\rop_x$. Hence, as advertised, the QFI is not necessarily linear in $\rop_x$, and can capture quantities that are inaccessible in the usual theory of quantum measurement. However, since the SLD ultimately depends on $x$, the optimal measurement strategy depends on the parameter that we want to estimate. Thus, the optimal strategy cannot be implemented without any prior knowledge of the unknown parameter. It is, however, possible to solve this problem through a two-stage measurement: first, a (perhaps suboptimal) measurement is made and repeated $\nbrep_0$ times to obtain a first estimate of $x$. Then this estimate is used to implement the optimal measurement strategy, which can itself be repeated $\nbrep$ times. By devoting only a small fraction of the total measurement rounds to the first suboptimal step (\textit{i.e.}, when $\nbrep\gg \nbrep_0$), the quantum CR bound \eqref{QCR} can be attained \cite{braunstein_statistical_1994,barndorff-nielsen_fisher_2000}.\\

Among others, the following properties of the QFI and SLD will be useful in the following \cite{demkowicz-dobrzanski_chapter_2015}:

\begin{itemize}
	\item The QFI is convex: $\pazocal{I}_x(\sum_i p_i \rop_x^i)\leq\sum_i p_i \pazocal{I}_x(\rop_x^i)$ with $\sum_i p_i=1$ and $p_i\geq0$: intuitively, mixing quantum state cannot increase estimation sensitivity. \footnote{More accurately, the QFI obeys an \textit{extended convexity} property \cite{alipour_extended_2015}. The state $\rop_x$ can be decomposed (in general, in a non-unique way) as a convex combination of density matrices: $\rop_x=\sum_i p_i(x) \rop_x^i$, and both the probability distribution $p_i(x)$ and $\rop_x^i$ can evolve during the parameter encoding. Then the QFI obeys the following extended convexity relation: $\pazocal{I}_x(\rop_x)\leq\sum_i p_i \pazocal{I}_x(\rop_x^i)+F_x({p^i})$. Intuitively, this expression contains both a quantum part which quantifies the sensitivity of each state $\rop_x^i$, and a classical part which quantifies the sensitivity of the classical probability distribution describing the mixture. Since the convex decomposition is in general non-unique, it is possible to optimize over the different decomposition to obtain a tighter bound. In the specific case where the distribution $p^i$ does not depend on $x$ (which is the case, in particular, when the SLD is independent of $x$), we find again the simple convexity relation.}
	\item The QFI does not increase under a channel which does not depend on the parameter $x$: $\pazocal{I}_x(\rop_x)\geq\pazocal{I}_x(\Lambda[\rop_x])$.
	\item In the eigenbasis of $\rop_x$, the SLD can be formally expressed: $\hat{\mathcal{L}}[\rop_x]=\sum_{i,j}\frac{2\bra{e_i(x)}\partial_x\rop_x\ket{e_j(x)}}{\mu_i(x)+\mu_j(x)}\ket{e_i(x)}\bra{e_j(x)}$, with $\rop_x=\sum_i\mu_i\ket{e_i(x)}\bra{e_i(x)}$.
\end{itemize}

\subsection{Specific cases}

We will now present the expression of the QFI in a few important specific cases. If the parameter is encoded by a unitary map $\Lambda_x[\rop]=U^\dagger_x(0\rightarrow \dur)\rop U_x(0\rightarrow \dur)$, the QFI is given by: 
\begin{equation}
	\pazocal{I}_x=\sum_{i,j}\frac{2(\mu_i-\mu_j)^2}{\mu_i+\mu_j}\lvert\bra{e_i}\mathcal{H}_x(\dur)\ket{e_j}\rvert^2,
	\label{QFIunit}
\end{equation}
with $\mu_i$ and $\ket{e_i}$ the eigenvalues and eigenstates of the input state $\rop_0$, and $\mathcal{H}_x(\dur)$ is the generator of the evolution:
\begin{equation}
	\mathcal{H}_x(\dur)=iU_x^\dagger(0\rightarrow \dur)\partial_xU_x(0\rightarrow \dur).
\end{equation}

If the input state is pure, $\rop_0=\ket{\psi_0}\bra{\psi_0}$, it will remain pure at all times of the evolution: $\rop_x(\dur)=\ket{\psi_{x}}\bra{\psi_x}$. Then Eq.\eqref{QFIunit} can be rewritten as \cite{demkowicz-dobrzanski_chapter_2015}:

\begin{align}
\pazocal{I}_x & =4\left[\langle\partial_x\psi_x|\partial_x\psi_x\rangle - \lvert\langle\psi_x|\partial_x\psi_x\rangle\rvert^2\right]=4\left[\langle\partial_x\psi_x|\partial_x\psi_x\rangle + \langle\psi_x|\partial_x\psi_x\rangle^2\right]\\ 
 & =4\text{Var}(\mathcal{H}_x(\dur))\lvert_{\ket{\psi_0}},
\label{QFIpure}
\end{align}

with the variance taken on the \textit{initial} state $\ket{\psi_0}$.

If the encoding map is a free evolution under a Hamiltonian which is both \textit{linear in} $x$ and \textit{time-independent}, \textit{i.e.}, $\Uop(0\rightarrow \dur)=e^{-ixt\Hop}$ \footnote{Or more generally, if we have a Hamiltonian which satisfies $[\partial_x\Hop(x,t),\Hop(x,t)]= [\partial_t\Hop(x,t),\Hop(x,t)]=0$.}, the QFI becomes:
\begin{equation}
	\pazocal{I}_x(\dur)=\dur^2\sum_{i,j}\frac{2(\mu_i-\mu_j)^2}{\mu_i+\mu_j}\lvert\bra{e_i}\Hop\ket{e_j}\rvert^2,
\end{equation}
 
and if the state is pure:

\begin{equation}
	\pazocal{I}_x(\dur)=4 \dur^2 \text{Var}^2(\Hop).
\end{equation}
This last expression leads to a particularly appealing relation: the error on the estimation of $x$ is bounded by 
\begin{equation}
	\delta x\geq\frac{1}{2 \sqrt{\nbrep} \hspace{3pt}\text{Var}(\dur\Hop)}.
	\label{VarH}
\end{equation}
This expression is strongly reminiscent of the usual Heisenberg uncertainty. Indeed, the operator $\dur\Hop$ generates the transformation associated with the parameter $x$. Hence, it can be interpreted as a generalized \textquote{conjugate variable} of the parameter $x$; then the fluctuations of $x$ and the variance of $\dur\Hop$ are linked by a Heisenberg-like inequality. To achieve a small uncertainty in $x$, it is necessary to use a probe with large uncertainty in $\Hop$. However, the generalized conjugate variable is in general protocol-dependent and different from the \textquote{true} conjugate variable of $x$.  
In particular, and as advertised earlier, this formalism can even describe situations when $x$ is a quantity with no true conjugate variable, for instance, if $x$ is the purity of a state. 
Finally, we may also want to estimate the duration $\dur$ itself; in this case, the encoding transformation is $e^{i\dur\Hop}$, and the inequality reads $\delta \dur \geq \frac{1}{2 \sqrt{\nbrep} \hspace{3pt} \text{Var}(\Hop)}$.

\section{Number and time resources}
\subsection{Standard Quantum Limit}

The QFI depends only on the probe state and encoding channel. In principle, it is still possible to reduce the quantum noise to arbitrarily small values, for instance by taking a Hamiltonian encoding and a pure state for which $\text{Var}(\Hop)\rightarrow\infty$. Such a probe, however, could in general only be prepared by spending an infinite amount of energy. Therefore, a central question in quantum metrology is to determine which precision can be reached with finite resources, and how to make the best possible use of said resources. Most of the literature has focused on two quantities: the number of probes, and the duration $\dur$ of the protocol. Indeed, both quantities are limited in most metrological protocols. The number of probes that can be controlled at once cannot grow indefinitely, both in atomic interferometry experiments \cite{pezze_quantum_2018} (because of collisions effects, among others), and in light interferometers (notably because of backaction noise). Similarly, the duration of a protocol is limited by decoherence or instability effects. \\

In this context, let us consider the class of protocols obeying the following properties:

\begin{itemize}
	\item The probe system is composed of a finite, fixed number $N$ of identical, distinguishable subsystems.
	\item Initially, the subsystems are in a factorizable state: $\rop_0=\bigotimes\rop_0^{(i)}$ (which means, in particular, that the subsystems are unentangled).
	\item The encoding map is a free evolution under a time-independent Hamiltonian $e^{-it\Hop(x)}$.
	\item The encoding Hamiltonian satisfies $[\partial_x\Hop(x),\Hop(x)]=0$. In particular, this is the case when the Hamiltonian is linear in $x$: $\Hop(x)=x\tilde{H}$.
	\item The Hamiltonian acts on each subsystem individually: $\Hop=\sum_i^{N} \hat{h}_i$, \textit{i.e.}, the encoding map is non-entangling.
\end{itemize}
Although it may seem restrictive, this set of hypotheses is satisfied by most atomic interferometric experiments performed today, in particular those used in state-of-the-art atomic clocks. 
Under these hypotheses, the state remains factorizable during the evolution: $\rop_x(\dur)=\bigotimes_i\rop_x^{(i)}(\dur)$. As a consequence, the QFI becomes additive in the number of components \cite{demkowicz-dobrzanski_chapter_2015}: $$\pazocal{I}_x\left(\bigotimes_i\rop_x^{(i)}\right)=\sum_i\pazocal{I}_x(\rop_x^{(i)}).$$

In particular, for pure states, Eq.\eqref{VarH} becomes:

\begin{equation}
	\delta x \geq\frac{1}{2\sqrt{\nbrep}\dur\sqrt{\sum_i^N\text{Var}^2(\hat{h}_i)}}=\frac{1}{2\dur\sqrt{\nbrep N}\text{Var}(\hat{h}_i)}\sim\frac{1}{\dur\sqrt{\nbrep N}}.
\end{equation}

Hence, in this situation, the precision scale linearly in the evolution time, and as the square-root of the number of components. Since the evolution acts on each subsystem individually, the protocol is equivalent to $N$ independent estimations running in parallel: hence the behavior in $\sqrt{N}$ can also be interpreted as a consequence of the central limit theorem. This bound is called the \textit{Standard Quantum Limit} (SQL), or sometimes the shot-noise limit \cite{demkowicz-dobrzanski_chapter_2015,pezze_quantum_2018}.

As we discussed, standard interferometric experiments are limited by the SQL bound. In the last two decades, considerable research has investigated how a precision beating the SQL bound could be obtained, without increasing the number of components or the protocol duration. In the remainder of this section, we will review several strategies that have been proposed to reach this goal.\\

Before this, however, let us mention that there is a second definition of SQL, which applies in the context of optomechanical sensors. As we discussed in the previous section, these devices are affected by both measurement  and backaction noises. When the device is driven by stronger light sources, more photons are used to probe the movable mirror. This reduces the measurement noise, but amplifies the backaction. Because of the trade-off between the two noises, there is an optimal drive power, for which a minimum error is reached. It is this error which is often referred to as the SQL in optomechanical sensing literature \cite{clerk_introduction_2010}. In short, in this case, the SQL arises from a trade-off between the measurement noise and the backaction noise, while the SQL in the case we consider arises solely from the measurement noise, when one has a fixed number of uncorrelated probes.

\subsection{Entangled state}

A first possibility to beat the SQL is to relax the separability assumption on the initial state. This is the strategy that has attracted the most attention so far, both theoretically and experimentally.

One of the most relevant examples is spin probes prepared in a spin-squeezed state. Without loss of generality, we will consider ensembles of spin polarized along the $y$-axis and squeezed in the $x$ direction. According to the Wineland criterion \cite{wineland_spin_1992}, such states obey the inequality $\xi_R^2=\frac{N\text{Var}^2(\Jx)}{\moy{\Jy}^2}<1$, with $\hat{J}_a$ the collective spin operators. Furthermore, the Heisenberg principle imposes \cite{pezze_quantum_2018}: $\text{Var}^2(\Jz)\geq \frac{N}{4\xi_R^2}$. Therefore, spin-squeezed states can achieve large variance for the $\Jz$ spin component. As such, they are useful probes for parameter estimation, according to the discussion of the previous section.
Indeed, if one of such states is used as a probe in a Ramsey interferometric experiment with encoding operator $e^{ixt\Jz}$, the achievable error is :
\begin{equation}
	\delta x=\frac{1}{\dur\sqrt{\nbrep N}\text{Var}(\Jz)}\leq\frac{\xi_R}{\sqrt{\nbrep N}\dur}<\frac{1}{\dur\sqrt{\nbrep N}}.
\end{equation}
Hence, using spin-squeezed states, it is possible to beat the SQL bound for this estimation task.

Spin-squeezed states are already available in many experiments, and as such are one of the most promising tools for quantum sensing tasks. It is, however, also interesting to study the $\textit{optimal}$ state, \textit{i.e.}, the state which allows us to maximize the achievable precision for a given number of probes. This optimal precision can be found through the following argument \cite{giovannetti_quantum_2006,boixo_generalized_2007}: let us consider an encoding Hamiltonian which acts independently on each subsystem, $e^{ixt\Hop}=e^{ixt\sum_i^N\hat{h}_i}$, and the associated QFI $4\text{Var}(\Hop)^2$. If the spectrum of $\Hop$ is bounded, then the following inequality holds: 

\begin{equation}
	4\text{Var}(\Hop)^2\leq\lvert\lvert\Hop\rvert\rvert^2,
	\label{seminormbound}
\end{equation}
with $\lvert\lvert\Hop\rvert\rvert$ a seminorm defined as: $\lvert\lvert\Hop\rvert\rvert=\mu_M-\mu_m$, where $\mu_M$ ($\mu_m$) is the maximum (minimum) eigenvalue of $\Hop$. Since the action on different subsystems are independent, $[\hat{h}_i,\hat{h}_j]=0$, and $\lvert\lvert\Hop\rvert\rvert=\sum_i^N\lvert\lvert\hat{h}_i\rvert\rvert$. Thus, if the encoding acts independently on each subsystem, the seminorm of $\Hop$ will scale linearly with the number of components, and the QFI will scale quadratically. In particular, if the action on each subsystem is identical, $\hat{h}_i=\hat{h}$, we have:
\begin{equation}
 	\delta x \geq \frac{1}{\sqrt{\nbrep}\dur N \text{Var}(\hat{h})}\sim\frac{1}{\sqrt{\nbrep}\dur N}.
 	\label{Heisenlimit}
 \end{equation}
This scaling of precision with $N$ is referred to as the \textit{Heisenberg limit} or \textit{Heisenberg scaling} in the literature. Note that despite its name, this limit is not a fundamental, unconditional limit like the Heisenberg uncertainty relations: this linear scaling is limiting only protocols involving a fixed number of probes, for which the encoding is made through a time-independent Hamiltonian linear in $x$, and, most importantly, acting independently on each subsystem. If these hypotheses are satisfied, the use of quantum correlations in the probe state allows to increase the precision from the SQL to the Heisenberg limit. \footnote{Note also that some works (for instance \cite{giovannetti_advances_2011}) have used the name \textit{Heisenberg limit} to refer instead to the optimal precision achievable with a Hamiltonian dynamic $\lvert\lvert\Hop\rvert\rvert^2$. Here, we will keep the more common definition and focus on the scaling of precision with respect to the number of probes.} 

The bound \eqref{Heisenlimit} can always be saturated by a state which is an equal superposition of the maximum and minimum eigenstates:
\begin{equation}
	\ket{\psi_0}=\frac{\ket{\mu_M}+\ket{\mu_m}}{2}.
\end{equation}
 For instance, for a set of qubits and an encoding operator $e^{ixt\Jz}$, the Heisenberg limit can be achieved by using Greenberger-Horne-Zeilinger (GHZ) states $\ket{\psi_0}=\frac{\ket{\downarrow\downarrow..\downarrow}+\ket{\uparrow\uparrow..\uparrow}}{\sqrt{2}}$.\\

More generally, the precision achievable can be linked to the entanglement within the state. For multipartite states, there is no single quantifier of entanglement. However, the notion of $k$-separability was shown to be related to the metrological power of a state. A pure state is $k$-separable if it can be written as a product of states $\ket{\psi}=\ket{\psi_1}\otimes\ket{\psi_2}\otimes..\otimes\ket{\psi_m}$, where each $\psi_i$ is an entangled state at most $k$ component \cite{guhne_multipartite_2005} (for a mixed state, a state is $k$-separable if it can be written as a convex sum of $k$-separable states.) For instance, the state $\frac{1}{2}\left(\ket{\uparrow\uparrow}+\ket{\downarrow\downarrow}\right)(\left\ket{\uparrow\uparrow}+\ket{\downarrow\downarrow}\right)$ is 2-separable but not 1-separable. The state $\left(\frac{\ket{\uparrow\uparrow\uparrow}+\ket{\downarrow\downarrow\downarrow}}{\sqrt{3}}\right)\ket{\uparrow}$ is 3-separable but not 2-separable. Hence, the notion of $k$-separability quantifies how many components are involved in the largest entangled cluster in the state. Two works, \cite{toth_multipartite_2012} and \cite{hyllus_fisher_2012}, have shown independently that if the probe is $k$-separable and the encoding map is non-entangling, we have

 $$\pazocal{I}_x\leq \dur([N/k]k^2+(N-[N/k]k)^2)\leq \dur Nk,$$ 

 with $[N/k]$ the integer part of $N/k$. For a separable state, $k=1$, we find the SQL $\pazocal{I}_x=TN$. For a GHZ state, $k=N$, and $\pazocal{I}_x=\dur N^2$. Between these two extremes, the precision is ultimately limited by $k$, the size of the largest entangled cluster in the system. \\

Finally, note also that a precision scaling linearly in $N$ can also be achieved in sequential (or multi-pass) protocols, where, instead of encoding once the parameter in $N$ probes, the parameter is encoded $N$ times in a single probe \cite{giovannetti_advances_2011,demkowicz-dobrzanski_chapter_2015}. Indeed, in that case, the duration of the protocol is multiplied by $N$, with a corresponding linear increase in precision. Hence, in the case described by Eq.\eqref{Heisenlimit}, it is equivalent to encode the parameter in parallel or sequentially.

\subsection{Entangling maps}

A second possibility to improve the precision is to encode the parameter through a map which creates correlations between subsystems, \textit{i.e.}, an entangling map. In \cite{boixo_generalized_2007}, Boixo and al. have considered Hamiltonians acting on $k$ subsystems at once, and symmetric under permutation. For instance, for a system of $N$ spins, the Hamiltonian $\Hop=\sigx\otimes\sigx\otimes\mathbb{1}_{N-2}+\sigx\otimes\mathbb{1}\otimes\sigx\otimes\mathbb{1}_{N-3}+...+\mathbb{1}_{N-2}\otimes\sigx\otimes\sigx$ satisfies the above conditions, for $k=2$. Such a Hamiltonian involves ${N\choose k}$ terms with an identical spectrum. The seminorm of the total Hamiltonian is bounded by the sum of the seminorm of each term: $\lvert\lvert\Hop\rvert\rvert\leq{N\choose k}\lvert\lvert\hat{h}^{(k)}\rvert\rvert$. Hence, the achievable precision is limited by $\delta x\geq\frac{1}{\sqrt{\nbrep}\dur{N\choose k}\lvert\lvert\hat{h}^{(k)}\rvert\rvert}$. In the case $k\ll N$, this reduces to:

\begin{equation}
	\delta x\geq\frac{k!}{\sqrt{\nbrep}\dur N^k\lvert\lvert\hat{h}^{(k)}\rvert\rvert}\sim\frac{1}{\sqrt{\nbrep}\dur N^k}.
\end{equation}

Hence, estimation protocols based on entangling maps can in principle achieve super-Heisenberg scaling in the number of particles. As earlier, the optimal state is an equal superposition of maximum and minimum eigenstates. Because the Hamiltonian is no longer separable, however, this state is not necessarily a massively entangled state like a GHZ state.
 With entangling maps, the input state itself does not need to be entangled to reach a high precision. It has been shown \cite{boixo_quantum-limited_2008} that, for the symmetric maps we have introduced above, separable input states could achieve a precision bounded by:

 \begin{equation}
 	\delta x\geq \frac{1}{\sqrt{\nbrep}\dur N^{k-1/2}}.
 \end{equation}
Hence, even using separable probe states, it is possible to achieve super-Heisenberg scaling in that case.
 These entangling maps can be implemented, for instance, using interaction processes between atoms in high-density clouds or BEC \cite{choi_bose-einstein_2008,braun_quantum-enhanced_2018}. Note that interaction strengths can not be measured directly and require indirect probing processes, which can be captured by the above formalism.

So far, we have considered $k\ll N$, ie, a protocol entangling only a small fraction of the total particle number. The best precision, however, will be achieved when $k$ is a significant fraction of $N$. In particular, when $k=N/2$, we will have 

\begin{equation}
	\delta x\geq\frac{1}{\sqrt{\nbrep}\dur2^N},
\end{equation}
that is, an \textit{exponential} scaling with $N$. Note however that to achieve this scaling, $k$ needs to increase with $N$; that is, the map needs to entangle more and more subsystems as the total system size grows. This is very unlikely to be realizable in practice. However, it remains a result of fundamental interest to understand the respective roles played by entangled probe states and entangling maps. \footnote{This kind of exponential scaling was also reported in \cite{roy_exponentially_2008}, with a slightly different protocol. In this case, the exponential scaling could even be achieved with a \textquote{trivial} input state {$\ket{\downarrow\downarrow..\downarrow}$}.}

Finally, another example of non-trivial maps concerns the preparation of a system near a critical point. This will be the focus of Chapter 5.

\subsection{Time-dependent Hamiltonians and quantum control}

So far we have shown how to make the best of a limited number of probes. We will now consider how similar scaling improvements can be achieved with respect to the duration $\dur$. When the encoding map is a time-independent Hamiltonian, the QFI scales like $\dur^2$. To improve this, it is necessary to consider time-dependent Hamiltonian. In that case, the encoding map is given by a unitary $U_x(0\rightarrow \dur)$, which can be expressed as a function of $\Hop_x(t)$ through time-ordering operators (in the specific case $[\partial_t\Hop_x(t),\Hop_x(t)]=0$ for all $t$, we have $U_x(0\rightarrow \dur)=e^{i\int_0^\dur\Hop_x(t)dt}$). If the probe state is pure, the QFI is given by $\pazocal{I}_x=4\text{Var}(\mathcal{H}_x(\dur))\lvert_{\ket{\psi_0}}$, according to Eq.\eqref{QFIpure}. For time-independent Hamiltonian, the QFI can be bounded by the difference between the maximum eigenvalues of $\Hop_x$, as we described in \eqref{seminormbound}. In \cite{pang_optimal_2017}, Pang and al. have found that a similar expression can be obtained in the time-dependent case:

\begin{equation}
	\pazocal{I}_x\leq\left[\int_0^\dur(\mu_M(t)-\mu_m(t))dt\right]^2,
	\label{controlbound}
\end{equation}
with $\mu_M(t)$ ($\mu_m(t)$) the instantaneous maximum (minimum) eigenvalues of $\partial_x\Hop_x(t)$. It is possible to saturate this bound if the two following conditions are met: 1) the state is initially prepared in the superposition $\frac{\ket{\mu_M(0)}+\ket{\mu_m(0)}}{\sqrt{2}}$, and 2) at all times, $\Uop(0\rightarrow \dur)\ket{\mu_m(0)}=\ket{\mu_m(\dur)}$ and $\Uop(0\rightarrow \dur)\ket{\mu_M(0)}=\ket{\mu_M(\dur)}$, meaning that the probe remain in a superposition of instantaneous maximum and minimum eigenvalue at all times. This last condition can be difficult to satisfy: in general, the eigenstates at time $t$ do not remain eigenstates during the evolution if the Hamiltonian is time-dependent (the case $[\partial_t\Hop,\Hop]=0$ being an exception). However, this problem can be circumvented by the use of quantum  control. In this case, the state evolves according to $\Hop_x(t)+\Hop_c(t)$, with $\Hop_c(t)$ a control Hamiltonian independent of the parameter to estimate $x$. By implementing an appropriate control, Pang and al. have shown that the bound \eqref{controlbound} can be saturated. If the eigenvalues are time-independent, we find again that the QFI scales like $\dur^2$. However, time-dependent eigenvalues also become possible, which leads to non-trivial time scalings. For example, this formalism has been applied to a qubit evolving under a uniformly rotating magnetic field \cite{pang_optimal_2017}, with an evolution described by a Hamiltonian $$\Hop(t)=-B(\cos(\omega t)\sigx+\sin(\omega t)\sigz).$$ This system can be used to evaluate the magnetic field frequency $\omega$. In this case, it was shown that the QFI was given by:
\begin{equation*}
	\pazocal{I}_\omega=B^2\dur^4,
\end{equation*}
that is, a \textit{quartic} time scaling. Further analysis of this process was made in \cite{gefen_control_2017}. This shows that control methods are promising to improve the performances of metrological protocols with limited encoding time.

\section{Metrology with photons}
So far, we have considered a fixed number of distinguishable particles, which is well adapted to describe most interferometric protocols involving atoms or ions. However, many relevant cases do not fall under this definition. The most important of these exceptions are protocols where light is used as a probe. Photons have two properties that justify a separate study. First, a photonic mode is described by a Hamiltonian unbounded from above, and can be populated by a superposition of different photon number. Second, the photons composing a state of light are, in general, indistinguishable. This last property is shared by other physical systems of metrological relevance, such as BECs. In this section, we will review some consequences of these two properties.

\subsection{Number superposition}

The possibility to superpose different numbers of particles can have drastic consequences on the achievable precision.

To illustrate this, let us consider a pure photonic state corresponding to a single mode of the electromagnetic field (for instance, a field propagating inside a waveguide), described by the creation operator $\adag$. This state is submitted to a phase shift $e^{ix\nop}$, with $\nop=\adag\aop$. The state is then measured (for instance through homodyne measurement), and the measurement results are used to estimate $x$. 
The QFI of this procedure is given by $\pazocal{I}_x=4\text{Var}^2(\nop)$. Hence, the optimal precision can be reached when maximizing the variance of $\nop$, with a fixed amount of resources. By analogy with the previous discussion, we will consider the average number of photons $\moy{\nop}$ as a resource quantifier. If the input state is a coherent state $\ket{\alpha}$, we have: $\moy{\nop}=\text{Var}(\nop)=\lvert\alpha\rvert$. Hence, the precision in this case is given by:

\begin{equation}
	\delta x \geq \frac{1}{\sqrt{\nbrep\moy{\nop}}}.
\end{equation}
Hence, we find a SQL-like bound in that case. Now, let us consider the following superposition of two Fock states: $\ket{\psi}=\sqrt{1-p}\ket{0}+\sqrt{p}\ket{K}$. Then $\moy{\nop}=pK$ and $\moy{\nop^2}=pK^2$. This means that we have $\pazocal{I}_x=4\moy{\nop}^2\frac{1-p}{p}$. Hence, by choosing an arbitrarily small $p$, it would appear that an infinite precision could be reached with a finite number of photons.\footnote{Additionally, an arbitrary \textit{scaling} may also be achieved: we can define a class of states, each with a different value of $K$. 
 If $p$ evolves like a power of $K$, $p=K^{-m}$, then we obtain $\pazocal{I}=4\nop^{1+\frac{1}{1-m}}$which can give an arbitrarily large scaling for $m\sim1$.} A similar reasoning can be made, for instance, for a superposition between the vacuum and a squeezed state \cite{rivas_sub-heisenberg_2012}.
There is, however, an important caveat in the argument above. To measure the state, we have assumed that we had access to any POVM, including homodyne measurements. However, such a measurement involves an external coherent state, which itself carries a vast number of photons in the measurement device. Formulated differently, a phase can only be defined with respect to a certain reference. A system, however, can only act as a perfect phase reference if it carries infinite energy. Hence, focusing only on $\moy{\nop}$ can lead us to underestimate the energetic resources involved in the estimation problem. This problem can be formulated in the language of quantum reference frame \cite{bartlett_reference_2007,jarzyna_quantum_2012,safranek_quantum_2015}. 
 It illustrates that giving a proper definition for the resources involved in a protocol can become a non-trivial issue when superposition between different number states are allowed. If no perfect phase reference is available, only a relative phase shift can be measured, for instance between the two arms of a Mach-Zehnder interferometer. 
\footnote{In \cite{benatti_sub-shot-noise_2013}, it was shown that a QFI {$\frac{\moy{\nop}^2}{p}$} can also be reached in a Mach-Zehnder-like interferometer, using a probe state {$\sqrt{1-p}\ket{00}+\sqrt{p}\ket{N0}$}. However, a careful analysis of the protocol reveals that the phase information is encoded both in the relative and the absolute phase of the state. Therefore, the precision is once more attributable to a global phase shift measured with a homodyne scheme. If the information is encoded only in the relative phase, the QFI limited by {$\moy{\nop}$}.}

In most experimental scenarios, however, energy shortage is not a limiting factor, and homodyne measurements are indeed available. By contrast, the average number of photons can generally not be increased arbitrarily, in particular because of backaction noise \cite{clerk_introduction_2010}. Hence, it is indeed relevant to optimize the scaling of precision versus $\moy{\nop}$. While keeping the above discussion in mind, we will keep $\moy{\nop}$ as the resource quantifier in the following.\\

\subsection{Indistinguishability}

In the case of distinguishable particles, we have shown that metrological performances are often (but not exclusively \cite{braun_quantum-enhanced_2018}) linked with entangled input states or entangling gates. For indistinguishable particles, however, the very notion of entanglement is very challenging to define. This is still an open and debated topic: see, among others, \cite{eckert_quantum_2002,ghirardi_general_2004,tichy_essential_2011,benatti_bipartite_2012,benatti_sub-shot-noise_2013,benatti_entanglement_2014,killoran_extracting_2014,demkowicz-dobrzanski_chapter_2015,sciara_universality_2017,braun_quantum-enhanced_2018} for some of the perspectives adopted on these issues. Here we will discuss a few important points in this debate that are related to quantum metrology.\\ 

To introduce the problem, let us start from a simple observation: for distinguishable particles, entanglement is typically difficult to produce. However, for indistinguishable bosons, there are many cases in which it apparently comes for free, due to the symmetrization principle \cite{benatti_sub-shot-noise_2013}. As an example, consider the Hong-Ou Mandel (HOM) experiment: two indistinguishable photons are sent in the two input ports of a balanced beam-splitter (BS). We will call $\ket{n_{a(b,c,d)}}$ the Fock basis describing the ports $a(b,c,d)$ of the BS, $a$ and $b$ being the two input ports, and $c$ and $d$ the output. Before the BS, the state of the system is $\ket{1_a,1_b}$. After the BS, the state is $\frac{\ket{2_c,0_d}+\ket{0_c,2_d}}{\sqrt{2}}$, which exhibits entanglement between the two output ports. More generally, if we send a Fock state $\ket{k,N-k}=\frac{(\adag)^k(\hat{b}^\dagger)^{N-k}}{\sqrt{k!(N-k)!}}\ket{0}$ in a BS with a reflexivity $\theta\neq0$, it will come out entangled. 
Hence, simply using linear optics elements on indistinguishable photons can generate entanglement.
It was shown in \cite{benatti_sub-shot-noise_2013} that this process is metrologically useful; if we want to evaluate the reflexivity $\theta$ of the BS, for instance, sending in a Fock state allows to reach a QFI $\pazocal{I}_\theta=4[N(2k+1)-2k^2]$, which is higher than $4N$ as soon as $k\neq0$ and $k\neq N$, and can reach $N^2$ scaling for $k=N/2$. If we send coherent states instead, the QFI will only scale like the average number of photons.\\

There are several ways in which this entanglement generation can be interpreted.
To study the entanglement properties of a system, we need first to define a partition of the system into several components. Entanglement then corresponds to quantum correlations between these subcomponents; or, more precisely, to the possibility to attribute well-defined properties to the whole system, but not to its individual components.
 For a fixed number of distinguishable particles, there is a natural partition of the system: each particle corresponds to a component. 
 For indistinguishable particles, the situation is more complex. For concreteness, we will discuss the case of electromagnetic field propagating in two waveguides, but the concepts presented here are very general. To describe the system, a first possibility is to switch to second-quantization formalism. In that case, the components of the system are the two waveguides, or more precisely the two \textit{modes} of the field inside these waveguides. Hence, we have a system composed of two components. The state of each component corresponds to the photon statistics within each mode, which is expressed in the two Fock bases $\ket{n}_a$ and $\ket{n}_b$. 
 Entanglement is then defined as nonseparability with respect to this partition into modes. States that can be expressed as $\ket{\psi}_a\otimes\ket{\phi}_b$ are \textit{mode-separable}. The states that cannot be expressed as (a statistical mixture of) mode-separable states are \textit{mode-entangled}.
 Now, how can we describe the HOM experiment in this framework? In the mode picture, the input state reads $\ket{1}_a\otimes\ket{1}_b$, which is clearly mode-separable. However, the action of the BS can be expressed by a unitary transformation $\Uop=e^{i\theta(\adag\bop+\aop\bdag)}$. This map does \textit{not} act independently on each mode; as a consequence, it can produce mode-entanglement. Hence, the HOM effect is interpreted as the application of an entangling encoding map on a separable state. When we send in a state $\ket{k,N-k}$ to evaluate the reflexivity of the BS, we still have only two modes, and therefore two components. Hence, the $N^2$ scaling of $\pazocal{I}_\theta$ is not linked with the number of components, but with the energy excitation within each component. \\

Alternatively, we may also keep the description at the first quantization level. In that case, the components of the system are the individual photons. Each component can be in either of two states $\ket{a}$ or $\ket{b}$, which corresponds to its presence in either of the two arms. 
 Due to the symmetrization principle, the input state in the HOM experience can be \textit{formally} written as $\frac{\ket{ab}+\ket{ba}}{\sqrt{2}}$. From the point of view of individual photons, the state appears non-separable, a property which we will refer to as \textit{formal entanglement} \footnote{We must emphasize that this entanglement is not linked with directly observable properties. If we consider two distinguishable qubits prepared in the state {$\frac{\ket{\uparrow\downarrow}+\ket{\downarrow\uparrow}}{\sqrt{2}}$}, the presence of entanglement can be witnesses by the possibility to assign definite properties to the whole system, but not to each subsystem. If we measure the spin state, we can predict with certainty that one particle will be found in {$\ket{\downarrow}$} state and the other in {$\ket{\uparrow}$}. However, if we measure one specific qubit, we cannot predict in advance the result.
 In other words, we can describe with certainty the state of the entire system, but not of each subsystem. By contrast, if the particles are indistinguishable, we cannot even make a measurement which addresses the state of a definite subsystem, since they are indistinguishable. The only statements that can be made are \textquote{One particle is in the state {$\ket{\uparrow}$}} and \textquote{One particle is in the state {$\ket{\downarrow}$}}, 
 each of which occurs with probability one. In other words, the results of the measurements we can perform on the system can be predicted with certainty. Since there is no physical way to address one definite particle, the statement: \textquote{one particle is in state {$\ket{\uparrow}$} and the other in state {$\ket{\downarrow}$}} is a complete description of the system. Contrary to usual entanglement or mode-entanglement, formal entanglement cannot be interpreted in terms of incomplete knowledge about the system. For more details about this interpretation of entanglement as a lack of information, see \cite{ghirardi_general_2004,benatti_entanglement_2014}.}.
  In this perspective, the BS acts on each particle individually, and therefore can not create entanglement. This provides a second interpretation of the HOM effect: the BS does not create entanglement. Rather, it reveals preexisting \textit{formal} entanglement in the input state (see in particular \cite{killoran_extracting_2014} for developments on this idea). When using the state $\ket{k,N-k}$ to evaluate the reflexivity of a BS, we are applying a separable encoding map on an entangled state made of $N$ components. In this interpretation, the $N^2$ scaling of $\pazocal{I}_\theta$ is indeed linked with the number of components.

The physical relevance of the notion of formal entanglement is controversial \footnote{ In addition to the absence of interpretation in terms of incomplete information, another critic one could make is that this notion is useless because it applies to every state. Indeed, it would appear at first sight that the symmetrization principle entails the existence of correlations between any pair of photons; and that \textit{any} particle-symmetric state should be formally entangled, except if all photons were to be in the same mode. This is, however, not the case. Consider, for example, two coherent states sent in both input ports of a BS, which reads {$\ket{\alpha}_a\ket{\beta}_b$} in the mode picture. In the particle picture, this state is written as {$\ket{\psi}=e^{-\frac{\lvert\alpha\rvert^2+\lvert\beta\rvert^2}{2}}\sum_{N=0}^\infty\frac{\sqrt{\lvert\alpha\rvert^2+\lvert\beta\rvert^2}^N}{\sqrt{N!}}\ket{\psi^{(N)}}$}, where each {$\ket{\psi^{(N)}}$} is a state of {$N$} photons which reads: {$\ket{\psi^{(N)}}=\left(\frac{\alpha}{\sqrt{\lvert\alpha\rvert^2+\lvert\beta\rvert^2}}\ket{a}+\frac{\beta}{\sqrt{\lvert\alpha\rvert^2+\lvert\beta\rvert^2}}\ket{b}\right)^{\otimes N}$}, that is, a product state where each photon is in the same quantum state. Hence, this state is not formally entangled, despite the fact that the photons are not in the same mode. If such a state is sent into a BS, the output will still be a coherent state in both arms; hence, sending this particle-separable state into a BS does not create entanglement, which is consistent with the discussion above.}. However, despite its formal nature, the particle picture can still provide useful hindsight to describe systems of indistinguishable particles and to give an interpretation of some forms of metrological advantages \cite{demkowicz-dobrzanski_chapter_2015}. We will further elaborate on these ideas in Chapter 6.

\subsection{Metrology with Gaussian states and channels}
In the last subsections, we have shown that the metrological performances of photonic states arise both as a consequence of indistinguishability and number statistics, making it more arduous to make a general study of metrologically useful states and channels. It is possible, however, to characterize some specific relevant cases. Gaussian states, in particular, are especially interesting because of their experimental availability.
The metrological properties of these states are discussed extensively in \cite{safranek_gaussian_2016}, we will recall a few key elements here. Gaussian states are best described in phase space, where they take a particularly simple form. We consider $q$ modes of the electromagnetic field, and collect the creation and annihilation operators in a vector $\mathbf{A}=(\hat{\textbf{a}},\hat{\textbf{a}}^{\dagger})^T$, with $\hat{\textbf{a}}=(\hat{a}_1,...,\hat{a}_q$). We will study the symmetric characteristic function $\text{Tr}[\rop e^{\mathbf{A}^{\dagger} K \xivect}]$, where $\xivect$ is in $\mathbb{C}^{2q}$ and can be written like $\xivect=\mathbf{x}\oplus\mathbf{x}^*$, and $K$ is the so-called symplectic form, which here reads as:

\begin{equation}
K=\left[ {\begin{array}{cc}
\mathbb{1}_q & 0\\
0 & -\mathbb{1}_q\\	
\end{array}} \right],
\end{equation}
where $\mathbb{1}_q$ is the $q\times q$ identity matrix. For Gaussian states, these functions take a Gaussian form, \textit{i.e.}
\begin{equation}
\text{Tr}[\rop e^{\mathbf{A}^{\dagger} K \xivect}]=\exp\Big(-\frac{\xivect^{\dagger}\sigma\xivect}{4}-i\ddagvect K\xivect\Big).
\end{equation}

Here $\sigma$ is the covariance matrix and $\dvect$ the displacement vector, defined as:
\begin{subequations}
\begin{equation}
	\dvect^i=\text{Tr}[\rop\mathbf{A}^i],
\end{equation}
\begin{equation}
	\sigma^{ij}=\text{Tr}[\rop\{\mathbf{A}^i-\dvect^i,\mathbf{A}^{j\dagger}-\dvect^{j\dagger}\}],
\end{equation}
\end{subequations}

with $\{A,B\}=AB+BA$. The simplest examples of Gaussian states are thermal states, whose density matrix are given by $\rop_{\text{th}}(T_1,...T_q)=\bigotimes_i^q\frac{1}{Z_i}e^{-\frac{\hbar\omega_i \adag_i\aop}{k_BT_i}}$. Here $k_B$ is the Boltzmann constant, $T_i$ is the temperature and $\omega_i$ the frequency of the $i$-th mode. In the phase space, these states satisfy $\dvect=0$ and $\sigma=D(\nu_1,...\nu_q)$, where $D$ is a $2q\times2q$ diagonal matrix:

\begin{equation}
	D(\{\nu_i\})=\text{diag}(\nu_1,\nu_2,...\nu_q,\nu_1,...\nu_q), \hspace{10pt} \nu_i=\text{cotanh}\left(\frac{\hbar\omega_i}{2k_BT_i}\right).
\end{equation}

\begin{figure}
	\centering
	\includegraphics[scale=0.3,angle=-90]{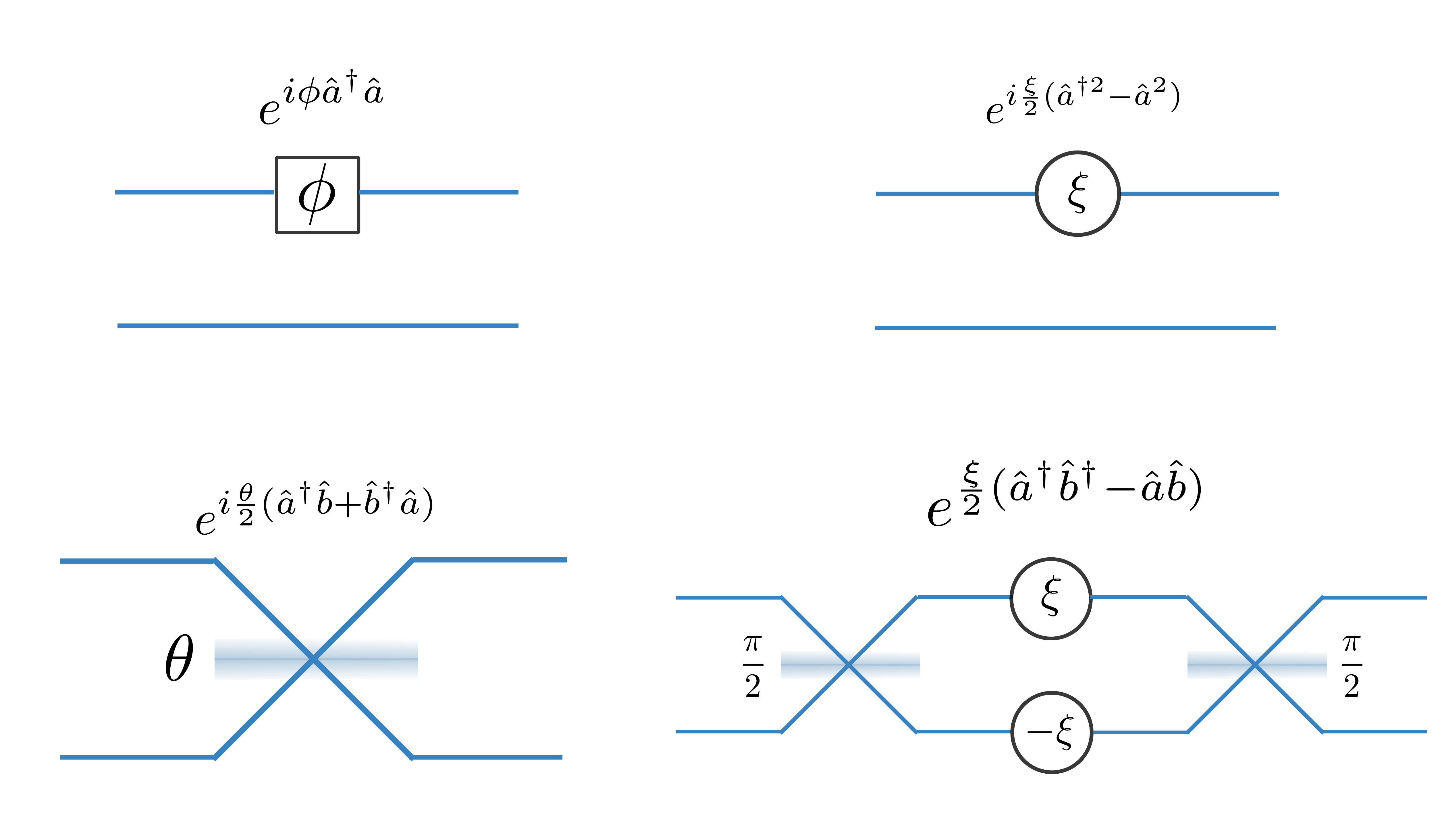}
	\caption[Schematic depiction of two-modes Gaussian operations.]{Pictorial representation of two-modes Gaussian operations. Top left: phase-shifting, top right: one-mode squeezing, bottom left: mode-mixing by a beam-splitter, bottom right: two-mode squeezing.}
	\label{Gaussianop}	
\end{figure}

Hence, a thermal state is uniquely defined by the set of parameters $\nu_i$. Every other Gaussian state can be obtained by applying unitary Gaussian operations on a thermal state:
$\rop=\hat{G} \hspace{5pt}\rop_{\text{th}}(\{T_i\}) \hspace{5pt} \hat{G}^\dagger$ \cite{ferraro_gaussian_2005}.
Gaussian operations $\hat{G}$ include displacement, squeezing, mode mixing, phase shifting, and compositions thereof (see Fig.\ref{Gaussianop}). In Hilbert space, these transformations are \textit{linear} transformations of the operators $\mathbf{A}$. In phase space, this translates into a linear transformation of both first and second moments according to: $\dvect\rightarrow \mathpzc{G} \dvect + \bm{\gamma}_\mathpzc{G}$ and $\sigma\rightarrow \mathpzc{G} \sigma \mathpzc{G}^{\dagger}$, where $ \bm{\gamma}_\mathpzc{G}$ is a vector and $\mathpzc{G}$ is a $2q\times2q$ \textit{symplectic matrices}, meaning that it verifies

\begin{equation}
 	\mathpzc{G} K \mathpzc{G}^\dagger=K.
 \end{equation} 

Therefore, the covariance matrix of every Gaussian state, pure or mixed, can be decomposed as:
$$\sigma=\mathpzc{G}_\sigma D(\nu_1,..\nu_q) \mathpzc{G}_\sigma^\dagger.$$ This decomposition is unique. The parameters $\{\nu_i\}$ are the \textit{symplectic eigenvalues} of $\sigma$. Note that they are not necessarily the usual eigenvalues of $\sigma$, since in general $\mathpzc{G}_\sigma^\dagger\neq\mathpzc{G}_\sigma^{-1}$; instead, they correspond to the eigenvalues of the matrix $\Mmat=K\sigma$.\\

If the input state \textit{and} encoding channels are both Gaussian, the state will remain Gaussian throughout the evolution (this is the case, for instance, when one uses coherent or squeezed state to evaluate a phase shift between the two arms of a Mach-Zehnder interferometer). In this case, we obtain parameter-dependent displacement vector $\dvect_x$ and covariance matrix $\sigma_x$, in terms of which the QFI can be expressed. In particular, it was shown in \cite{monras_phase_2013,banchi_quantum_2015} that the QFI can be written on the following implicit form:

\begin{equation}
	\pazocal{I}_x=\frac{1}{2}\text{Tr}[(\dot{\sigma}_x\mathcal{Y})^2] + 2\dot{\dvect}_x^\dagger\sigma^{-1}\dot{\dvect}_x,
\end{equation}

where $\mathcal{Y}$ is a solution of the equation $\dot{\sigma}_x=\sigma_x \mathcal{Y}\sigma_x-K\mathcal{Y}K$, and the dot means derivative with respect to $x$. If all the symplectic eigenvalues $\{\nu_i\}$ are larger than $1$, it was shown in \cite{safranek_quantum_2015-1} that the QFI can be rewritten explicitly as an infinite series:

\begin{equation}
	\pazocal{I}_x=\frac{1}{2}\sum_{n=1}^\infty \text{Tr}[(\Mmat_x^{-n}\dot{\Mmat}_x)^2] + 2\dot{\dvect}_x^\dagger\sigma_x^{-1}\dot{\dvect}_x.
\end{equation}

In the following, we will be particularly interested in \textit{isotropic} states, which satisfy $\nu_i=\nu$ for all $i$. Furthermore, we will assume that the symplectic eigenvalues are independent of $x$: $\dot{\nu}=0$ (this is the case, in particular, when the encoding channel involves only passive linear elements such as beam-splitter or phase shifters). With these assumptions, the QFI can be further simplified (see Appendix B for details):

\begin{equation}
	\pazocal{I}_x=\frac{1}{2(\nu^4-1)}\left[\nu^4Tr[(\sigma_x^{-1}\dot{\sigma}_x)^2]+Tr[(K\dot{\sigma}_x)^2]\right]+ 2\dot{\dvect}_x^\dagger\sigma_x^{-1}\dot{\dvect}_x.
\end{equation}

Furthermore, under these assumptions, we also have $\sigma_x^{-1}\dot{\sigma}_x\sigma_x^{-1}=-\frac{K\dot{\sigma}_xK}{\nu^2}$ , which allows one last simplification:

\begin{equation}
	\pazocal{I}_x=-\frac{Tr[(K\dot{\sigma}_x)^2]}{2(1+\nu^2)}+2\dot{\dvect}_x^\dagger\sigma_x^{-1}\dot{\dvect}_x=\frac{\nu^2Tr[(\sigma_x^{-1}\dot{\sigma}_x)^2]}{2(1+\nu^2)}+2\dot{\dvect}_x^\dagger\sigma_x^{-1}\dot{\dvect}_x.
\end{equation}

By exploiting the formulas above, it is possible to fully characterize the precision achievable with Gaussian input probe, for various encoding channels. This was done for single-mode states in \cite{pinel_quantum_2013} (with refinements in \cite{safranek_optimal_2016}), and for two-mode states (both isotropic and anisotropic) in \cite{safranek_optimal_2016}. In this last work, several Gaussian channels were considered: phase-shifting, squeezing, mode-mixing, and two-mode squeezing. For all these protocols, it was shown that the best achievable QFI scales like $\moy{\nop}^2$.
Attaining this precision requires using squeezed input states; coherent input probes give a QFI linear in $\moy{\nop}$. By contrast, mode entanglement does not seem to play a significant role.

\subsection{Non-Gaussian channels}
We have seen that, for distinguishable particles, the precision could be increased by using non-trivial channels. For indistinguishable particles, we have so far focused on Gaussian channels, which act linearly on the set of creation and annihilation operators. To conclude this section, we will now briefly discuss the effect of nonlinear channels. The simplest example is given by Kerr nonlinearity, which can be expressed by the unitary map: $e^{ix(\adag\aop)^2}$. More generally, let us consider an encoding channel $e^{ix(\adag\aop)^k}$: how precisely can we estimate the parameter $x$? It was argued, on general grounds, that a scaling $\pazocal{I}_x\propto \moy{\nop}^{2k}$ could be achieved in that case \cite{luis_quantum_2007}. In \cite{tilma_entanglement_2010}, several input states which achieve this scaling have been exhibited. It was shown that mode entanglement was again unnecessary to achieve this precision, and that separable probes could beat non-separable ones in some cases.
Similarly, in \cite{berrada_quantum_2013}, it was shown that a higher precision could be reached by increasing the order of nonlinearity.\\

\section{Effect of dissipation}

So far we have considered perfect protocols, with no error outside of the fundamental measurement noise. However, every quantum estimation protocol is affected by decoherence, which can have a dramatic effect on the achievable precision. Although the effect of decoherence will depend on the specifics of each system, its importance can be illustrated by a simple example. Let us consider $N$ qubits, each prepared in the superposition $\frac{\ket{0}+\ket{1}}{\sqrt{2}}$. Each qubit evolves independently under to a channel $e^{ix\sigz \dur}$, with $x$ the parameter to be evaluated. In the absence of dissipation, the precision of the protocol is bounded by the limit $\delta x=\frac{1}{\sqrt{N}\dur}$. If, by contrast, the qubits are also experiencing dephasing noise, we obtain a precision \cite{huelga_improvement_1997}: $$\delta x=\sqrt{\frac{1-\cos^2(x \dur)e^{-2\Gamma \dur}}{N\dur^2e^{-2\Gamma \dur}\sin^2(x \dur)}},$$ where $\Gamma$ is the dephasing rate. Contrary to the dissipationless case, this error does not decrease monotonically with $\dur$; there is a trade-off between the acquisition of information and the blurring effect of dephasing. 
The maximum precision is achieved for a duration $\dur_c=1/(2\Gamma)$, and is equal to $\delta x_{\text{opt}}=\sqrt{\frac{2e\Gamma}{N \dur_c}}$, which has again a square-root scaling in $N$ and $\dur_c$. If, instead of a separable state, the probes are initially in a GHZ-like state $\frac{\ket{00..0}+\ket{11..1}}{\sqrt{2}}$, a precision $\delta x=\frac{1}{N\dur}$ could in theory be achieved. However, in the presence of dissipation, the GHZ state is dephased at an improved rate $N\Gamma$, much faster than the separable state. As a consequence, the optimal precision will be reached with a much shorter time $\dur_c=\frac{1}{2N\Gamma}$. The associated error is $\delta x_{\text{opt}}=\sqrt{\frac{2e\Gamma}{N \dur_c}}$, which is exactly the scaling we obtained with separable input. Since the evolution protocol is much shorter in this case, we could improve the precision by repeating the experiment many times. However, if we repeat the experiment $\nbrep$ times, we will obtain an error 

\begin{equation}
\delta x_{\text{opt}}=\sqrt{\frac{2e\Gamma}{\nbrep N \dur_c}}=\sqrt{\frac{2e\Gamma}{N \dur}},
\label{GHZdephas}
\end{equation}

where $\dur=\nbrep \dur_c$ is the total duration of the protocol. In particular, if $\nbrep=N$, we have $\dur=\Gamma/2$. Compared to the separable input case, we achieve exactly the same precision in the same amount of time. In other words, when dephasing is present, we can either consider a long experimental sequence with $N$ uncorrelated probes, or a series of short rounds with $N$ entangled probes; when the duration of each round is optimized, both protocols give the same results.

Hence, the presence of dephasing can completely destroy the quantum advantage offered by entanglement and restore shot-noise scaling. Note also that this result is independent of the value of $\Gamma$, meaning that even an infinitesimally small amount of dissipation is sufficient to destroy the scaling advantage. This insight has been generalized by several works. In particular, a very general bound was derived in \cite{escher_general_2011}, and further developed in \cite{demkowicz-dobrzanski_elusive_2012}. During the evolution stage, the probe evolves according to a channel $\Lambda_x$, which describes the effect of both the encoding procedure and the dissipation. This quantum channel can be decomposed in a non-unique way in terms of a Kraus map:

\begin{equation}
	\rop_x=\Lambda_x[\rop_0]=\sum_j\hat{K}_j(x)\rop_0\hat{K}_j^\dagger(x)
\end{equation}

with the Kraus operators satisfying the condition $\sum_j\hat{K}_j^\dagger(x)\hat{K}_j(x)=1$. Then the QFI is bounded by the following expression: $\pazocal{I}_x\leq[\moy{\hat{\alpha}_K}-\moy{\hat{\beta}_K}^2]$, with:

\begin{align}
	\hat{\alpha}_K=4\sum_j\frac{d\hat{K}^\dagger_j(x)}{dx}\frac{d\hat{K}_j(x)}{dx}\\ \nonumber
	\hat{\beta}_K=2i\sum_j\frac{d\hat{K}^\dagger_j(x)}{dx}\hat{K}_j(x)
\end{align}

In general, the evolution channel can be decomposed into Kraus operators in several ways; therefore, an upper bound for the precision can be obtained by minimizing over all possible decompositions. It was actually shown in \cite{escher_general_2011} that this bound is attainable, which implies:

\begin{equation}
	\pazocal{I}_x=\text{min}_{[\hat{K}_j(x)]}[\moy{\hat{\alpha}_K}-\moy{\hat{\beta}_K}^2]
\end{equation}

When no dissipation is present and the evolution is unitary, we have a single Kraus operator $\hat{K}(x)=\Uop_x$, and we retrieve the expression given in \eqref{QFIpure}. Let us now consider $N$ probes, and assume that the encoding process and the dissipation act independently on each probe. Then we can write the Kraus map as a direct product of operators acting on each component. In this case, the term $\moy{\hat{\alpha}_K}$ scales like $N$, while the term $\moy{\hat{\beta}_K}^2$ can scale like $N^2$. Therefore, an interplay between the SQL and the Heisenberg scaling emerge from this expression. Depending on the shape of the dissipation channel, one may retrieve either scaling. This formalism was applied in several paradigmatic cases. The first situation is the estimation of a phase-shift in a photonic interferometer, in the presence of photon loss. In this case, it was found that the QFI is bounded by:

\begin{equation}
	\pazocal{I}_x\leq\frac{4\gamma\moy{\nop}\moy{\Delta\nop^2}}{\moy{\Delta\nop^2}(1-\gamma)+\gamma\moy{\nop}}
	\label{dissipphotonQFI}
\end{equation}
with $\moy{\nop}$ $\moy{\Delta\nop^2}$ the mean and square variance of the total photon number, respectively, and $\gamma$ quantifies the photon loss (from $\gamma=1$, lossless case, to $\gamma=0$, complete absorption). For $\frac{\moy{\Delta\nop^2}}{\moy{\nop}}\ll\frac{\gamma}{1-\gamma}$, the QFI is given by the Heisenberg limit $\pazocal{I}_x\leq4\moy{\Delta\nop^2}$. However, for $\frac{\moy{\Delta\nop^2}}{\moy{\nop}}\gg\frac{\gamma}{1-\gamma}$, we obtain the linear scaling $\pazocal{I}_x\leq4\frac{\gamma}{1-\gamma}\moy{\nop}$. For a GHZ input state, we have $\frac{\moy{\Delta\nop^2}}{\moy{\nop}}\sim\moy{\nop}$; this means that, as the number of photons increases, the precision will go from Heisenberg to SQL scaling. No matter how small $1-\gamma$ is, the Heisenberg scaling will always disappear for large photon numbers. However, the transition takes place for $\moy{\nop}\sim N_c=\frac{\gamma}{1-\gamma}$. As the dissipation rate is reduced, the Heisenberg scaling will remain valid for higher and higher photon numbers. Furthermore, even when the linear scaling is recovered, it comes with a prefactor $N_c$, which can become arbitrarily large as the absorption becomes less important. Therefore, the systematic downfall of the Heisenberg limit does not mean that the precision cannot be increased by reducing the dissipation to extremely small values. In \cite{demkowicz-dobrzanski_elusive_2012}, these ideas were also applied to atomic interferometry in the presence of depolarisation, dephasing, or spontaneous emission. In each case, it was found that the precision achieved a linear scaling in the number of probes, the prefactor being equal to $0$ for a complete loss, and infinite for a vanishing dissipation.

Going back to \eqref{dissipphotonQFI}, we can also make the following remark: instead of sending all the photons at once, we can decompose the probes into several packets containing on average $\moy{\nop}_i\lessapprox N_c$ photons. Each packet will provide a Heisenberg-limited QFI $\moy{\nop}_i^2$. Since we have in total $\moy{\nop}_i/\moy{\nop}$ packets, the final precision will be given by:

\begin{equation}
	\pazocal{I}_x\sim \moy{\nop}_i\moy{\nop}\sim N_c\moy{\nop}
\end{equation}
and we find again the result of \eqref{dissipphotonQFI} for a single packet with $\frac{\moy{\Delta\nop^2}}{\moy{\nop}}\gg N_c$. This echoes to the beginning of the paragraph, when we showed how noisy evolution could be split into several shorter steps. 

Let us now make the argument a bit more precise: with a time-independent Hamiltonian, we achieve a scaling $\delta x=\frac{1}{\dur}$. If the duration of the protocol is limited to some duration $\dur_c$ (because of decoherence, or for any other reason), the best we can do is a series of $\nbrep$ independent rounds of measurement of duration $\dur_c$. This gives us a precision $\delta x=\frac{1}{\sqrt{\nbrep}\dur_c}=\frac{1}{\sqrt{\dur\dur_c}}$ with $\dur$ the total protocol duration. Now, as we explained earlier, the use of coherent and time-dependent protocols allow to achieve a precision with improved scaling in time: $\delta x\propto\frac{C^a}{\dur_c^{1+a}}$; where the constant $C$ must be homogeneous to a time to ensure the correct dimensionality of the expression. If the limiting duration is the same as in the time-independent protocol (which of course is not guaranteed to occur, as the discussion at the beginning of this section has shown), we can still consider a series of short cycles of duration $\dur_c$. We will then achieve a precision:
\begin{equation}
	\delta x\propto\frac{C^a}{\sqrt{\dur \dur_c}\dur_c^{a}}.
	\label{tlTscaling}
\end{equation}
 In other words, if the duration of the measurement cycle is limited, \textit{the error will always scale like the square root of the total experimental time $\dur$}, even if the time scaling in each cycle is arbitrarily high. Compared to the time-independent protocol, the difference is entirely contained in the \textit{prefactor} $\left(\frac{C}{\dur_c}\right)^a$. This is not just to say that experimental imperfections should be kept in mind when deriving a theoretical model; this remark is more fundamental. We studied the evolution of precision with time in the first place because we assumed that time was a limited resource; in other words, that there exists some typical time scale that the experimental cycle should not exceed. Therefore, we can expect this square-root scaling to be ubiquitous in estimation protocols (this is indeed the case, for instance, in atomic clocks \cite{pezze_quantum_2018}). Improved time-scalings (outside of course of their fundamental interest) are only interesting from a practical perspective insofar as they lead to an improved prefactor in the final expression. The same is true for the scaling with particle number. Let us assume that the precision for the non-dissipative protocol is given by $\pazocal{I}\propto N^a$, and that the dissipation imposes a maximum number $N_c$ that can be used simultaneously in the experiment. Then by splitting the probes into packets of size $N_c$, we obtain a precision $\pazocal{I}\propto N_c^{a-1}N$. Whatever the original scaling $a$, the final QFI scale linearly with the total particle number. However, $a$ can still be found in the prefactor $N_c^{a-1}$. As the dissipation is reduced, this prefactor can grow and becomes infinite in the limit of vanishing dissipation.

Thus, when writing an expression for the QFI which involves a scaling with any parameter, it should be kept in mind that the parameter in question will be limited by a certain factor. To increase the QFI beyond this limiting value, the estimation should be decomposed into shorter steps, which will always restore a linear scaling for the QFI (which means a square-root scaling for the error). The improved precision can only take the form of a prefactor. For practical applications, two things should be checked: 1) how the use of a quantum protocol affects the maximum value for particle number or time (as we saw, using entangled probes can greatly reduce the possible duration cycle) and 2) what are the prefactors involved in the expression of the QFI, and how do they compare with the limiting values defined above. These remarks will be of interest in Chapter 5.

In this section, we have focused only on the precision bound which arise in the presence of decoherence. Several works have been put forward to compensate for these effects and recover Heisenberg scaling, such as error correction \cite{demkowicz-dobrzanski_adaptive_2017,zhou_achieving_2018}, or the use of decoherence-free subspaces; however, we will not discuss these approaches here. The interested reader may find reviews in \cite{haase_precision_2016,pezze_quantum_2018}, which discuss in detail the general bounds we have described, as well as the strategies that can be implemented to sidestep them.

\section{Experimental realizations}

So far the discussion has been completely theoretical. In this last section, we will review some recent progress in the practical implementation of quantum metrology protocols. Most of the results we will discuss have been achieved in the context of atom interferometry: a comprehensive discussion with this platform can be found in \cite{pezze_quantum_2018}.\\

Much effort has been devoted to the study of atomic clocks improved by spin squeezing. As we have discussed in the first section, standard atomic clocks work by applying an encoding map $e^{i(\Omega-\Omega_{LO})\dur_R\Jz}$ on a separable cloud of atoms, $\Omega$ being the frequency of the atoms, $\Omega_{LO}$ the frequency of the local oscillator, and $\dur_R$ the interrogation time. The estimation results are used to lock the local oscillator on the atomic frequency. In practice, several independent cycles are repeated over a given period of time $\dur$. The relative frequency $\Delta(t)=\frac{\Omega-\Omega_{LO}(t)}{\Omega}$ fluctuates in time: the goal then is to achieve a relative frequency centered on $0$, with the smallest fluctuations possible. These fluctuations are ultimately determined by the accuracy of the estimation procedure. They are characterized by the Allan standard deviation, which quantifies the variance of $\Delta(t)$ averaged over a bin of duration $\dur$ \cite{riehle_frequency_2006}. If the estimation protocol reaches the SQL, the Allan deviation is:

\begin{equation}
	\frac{1}{\Omega \dur_R\sqrt{N}}\sqrt{\frac{\dur_{\text{cycle}}}{\dur}},
\end{equation}
with $\dur_{\text{cycle}}\geq \dur_R$ is the total duration of a cycle (which includes the interrogation itself, the data acquisition, etc.), and $\frac{\dur}{\dur_{\text{cycle}}}$ is the total number of cycles performed in a time $\dur$. Current state-of-the-art atomic clocks are almost, or already, limited by this bound.
This limit can be improved by increasing the number of atoms $N$, the interrogation time $\dur_R$, or the atomic frequency $\Omega$. All three strategies, however, are limited in practice: for instance, the interrogation time is limited by the stability of the local oscillator, while atomic collisions place an upper bound on the number of atoms \cite{pezze_quantum_2018}.
As a consequence, much effort has been devoted to improving atomic clocks using squeezed atomic states as probes. The current record is held by a clock which uses a squeezed Rb atoms with a transition frequency of a few GHz \cite{hosten_measurement_2016}. The protocol uses a squeezed cloud of $10^5$ atoms and an interrogation time of $\dur_R\sim 200\mu$s. The Allan deviation achieved is $9.7*10^{-11}$ for $\dur=1s$, which represents an improvement of one order of magnitude ($10.5$dB) compared to a clock with the same $N$ and $\dur_R$ but no squeezing. This setup, however, does not yet reach the performances of state-of-the-art fountains clocks with Cs atoms, which can reach Allan deviation of $10^{-13}$ or $10^{-14}$ for $\dur=1s$ \cite{wynands_atomic_2009} without using squeezing. This is mostly because these clocks operate with a much higher interrogation time $\dur_R\sim 1$s and a larger number of atoms $N\sim 10^7-10^8$. Even higher sensitivities can be reached with clocks exploiting optical transitions in \textit{Sr}; the much higher frequency $\Omega\sim 10^{14}$Hz, despite a lower $N\sim 10^4$ atoms, allows to reach Allan deviations of order $10^{-16}$ for $\dur=1s$ (or even $10^{-18}$ with longer integration time) \cite{nicholson_systematic_2015}. However, so far there are no experiments exploiting spin-squeezed states in this regime of frequency. To summarize, it has been experimentally proven that, all things being equal, squeezing does allow to improve the precision of quantum clocks. However, practical squeezed clocks have not reached yet the precision of their unsqueezed counterpart because they are still operated with a smaller number of atoms and shorter interrogation time. \\

A second application concerns magnetometry, in particular scanning-probe magnetometers which utilize BEC. The goal is to evaluate a (potentially inhomogeneous) magnetic field on a surface. This can be done by positioning a levitating BEC near the surface. The magnetic field induces a Zeeman shift between two hyperfine levels of the BEC; this shift can then be studied through a Ramsey interferometric protocol or Loschmidt echo, allowing to reconstruct the magnetic field.
For a system operating at SQL, the error is given by $\delta B=\frac{\hbar}{\gamma}\frac{1}{\dur_R\sqrt{N}}\sqrt{\frac{\dur_{cycle}}{\dur}}$, with $\gamma$ the gyromagnetic ratio, and the other quantities being the same as before. In \cite{ockeloen_quantum_2013}, a protocol beating this limit was demonstrated, using an ensemble of $N=1400$ atoms in a squeezed state. With an interrogation time $\dur_R\sim 10$ms, a precision $\delta B=23$pT was achieved for a total protocol duration $\dur=11$s. This represents a sensitivity $\delta B \sqrt{\dur}=77$pT/$\sqrt{\text{Hz}}$, which is $4$dB below the SQL. The BEC probe has a small volume $V=20\mu\text{m}^3$, which allows for a good spatial resolution. A similar experiment was performed in \cite{muessel_scalable_2014}, using $\sim 30$ independent BEC with $400$ atoms each. With an interrogation time $\dur_R\sim 10^{-4}$s, a sensitivity of $1.9$nT/$\sqrt{\text{Hz}}$ has been achieved, which is $3.4$dB below the SQL. By comparison, experiments which do not exploit squeezing \cite{vengalattore_high-resolution_2007,eto_spin-echo-based_2013} can reach better sensitivities, around $10$pT/$\sqrt{\text{Hz}}$, notably because they use more atoms; this, however, comes at the cost of a much larger probe volume. Hence, these experiments demonstrate that squeezing-enhanced interferometry can be useful for spatially resolved magnetometry.

The experiments above exploit squeezed probes and channels acting independently on each probe. A protocol exploiting  non-linear encoding maps have also been implemented \cite{napolitano_nonlinear_2010,napolitano_interaction-based_2011,sewell_ultrasensitive_2014}. In this setup, a cloud of atoms interacts with polarized photons. The interaction Hamiltonian is $\Hop=c_1\Jz\hat{S}_z+c_2(\Jx\hat{S}_x+\Jy\hat{S}_y)$, with $\hat{J}_a$ and $\hat{S}_a$ the collective polarization operators of the spins and the light, respectively. Both atoms and light are in a separable coherent state. The goal is to evaluate the average spin polarization $\moy{\Jz}$. In a suitable regime of parameters, the system is described by an effective dynamics $\Hop_{\text{eff}}=c_1\Jz\hat{S}_z+c_2\Jz\hat{S}_z\hat{S}_0$, with $\hat{S}_0=\sum_i^{N_P}\hat{s}_0^{(i)}=\frac{N_P}{2}\mathbb{1}$, with $N_P$ the number of photons \cite{napolitano_nonlinear_2010}. The average spin polarization $\moy{\Jz}$ induces a rotation of the light polarization, which is then measured. The rotation angle, however, evolves non-linearly with the photon number, which is very similar to the non-linear photonic channels which we have discussed in the previous section. Thanks to this behavior, a precision $\delta\moy{\Jz}\sim\frac{1}{N_P^{3/2}}$ was achieved, demonstrating that nonlinear encoding channels allow indeed for scaling improvement of sensing protocols. \\

Optomechanical sensors have also generated an important experimental effort. Since the SQL in this case arises from the trade-off between measurement and backaction noise, several strategies can be implemented to beat this limit. A first possibility is to inject squeezed light in the device to reduce the measurement noise. This has been achieved at a large scale in the GEO \cite{ligo_scientific_collaboration_gravitational_2011} and LIGO \cite{ligo_scientific_collaboration_enhanced_2013} gravitational wave detectors. A sensitivity improvement of up to $2$dB was achieved in these experiments.

A second strategy is to exploit the correlations between measurement and back-action noises. Indeed, the SQL is derived by assuming that the two noises are uncorrelated, which is not always the case. By measuring carefully-chosen quadratures of the outgoing field, it is possible to exploit noise correlation to beat the SQL. This was achieved in a recent experiment \cite{mason_continuous_2019}, where a sensitivity of $1.5$ dB below the SQL was demonstrated.

Finally, another way to improve the precision is to implement so-called \textit{back-action evading measurements}, which reduces the backaction by transferring it from the degree of freedom under scrutiny to another one. This idea has been implemented in many experiments (see for instance 
\cite{ockeloen-korppi_quantum_2016,moller_quantum_2017} and references therein). Note however that these experiments are affected by additional, non-fundamental sources of noise such as thermal noise, which can prevent to reach the SQL despite the reduction of quantum noise.

\chapter{Study of the two-photon Dicke model}


\epigraph{\textit{The most important thing is: don't fool yourself. And you're the easiest person to fool.}}{Richard Feynman}

Thanks to our improving control over quantum light-matter interaction, dynamics that go beyond the standard Dicke picture are now within experimental reach. This opens perspectives to engineer quantum states and study new effects. For instance, in the anisotropic Dicke model, the rotating- and counter-rotating terms have independent, tunable strengths. This leads to parity-symmetry breaking, and a rich phase diagram in the ultrastrong coupling (USC) regime \cite{baksic_controlling_2014,xie_anisotropic_2014}. The two-photon Rabi model \cite{emary_bogoliubov_2002,dolya_quadratic_2009,travenec_solvability_2012,maciejewski_comment_2015,travenec_reply_2015,chen_exact_2012,felicetti_spectral_2015,duan_two-photon_2016,peng_dark-like_2017} is another example of generalized light-matter coupling. In the USC regime, it exhibits exotic spectral features, such as a coalescence of energy levels known as spectral collapse. However, few results are known about its many-body version. 

In this Chapter, we study the existence of phase transitions in the two-photon Dicke model. First, for a system at equilibrium, we prove that the two-photon coupling can be mapped to an effective one-boson coupling in the thermodynamic limit. Using this mapping, we show the emergence of a second-order phase transition, and study its interplay with the spectral collapse. Next, we study the behavior of the system in the presence of dissipation. At the mean-field level, the second-order transition is replaced by a first-order one. The spectral collapse is replaced by an instability, which can be removed by increasing the dissipation rate.  

This Chapter is divided into three parts. In the first section, we introduce the two-photon Dicke Hamiltonian and its spectral properties. In the second part, we present the mapping to the one-photon coupling and the emergence of phase transition in an isolated system. In the last section, we discuss the behavior of the model in a dissipative setting. 

The first part reviews previous results, while the second and third sections are original contributions, published in Ref.\cite{garbe_superradiant_2017} and \cite{garbe_dissipation-induced_2020}.

\section{Introduction to the two-photon Dicke model}

\subsection{Two-photon coupling and spectral collapse}

\begin{figure}[h!]
	\centering
	\includegraphics[angle=-90,width=0.7\linewidth]{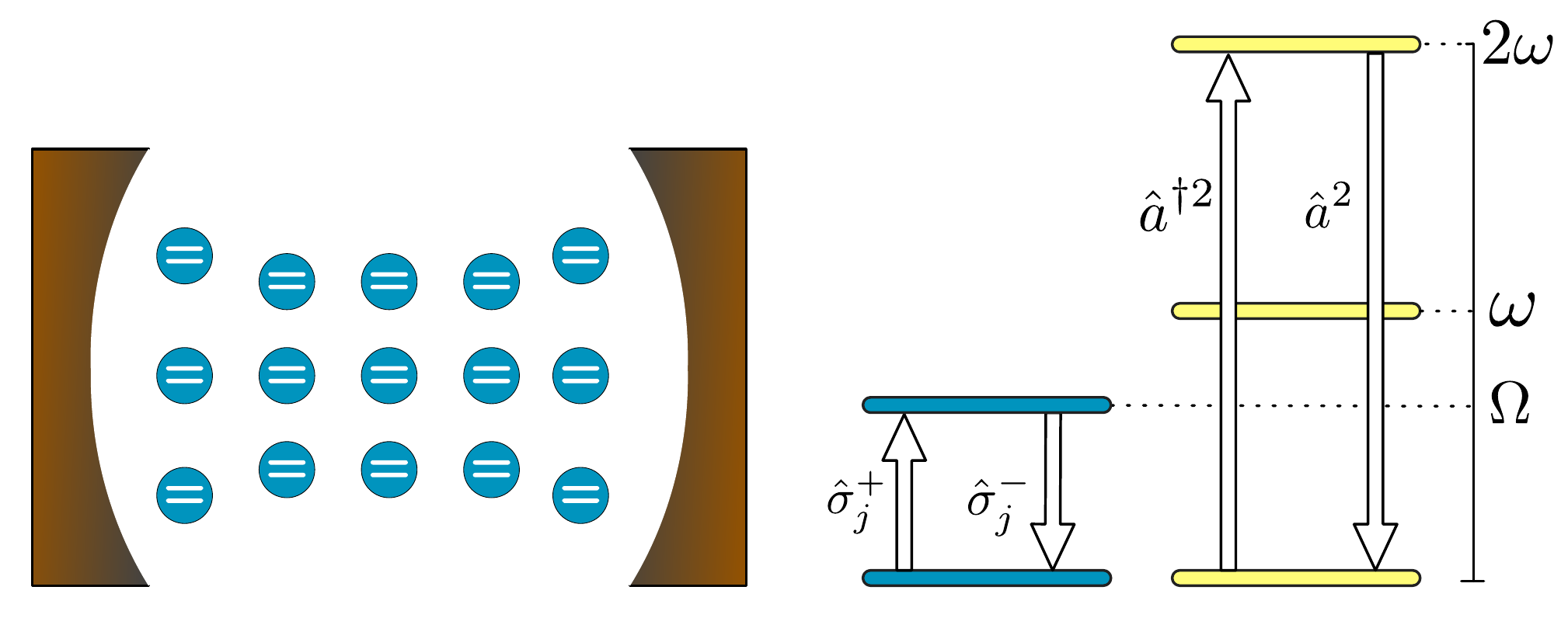}
	\caption[Sketch of an ensemble of qubits interacting with a single bosonic mode via two-photon coupling]{(Left) Sketch of an ensemble of qubits interacting with a single bosonic mode via two-photon coupling terms, as in Eq.\eqref{Htwophoton}. (Right) Energy levels of the uncoupled systems. Arrows represent rotating and counter-rotating two-photon transitions. Figure from Ref.\cite{garbe_superradiant_2017}. \label{sketch}}
\end{figure}

We consider an ensemble of $N$ qubits interacting with a bosonic mode via two-photon interaction, as sketched in Fig.\ref{sketch}. Each qubit excitation is associated with the simultaneous creation or destruction of two bosonic excitations. The behavior of the system is described by the following Hamiltonian:

\begin{eqnarray}
\hat{H} = \omega \adag \aop + \Omega\Jz + 2\frac{\gtwo}{N}\Jx(\aop^2+\adagsq) = \omega \adag \aop + \Omega\Jz + 2\frac{\gcoll}{\sqrt{N}}\Jx(\aop^2+\adagsq).
\label{Htwophoton}
\end{eqnarray}

\begin{figure}[h!]
	\centering
	\includegraphics[angle=-90]{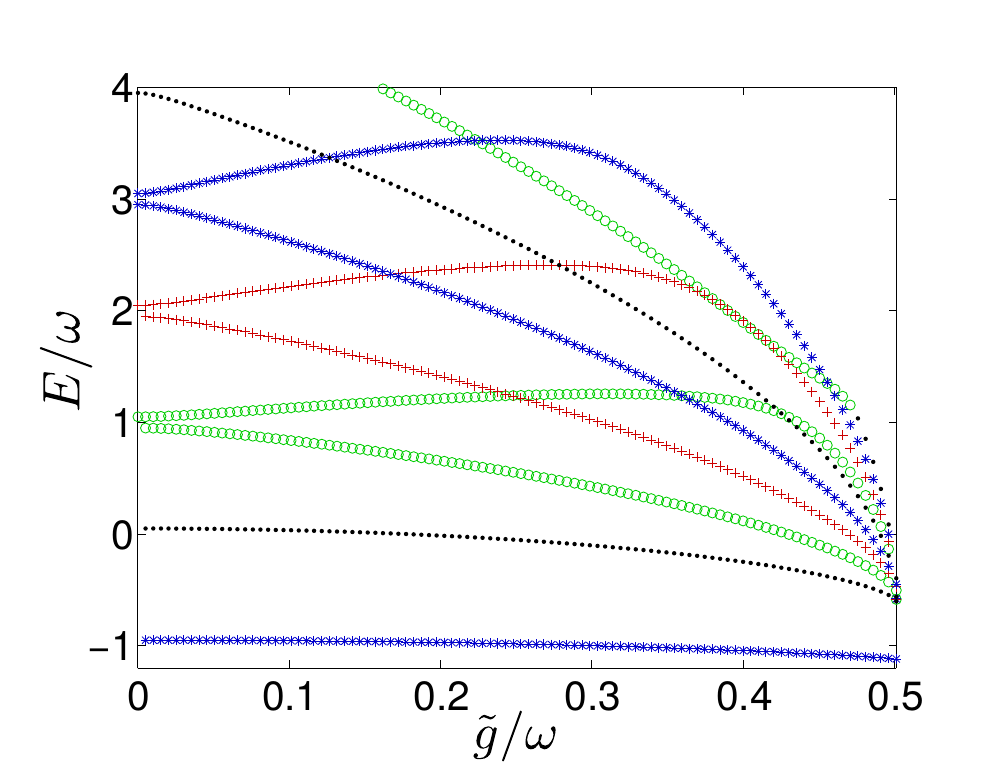}
	\caption[Spectrum of two-photon Rabi model, showing the spectral collpase.]{Spectrum of \eqref{Htwophoton}, for $N=1$ and $\Omega=1.9$, as a function of $\gtwo$. For $\gtwo=\omega/2$, the energy levels coalesce and the spectral collapse occurs. Reproduced from \cite{felicetti_spectral_2015}. \label{Spectralcollapse}}
\end{figure}

Where we have defined the collective spin operators $\hat{J}_a=\frac{1}{2}\sum_{j=1}^N{\hat{\sigma}_j}^a$, with $a=\{x,y,z\}$. Here, $\gtwo$ and $\gcoll$ represent different choices of normalization for the collective coupling. The first choice is best suited to discuss the spectral properties of the model. These properties have been studied in the literature with different techniques \cite{emary_bogoliubov_2002,dolya_quadratic_2009,travenec_solvability_2012,maciejewski_comment_2015,travenec_reply_2015,chen_exact_2012,felicetti_spectral_2015,duan_two-photon_2016,peng_dark-like_2017}; in particular, the method developed by Braak to solve exactly the one-photon Rabi model \cite{braak_integrability_2011} has been extended to the two-photon case for $N=1$ \cite{travenec_solvability_2012,chen_exact_2012,duan_two-photon_2016}. The most salient feature which has been highlighted by these studies is the presence of a \textit{spectral collapse}. When $\gtwo=\frac{\omega}{2}$, the energy levels coalesce into a continuous band, as shown in Fig.\ref{Spectralcollapse}. For higher values of $\gtwo$, the Hamiltonian is not bounded from below and the model is no longer well-defined. Previous studies of the spectral collapse \cite{travenec_solvability_2012,felicetti_spectral_2015,duan_two-photon_2016} have confirmed that, at least for few qubits, the collapse occurs for $\gtwo=\frac{\omega}{2}$ regardless of the value of $N$ and $\Omega$.

The origin of this dramatic behavior may be grasped by the following argument \cite{felicetti_spectral_2015}: when the coupling $\gtwo$ becomes large the dynamic is dominated by the coupling term: $\hat{H}\sim\omega \adag \aop +  \frac{\gtwo}{N}\Jx(\aop^2+\adagsq)=(\frac{\omega}{2}+2\gtwo\frac{\Jx}{N})\xop^2 + (\frac{\omega}{2}-2\gtwo\frac{\Jx}{N})\pop^2$, with $\xop$ and $\pop$ the field quadratures. This effective potential is only stable as long as $(\frac{\omega}{2}\pm 2\gtwo\frac{\Jx}{N})$ is positive. Since $2\frac{\Jx}{N}$ takes value in the interval $[-1,1]$, this is always the case for $\gtwo<\omega/2$. For $\gtwo=\omega/2$, however, there are spin states which give a flat potential. This corresponds to unbounded excitation modes, whose spectrum is described by a continuous band. For $\gtwo>\omega/2$, the potential is no longer bounded from below. Since the free spin term $\Omega\Jz$ is bounded, it cannot compensate the diverging behavior of the interaction. By contrast, in the one-photon Dicke model, the interaction potential is only linear, and can never compensate the free (quadratic) potential $\frac{\omega}{2}(\xop^2+\pop^2)$. Hence, the spectral collapse is a feature specific to the two-photon coupling. Finally, let us say a word about the symmetry of the model: while the usual Dicke model has a $\mathbb{Z}_2$ symmetry (see Chapter 1 and 2), the two-photon version has a four-folded symmetry, which corresponds to the simultaneous exchange: \begin{equation}\label{symmetry}
\left\lbrace
\begin{split}
&\aop \to i\aop, \\
&\hat{\sigma}_j^x\rightarrow -\hat{\sigma}_j^x \quad \forall j.
\end{split}
\right.
\end{equation}

\subsection{Proposals of implementation}

Although the spectral collapse has not been observed experimentally yet, several proposals have shown how it could be realized with present-day technology. The main challenge is to implement an ultrastrong two-photon coupling; this is difficult with ordinary dipole coupling, since two-photon processes arise only as a second-order correction in this case. One solution around this problem is provided by quantum simulation. For instance, in \cite{felicetti_spectral_2015}, it was shown that an effective two-photon dynamic could be implemented in trapped ions platforms, in which the internal and vibrational states of the ions are used to simulate the qubits and bosonic degrees of freedom, respectively. Lasers couple the internal state of the ions to their motional degrees of freedom via sideband coupling. Let us assume that the properties of the trap allow us to single out a single vibrational mode of frequency $f$. Then if the detuning between the laser and the ion internal transition is close to $-2f$ (red-detuned laser), the laser excites a process in which two phonons are emitted while a qubit excitation is destroyed (or vice-versa). If the detuning is close to $2f$ (blue-detuned laser), the energy brought by the laser creates two phonons while exciting the qubit. By using both red- and blue-detuned lasers, it is possible to engineer an effective phonon-qubit coupling as described by \eqref{Htwophoton}. Furthermore, the vibrational states are modulated by the light pump, which renormalize their frequencies. This allows us to tune both the effective phonon-qubit coupling and the phonon frequency $\omega$, and to reach the USC regime, even if the underlying laser-qubit coupling is weak.

Recently \cite{felicetti_two-photon_2018}, it was also pointed out that \textit{genuine} two-photon interactions could also be implemented in superconducting circuits. Here, an artificial atom is inductively coupled to a superconducting quantum interference device (SQUID). In this case, the linear coupling is suppressed, while the two-photon coupling emerges as the natural light-matter interaction in an undriven system and not as a result of a quantum simulation scheme. It has been predicted that a spectral collapse can be observed in this system by analyzing the energy level structure through a pump-probe experiment \cite{felicetti_ultrastrong-coupling_2018}. This procedure could also be extended to include more than one qubit.

\section{Phase transition in the two-photon Hamiltonian}

\subsection{Mapping to the one-photon coupling}
In this section, we will study the two-photon Dicke model in the thermodynamic limit, $N\rightarrow\infty$. As we mentioned earlier, exact results can be obtained for both the one-photon and two-photon Rabi model. Several works have studied how to extend these results to the multi-qubit case \cite{braak_solution_2013,he_exact_2015}. However, these approaches are challenging to use in practice for a large number of qubits, and are not suited to develop physical intuition.
Furthermore, and contrary to the one-photon version, it is not possible to directly solve the model through HP and Bogoliubov transformation, since the resulting coupling term $(\bop+\bdag)(\aop^{\dagger2}+\aop^2)$ is not quadratic.
Therefore, we will use another approach, which will be more suited to describe the phenomenology of the system in the thermodynamic limit. This approach is based on a mapping between the two-photon and one-photon couplings, which is achieved with the two following steps. First, we use HP transformation: 
\begin{equation}
\Jp=\bdag \bsqr,\hspace{5pt} \Jz=\bdag \bop - \frac{N}{2},
\label{Hosltein-Primakoff}
\end{equation}
with $[\bop, \bdag]=1$. Second, since only quadratic combinations of bosonic operators are present in \eqref{Htwophoton}, we define the following operators:
\begin{eqnarray}
\Ko=\frac{1}{2}\left(\adag \aop + \frac{1}{2}\right), \hspace{5pt} \Kp=\frac{1}{2} \adagsq,\hspace{5pt} \Km=\frac{1}{2} \aop^2,
\label{Kdef}
\end{eqnarray}
which obey SU(1,1) commutation relations: $[\Ko,\hat{K}_{\pm}]=\pm \hat{K}_{\pm}$, and $[\Kp, \Km]=-2\Ko$. Thus the Hamiltonian \eqref{Htwophoton} may be rewritten as:

\begin{equation}
 	\Hop = 2\omega\Ko + \Omega \bdag \bop + \frac{2\gtwo}{\sqrt{N}}(\Kp + \Km)\left(\sqrt{1-\frac{\bdag\bop}{N}}\bop+\bdag\sqrt{1-\frac{\bdag\bop}{N}}\right) -\frac{\omega}{2} -\Omega\frac{N}{2}.
 	\label{HwithHP}
 \end{equation} 

So far we have not made any approximation. Let us now make two informal remarks which will serve as the starting point of a more rigorous analysis. First, in the thermodynamic limit, we expect that spin fluctuations will be small with respect to $N$: hence we can develop $\Jp\sim\sqrt{N}\bdag$, with nonlinear bosonic terms acting only as a perturbation. Second, we point out the formal similarity between the operators defined in \eqref{Kdef}, and standard spin operators which obey SU(2) commutation relations. Hence, we can rewrite the Hamiltonian \eqref{HwithHP} as: 

\begin{equation}
\Hop \sim 2\omega \Ko + \Omega \bdag \bop + \frac{2\gtwo}{\sqrt{N}}(\Kp + \Km)(\bop+\bdag) + O\left(\frac{1}{N\sqrt{N}}\right).
\label{Hmappe}
\end{equation}

This Hamiltonian involves only a linear coupling between a bosonic mode $\bop$ and a spin-like operator $\Ko$. Nonlinear coupling terms are also present, but only as a higher-order perturbation. Based on this observation, we expect that standard methods used to treat the one-photon Dicke Hamiltonian will also be effective to describe this model; in the next subsection, we show that this is indeed the case. Consistent with this intuition, we will use the following notation: $\hat{K}_\pm=\Kx\pm i\Ky$. These operators verify $[\Ko,\Kx]=i\Ky$ and $[\Ky,\Ko]=i\Kx$, but $[\Kx,\Ky]=-i\Ko$.

\subsection{Mean-field analysis}
We will now ignore the higher-order terms in \eqref{Hmappe} and study the behavior of the system by a mean-field ansatz. We assume the mode $\bop$ to be in a coherent state $\ket{\beta}$, $\bop\ket{\beta}=\sqrt{N}\beta\ket{\beta}$. In this state, the spin-like operator $\Ko$ experience an effective Hamiltonian:

\begin{equation}
	\Hop(\beta) = 2\omega\Ko + 4\gprime\Kx + N\Omega\lvert\beta\rvert^2 -\frac{\omega}{2} - \frac{\Omega N}{2},
\end{equation}
with $\gprime=\gtwo(\beta + \beta^*)\betasqr$. This Hamiltonian may be diagonalized easily by the following transformation, which preserves the commutation relations:
\begin{eqnarray}
\label{BogoMF}
\Kop(\beta) =\cosh(2\xi_b)\Ko + \sinh(2\xi_b)\Kx,\\ \nonumber
\Kx^{\prime}(\beta) = \cosh(2\xi_b)\Kx+ \sinh(2\xi_b)\Ko,
\end{eqnarray}

with $\xi_b = \frac{1}{2}\arctanh\left(\frac{2\gprime}{\omega} \right)$. Then the ground state is just the first eigenstate of $\Kop(\beta)$, and the ground state energy is given by \begin{equation}
E_G = 4\sqrt{\frac{\omega^2}{4} - \gprime^2} + N\Omega\lvert\beta\rvert^2 - \Omega\frac{N}{2} - \frac{\omega}{2}.
\end{equation} 
The final step is to minimize $E_G$ as a function of $\beta$ to find which value is adopted by the systems. $\beta$ then plays the role of order parameter for our system: a change in its behavior leads to a modification of the qualitative properties of the ground state, indicating a phase transition. We find that if $\gtwo<\gtwo_p=\sqrt{\frac{\omega\Omega N}{4}}$, $E_G$ is minimal for $\beta=\beta^*=0$. For $\gtwo>\gtwo_p$, we have two degenerate minima: 
\begin{equation}
\beta=\beta^*= \pm \sqrt{\frac{1}{2}} \left(1-\sqrt{\frac{1-(\frac{\gtwo}{\gtwo_c})^2}{(\frac{\gtwo}{\gtwo_p})^4-(\frac{\gtwo}{\gtwo_c})^2}}\right)^{1/2}= \pm \beta_\gcoll,
\end{equation}
where $\gtwo_c=\frac{\omega}{2}$ is the coupling value at which the spectral collapse occurs. \\

Let us now describe the state of the system in terms of the original operators. We will consider the following quantities: $\Jx$, $\Jy$, $\Jz$, $\adag\aop$, $\Xop=\aop^2+\adagsq$, $\Yop=\aop^2-\adagsq$. If the four-folded symmetry \eqref{symmetry} is unbroken, these quantities are all equal to $0$; therefore, we can expect that they will be good order parameters to witness the transition.
 
 In terms of the field operator $\aop$, the transformation \eqref{BogoMF} is a Bogoliubov transformation. Therefore, in the ground state, the bosonic field is squeezed, with squeezing parameter $\xi_b$. For $\gtwo<\gtwo_p$, the squeezing parameter vanishes, and the field is in a vacuum state: $\moy{\adag\aop}=\moy{\Xop}=\moy{\Yop}=0$. The collective spin is polarized along the $z$-axis: $\moy{\Jz}=N\lvert\beta\rvert^2-\frac{N}{2} = -\frac{N}{2}$, $\moy{\Jx}=\moy{\Jy}=0$.  For $\gtwo>\gtwo_p$, however, the field enters a squeezed vacuum state, and the spins experience a collective rotation in the $x-z$ plane. The direction of both the squeezing and the rotation is given by the sign of $\beta$: for $\beta=+\beta_\gcoll$, the field is squeezed along the $\xop$ quadrature, and $\moy{\Jx}\geq0$. For $\beta=-\beta_\gcoll$, the field is squeezed along the $\pop$ quadrature, and $\moy{\Jx}\leq0$. Since, in this second phase, we have $\moy{\Jx}\neq0$, $\moy{\Xop}\neq0$ and $\moy{\Yop}\neq0$, the symmetry \eqref{symmetry} is (at least partially) broken. The order parameters $\moy{\Jx}$ $\moy{\Xop}$ remain continuous at $\gtwo=\gtwo_p$, indicating a second-order transition.

Interestingly, for $\gtwo=\gtwo_c$, the field becomes infinitely squeezed, and the spins are polarized along the $x$ direction. For $\gtwo>\gtwo_c$, $\beta$ is no longer well-defined. Hence, in addition to predicting a phase transition, this analysis also correctly identifies that a qualitative change takes place at the collapse point. Furthermore, if we want the transition to occur before the spectral collapse, we need to have $\gtwo_p < \gtwo_c$, which implies $\Omega N < \omega$.  

To summarize, the two-photon Dicke model exhibits a phase transition when the field-to-spin frequency ratio scales at least like the number of spins. During this transition, the field experiences squeezing, correlated with a collective rotation of the two-level systems. This is similar to the one-photon Dicke model, in which the field experiences displacement instead. In both cases, the field acquires a non-zero population in the ground state.\\

\subsection{Scaling of the coupling}

There is, however, an important difference between the one- and two-photon model, in the scaling of the bosonic population. In the one-photon Dicke model, the  coupling may be expressed in terms of individual coupling strength like $\gind(\adag+\aop)\Jx$. For $\gind=\gind_p=\sqrt{\frac{\Omega\omega}{2N}}$, the system enters a phase in which the ground-state average photon number is non-zero. This number depends in particular on the coupling and the qubit number $N$. In this context, the name "superradiant" has the following meaning: the photon population scales like $N$, \textit{when the ratio $\frac{\gind}{\gind_p}$ is kept constant}. This last condition is necessary because $\gind_p$ itself depends on $N$. If we increase $N$ by keeping $\gind$ constant, the system will effectively drift away from the critical point, and it will be ambiguous whether the increase in the photon number is solely due to the increase in qubit number or because we are operating further away from the critical point. This condition is best expressed in terms of the renormalized coupling $\gcoll=\gind\sqrt{N}$; since $\gcoll_p=\sqrt{\frac{\Omega\omega}{2}}$ is independent of $N$, we may simply increase $N$ while keeping $\gcoll$ constant.

In the two-photon Dicke case, however, we have $\textit{two}$ coupling scales, corresponding to the transition and collapse, respectively. And since $\gtwo_p$ depends on $N$ but $\gtwo_c$ does not, these two points will move respectively to one another when $N$ increases. Therefore, there is no single best choice of coupling normalization: $\gtwo$ is most suited to describe the collapse, and $\gcoll=\frac{\gtwo}{\sqrt{N}}$ allows to efficiently describes the transition. As a consequence, if we wanted to describe the "superradiant" character of the two-photon Dicke, it is ambiguous whether we should increase $N$ while keeping constant$\frac{\gtwo}{\gtwo_p}$, $\frac{\gtwo}{\gtwo_c}$, or some hybrid quantity like $\frac{\gtwo}{\sqrt{\gtwo_p\gtwo_c}}$. This ambiguity shows that the properties of the one- and two-photon Dicke Hamiltonians are very different in that respect and cannot be directly compared. \\

\begin{figure}[h]
	\centering
	\includegraphics[angle=-90,width=.5\linewidth]{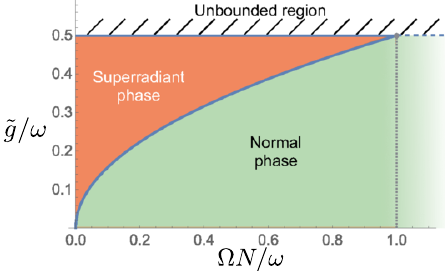}
	\caption[Phase diagram of the two-photon Dicke model.]{Phase diagram of our model in the mean-field approximation. In the unbounded region, the model is no longer valid. For $N\Omega>\omega$, the phase transition can no longer happen before the spectral collapse and only the normal phase exists.}
	\label{phasediag}
\end{figure}

To sidestep this difficulty, we will assume that the qubit frequency is rescaled with the qubit number, by imposing 
$$\ratbis=\frac{N\Omega}{\omega}=\text{cst}.$$ Since $N$ is large, this means we are now considering that the frequency ratio $\frac{\Omega}{\omega}$ is now very \textit{small}, which is the regime opposite to the one in which we studied the Rabi model. With this scaling, both $\gtwo_p$ and $\gtwo_c$ become independent of $N$. Furthermore, if $\ratbis\leq1$, $\gtwo_p=\sqrt{\ratbis}\gtwo_c\leq \gtwo_c$, and the transition always happens before the collapse. Between $\gtwo_p$ and $\gtwo_c$, the photon number increases from $0$ to infinity, but is independent of $N$. For this reason, we will restrain from using the name "superradiant phase" to describe this regime, and rather talk about the "squeezed phase". The phase diagram of the model is given in Fig.\ref{phasediag}. From an experimental perspective, the scaling $\ratbis=\text{cst}$ could be achieved through a quantum simulation scheme. In the proposal of \cite{felicetti_spectral_2015}, for instance, both qubit and field frequencies can in principle be tuned to arbitrarily low values. 

\subsection{Finite-size analysis}

The previous study becomes exact in the thermodynamic limit, when the fluctuations of $\bop$ field are negligible. We will now study the case of a large but finite number of spins. We will assume that the frequency condition obtained through the mean-field analysis is satisfied: $\ratbis=\frac{N\Omega}{\omega}\leq1$.
 
\begin{figure}
	\centering
	\includegraphics[angle=-90,width=0.5\linewidth]{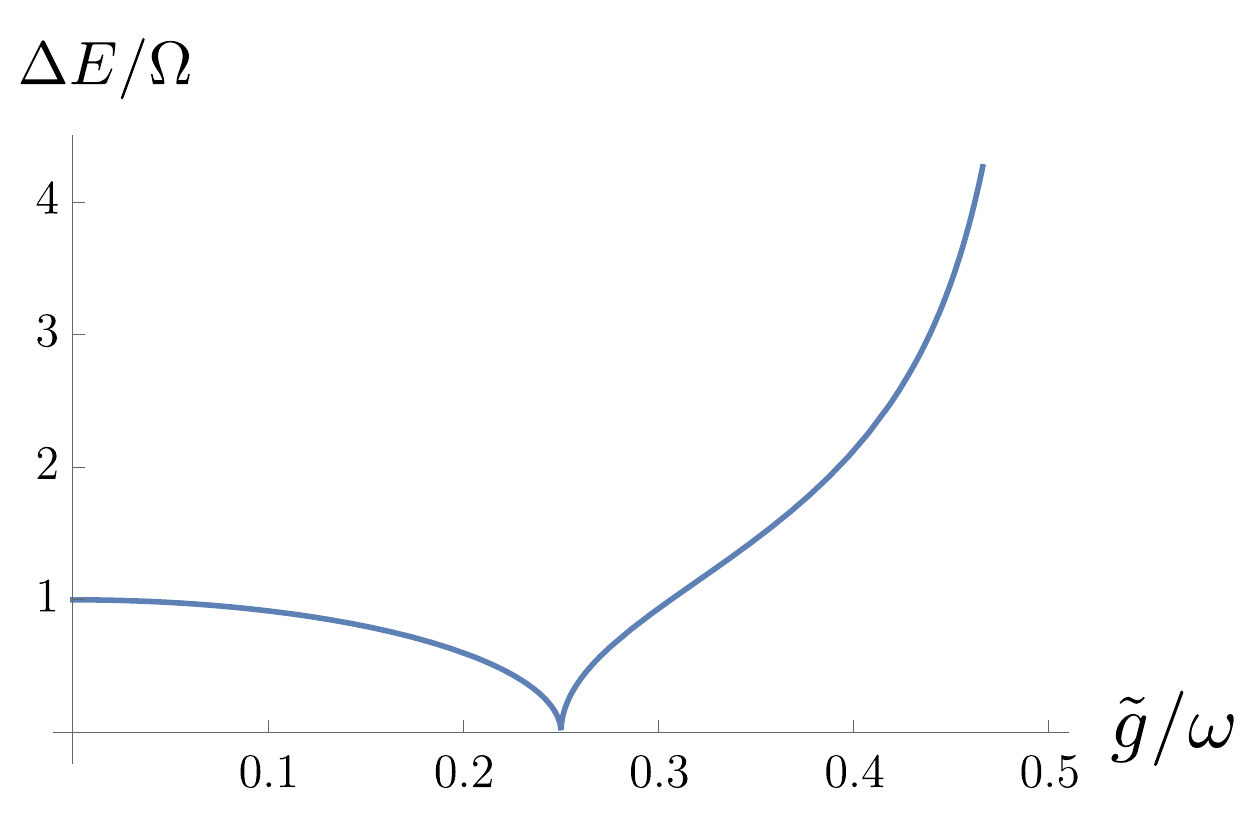}
	\caption[Energy gap in the two-photon Dicke model.]{Excitation energy $\Delta E=E_1-E_0$ versus $\gtwo/\omega$, for $\omega=4\Omega N$. In the regime of parameters considered, the excitation energy is independent of the qubit number $N$. It vanishes for $\gtwo=\gtwo_p$, indicating the phase transition. Near the collapse point $\gtwo=0.5\omega$ however, the gap diverges instead of vanishing, which is incompatible with previous results and shows the limitations of this treatment.}
	\label{Egap}
\end{figure}

We will study the fluctuations of the $\bop$ field around its mean value: $\bop=\sqrt{N}\beta+\dop$. In the normal phase, $\beta=0$, and the Hamiltonian \eqref{HwithHP} may be rewritten as (dropping a constant additional factor):

\begin{equation}
\Hop = 2\omega \Ko + \omega\frac{\ratbis}{N} \opddag \dop + 2\omega\sqrt{\frac{\ratbis}{N}}\lambda\Kx(\dop+\opddag),
\end{equation}
where we have again defined the normalized coupling strength $\lambda=\frac{\gtwo}{\gtwo_p}$. This is almost the one-photon Dicke model, with spins replaced by the $\hat{K}$ operators. Furthermore, the frequency of the $\dop$ mode is very weak with respect to the frequency of $\Ko$. This is precisely the regime in which the model can be solved by SW transformation. The full computations can be found in Appendix A, here we just give the main steps.
First, we apply the following unitary transformation: $\Uop=e^{\sqrt{\frac{\ratbis}{N}}\Sop}$ with $\Sop=i\lambda\Ky\frac{\Jx}{N}\sim i\lambda\Ky\frac{\dop+\opddag}{2}$. This yields the following Hamiltonian:

\begin{equation}
	e^{i\Sop}\Hop e^{-i\Sop}=2\omega\Ko + \omega\frac{\ratbis}{N}\opddag\dop -\omega\frac{\ratbis}{N}\lambda^2\Ko(\dop+\opddag)^2, 
\end{equation}

plus some additional terms of order $O(\omega\frac{1}{N\sqrt{N}})$. Up to this order, the Hamiltonian now commutes with $\Ko$. Thus, we can study the eigenspaces of $\Ko$ independently. We project the Hamiltonian in the lowest of these subspaces: $\Ko\rightarrow \frac{1}{4}$. This gives an effective low-energy field dynamic:

\begin{equation}
	\frac{\Hop_f^{(2)}}{\omega}=-\frac{1}{2} + \frac{\ratbis}{N} \left(\opddag\dop - \lambda^2(\dop+\opddag)^2\right).
	\label{order2SW}
\end{equation}
This is a quadratic field Hamiltonian, that can be solved through Bogoliubov transformation. The eigenstates are squeezed Fock states $\ket{m,\xi_b^{(N)}}$, with squeezing parameter $\xi_b^{(N)}=\frac{1}{4}\text{ln}(1-\lambda^2)$. The associated eigenvalues are $E^{(N)}_m=m \Omega \sqrt{1-\lambda^2}$. In terms of the original degrees of freedom, the system experiences spin squeezing near the transition. This can be measured by the standard Wineland's criterion \cite{wineland_spin_1992}: $\frac{2\Delta^2(\Jy)}{\moy{\Jz}}=-\moy{(\dop-\opddag)^2}=e^{2\xi_b^{(N)}}\leq1$. As one gets closer to the transition, the amount of squeezing diverges and the gap $\Delta E=E_1^{(N)}-E_0^{(N)}$ vanishes (as shown in Fig.\ref{Egap}), indicating the transition.\\

A similar analysis can be performed in the squeezed phase. The detailed calculations can be found in Appendix A. Once more, we obtain a ladder of squeezed Fock states for the $\dop$ field. The squeezing parameter and eigenvalues are, respectively:

$$\xi_b^{(S)}=-\frac{1}{4}\text{ln}\left(\frac{\lambda^4}{4(\lambda^4-1)}\left(1+\sqrt{\frac{1-\frac{\gtwo^2}{\gtwo_c^2}}{\lambda^4-\frac{\gtwo^2}{\gtwo_c^2}}}\right)^2\right),$$ 
$$E^{(S)}_m=m\Omega\sqrt{\frac{\left(\lambda^4-\frac{\gtwo^2}{\gtwo_c^2}\right)\left(1-\frac{1}{\lambda^4}\right)}{\left(1-\frac{\gtwo^2}{\gtwo_c^2}\right)}}.$$

\begin{figure}[h!]	
	\begin{minipage}{\linewidth}
		\begin{center}
		\hspace{10pt}\includegraphics[width=0.4\linewidth]{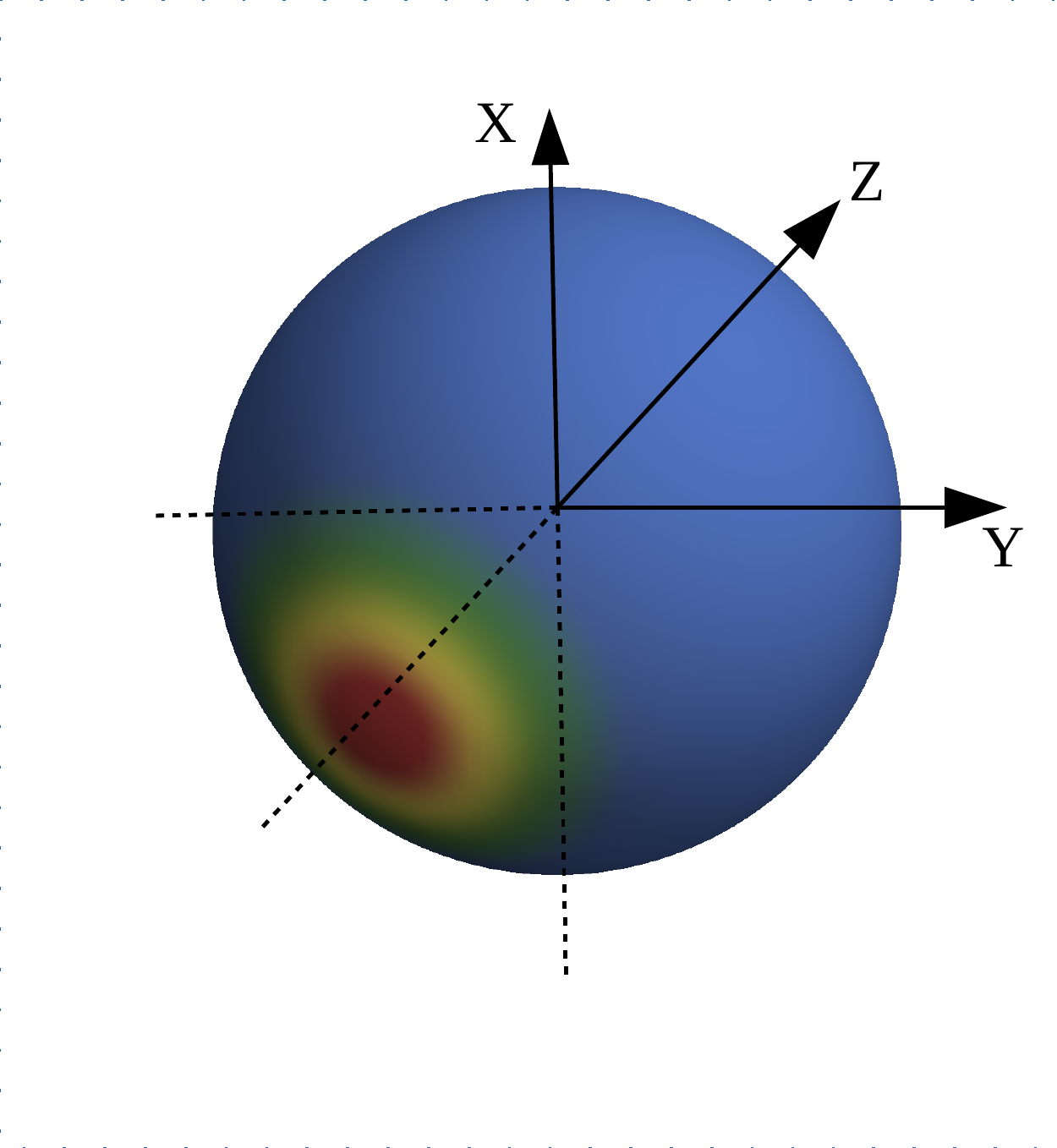} \hspace{10pt}
		\includegraphics[width=0.4\linewidth]{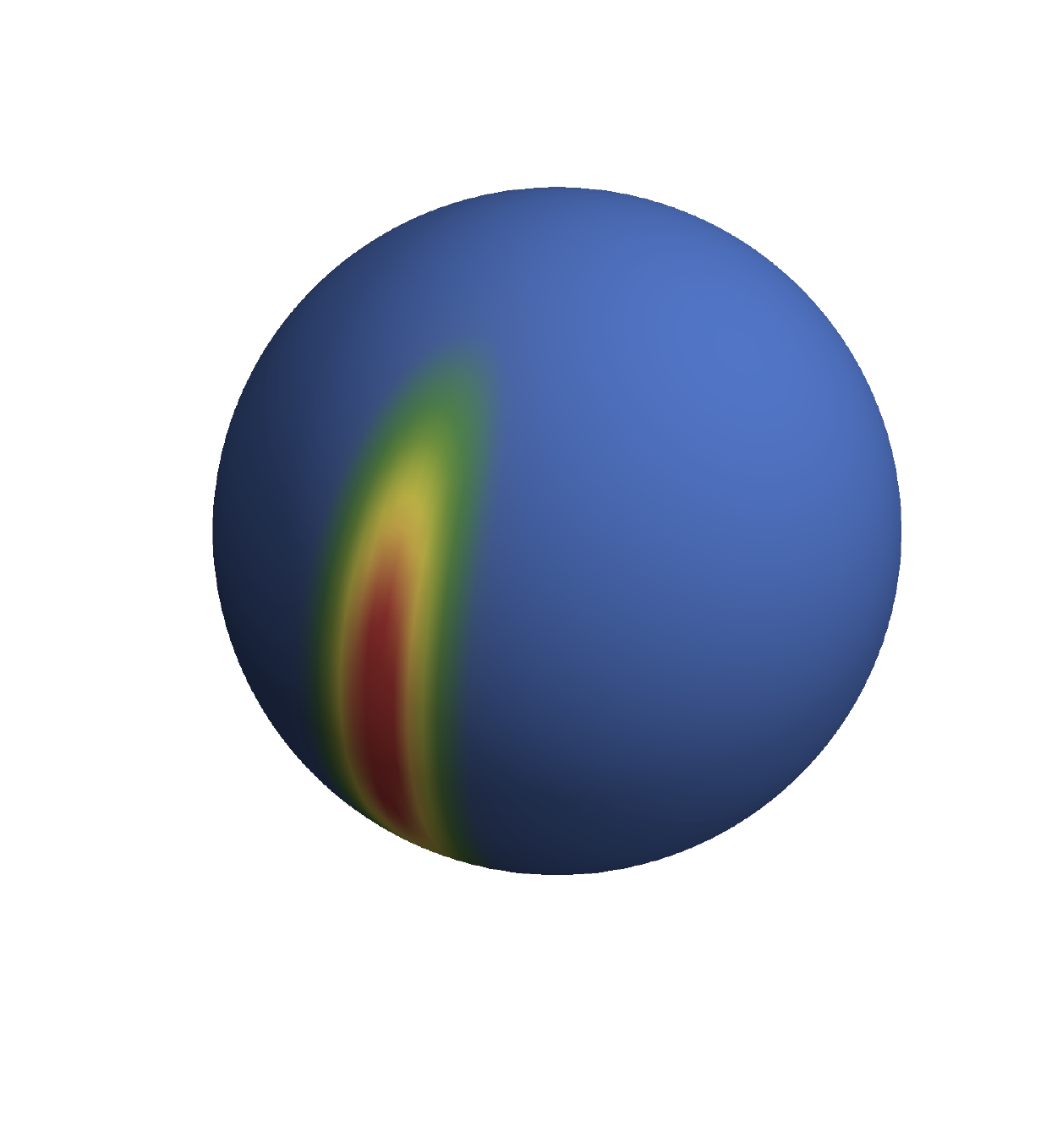}
		\end{center}
	\end{minipage}
	\newline
	\begin{minipage}{\linewidth}
		\begin{center}
		\hspace{10pt} \includegraphics[width=0.4\linewidth]{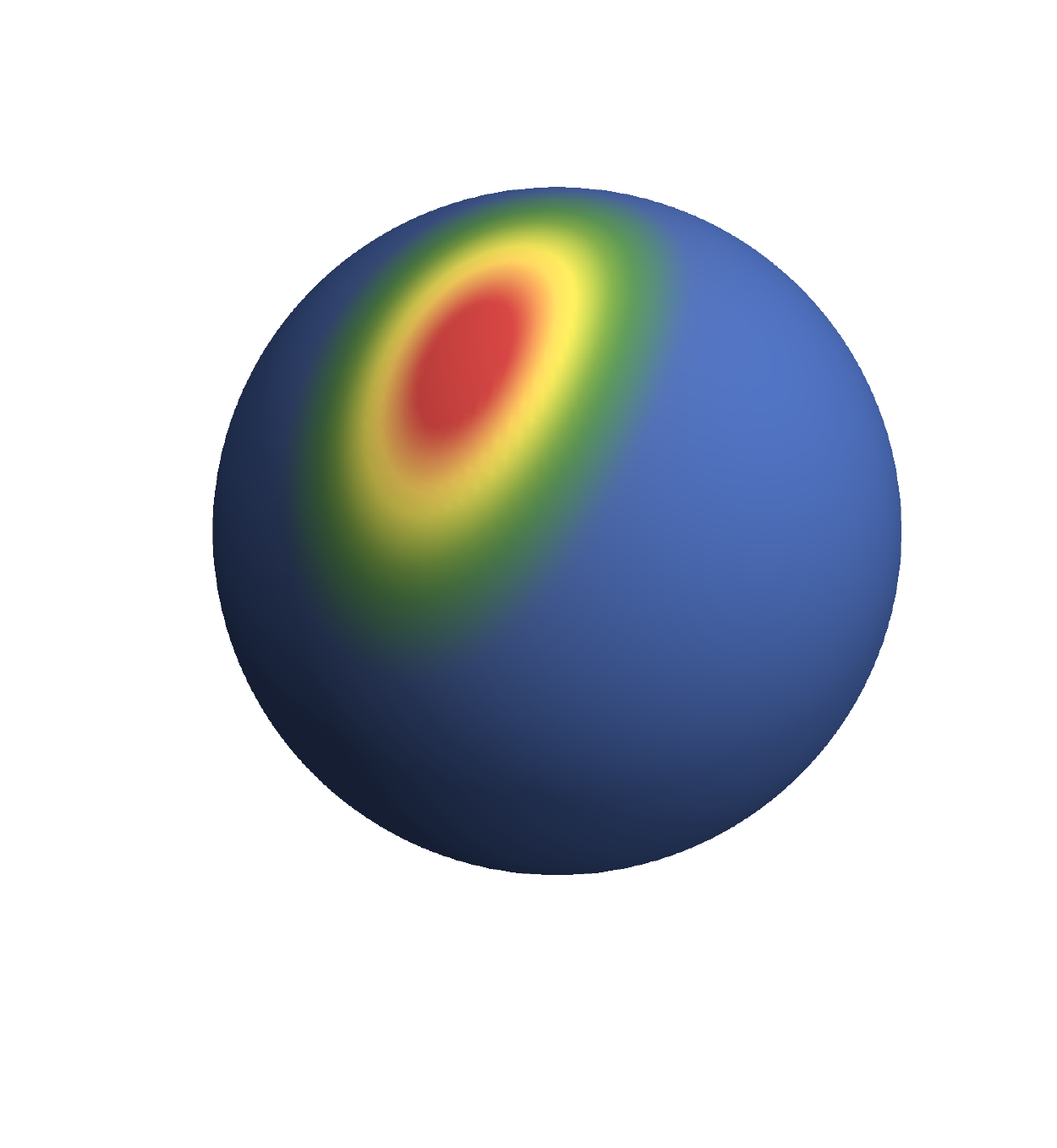} \hspace{10pt}
		\includegraphics[width=.4\linewidth]{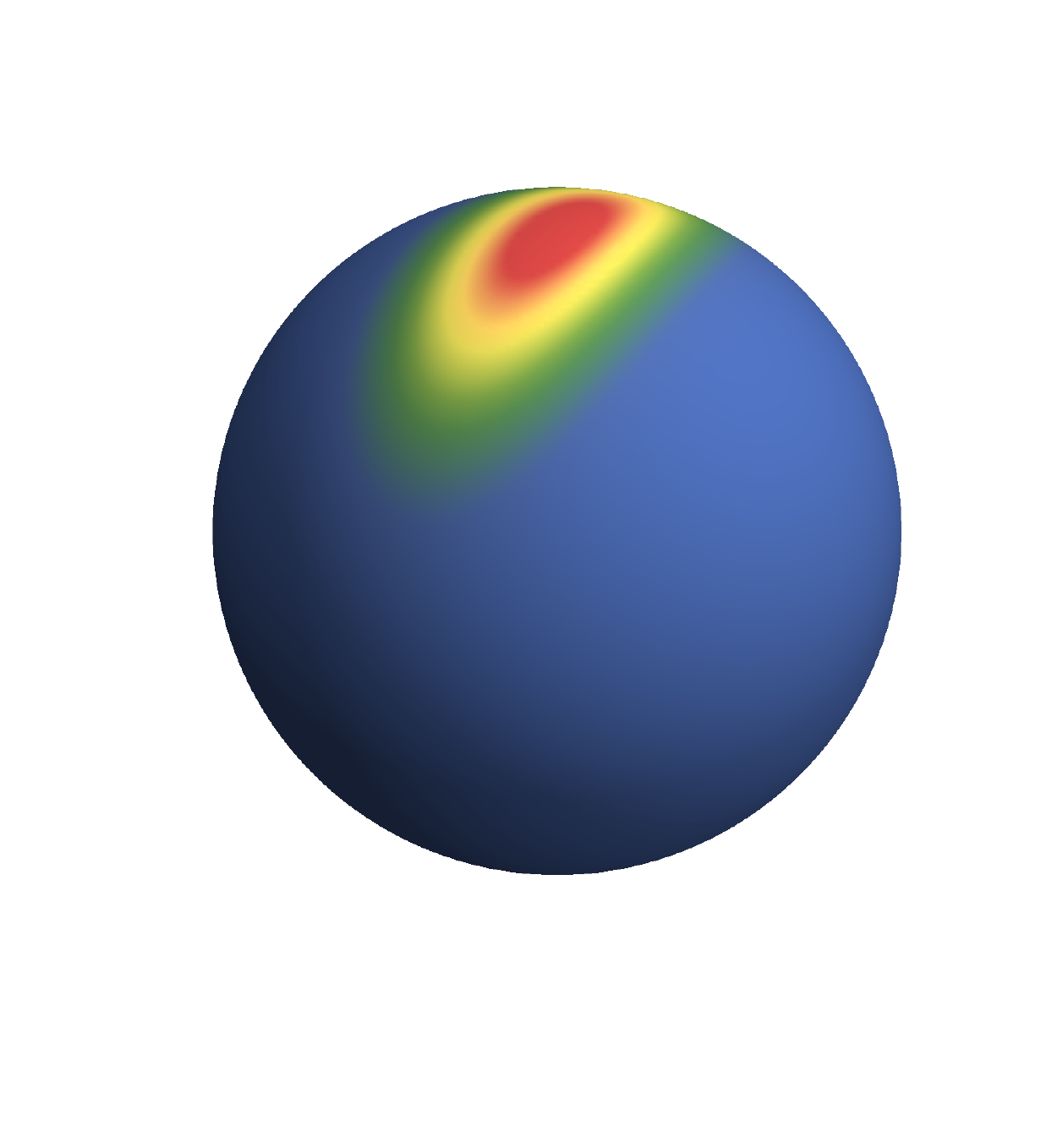}
		\end{center}
	\end{minipage}

	\caption[Schematic description of the collective spin behavior in the two-photon Dicke model.]{Schematic description of the collective spin behavior in a generalized Bloch sphere. Top left: $\gtwo<\gtwo_p$, top right: $\gtwo\lesssim \gtwo_p$, bottom left: $\gtwo_p<\gtwo<\gtwo_c$, bottom right: $\gtwo\lesssim \gtwo_c$. In the first phase, the spins fluctuate around the $z$-axis. The fluctuations are squeezed along the $y$ direction, a behavior that becomes more pronounced near the transition. In the second phase, the spins experience collective rotation in the $x-z$ plane.}
	\label{Spinbehavior}
\end{figure}

Let us summarize the behavior of the spins across the parameter space. For $\gtwo\leq \gtwo_p$, the spins are polarized along the $z$-axis. When $\gtwo$ increases, the fluctuations of $\Jy$ are reduced, while those of $\Jx$ are amplified proportionally. In the second phase, the spin polarization will gradually evolve from the $z$-axis to the $x$-axis as $\gtwo$ increases. Near the spectral collapse $\gtwo=\gtwo_c$, the spins are polarized along the $x$-axis. This behavior is schematically depicted in Fig.\ref{Spinbehavior}.
As for the $\aop$ field, we have a coherent vacuum state in the first phase. Even when the spin fluctuations become large close to the critical point, the field state is only slightly modified, since the field frequency is much larger than the characteristic spin energy. At the transition, the field acquires squeezing properties, which increase with $\gtwo$ and diverge at the spectral collapse. We can also compute the critical exponents associated with the transition. The square spin variance diverges as $\text{Var}^2(\bop+\bdag)\propto\lvert \lambda-1\rvert^{-1/2}=\lvert \gtwo-\gtwo_p\rvert^{-1/2}$, and the gap vanishes as $\Delta E\propto\lvert \lambda-1\rvert^{1/2}$. These exponents are identical to the ones of the one-photon Rabi and Dicke model.\\ 

We have described the system away from the critical point. Close to the transition, higher-order terms need to be taken into account by pushing the SW at the next order, similarly to the treatment of the Rabi model. The detailed calculations can again be found in Appendix A. After projecting out the $\Ko$ degree of freedom, we obtain the effective Hamiltonian:

\begin{align}
	\nonumber
	\frac{\Hop_f^{(4)}}{\omega}=&-\frac{1}{2} + \frac{\ratbis}{N} \left(\opddag\dop - \frac{\lambda^2}{4}(\dop+\opddag)^2\right)-\frac{\ratbis^2}{N^2}\frac{\lambda^2}{4}-\frac{\ratbis^2}{16N^2}\lambda^4(\opddag+\dop)^4\\
&+\frac{\ratbis}{8N^2}\lambda^2[(\dop+\opddag)({\dop}^{\dagger2}\dop+\opddag\dop^2)+({\dop}^{\dagger2}\dop+\opddag\dop^2)(\dop+\opddag)].
	\label{order2SW}
\end{align}

Although this expression is too complicated for an exact analytical study, it can be considerably simplified by the argument we already used in Chapter 2 for the Dicke model. The last term $(\dop+\opddag)({\dop}^{\dagger2}\dop+\opddag\dop^2)+({\dop}^{\dagger2}\dop+\opddag\dop^2)(\dop+\opddag)$ involves quartic products of quadratures for the $\dop$ field. Near the critical point, we expect that the fluctuations of $\dop+\opddag$ will be large, while those of $i(\dop-\opddag)$ will be suppressed. Therefore, we expect that the dominant term in this product of quadrature will be the one involving only the dominant quadrature, which is proportional to $(\dop+\opddag)^4$. A straightforward calculation gives $(\dop+\opddag)({\dop}^{\dagger2}\dop+\opddag\dop^2)+({\dop}^{\dagger2}\dop+\opddag\dop^2)(\dop+\opddag)\sim \frac{(\dop+\opddag)^4}{2}$. Then we obtain (dropping constant terms):

\begin{equation}
	\frac{\Hop_f^{(4)}}{\omega}\sim \frac{\ratbis}{N} \left(\opddag\dop - \frac{\lambda^2}{4}(\dop+\opddag)^2\right)+\frac{\ratbis^2}{16N^2}\lambda^2\left[\frac{\gtwo_c^2}{\gtwo_p^2}-\lambda^2\right](\opddag+\dop)^4.
	\label{order2SW}
\end{equation}

Two regimes of parameters are apparent in this expression: for $\gtwo_p<\gtwo_c$, we will have a transition at $\lambda=1$ in which the quadratic term vanishes and the quartic term is positive. This is exactly the same behavior as the Rabi and Dicke model. The finite-size scaling exponent $\text{Var}^2(\dop+\opddag)\propto N^{1/3}$ is also identical.
By contrast, if $\gtwo_c<\gtwo_p$, the quartic term becomes unstable before the phase transition, when $\gtwo$ reaches $\gtwo_c$, and the spectral collapse occurs.

\section{Driven-dissipative case}

\subsection{Introduction}
So far, we have focused on the Hamiltonian model. We will now consider the case of an open system. We will include three dissipative channels: individual qubit decay, individual qubit dephasing, and photon loss. The behavior of the system is  described by the following Lindblad equation: 

\begin{equation}
\label{Lindblad}
\partial_t\hat{\rho}=-i[\hat{H},\rho]+\kappa \mathcal D[\aop](\hat{\rho}) + \sum_{j=1}^N \Gamma_{\downarrow}\mathcal D[\hat{\sigma}^j_-](\hat{\rho}) + \Gamma_{\phi}\mathcal D[\hat{\sigma}^j_z](\hat{\rho}), 
\end{equation}
where $\hat{\rho}(t)$ is the density matrix of the system at time $t$, $\Hop$ is given by \eqref{Htwophoton} (we will use $\frac{\gcoll}{\sqrt{N}}$ rather than $\frac{g}{N}$ for the coupling convention), and $\mathcal D[\hat{A}]$ are the Lindblad dissipation superoperators.
\begin{equation*}
\mathcal D[\hat{A}]\hat{\rho}=\hat{A}\hat{\rho}\hat{A}^{\dagger}-\frac{1}{2}\hat{\rho}\hat{A}^{\dagger}\hat{A}-\frac{1}{2}\hat{A}^{\dagger}\hat{A}\hat{\rho}.
\end{equation*}

Although the symmetry of a Hamiltonian may not be directly translated into one of a Lindbladian, Eq.\eqref{Lindblad} also remains invariant under the transformation \eqref{symmetry}. In this regard, the Lindblad master equation presents a four-folded symmetry similar to that of the Hamiltonian case. Let us note that this equation is a phenomenological one, which is suited to describe a quantum simulation setup. To describe dissipation in a genuine system, it is necessary to use dressed operators, which lead towards the true polaritonic ground state. By contrast, the bare operators which we consider here effectively pump energy into the system, forcing the system away from its ground state and towards a different steady-state. Therefore, this analysis aims to describe effective implementations of the model, where the incoherent processes can be tuned by bath-engineering techniques. Note also that, since quantum simulation allows to renormalize both the effective frequencies $\omega$ and $\Omega$ and the coupling constant $\gcoll$, the dissipation constants may be large compared with $\omega$, $\Omega$ and $\gcoll$, while remaining small compared with the actual frequencies of the system. \\

\subsection{Cumulant expansion}

Because the dissipation acts on each qubit individually, it is no longer possible to treat the qubits as a collective bosonic mode. Informally, we may say that we can no longer ignore the fact that the system is many-body. The complete quantum behavior of the system becomes very difficult to obtain in this case. However, we can still study the average values of a few well-chosen quantities. 

We will be interested again in the average values of $\Jx$, $\Jy$, $\Jz$, $\adag\aop$, $\Xop=\aop^2+\adagsq$, $\Yop=i(\adagsq-\aop^2)$. Starting from  Eq.\eqref{Lindblad}, we obtain the following system of equations:
\begin{equation}\label{Eq:observables_not_closed}
    \begin{cases}
    \partial_t \moy{\hat{X}}=-\kappa \moy{\hat{X}}+2\omega \moy{\hat{Y}}, \\
    \partial_t \moy{\hat{Y}}=-\kappa \moy{\hat{Y}}-2\omega \moy{\hat{X}} - 8\gcoll\frac{\moy{\hat{J}_x}}{\sqrt{N}} -16\gcoll\frac{\moy{\hat{J}_x \adag\aop}}{\sqrt{N}},\\ 
    \partial_t \moy{\adag\aop}=-4 \gcoll \frac{\moy{\hat{J}_x\hat{Y}}}{\sqrt{N}}-\kappa \moy{\adag\aop},\\
    \partial_t \moy{\hat{J}_x}=-2\Omega\moy{\hat{J}_y} - \Gamma^\prime \moy{\hat{J}_x},\\
    \partial_t \moy{\hat{J}_y}=2\Omega\moy{\hat{J}_x}-\Gamma^\prime \moy{\hat{J}_y}-\frac{2\gcoll}{\sqrt{N}}\moy{\hat{J}_z \hat{X}},\\
    \partial_t \moy{\hat{J}_z}=\frac{2\gcoll}{\sqrt{N}}\moy{\hat{J}_y \hat{X}}-\Gamma_{\downarrow}\moy{\hat{J}_z}-\Gamma_{\downarrow}\frac{N}{2},
    \end{cases}
\end{equation}

where we have defined $\Gamma^\prime =2\Gamma_{\phi}+\frac{\Gamma_{\downarrow}}{2}$. We will study the steady-state solutions of the system, defined by $\partial_t\moy{\hat{A}}=0$ for all $\hat{A}$. We can immediately see that the normal phase defined in the previous section, characterized by $\moy{\Xop}=\moy{\Yop}=\moy{\adag\aop}=\moy{\Jx}=\moy{\Jy}=0$ and $\moy{\Jz}=-\frac{N}{2}$, is one of these solutions.

However, other solutions are challenging to obtain in general. Indeed, the system \eqref{Eq:observables_not_closed} involves higher-order correlation functions, which results in an infinite hierarchy of coupled equations. To truncate this hierarchy, we will use a first-order cumulant expansion method. We assume that the system density matrix $\rop$ can be factorized as a tensor product between the qubit and photonic part, meaning we neglect all quantum correlations between qubits and field. For instance, $\moy{\Jx\adag\aop}=\moy{\Jx}\moy{\adag\aop}$. In the context of the one-photon Dicke model, this method can predict qualitatively correct results for quantities such as $\moy{\adag\aop}$ or $\moy{\Jz}$, and becomes accurate in the thermodynamic limit $N\rightarrow\infty$ \cite{shammah_superradiance_2017,kirton_suppressing_2017,kirton_superradiant_2018}. Under this approximation, we obtain a closed set of equations:

\begin{equation}\label{Eq:observables_closed}
    \begin{cases}
    \partial_t \moy{\hat{X}}=-\kappa \moy{\hat{X}}+2\omega \moy{\hat{Y}}, \\
    \partial_t \moy{\hat{Y}}=-\kappa \moy{\hat{Y}}-2\omega \moy{\hat{X}} - 8\gcoll\frac{\moy{\hat{J}_x}}{\sqrt{N}} -16\gcoll\frac{\moy{\hat{J}_x}}{\sqrt{N}} \moy{\adag\aop},\\ 
    \partial_t \moy{\adag\aop}=-4 \gcoll \frac{\moy{\hat{J}_x}}{\sqrt{N}}\moy{\hat{Y}}-\kappa \moy{\adag\aop},\\
    \partial_t \moy{\hat{J}_x}=-2\Omega\moy{\hat{J}_y} - \Gamma^\prime \moy{\hat{J}_x},\\
    \partial_t \moy{\hat{J}_y}=2\Omega\moy{\hat{J}_x}-\Gamma^\prime \moy{\hat{J}_y}-\frac{2\gcoll}{\sqrt{N}}\moy{\hat{J}_z} \moy{\hat{X}},\\
    \partial_t \moy{\hat{J}_z}=\frac{2\gcoll}{\sqrt{N}}\moy{\hat{J}_y} \moy{\hat{X}}-\Gamma_{\downarrow}\moy{\hat{J}_z}-\Gamma_{\downarrow}\frac{N}{2}.
    \end{cases}
\end{equation}

The normal phase is still a solution to this set of equations. However, there are two other solutions, namely:

\begin{equation}\label{Eq:solutionsuperrad}
    \begin{cases}
    \moy{\hat{X}}_{\rm ss}= \frac{4 \gcoll \omega \sqrt{N} \moy{\hat{J}_x}_{\rm ss}}{-\frac{N}{4}(\kappa^2 + 4 \omega_c^2) + 16 \gcoll^2 \moy{\hat{J}_x}_{\rm ss}^2}, \\
    \moy{\hat{Y}}_{\rm ss} = \frac{\kappa}{2 \omega} \moy{\hat{X}}_{\rm ss}, \\
     \moy{\adag\aop}_{\rm ss}= \frac{-4 \gcoll }{\kappa\sqrt{N}}  \moy{\hat{J}_x}_{\rm ss} \moy{\hat{Y}}_{\rm ss},\\
    \moy{\hat{J}_x}_{\rm ss} = \pm \sqrt{\frac{N}{4}\frac{\kappa^2 + 4 \omega_c^2}{16 \gcoll^2} + \frac{\omega \Omega}{ \left(4 \Omega^2 + \Gamma'^2\right)} \moy{\hat{J}_z}_{\rm ss}},\\    
    \moy{\hat{J}_y}_{\rm ss} = - \frac{\Gamma'}{2 \Omega}  \moy{\hat{J}_x}_{\rm ss},\\
    \moy{\hat{J}_z}_{\rm ss} = \frac{N}{2}\left(- \frac{1+ Z}{2}+ \sqrt{\left(\frac{1+ Z}{2}\right)^2 - Z \left(\frac{\gcoll_p^D}{\gcoll}\right)^2}\right),
    \end{cases}
\end{equation}
where we have introduced $Z = \frac{\omega \Gamma'}{2 \Omega N \Gamma_\downarrow}$ and 

\begin{equation}
	\gcoll_p^D=\sqrt{\frac{1}{8}\left(2 \omega + \frac{\kappa^2}{2 \omega} \right)\left(2 \Omega + \frac{\Gamma'^2}{2 \Omega}\right)}
\end{equation} 
is the critical coupling constant in the dissipative case. We can already notice that the superradiant phase is well-defined only for $\gcoll\geq\gcoll_p^D$. To determine whether the system will reach this phase, however, we also need to study the stability of each solution. We have done this by considering linear perturbation around the steady-state values:

\begin{equation}
    \vec{A}=  \left\{ \moy{\hat{X}},\moy{\hat{Y}}, \moy{\adag\aop} , \moy{\hat{J}_x} ,  \moy{\hat{J}_y},  \moy{\hat{J}_z}\right\}^{\operatorname{T}} = \vec{A}_{\rm ss} + \delta\vec{A}.
\end{equation}
For the normal solution, we have
\begin{equation}
    \partial_t \vec{A}=\partial_t(\delta \vec{A})=M_\text{N} \delta \vec{A} = \begin{bmatrix}-\kappa & 2\omega & 0 & 0 & 0 & 0 \\ 
    -2\omega & -\kappa & 0 & -\frac{8\gcoll}{\sqrt{N}} & 0 & 0 \\
    0 & 0 & -\kappa & 0 & 0 & 0  \\
    0 & 0 & 0 & -\Gp & -2\Omega & 0\\
    \gcoll\sqrt{N} & 0 & 0 & 2\Omega & -\Gp & 0 \\
    0 & 0 & 0 & 0 & 0 & -\Gamd \end{bmatrix} \delta \vec{A},
\end{equation}
while for the two other solutions 
\begin{equation}
    \partial_t \vec{A}=\partial_t(\delta \vec{A})=M_\text{S} \delta \vec{A} =\begin{bmatrix}-\kappa & 2\omega & 0 & 0 & 0 & 0 \\ 
    -2\omega & -\kappa & -\frac{16\gcoll}{\sqrt{N}}\moy{\Jx}_s & -\frac{8\gcoll}{\sqrt{N}}(1+2\moy{\adag\aop}_s) & 0 & 0 \\
    0 & -\frac{4\gcoll}{\sqrt{N}}\moy{\Jx}_s & -\kappa & -\frac{4\gcoll}{\sqrt{N}}\moy{\Yop}_s & 0 & 0  \\
    0 & 0 & 0 & -\Gp & -2\Omega & 0\\
    -\frac{2\gcoll}{\sqrt{N}}\moy{\Jz}_s & 0 & 0 & 2\Omega & -\Gp &  -\frac{2\gcoll}{\sqrt{N}}\moy{\Xop}_s \\
    \frac{2\gcoll}{\sqrt{N}}\moy{\Jy}_s & 0 & 0 & 0 & \frac{2\gcoll}{\sqrt{N}}\moy{\Xop}_s & -\Gamd \end{bmatrix} \delta \vec{A}.
\end{equation}

The first (second) solution is stable if and only if all the eigenvalues of $M_N$ ($M_S$) are negative. These eigenvalues have been computed numerically.

We have also attempted to simplify the problem through adiabatic elimination by considering $\kappa\gg\Gamd\sim\Gamma_\phi$, i.e., that the bosonic field reaches a steady-state long before the qubits do. Although this technique proved successful to describe the first solution, it failed to capture the second solution.

\subsection{Nature of the transition}

\begin{figure}
	\begin{minipage}{\linewidth}
		\begin{center}
		\includegraphics[angle=-90,width=0.4\linewidth]{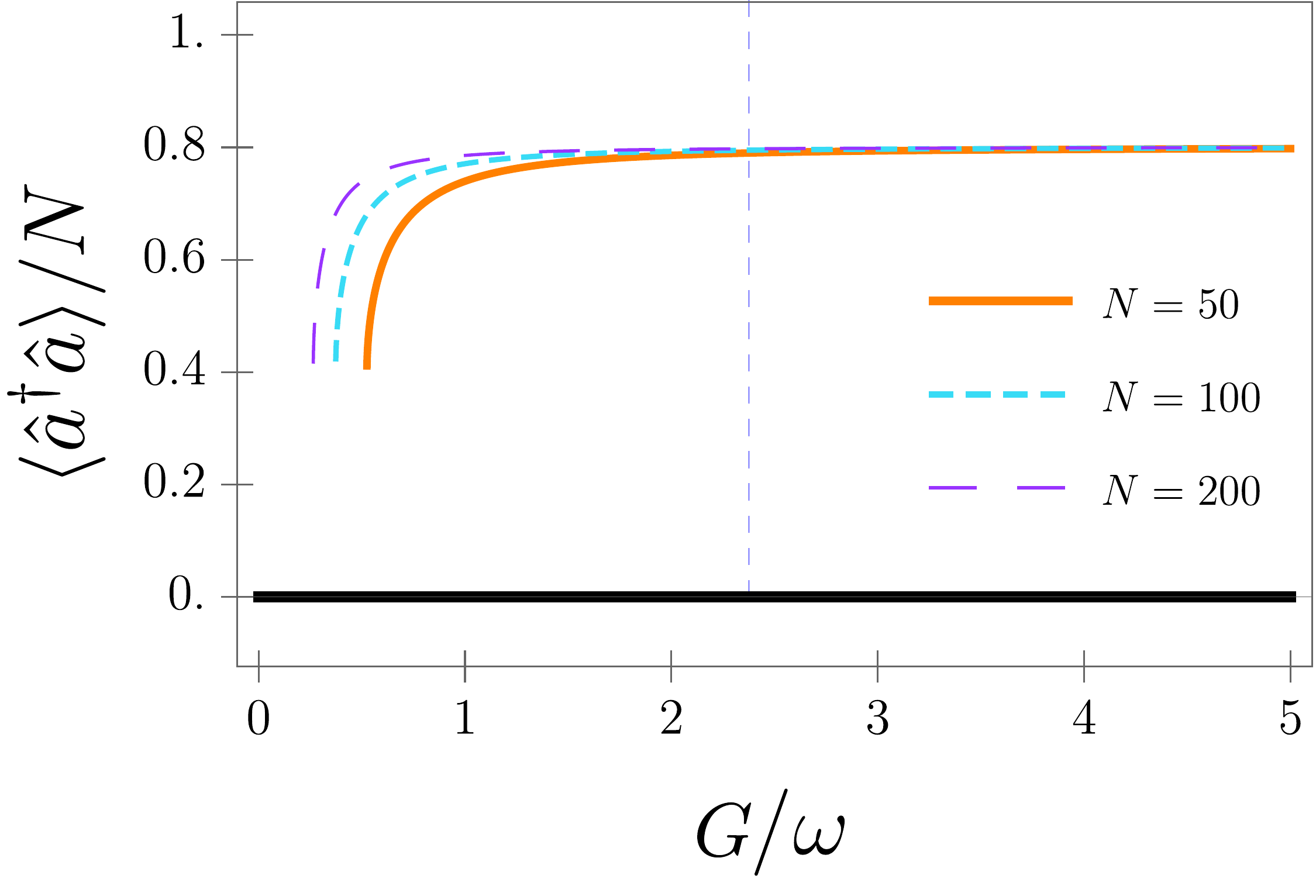}\hspace{10pt}
		\includegraphics[angle=-90,width=.4\linewidth]{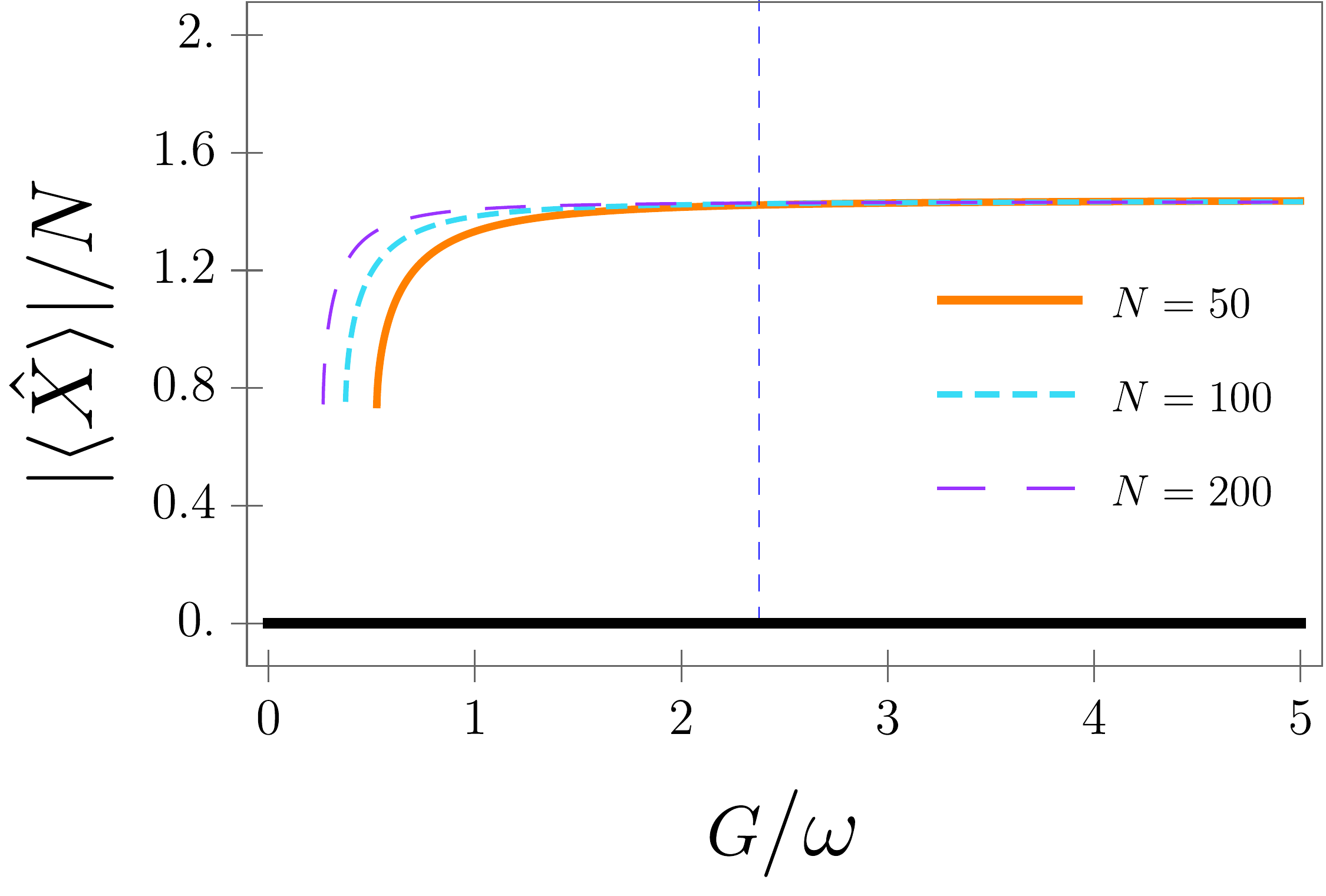}
		\end{center}
	\end{minipage}
	\newline
	\begin{minipage}{\linewidth}
		\begin{center}
		\includegraphics[angle=-90,width=0.4\linewidth]{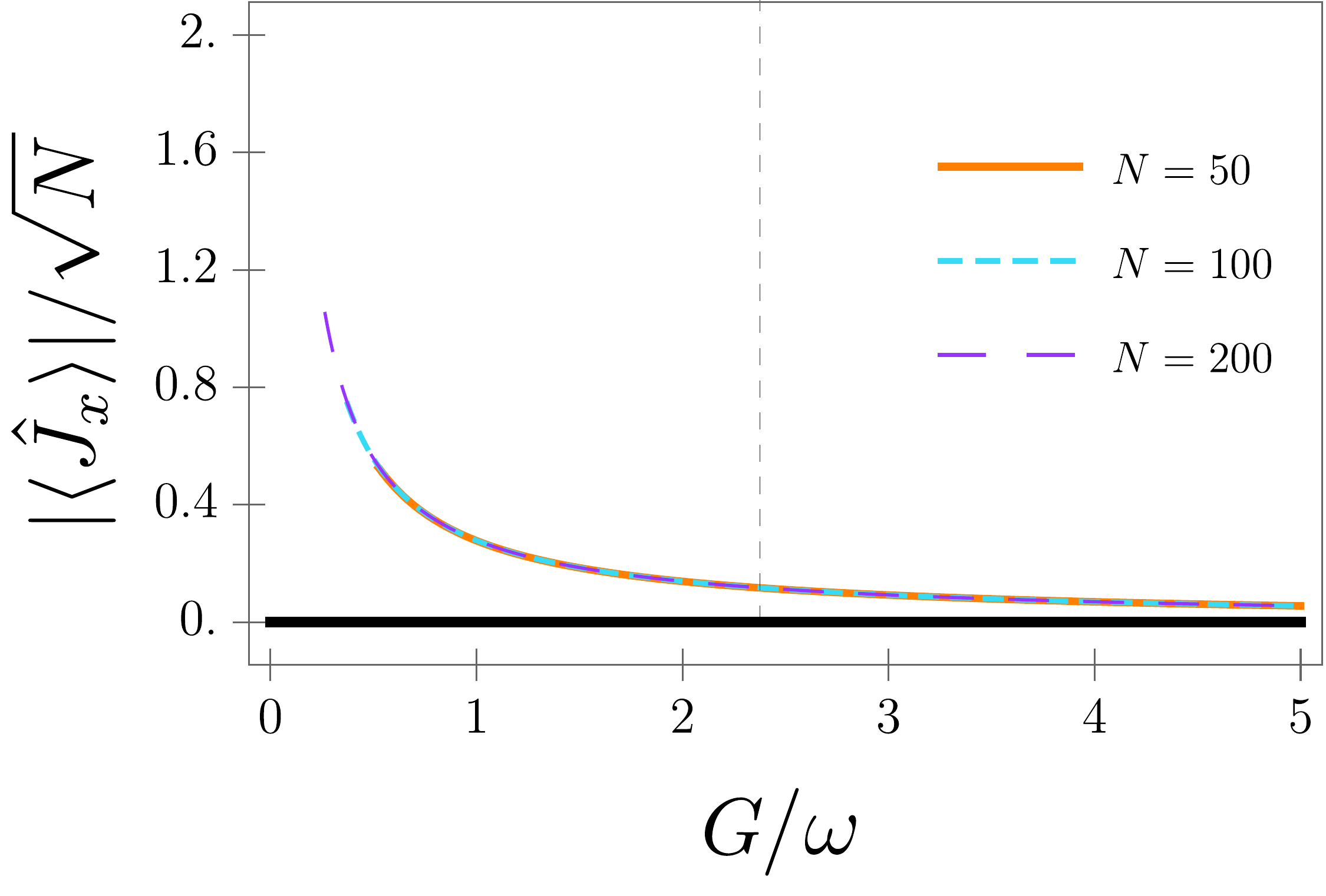}\hspace{10pt}
		\includegraphics[angle=-90,width=.4\linewidth]{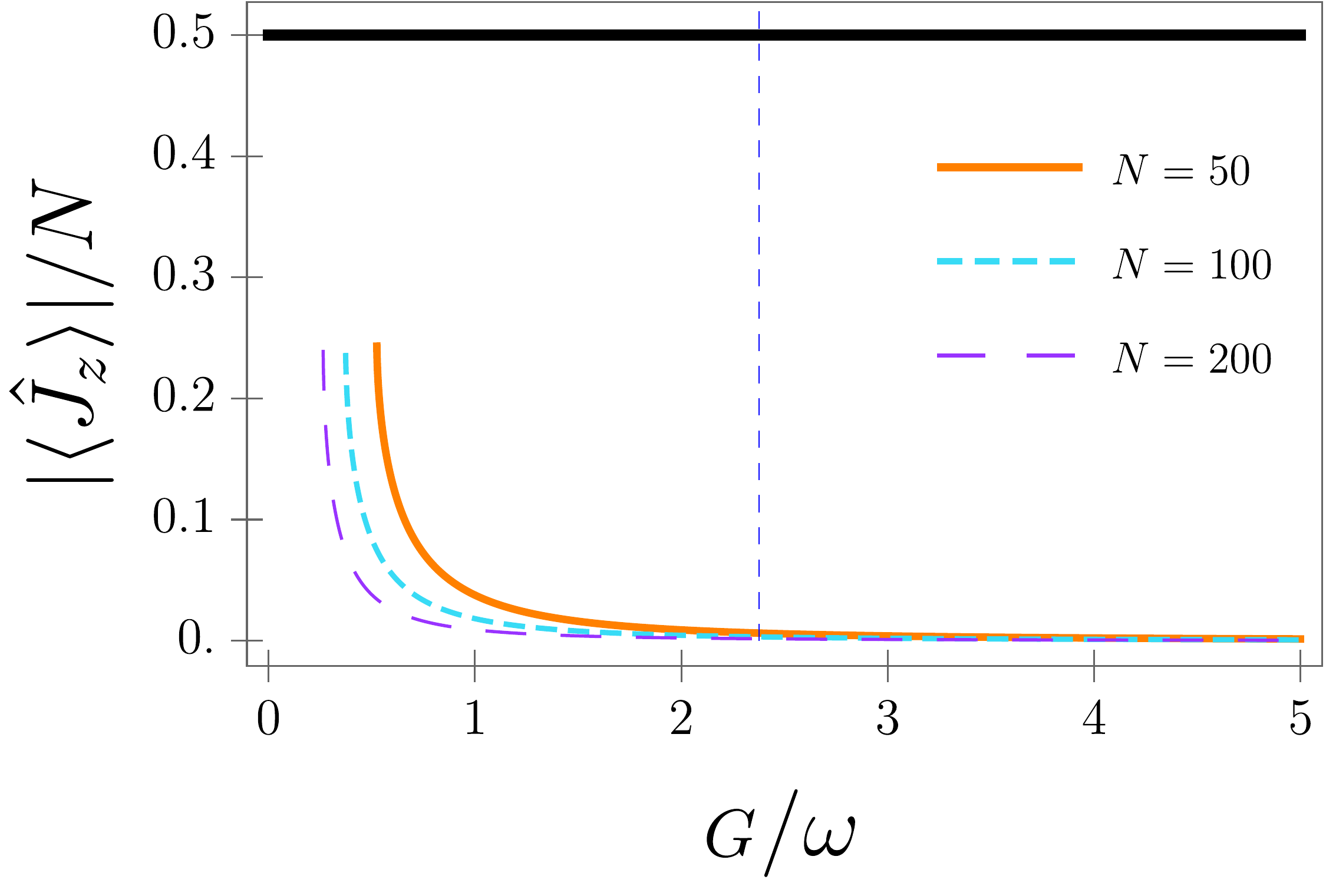}
		\end{center}
	\end{minipage}
	\caption[Various observables in the superradiant phase of the dissipative two-photon Dicke model.]{Various observables in the superradiant phase versus normalized light-matter coupling, for several qubit numbers $N$. We have set $\kappa=\omega$ and $\Gamd=\Gamp=3\omega$. For $\gcoll\leq\gcoll_p$, the superradiant phase becomes unphysical. For $\gcoll\geq\gcoll_p$, it is both physical and stable. The horizontal black line is a guide to the eye indicating the value adopted by the different observables in the normal phase. The vertical dashed line indicates the point where the normal phase becomes unstable (this point is independent of $N$). In all cases, the superradiant phase becomes stable before the normal phase becomes unstable, indicating bistability.}
	\label{plot_observables}
\end{figure}

At the mean-field level, the system has several qualitatively different steady-state. The stability analysis reveals that those solutions are stable for different parameter values, which hints that a phase transition is taking place.
In Fig.\ref{plot_observables}, we show the value of several observables in one of the solutions \eqref{Eq:solutionsuperrad}. These solutions closely resemble what we have in the squeezed phase of the Hamiltonian $\Hop$: the field $\aop$ is squeezed, while the spins experience collective rotation in two possible directions (although this time the spin also has a nonzero $\Jy$ component). The squeezing is again correlated with the spin rotation direction. Finally, these solutions break the four-folded symmetry of the Liouvillian. 

However, the plots of Fig.\ref{plot_observables} also reveals several essential differences between the two phases. First, the dissipative solutions are well-defined for $\textit{all}$ values of $\gcoll\geq \gcoll_p^D$.  
We have also studied the stability of these solutions for very large values of $\gcoll$ (up to $10^3$), and found no equivalent of the spectral collapse. Therefore, the dissipative two-photon Dicke model is characterized by a single coupling constant scale $\gcoll_p$, instead of two scales $\gcoll_p$ and $\gcoll_c$. Moreover, both $\moy{\adag\aop}$ and $\moy{\Xop}$ scale linearly with the number of qubits. Hence, this time, the transition can be properly called "superradiant". Moreover, the disappearance of the spectral collapse also means the frequency condition $\Omega=\frac{\omega}{N}$ is now lifted. Indeed, the transition may even take place in the resonant regime $\Omega=\omega$, which could strongly simplify experimental implementations.

Two other important facts need to be discussed: first, the number of photons does not go to zero when one approaches the limit of stability from above. Second, the point at which the normal phase becomes unstable and the superradiant phase becomes stable do not coincide. Therefore, the driven-dissipative two-photon Dicke model exhibits bistability at the mean-field level. 
This is very similar to what happens in the one-photon Dicke model. In the absence of dissipation, the one-photon Dicke Hamiltonian is characterized by a continuous, second-order transition; by contrast, the driven-dissipative Dicke model exhibits bistability at the mean-field level \cite{bowden_first-_1979,gelhausen_dissipative_2018}. When the quantum fluctuations are taken into account, the two stable branches collapse into a single quantum solution which experience a sharp, first-order solution \cite{gelhausen_dissipative_2018}. 
The presence of mean-field bistability associated with a first-order transition at the full quantum level has been observed in many other open quantum systems, such as Kerr resonators \cite{bartolo_exact_2016}, driven spin systems \cite{jin_phase_2018,landa_multistability_2020}, or the Bose-Hubbard model \cite{le_boite_bose-hubbard_2014,savona_spontaneous_2017,foss-feig_emergent_2017,vicentini_critical_2018}.

\subsection{Effect of dissipation}

\begin{figure}
	\includegraphics[angle=-90,width=\linewidth]{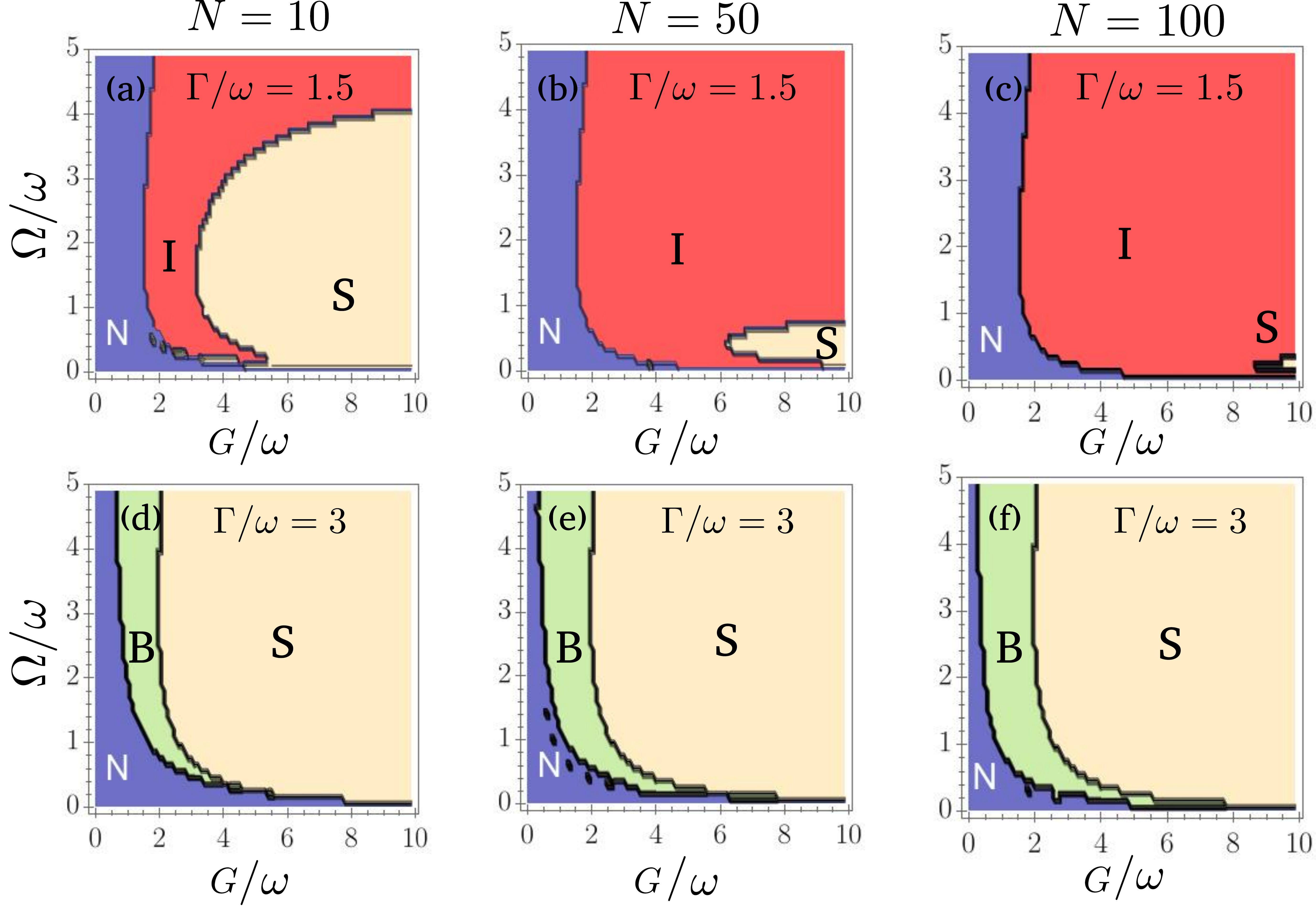}
	\caption[Phase diagram of the dissipative two-photon Dicke model.]{Phase diagram of the model for various $\Gamd=\Gamp=\Gamma$ and $N$, with $\kappa=\omega$. Upper row: $\Gamma/\omega=1.5$, lower row: $\Gamma/\omega=3$. N: normal phase, S: superradiant phase, B: bistable phase, I: instability. For $\Gamma/\omega=1.5$, the region of stability for the superradiant phase shrinks when the number of qubits increases. For $\Gamma/\omega=3$, a bistable behavior is observed, and the phase diagram is almost invariant when the number of qubits is increased beyond a few dozens.}
\label{Phasediag_gw}
\end{figure}

By studying the stability of both phases for a broad range of parameters, we can produce the phase diagram of the model. The analysis of these diagrams reveals the existence of two regimes of dissipation. In Fig.\ref{Phasediag_gw}, we display the phase diagram in the $\gcoll-\omega$ plane, for two values of $\Gamd=\Gamp=\Gamma$: $\Gamma=1.5\omega$ and $\Gamma=3\omega$, and for various qubit number $N$. For $\Gamma=1.5\omega$ and the smaller value $N=10$ qubits, we observe that the mean-field equations predict the existence of a zone where the superradiant phase is stable. However, the size of this zone shrinks when $N$ increases. Since the mean-field description becomes correct only for $N\rightarrow\infty$, \textit{no phase transition can happen in the mean-field limit for this value of dissipation}. The system will either reach the normal steady-state or be unstable. 
For $\Gamma=3\omega$, however, we observe that bistability becomes possible. In the thermodynamic limit, the region of stability becomes independent of the number of qubits, meaning that \textit{a phase transition can take place in the $N\to\infty$ limit}. 
An examination of other values for $\Gamma/\omega$ confirms that there are two regimes of dissipation: a large dissipation regime in which a phase transition is possible, and a low dissipation regime in which only the normal phase is stable in the thermodynamic limit.

\begin{figure}[ht]
\includegraphics[angle=-90,width=\linewidth]{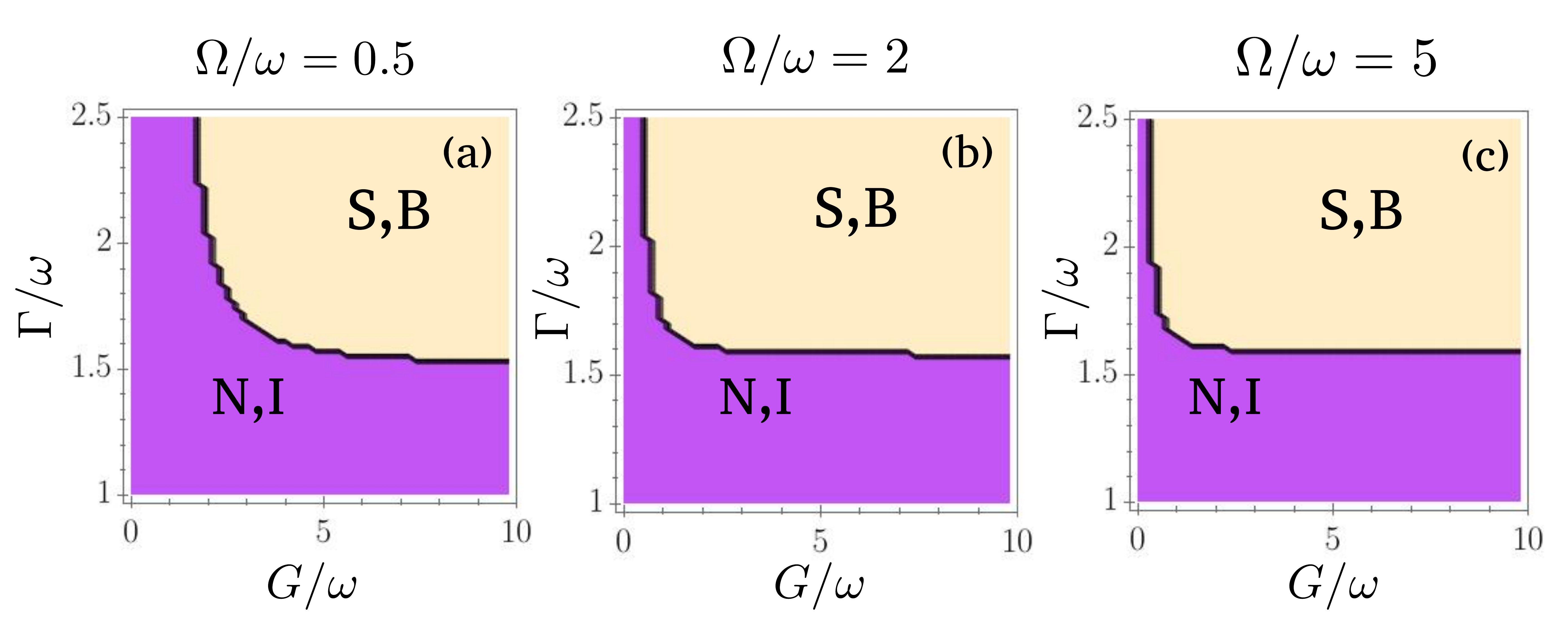}
\caption[Stability of the two-photon Dicke transition versus coupling and dissipation.]{Stability of the superradiant phase versus coupling and dissipation, for various frequencies. For all plots, $\kappa=\omega$ and $N=100$ qubits. S,B: superradiant or bistable phase; N,I: normal phase or instability. Here, we do not distinguish between the normal and unstable phase or between superradiant and bistable phase because we want to focus only on the stability of the superradiant phase. Except for small values of $\Omega/\omega$ or $\gcoll$, the value of $\Gamma/\omega$ for which the transition occurs is almost the same for all parameters, around $1.6$. This means the phase diagram as a whole changes drastically when $\Gamma$ goes across this value.}
\label{Phasediag_gGamma}
\end{figure}

Interestingly, the transition between these two regimes of parameters when $\Gamma$ increases is quite sharp, especially in the thermodynamic limit. To visualize this, we study the stability of the superradiant phase versus both $\gcoll$ and $\Gamma$, for $N=100$ qubits, and for various values of $\Omega$, the other parameters being the same (this amounts to taking horizontal slices in Figure~\ref{Phasediag_gw} and study their evolution when $\Gamma$ changes).
The results are displayed in Fig.~\ref{Phasediag_gGamma}: for $\Gamma/\omega\approx 1.6$, the instability disappears and the superradiant phase becomes stable for almost all values of $\Omega$ and $\gcoll$. Hence, the phase diagram as a whole changes drastically when $\Gamma/\omega$ goes across this threshold.

Hence, we have established that spin dissipation is essential in stabilizing this superradiant transition. We propose the following interpretation for this behavior: as we discussed in the previous section, the spectral collapse emerges from the competition between the positive field term $\omega\adag\aop$ and the interaction term, which is an inverse quadratic potential as long as $\Jx<0$. The spin dissipation, however, prevents the value of $\lvert\moy{\Jx}\rvert$ from growing too much, and thus tends to remove this instability. Since the photon loss term acts as an effective quadratic potential, we may also expect that it will tend to stabilize the dynamic. However, it also drives the system towards the vacuum of photon, increasing the value of $\gcoll_p$, and thus making the superradiant phase more difficult to produce.\\

\begin{figure}[ht]
    \centering
    \includegraphics[angle=-90,width=\linewidth]{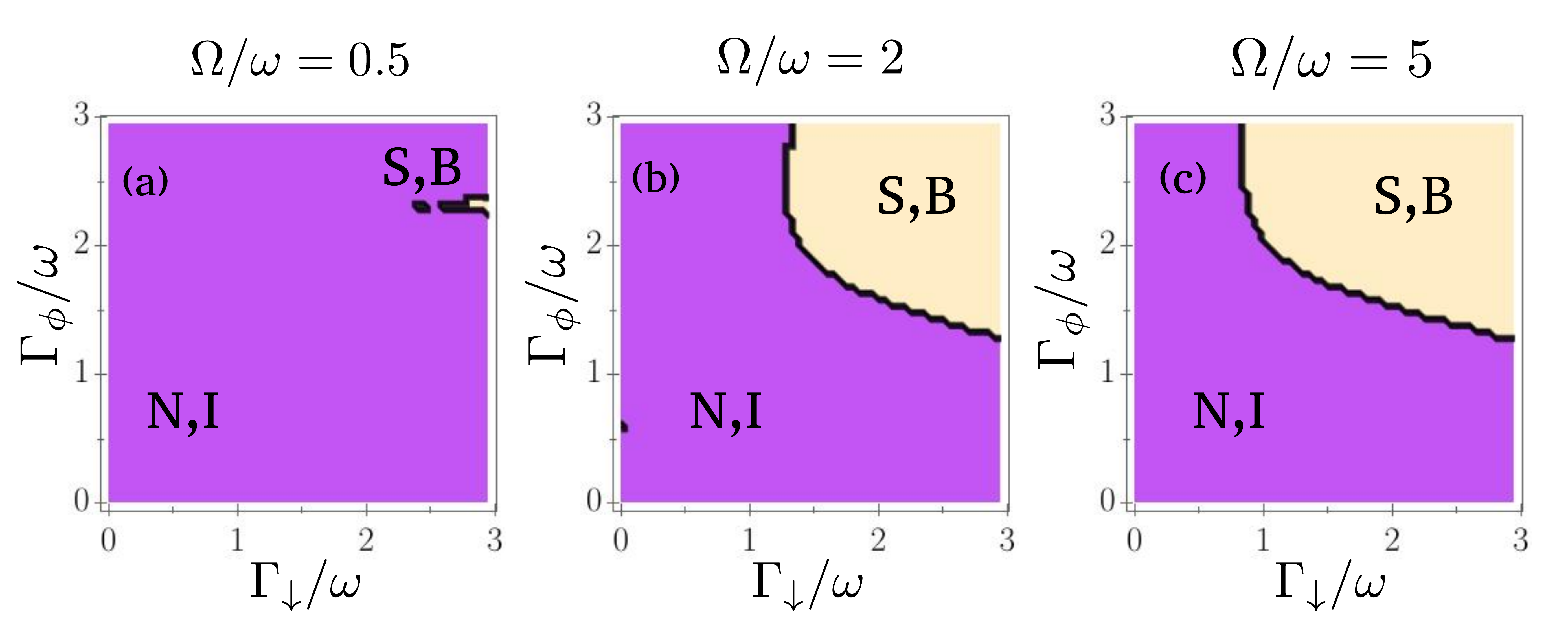}
    \caption[Stability of the two-photon Dicke transition versus qubit dephasing and decay.]{Stability of the superradiant phase versus qubit dephasing and decay, for various frequencies. For all plots, $\kappa/\omega=1$, $\gcoll/\omega=1$, and $N=100$ qubits. S,B: superradiant or bistable phase; N,I: normal phase or instability. Here, both dephasing and decay contribute positively to the stabilization of the superradiant phase.}
    \label{Phasediag_GG}
\end{figure}

To conclude this analysis, we have tried to discriminate the respective roles played by spin dephasing and spin decay. 
In the one-photon Dicke model, it was shown that the superradiant transition is destroyed in the presence of spin dephasing alone, but restored in the presence of both dephasing and decay \cite{kirton_suppressing_2017}. We did not find any equivalent of this phenomenon in our case. If the decay rate $\Gamd$ is identically zero but $\Gamp\neq0$, the superradiant solution \eqref{Eq:solutionsuperrad} is still well-defined. Furthermore, both decay and dephasing can contribute positively to the stabilization of the phase. This is illustrated in Fig.\ref{Phasediag_GG}, where we show the stability of the superradiant phase with respect to both $\Gamd$ and $\Gamp$. (Note that in both Fig.\ref{Phasediag_gGamma} and Fig.\ref{Phasediag_GG}, we have aggregated the normal phase and instability and the superradiant and bistable phase, because we only want to know whether the superradiant phase is stable or not, independently of the stability of the normal phase.)  

\section{Conclusion}

In this Chapter, we have presented some new results concerning the presence of a phase transition in a many-body system with two-photon coupling, both in equilibrium and driven-dissipative settings.

In the former case, the competition between the field potential and interaction term leads to both a second-order phase transition and a spectral collapse. During the transition, the field experiences squeezing correlation with a collective spin rotation, a phenomenology  similar to the superradiant transition of the one-photon Dicke model. A hallmark of one-photon superradiance is the linear scaling of the photon population with the number of spins $N$. In the two-photon case, however, the situation is much more complex. Because the transition and collapse point scale differently with $N$, a meaningful analysis can only be performed when the frequency ratio $\frac{\Omega}{\omega}$ itself is reduced when $N$ increases.

In the presence of dissipation, the spectral collapse is removed, at least at the mean-field level. The frequency condition is lifted, and the transition becomes a \textit{bona fide} superradiant transition. The system exhibits bistability, which hints that the transition is now first-order. The interaction term, however, can still produce an instability in this case. Although the different parameters compete in a highly non-trivial way, we showed that spin dissipation is key to stabilizing the system and enabling the transition.

These results can be extended in several directions. Although the SW method proved performant to study the phase transition in the equilibrium case, it predicts a diverging gap near the spectral collapse, a result that seems at odds with results obtained in the few-qubits limit. A study of the collapse point through different methods (such as the polaron picture \cite{cong_polaron_2019}) should elucidate this point.

In the dissipative case, including the quantum fluctuations would allow us to confirm that the mean-field bistability leads to a first-order transition at the full quantum level, similar to what has been observed in other quantum optical models. This is challenging, however, due to the fast-increasing size of the Liouvillian space. Two strategies could be simultaneously implemented to overcome this limitation: first, the spin subspace may be reduced by exploiting the permutational invariance of the spins \cite{shammah_open_2018}. Second, the use of quantum trajectories, which reduces the computational overhead from being that of the Liouvillian space to just that of an effective Hilbert space. Although a full quantum solution necessitates averaging over many runs, the intermittent dynamics characterizing a bi-stable phase can be grasped even by single quantum trajectory simulations.

\chapter{Critical metrology with a finite-size quantum system}


\epigraph{\textit{Je me suis gardé de faire de la vérité une idole, préférant lui laisser son nom plus humble d'exactitude.}}{Marguerite Yourcenar}

Physical systems near a phase transition exhibit a diverging susceptibility, suggesting that they could be used as probes for sensing tasks. This idea has already been put to use several decades ago with \textit{classical} phase transitions, in particular for particle detection.
However, our new ability to access \textit{quantum} phase transitions opens new perspectives in this respect. A growing body of work has been devoted to these questions, with many different approaches. Several important questions, however, still need to be resolved. In particular, near the transition, the closure of the energy gap  is expected to lead to a slowing down of the dynamics and an inevitable growth of the protocol duration. 

In this Chapter, we propose a metrology protocol exploiting the critical behavior of a \textit{finite-size} system. More specifically, we show how the critical behavior of the Rabi model could be used to measure the qubit frequency with accuracy. We study the time needed to implement the protocol, and show that the Quantum Fisher information achieves a quartic time scaling in the protocol duration, while ordinary interferometric protocols scale quadratically. We also show that non-trivial scaling behavior can still be achieved in the presence of dissipation.

This Chapter is divided into four sections. In the first section, we introduce the concept of quantum critical metrology, and review previous results. The second and third sections are original contributions, published in \cite{garbe_critical_2020}. In the second section, we introduce a sensing protocol based on the critical behavior of the Rabi Hamiltonian, and study the scaling of precision with time. In the third section, these results are extended to the dissipative case. Finally, some conclusions and perspectives are given in the last section.

\section{Introduction}

In a system close to a phase transition, small perturbations can lead to large, observable change. It is thus intuitive that such systems could be used for sensing tasks. This principle has already been applied with classical phase transition, most notably for particle detection. Bubble chambers are an example: when a particle goes through a liquid superheated in a metastable state, it deposits energy, which makes the liquid vaporize locally and create an observable trail. Similarly, superconducting circuits just below the superconducting-metal phase transition can be used to detect single photons with excellent efficiency \cite{irwin_transition-edge_2005,lita_counting_2008}.

\begin{figure}
	\begin{center}
	\begin{minipage}{.48\linewidth}
		\includegraphics[height=\linewidth,angle=90,origin=b]{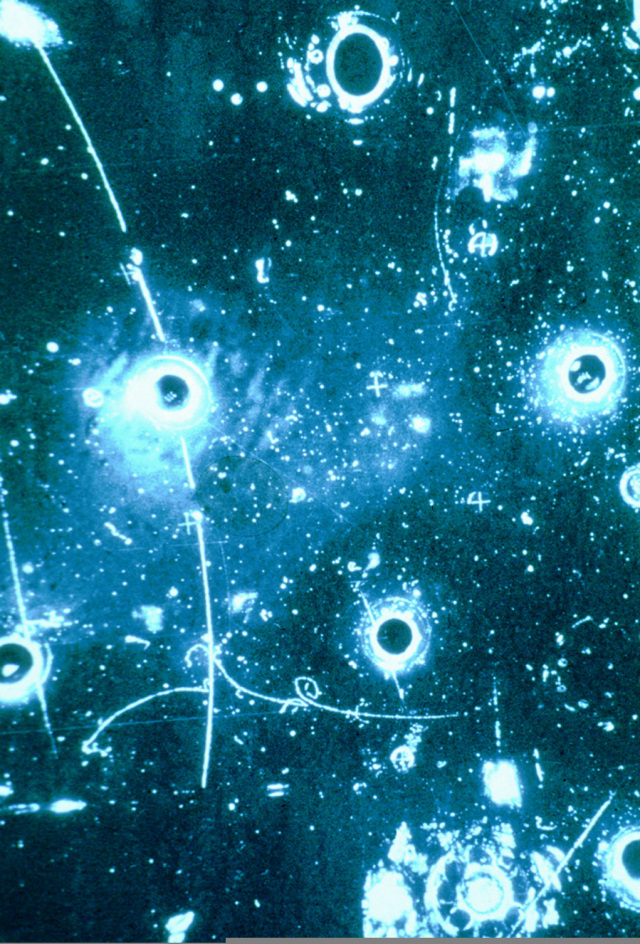}	
	\end{minipage}
	\begin{minipage}{.4\linewidth}
		\centering\includegraphics[width=.85\linewidth]{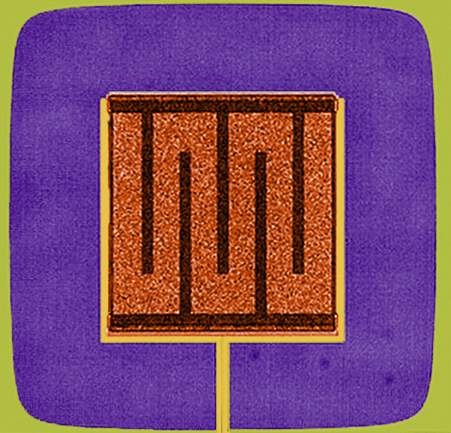}
	\end{minipage}
	\end{center}
	\caption[Two examples of sensors exploiting phase transitions: bubble chambers and superconducting photon detector.]{Two examples of sensors exploiting phase transitions. Left: trails left by a neutrino interacting with an electron in a bubble chamber (CERN image archives). Right: Superconducting circuit used for single-photon detection (NIST image archives).}
\end{figure}

In this context, quantum technologies carry two interesting perspectives. First, studying quantum phase transition in addition to classical ones extends the scope of phenomena available; in particular, quantum critical effects can also be observed at zero temperature.
Second, thanks to the development of quantum metrology, it is possible to derive both ultimate bounds on the achievable precision, and indications about how to reach this precision. Indeed, if the symmetric logarithm derivative is known, it is possible to compute which observable allows to saturate the CR bound. More generally, the tools of quantum metrology give us indications to "dissect" experimental signatures and extract the relevant information from it.\\

In the last few years, a growing number of works have studied phase transitions from a quantum sensing perspective. For instance, several protocols have been proposed to amplify a weak input signal by crossing a first-order phase transition, or by exploiting symmetry breaking effects \cite{gammelmark_phase_2011,ivanov_quantum_2015,fernandez-lorenzo_quantum_2017,yang_quantum_2019}. This is the quantum equivalent of the existing protocols presented above.

Other works have focused instead on the correlations near a critical point, with different concepts and languages. For the sake of this presentation, it is possible to roughly divide them into two classes. 

The first approach, which we will call the \textit{dynamical} paradigm, focus on the time evolution induced by a perturbation of a Hamiltonian close to a critical point \cite{wang_quantum_2014,tsang_quantum_2013,rams_at_2018,frerot_quantum_2018}. Let us take a simple example, based on the Ising Hamiltonian with transverse field $\Hop_I=-J\sum_{i,j}\sigmaiz\sigmajz - h\sum_i\sigmaix$ at zero temperature. The model exhibits a critical point at $J=h$. Near this point, the fluctuations of the collective spin $\Jz=\sum_i\sigmaiz$ become very large \cite{frerot_quantum_2018}. Then according to the discussion of Chapter 2, such a spin state can be used as an input in an interferometric experiment to accurately estimate a small magnetic field perturbation in the $z$ direction. To be precise, if we prepare the system in the ground state $\ket{GS}$ of $\Hop_{I}$, apply (instantaneously) a small magnetic field perturbation $\epsilon$ and let the system evolve under the Hamiltonian $\epsilon\Jz$ for a time $\dur$, we can evaluate $\epsilon$ with a QFI given by $\dur^2 \bra{GS}(\Jz-\langle\Jz\rangle)^2\ket{GS}$. Here the ground state evolves under an external perturbation; however, we may also consider that we change one of the parameters describing the Hamiltonian. Starting from the ground state of $\Hop(x=x_0)$, we apply a small perturbation $\epsilon$ and to let the system evolve according to $\Hop(x_0+\epsilon)$. In general, high fluctuations near the critical point can be related to accurate sensing of a "conjugate" variable, a link which can also be expressed in the language of dynamic susceptibilities \cite{hauke_measuring_2016}.\\

The second approach, which we will call the \textit{static} paradigm, instead focus on the steady-state properties of the Hamiltonian \cite{zanardi_quantum_2008,salvatori_quantum_2014,rams_at_2018,pezze_adiabatic_2019,mehboudi_thermometry_2015,invernizzi_optimal_2008,bina_dicke_2016,bin_mass_2019,wald_and_2020}. Let us take again the example of the Ising model. The idea here is that, close to the critical point, the ground state of the model for, say, a coupling value $J+\delta J$, is very different from the ground state for a coupling value $J$. Hence, if the system is prepared in its ground state (for instance by adiabatically evolving the system from $J=0$ to a value close to the critical point), all parameters except $J$ being known, it is possible to estimate $J$ directly by measuring the ground state. The associated QFI is proportional to $\langle\partial_J\psi_J|\partial_J\psi_J\rangle-(\langle\partial_J\psi_J|\psi_J\rangle)^2$, with $\ket{\psi_J}$ the ground state for a given $J$. This idea can be generalized to finite temperature and out-of-equilibrium phase transitions.
Note that the QFI can not always be attained by measuring only intuitive, macroscopic observables (such as the global magnetization of a sample). In general, extracting the relevant information may require measuring non-trivial correlation functions \cite{eckert_quantum_2008,mehboudi_thermometry_2015}. In this study, we will focus on this last approach.

Two points need to be made here. First, these two approaches do \textit{not} require to actually cross the transition, but only to get close to it. Second, for pedagogical purposes, we have presented these works from the perspective of designing sensing protocols. The use of quantum metrology to study critical phenomena, however, has a fundamental interest that goes beyond possible practical applications. In particular, the use of these new tools allows us to better understand the quantum correlations in critical systems \cite{chiara_genuine_2018}. \\

Despite all these studies, several important questions still need to be resolved. One of these questions concerns the duration of the protocol, especially in what we have called the static scenario. Indeed, adiabatically preparing the system closer and closer to the critical point while maintaining it in its ground state will require an increasing amount of time, due to the closure of the Hamiltonian gap. Similarly, if we try to prepare the system by cooling it down (or more generally by letting it interact with a dissipative bath), the time needed to reach the ground state will diverge at the critical point, an effect known as critical slowing down.  
 Therefore, it is important to carefully assess the time resources needed to implement the protocol. Although a few studies have considered the scaling of precision with time \cite{tsang_quantum_2013,macieszczak_dynamical_2016,rams_at_2018}, this question is still open in many cases.\\

Second, most of the previous studies have focused on many-body systems, where the phase transition occurs in the limit of an infinite number of components. In particular, several studies have explicitly studied the scaling of precision with the number of components, sometimes looking for super-Heisenberg scaling \cite{gammelmark_phase_2011,frerot_quantum_2018,rams_at_2018,pezze_adiabatic_2019}. However, critical effects may also be observed in finite-size systems, notably in the Rabi model \cite{bakemeier_quantum_2012,ashhab_superradiance_2013,hwang_quantum_2015,hwang_dissipative_2018,peng_unified_2019,felicetti_universal_2020}. These systems, which have so far attracted less attention in a critical metrology context, nonetheless open very interesting perspectives.
As we discussed in Chapter 2, these systems are the most natural platform in which concepts of quantum information and metrology can be studied. A question we ask here is how these quantum properties would be modified in the presence of critical behavior.
Moreover, the characteristics of these few-body systems could also be interesting from a practical perspective. For instance, compared to many-body systems, the size of the device could be reduced, which is very relevant in certain tasks. An example is scanning-probe magnetometry, in which a small probe is used to sense a local magnetic field. The use of quantum probes for this task has already been demonstrated experimentally \cite{ockeloen_quantum_2013,muessel_scalable_2014}. The use of critical correlations might improve further the precision without increasing the size of the probe, which is a key requirement to perform space-resolved magnetometry.
Using a finite-size system also removes the requirement to implement complex non-local operations \cite{eckert_quantum_2008,mehboudi_thermometry_2015} to manipulate the system and extract the relevant information from it. Thus, these systems could lead to better controllability and readability. The price to pay is to access an unusual regime of parameters, as in the Rabi model, where the critical behavior is predicted to occur when the spin frequency is much larger than the field frequency. This exotic regime, however, has already been achieved experimentally \cite{dareau_observation_2018}.\\

In this Chapter, we address these two topics, by designing a parameter-estimation protocol based on the critical steady-state properties of the Rabi model. By contrast with most previous studies, we explicitly take into account the time needed to implement the protocol, and focus on the scaling of precision versus protocol duration, instead of precision versus system size.

\section{Hamiltonian case}

\subsection{Presentation of the protocol}

Let us consider a qubit interacting with a single bosonic mode according to the quantum Rabi Hamiltonian: 

\begin{equation}
	\Hop=\Of \adag\aop + \Oq\sigz + \gind(\adag+\aop)\sigx.
	\label{Hrabi}
\end{equation}

As discussed in Chapter 2, such models can be implemented with many different systems, ranging from ions interacting with their vibrational degree of freedom \cite{dareau_observation_2018}, to atoms trapped in cavities.
In the limit $\rat=\frac{\Oq}{\Of}\rightarrow\infty$, the phase transition takes place for $\gind=\gind_p=\frac{\sqrt{\Of\Oq}}{2}$. We will study how this critical behavior could be used to accurately evaluate several parameters of the model. First, we will be interested in measuring the spin frequency $\Oq$, assuming all the other parameters are known. This task is relevant for several applications \cite{pezze_quantum_2018}. An example is magnetometry: if the qubit is realized by two Zeeman sublevels of an atom or trapped ion, magnetic field perturbations  result in changes of $\Oq$ which can then be measured. Atomic clocks are another field where spin frequency measurement is an essential task. 

\begin{figure}
	\centering \includegraphics[width=.8\linewidth]{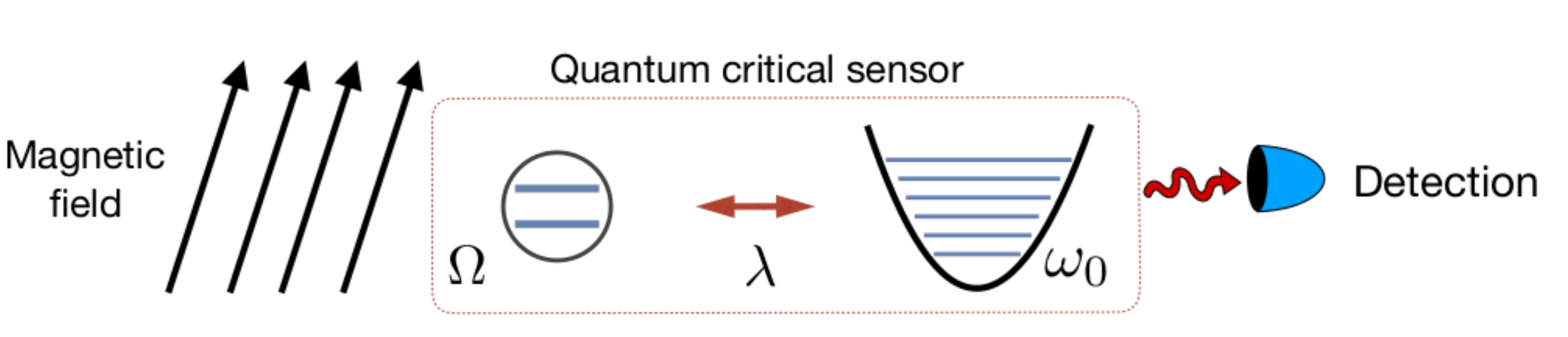}
	\caption[Schematic representation of our critical estimation protocol.]{Schematic representation of the estimation protocol: the quantum critical sensor is used to estimate the intensity of an external magnetic field.}
	\label{Protocol}
\end{figure}

Let us first outline the general procedure. The spin frequency is measured through a two-stage procedure \cite{barndorff-nielsen_fisher_2000}: first, one performs a generic, potentially suboptimal, measurement to get a rough estimate of $\Oq$. The second stage is the critical protocol  proper. The system is initialized in its ground state for $\gind=0$. Then one adiabatically tune the coupling parameter from $0$ to some value very close to $\gind_p$. 
Finally, the measurement of a relevant observable is performed, and a more accurate estimation of $\Oq$ is extracted from the measurement results, as sketched in Fig.\ref{Protocol}.\\

To evaluate the performances of this protocol, let us first characterize the ground state of the system. We recall the results of Chapter 2: the Hamiltonian \eqref{Hrabi} can be diagonalized by a SW transformation. In the normal phase, this gives:

\begin{equation}
	\Uop^{\dagger}\Hop\Uop=\Oq\sigz + \Of\adag\aop + \frac{\Of}{2}\frac{\gind^2}{\gind_p^2}\sigz (\aop+\adag)^2-\frac{1}{\rat}\frac{\Of}{16}\frac{\gind^4}{\gind_p^4}\sigz(\adag+\aop)^4,
	\label{Hrabi4}
\end{equation}

up to terms $O(\Of\rat^{-3/2})$. This effective Hamiltonian provides a description of the system behavior for all values of $\gind$ around the transition. Even at the critical point $\gind=\gind_p$, higher-order terms remain subleading. There is, however, a non-trivial interplay between the quadratic and quartic terms. Two regions of parameters can be distinguished: for $1-\frac{\gind}{\gind_p}\gg\rat^{-2/3}$, the quartic term is negligible. For $1-\frac{\gind}{\gind_p}\ll\rat^{-2/3}$, it becomes dominant. In this analysis, we will mostly focus on the first region, and keep only quadratic field terms.
We can diagonalize the Hamiltonian \eqref{Hrabi4} by projection into the lower spin eigenspace and Bogoliubov transformation. For a given value of $\gind$, the ground state is given by $$\ket{\psi^\gind_\Oq}=\ket{\xi_\Oq^\gind}\otimes\ket{\downarrow} + O(\rat^{-1/2}),$$ with $\ket{\downarrow}$ the spin eigenstate corresponding to the lowest $\sigz$ eigenvalue, and $\ket{\xi_\Oq^\gind}$ a squeezed state with squeezing parameter $\xi_\Oq^\gind=-\frac{1}{4}\text{ln}(1-\frac{\gind^2}{\gind_p^2})$. Close to the critical point, the field fluctuations become large, while the spin fluctuations remain small, due to the much larger spin frequency which suppresses the probability of exciting the spin. In turn, the excitation energy $\Of\sqrt{1-\frac{\gind^2}{\gind_p^2}}$ becomes small near the transition.\\

During the adiabatic evolution, the coupling parameter $\gind$ will evolve continuously from $\gind_0=0$ to $\gind_\dur$, with $\dur$ the total evolution time. At each time $t$, the state of the system will be given by $\ket{\psi^{\gind_t}_\Oq}$. This evolution can be seen as a quantum control procedure, during which the state evolves according to some effective (time-dependent) Hamiltonian $\tilde{H}_\Oq(t)$ (note that this Hamiltonian is very different from $\Hop$ in general). 
For two different values of $\Oq$, the system will evolve along a different path, and end up in a different final state. Our ability to distinguish between the two values $\Oq_1$ and $\Oq_2$ will then be given by the overlap between the two final states $\lvert\langle\psi^{\gind_\dur}_{\Oq_1}|\psi^{\gind_\dur}_{\Oq_2}\rangle\rvert^2$, or, for two very close values, by the QFI:
$$\pazocal{I}_\Oq(\dur)=4[\langle\partial_\Oq\psi^{\gind_\dur}_{\Oq}|\partial_\Oq\psi^{\gind_\dur}_{\Oq}\rangle + (\langle\partial_\Oq\psi^{\gind_\dur}_{\Oq}|\psi^{\gind_\dur}_{\Oq}\rangle)^2].$$ A direct calculation yields the QFI: 

\begin{eqnarray}\nonumber
\pazocal{I}_\Oq(\dur) & = & (\partial_\Oq\xi_\Oq^{\gind_\dur})^2\bra{\xi_\Oq^{\gind_\dur}}(\aop^2-\adagsq)(\adagsq-\aop^2) \ket{\xi_\Oq^{\gind_\dur}}\\
 & = & \frac{1}{8\Oq^2}\left[\frac{\gind_p^2}{\gind_p^2-\gind_\dur^2}\right]^2.
	\label{QFInormal phase}
\end{eqnarray}
The last equality is valid when we stop the evolution near the critical point: $\gind_\dur\sim\gind_p$. The minimal uncertainty associated with the measure of $\Oq$ is $1/\sqrt{\pazocal{I}_\Oq}$; the associated signal-to-noise ratio is given by:

\begin{equation}
	\pazocal{Q}_\Oq=\Oq\sqrt{\pazocal{I}_\Oq}\sim\frac{1}{2\sqrt{2}}\left[\frac{\gind_p^2}{\gind_p^2-\gind_\dur^2}\right].
\end{equation}

An important comment needs to be made here. Near the transition, the squeezing value $\xi_\Oq^\gind$ becomes large. However, the expression of the QFI \eqref{QFInormal phase} does not involve $\xi_\Oq^\gind$, but its derivative with respect to $\Oq$. In other words, the accuracy of this protocol does not come from the large amount of squeezing, but from the fact that this amount is very sensitive to the value of $\Oq$. This is a general feature of what we have called static protocols. In the dynamical paradigm, by contrast (for instance \cite{frerot_quantum_2018}), the Hamiltonian ground state is used as an input for an interferometric protocol. In this case, the precision is indeed a consequence of the large amount of squeezing.

\begin{figure}
	\centering\includegraphics[angle=-90,width=.7\linewidth]{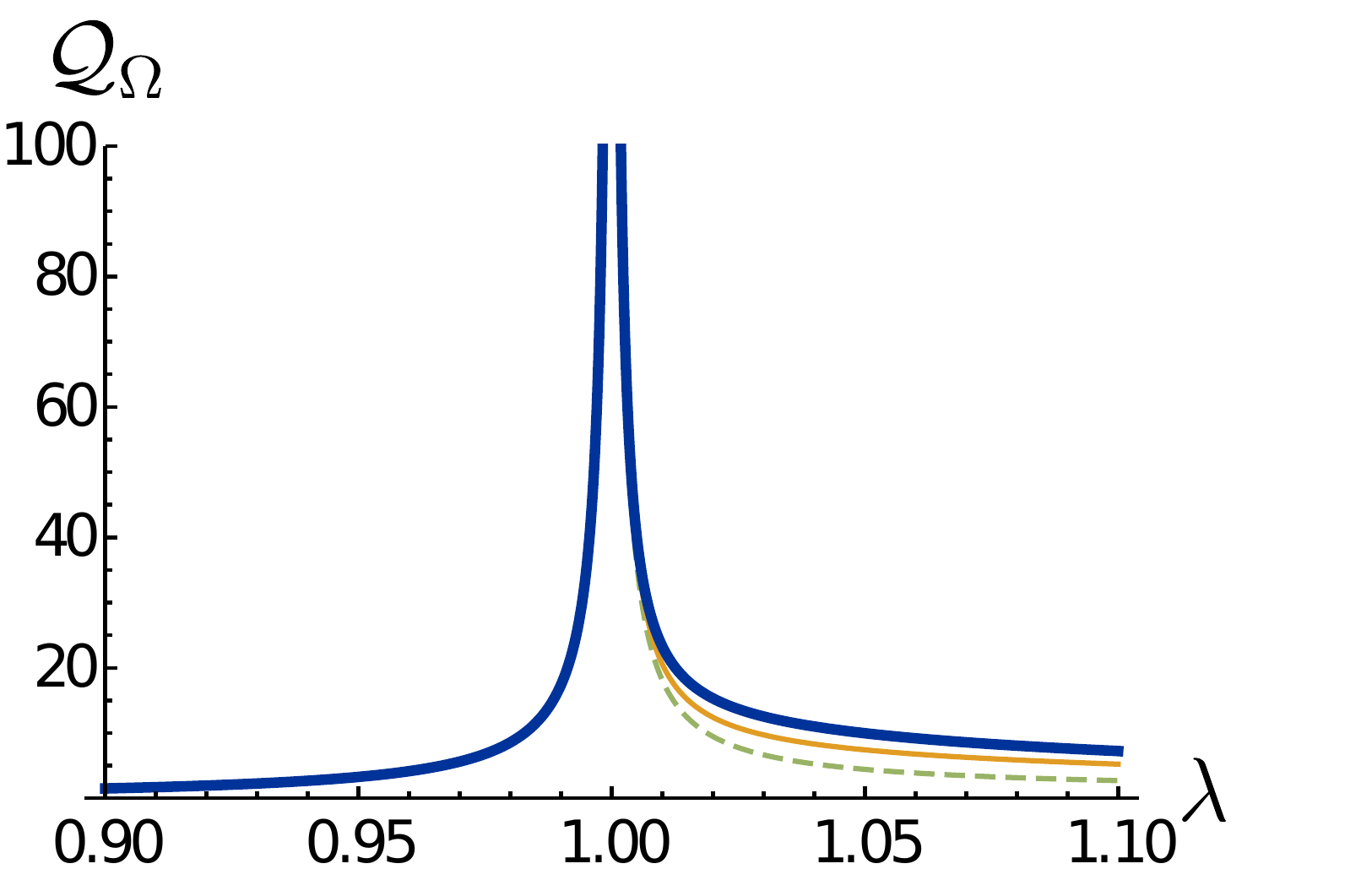}
	\caption[Optimal signal-to-noise ratio achievable with our critical protocol.]{Signal-to-noise ratio $Q_\Oq$ versus $\lambda=\gind/\gind_p$, with $\frac{1}{\rat}=0.1$ (thin dashed line), $0.2$ (thin full line), and $0.01$ (thick line). The QFI in the superradiant phase was computed assuming the system is either in the state $\ket{\psi_S^{+}}$ or $\ket{\psi_S^{-}}$. In the normal phase, $\pazocal{Q}_\Oq$ is independent of $\rat$. In the superradiant phase, there is a small $\rat$-dependent correction which becomes negligible near the critical point.}
	\label{QFI}
\end{figure}

So far we have focused on the estimation of the spin frequency $\Oq$ in the normal phase, assuming $\Of$ and $\gind$ are known. 
This analysis can be extended in two directions. 
First, it would be possible, in principle, to prepare the system in the superradiant phase, then to adiabatically reduce $\gind$. In this case, the system can adopt two possible states $ \ket{\psi_S^{\pm}}=\hat{D}(\pm\alpha_g){\hat{S}}(\xi^{(S)})\ket{0}\otimes\ket{\downarrow^{\pm}}$, with $\hat{D}$ the displacement operators, and $\ket{\downarrow^{\pm}}$ are spin states which belong to the $z-x$ plane (see Chapter 2 for details and exact expressions). For both of these states, the QFI can be computed as $\pazocal{I}_{\Oq}\sim \frac{1}{2\Oq^2}\frac{\gind_p^8}{(\gind_\tau^4-\gind_p^4)^2} + \frac{1}{\Oq\Of}\frac{\gind_p^6}{\gind^4\sqrt{\gind_\tau^4-\gind_p^4}}$, the first term being dominant when $\gind_\tau\sim\gind_p$. Again, pushing $\gind_\tau$ closer and closer to the transition increases the signal-to-noise ratio, as shown in Fig.\ref{QFI}. However, in general, it is unclear whether the system will reach $\ket{\psi_S^{+}}$, $\ket{\psi_S^{-}}$, a superposition of the two, or a statistical mixture. This will depend on the specifics of the preparation procedure, as well as the presence or absence of dissipation in the system. These various cases will yield different precision; in particular, a statistical mixture of the two states will reduce the metrological performances of the protocol. Therefore, in the following, we will focus only on the normal phase, where a clear picture of the ground state can be obtained. Furthermore, the initialization of the system is much easier in that case, since one can start from $\gind=0$ and separately prepare the field and spin in their ground state, before tuning up the coupling. \\

The second extension concerns the estimation of other parameters. For instance, if we assume that $\Oq$, this time, is known, the same protocol could be used to estimate the bosonic frequency $\Of$. If the Rabi model is implemented by coupling an atom or a spin to its vibrational degree of freedom \cite{dareau_observation_2018} for instance, this could be used to measure a frequency of vibration.
In this case, the QFI can be obtained by a similar calculation: we find $\pazocal{I}_\Of(\tau)=\frac{1}{8\Of^2}\left[\frac{\gind_p^2}{\gind_p^2-\gind_\tau^2}\right]^2$, which yields a signal-to-noise ratio $\pazocal{Q}_\Of=\pazocal{Q}_\Oq$.

\subsection{Homodyne and photon-number measure}

\begin{figure}
	\centering\includegraphics[angle=-90,width=.7\linewidth]{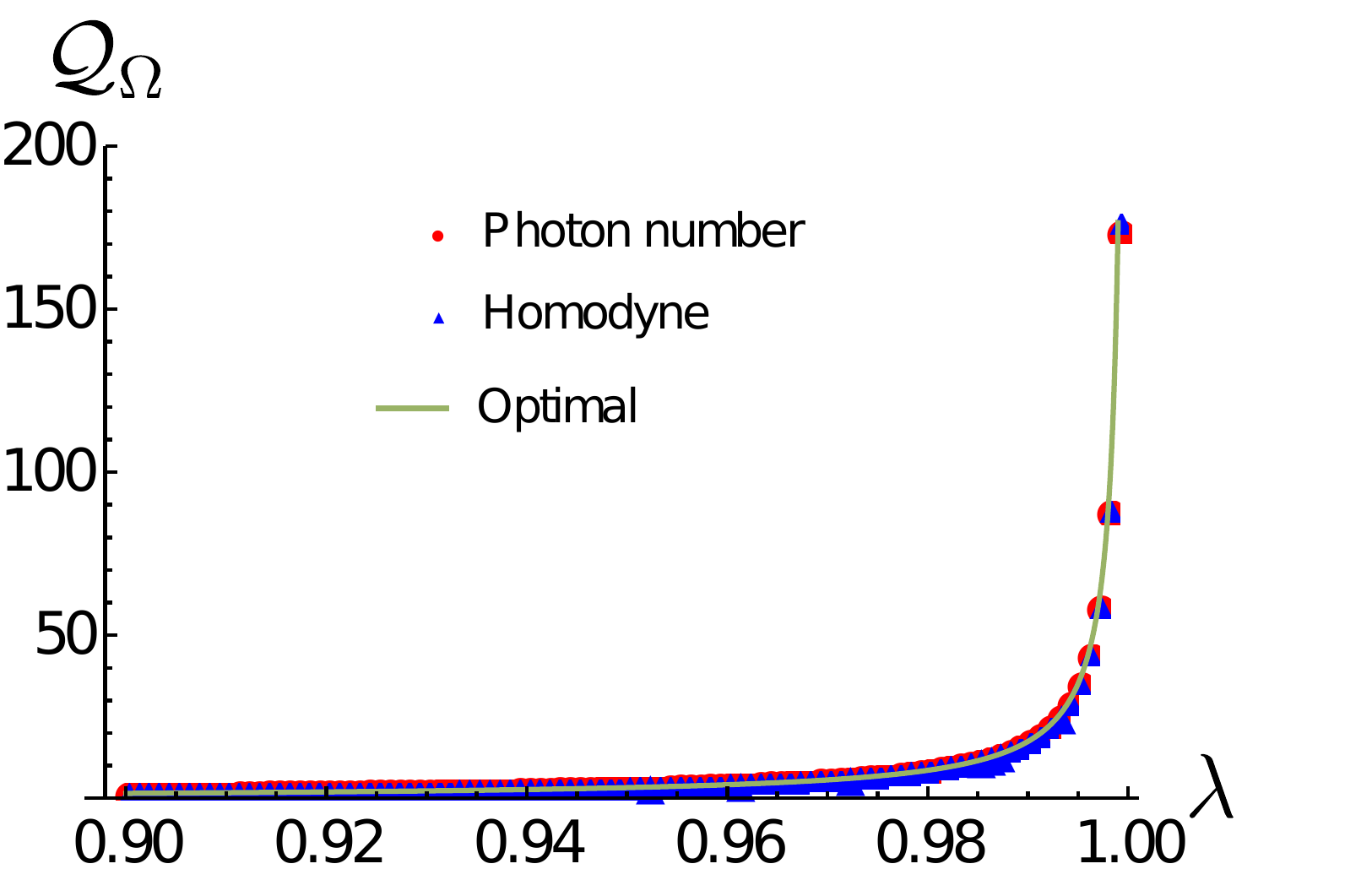}
	\caption[Signal-to-noise ratio achievable in our protocol with photon-number or homodyne measurement, compared with the optimal value.]{Signal-to-noise ratio achievable with photon-number measurement or homodyne measurement along the $\xop$ quadrature, compared with the optimal value in the normal phase. Both measurement strategies allow us to attain the optimal precision.}
	\label{ComparisonFIQFI}
\end{figure}

Having established the optimal sensitivity of the protocol, we need to consider which measure should be implemented to saturate this bound. We have considered both photon-number and homodyne measurement of the bosonic field. For each measurement, the achievable precision is given by the corresponding FI. The homodyne detection can be implemented along any field quadrature $\xop_\phi=\cos(\phi)\xop+\sin(\phi)\pop$. First, we have considered $\phi=0$. The FI and the corresponding signal-to-noise ratio were computed 
numerically, and compared to the optimal value given by the QFI. Results are shown in Fig.\ref{ComparisonFIQFI}: both photon-number measurement and homodyne measurement along the $\xop$ quadrature allow to saturate the CR bound.\\

For photon-number measurement, this can be confirmed analytically. At the end of the evolution, the probability distribution of the ground state $\ket{\xi_\Oq^{\gind_\dur}}$ over the Fock basis reads: $p(2m,\xi_\Oq^{\gind_\dur})=\lvert\langle2m|\xi_\Oq^{\gind_\dur}\rangle\rvert^2=\frac{\tanh(\xi_\Oq^{\gind_\dur})^{2m}}{\cosh(\xi_\Oq^{\gind_\dur})}\frac{(2m)!}{4^m(m!)^2}$ and $p(2m+1,\xi_\Oq^{\gind_\dur})=0$ for all $m$. The FI reads:
$$\sum\limits_{m=0}^\infty \frac{1}{p(m,\xi_\Oq^{\gind_\dur})}\left(\frac{\partial p(m,\xi_\Oq^{\gind_\dur})}{\partial\Oq}\right)^2=(\partial_\Oq\xi_\Oq^{\gind_\dur})^2 \sum\limits_{m=0}^\infty \frac{\tanh(\xi_\Oq^{\gind_\dur})^{2m}}{\cosh(\xi_\Oq^{\gind_\dur})^3\sinh(\xi_\Oq^{\gind_\dur})^2}\frac{(2m)!}{4^m(m!)^2}[2m-\sinh(\xi_\Oq^{\gind_\tau})^2]^2=2(\partial_\Oq\xi_\Oq^{\gind_\dur})^2,$$

where we have used $\sum\limits_{m=0}^\infty x^m\frac{(2m)!}{4^m(m!)^2}=\frac{1}{\sqrt{1-x}}$ and $\sum\limits_{m=0}^\infty m  x^m\frac{(2m)!}{4^m(m!)^2}=x\frac{d}{dx}\left(\frac{1}{\sqrt{1-x}}\right)$. After straightforward manipulation, this yields the QFI expression given in Eq.\eqref{QFInormal phase}, showing that photon-number measurement allows us to reach the QFI. \\

\begin{figure}[h]
	\begin{center}
	\includegraphics[angle=-90,width=.24\linewidth]{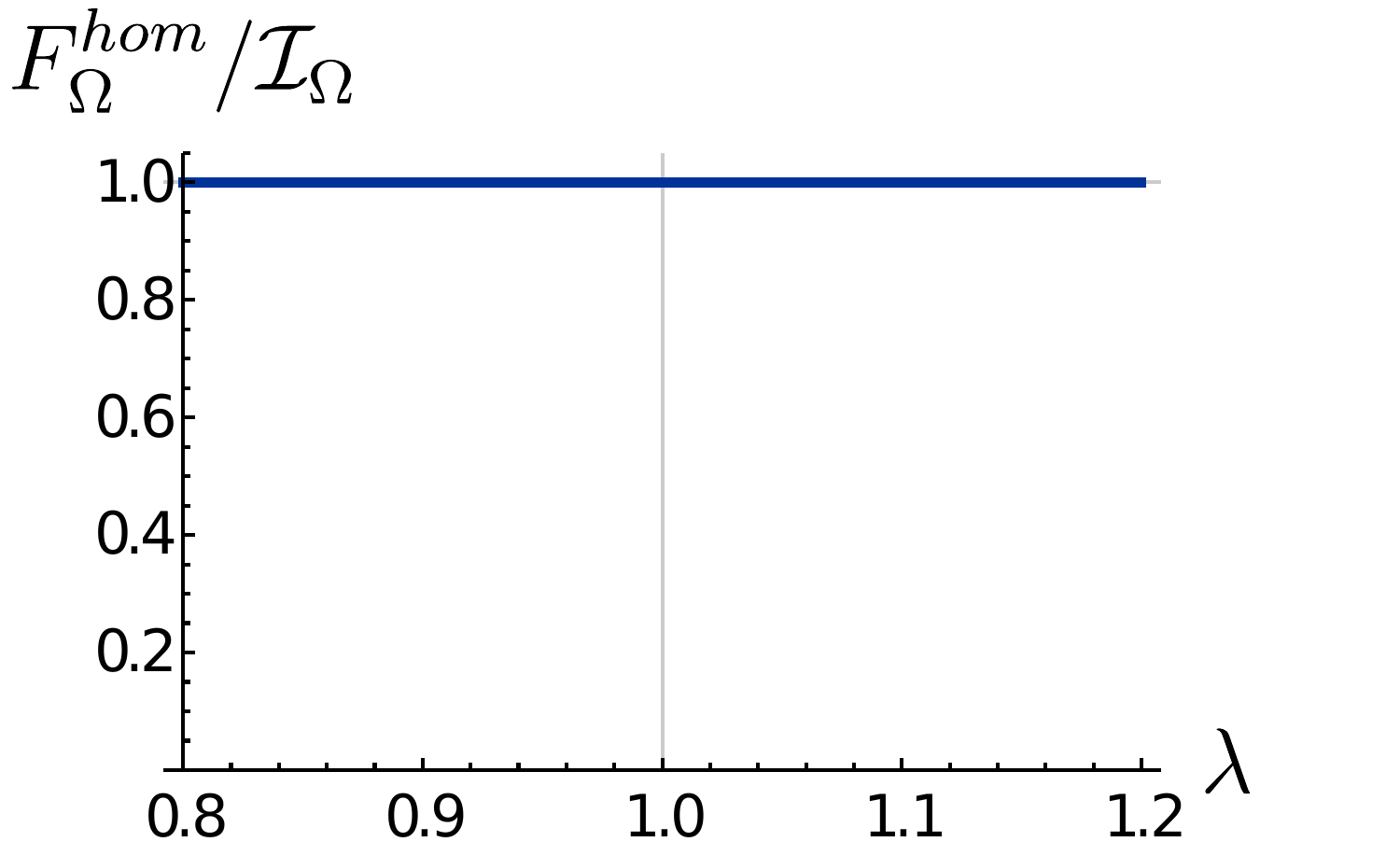}
	\includegraphics[angle=-90,width=.24\linewidth]{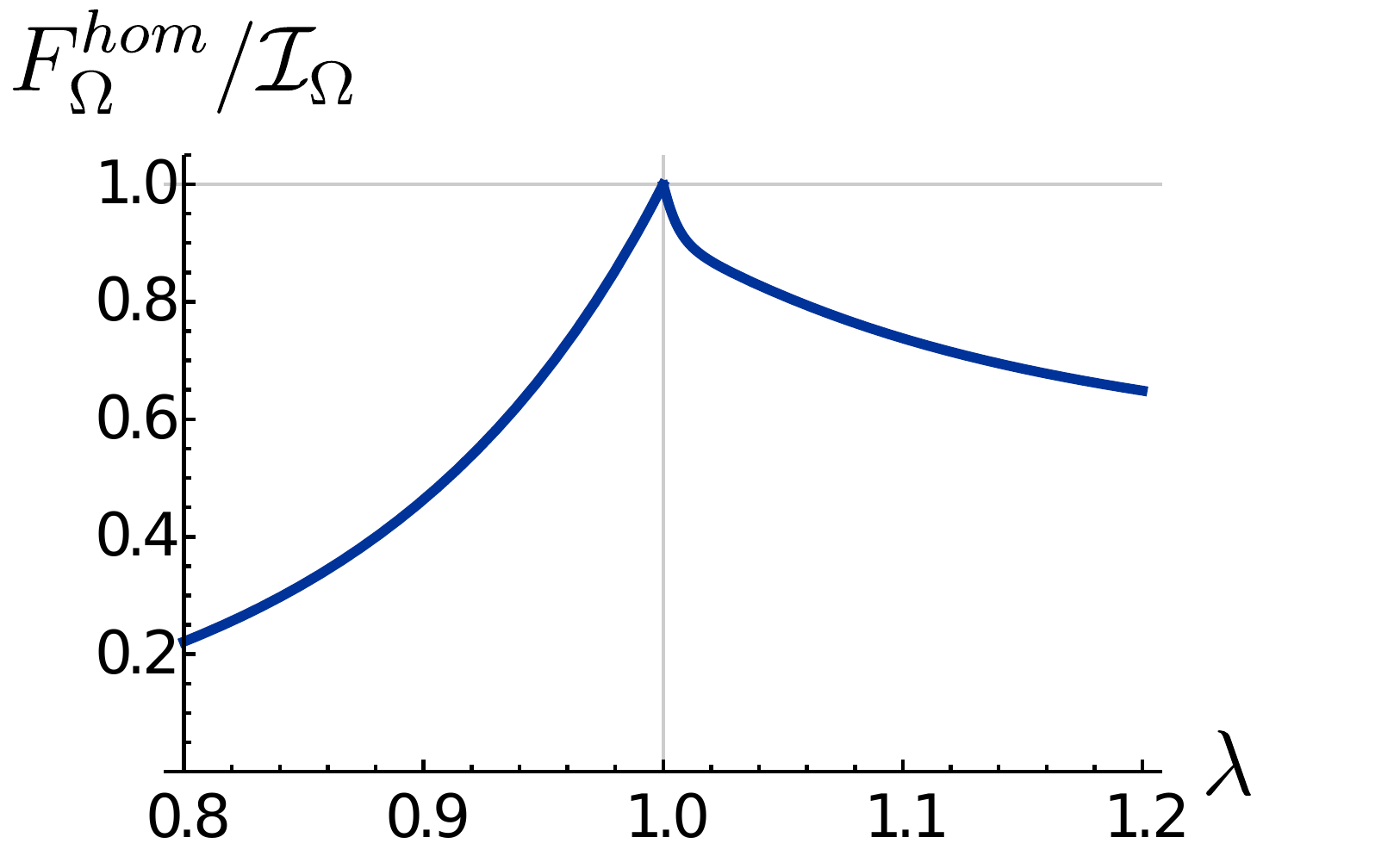}
	\includegraphics[angle=-90,width=.24\linewidth]{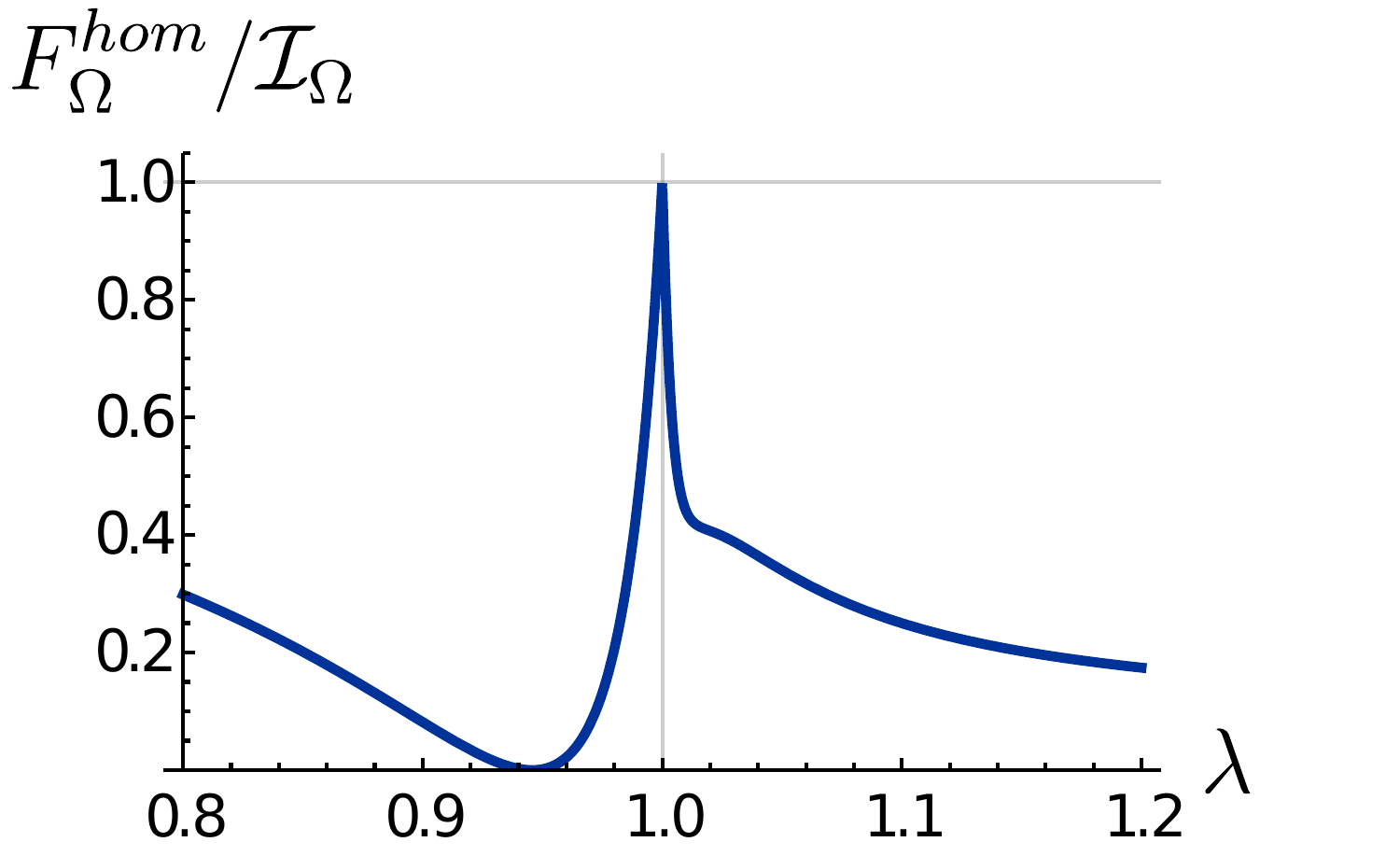}
	\includegraphics[angle=-90,width=.24\linewidth]{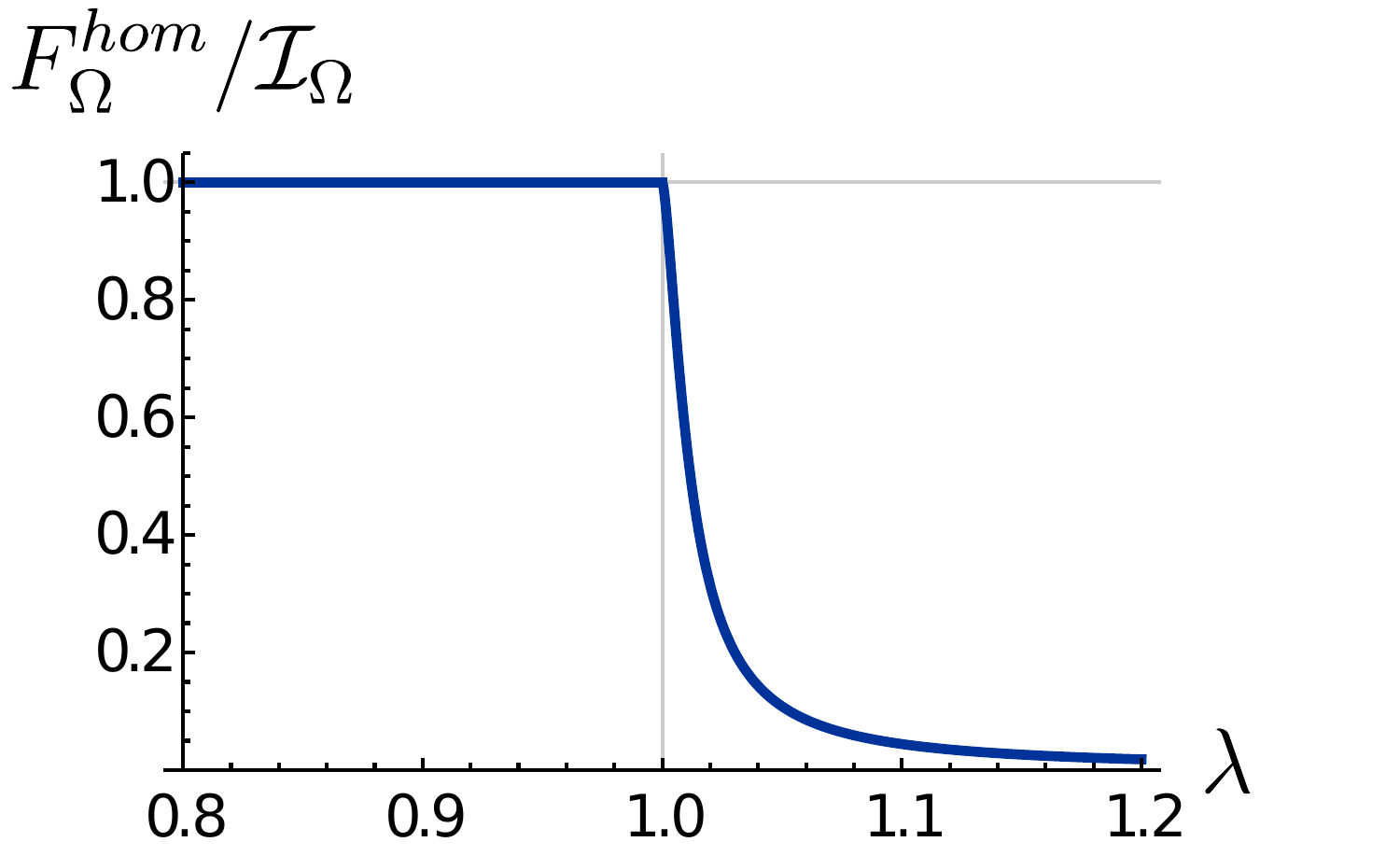}
	\end{center}
	\caption[FI-to-QFI ratio achievable with our protocol with homodyne measurement along different quadratures.]{FI-to-QFI ratio versus coupling constant, for a homodyne measurement performed along various quadratures $\xop_\phi$. From left to right: $\phi=0$, $\phi=\frac{1}{4}\pi$, $\phi=\frac{2}{5}\pi$, $\phi=\frac{1}{2}\pi$. In general, the precision depends on $\gind$ and it becomes optimal in close proximity of the critical point $\gind=\gind_p$. For $\phi=0$ and $\phi=\frac{1}{2}\pi$, the optimal precision is reached for all values of $\gind\leq\gind_p$, hence the QFI is saturated by homodyne measurement along the quadrature $\hat x$ or $\hat p$.}
	\label{FIQFIratio}
\end{figure}

Finally, we have also considered homodyne detection of different quadratures. In Fig.\ref{FIQFIratio}, we show the FI-to-QFI ratio for several quadratures. In general, the precision depends on $\frac{\gind}{\gind_p}$, and therefore on $\Oq$, as it is usually the case in quantum parameter estimation. In the proximity of the critical point $\gind=\gind_p$, most quadratures allow to saturate the CR bound. Away from this point, the precision may drop steeply or remain quasi-constant. In the normal phase $\gind<\gind_p$, the best choices of quadratures are $\xop_0=\xop$ and $\xop_{\frac{\pi}{2}}=\pop$, which both saturate the CR bound for every value of $\gind$. However, it is also possible to reach a very significant fraction of the optimal precision with the other quadratures.

One comment needs to be made about the homodyne detection. The homodyne signature of the ground state is a random noise centered around $0$. Its variance is linked with the amount of squeezing, which, as we discussed in the previous section, encodes the relevant information. Hence, this random noise should not be regarded as a nuisance that needs to be averaged out. Instead, the goal is to reconstruct its statistics. This also explains why the optimal precision may be reached with both $\xop$ and $\pop$. Although $\xop$ is considerably more noisy than $\pop$, the noise statistics is equally sensitive to the value of $\Oq$ in both cases, meaning that the optimal precision will be the same in both cases. The usefulness of homodyne detection to extract a parameter encoded in a squeezing channel was also pointed out in previous works, such as \cite{gaiba_squeezed_2009}.

\subsection{Duration of the protocol}

We will now consider the time needed to implement the adiabatic preparation of the initial state. The coupling constant evolves in time according to some profile $\gind_t$, from the initial value $\gind_0=0$ to the final one $\gind_\tau < \gind_p$. In general, we allow for a non-linear, adaptative evolution, during which the speed of evolution $\mathcal{V}=\frac{d\gind_t}{dt}$ is continuously reduced as the gap decreases. We expect that such an adaptative evolution will be easier to implement in a finite-size system like the one we are considering. At each moment $t$, the instantaneous eigenstates can be divided among the $\ket{\downarrow}$ and $\ket{\uparrow}$ sectors, which correspond to different eigenstates of $\sigz$. These two sectors remain well-separated in energy, even at the critical point; therefore, the high-energy sector associated with $\ket{\uparrow}$ may be safely ignored in the adiabatic procedure. In the low-energy sector, the eigenstates are squeezed Fock states $\ket{n,\xi_\Oq^{\gind_t}}\otimes\ket{\downarrow}+O(\rat^{-1/2})$, and the corresponding eigenenergies are $E_n(t)=n\Of\sqrt{1-\frac{\gind_t^2}{\gind_p^2}}$. The state of the spin will not change (at dominant order) during the adiabatic procedure and will be dropped henceforth. At each time, the state of the field can be decomposed in the squeezed Fock state eigenbasis as:

\begin{equation}
 	\ket{\psi(t)}=\sum_{n=0}^\infty c_n(t)e^{-i\Theta_n(t)}\ket{n,\xi_\Oq^{\gind_t}}, \hspace{10pt} \Theta_n(t)= \int_0^t E_n(t^\prime) dt^\prime.
\end{equation} 

The goal then is to ensure that the system will remain in its ground state during the evolution, \textit{i.e.}, $c_n(t)\sim0$ for $n\neq0$ and for all $t$. The evolution of the $c_n$ coefficients can be computed using Schrödinger equation:

\begin{equation}
\frac{dc_n(t)}{dt} = - \sum_{m=0}^\infty c_m(t) e^{-i\left[\Theta_m(t) - \Theta_n(t) \right] } \bra{n,\xi_\Oq^{\gind_t}} \frac{\partial}{\partial t}\ket{m,\xi_\Oq^{\gind_t}},
\end{equation}

which can be formally solved in time, and rewritten using the integration variable $d\gind=\mathcal{V}dt$:

\begin{equation}
 	c_n(\gind_t)=-\sum_{m=0}^\infty\int_0^{\gind_t} c_m(\gind^\prime) e^{-i\left[\Theta_m(\gind^\prime) - \Theta_n(\gind^\prime) \right]  }  \bra{n,\xi_\Oq^{\gind^\prime}}\frac{\partial}{\partial \gind^\prime}\ket{m,\xi_\Oq^{\gind^\prime}}d\gind^{\prime}.
 \end{equation} 

To ensure that the system will remain in the ground state, we will use an ansatz-verification procedure. We assume that the adiabatic condition is met at all times before a certain moment $t$, \textit{i.e.}, $c_m(\gind^\prime)\sim\delta_{m0}$ for $\gind^\prime < \gind_t$. Then the goal is to ensure that the condition is met at time $t$, provided it was at all earlier times. We can write: 

\begin{equation}
	c_n(\gind_t)=-\int_0^{\gind_t} e^{-i\left[\Theta_0(\gind^\prime) - \Theta_n(\gind^\prime) \right]  }  \bra{n,\xi_\Oq^{\gind^\prime}}\frac{\partial}{\partial \gind^\prime}\ket{0,\xi_\Oq^{\gind^\prime}}.
\end{equation}
We can compute directly the matrix elements involved:

\begin{equation}
	\bra{n,\xi_\Oq^{\gind^\prime}}\frac{\partial}{\partial \gind^\prime}\ket{0,\xi_\Oq^{\gind^\prime}}=\bra{n}\hat{S}^\dagger(\gind^\prime)\frac{\partial}{\partial \gind^\prime}\hat{S}(\gind^\prime)\ket{0} = \frac{\sqrt{2}}{4}\frac{\gind^\prime}{\gind_p^2-{\gind^\prime}^2} \delta_{n,2},
\end{equation}

so at this order, only transitions to the second excited-state $\ket{2,\xi_\Oq^{\gind^\prime}}$ should be taken into account. We obtain:
\begin{equation}
	c_2(\gind_t) = -\frac{1}{2\sqrt{2}}\int_0^\gind f(\gind^\prime)e^{iR(\gind^\prime)}d\gind^\prime,
	\label{eqalpha}
\end{equation}
where we defined $f(\gind^\prime)=\frac{\gind^\prime}{\gind_p^2-{\gind^\prime}^2}$ and
$R(\gind^\prime) = \Theta_2(\gind^\prime) - \Theta_0(\gind^\prime) = \frac{2\Of_0}{\gind_p}\int_0^{\gind^\prime}  \frac{\sqrt{\gind_p^2- {\gind''}^2}}{\mathcal{V}(\gind'')} d\gind''.$ 

The goal now is to ensure that this expression is small. This is achieved when the speed $\mathcal{V}$ is small: indeed, in that case, the exponential $e^{iR(\gind^\prime)}$ is a fast-oscillating term which cancels the integral. We can propose a more precise value through the following hand-waving argument: the exponential term must oscillate faster than the evolution of $f$, in order to cancel the changes in $f$ before they have the possibility to build up. We can express this condition as $\frac{\dot{f}}{f}\ll\dot{R}$, where here the dot means derivative with respect to $\gind$. Based on this intuition, we propose the following ansatz for the speed:
\begin{equation*}
	\mathcal{V}(\gind)=\mesc\Of\gind_p\left(\frac{\gind_p^2-\gind^2}{\gind_p^2}\right)^{3/2},
	\label{ansatzv}
\end{equation*}
with $\mesc$ a small constant. We can now verify that this ansatz gives the desired results. We rewrite \eqref{eqalpha} using integration by part, and we obtain:

\begin{align} 
\nonumber
c_2(t) &= +\frac{i}{2\sqrt{2}}\int_0^{\gind_t}  \frac{\gind^\prime}{\left(\gind_p^2-{\gind^\prime}^2\right)\dot R(\gind^\prime)} \frac{\partial}{\partial \gind^\prime} e^{iR(\gind^\prime)} d\gind^\prime \\ \nonumber
 & =\frac{i\gind_p}{4\sqrt{2}\Of}\left(e^{iR(\gind_t)}\frac{\mathcal{V}(\gind_t) \gind_t }{(\gind_p^2-\gind_t^2)^{3/2}} -\int_0^{\gind_t}  e^{iR(\gind^\prime)} \frac{\partial}{\partial \gind^\prime} \left[\frac{\mathcal{V}(\gind^\prime) \gind^\prime}{(\gind_p^2-{\gind^\prime}^2)^{3/2}}\right] d\gind^\prime \right)\\
 & = \frac{i\mesc \gind_t}{4\sqrt{2}\gind_p} e^{iR(\gind_t)} -\frac{i}{4\sqrt{2}\gind_p}\int_0^{\gind_t} \mesc e^{iR(\gind^\prime)}d\gind^\prime.
\end{align}

For all values of $\gind$, $e^{iR(\gind')}$ is a fast-oscillating term. Therefore, we have $\int_0^{\gind_t} \mesc e^{iR(\gind^\prime)}d\gind^\prime\ll\mesc \gind_t  e^{iR(\gind_t)}$, and we can omit the second factor in the expression of $c_2(t)$. Finally, we can write the probability of exciting the system during the adiabatic sweep as $\vert c_2 (\gind_t) \rvert^2=\frac{\mesc^2}{32}\frac{\gind_t^2}{\gind_p^2}$, which is small as long as $\mesc$ is. We have thus proven the validity of the ansatz \eqref{ansatzv}, and we can now evaluate the total time needed to perform the adiabatic evolution

\begin{equation}
	\dur=\int_0^{\gind_\dur}\frac{1}{\mathcal{V}(\gind^{\prime})}d\gind^{\prime}=\frac{1}{\mesc\Of}\frac{\gind_\tau}{\gind_p}\left(\frac{\gind_p}{\gind_p-\gind_\dur}\right)^{1/2},
\end{equation}

\begin{figure}
	\includegraphics[angle=-90, width=\linewidth]{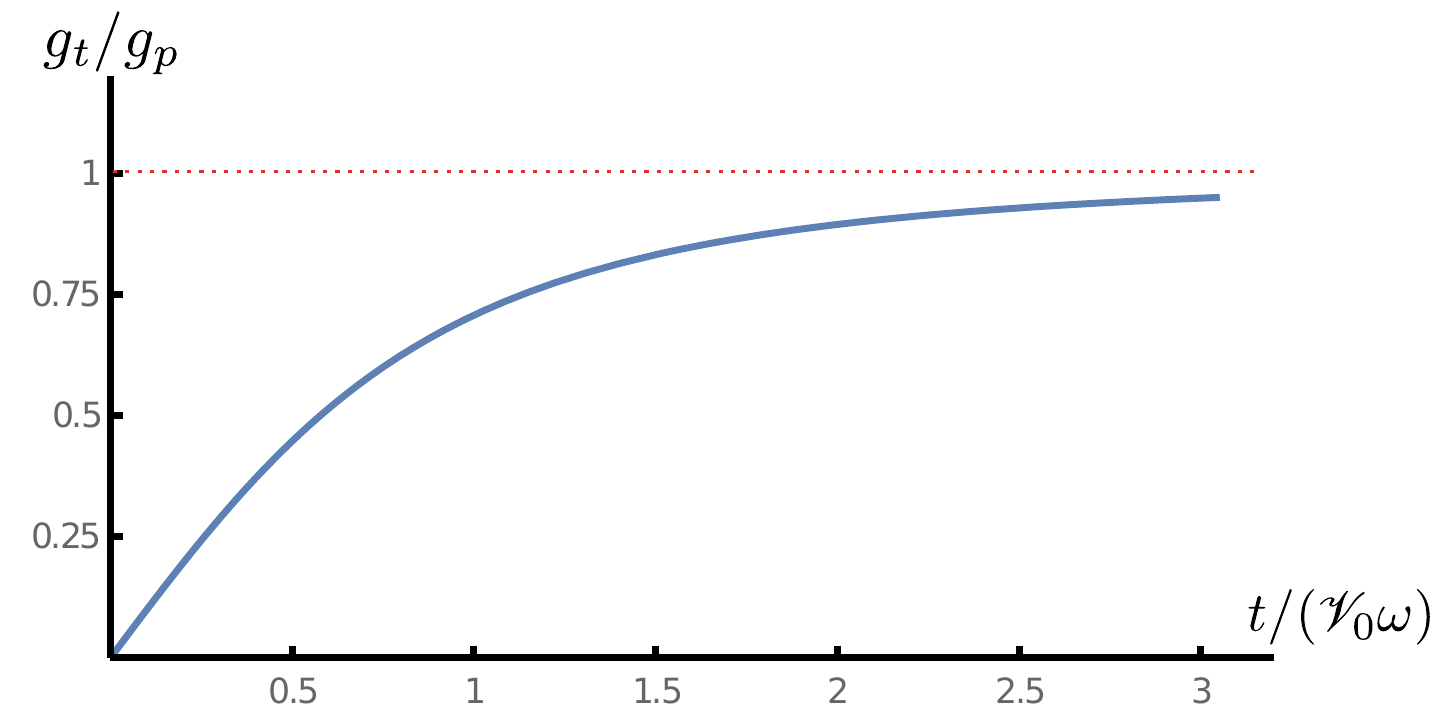}
	\caption{Time profile of the coupling during the adiabatic ramp. As $g$ approaches the critical value $g_p$, the speed is reduced in order to maintain the system in its ground state.}
\end{figure}

Finally, we can inject this expression into \eqref{QFInormal phase}, and we obtain near the transition: 

\begin{equation}
 	\pazocal{I}_\Oq\sim\frac{\mesc^4\Of^4}{8\Oq^2}\dur^4=\frac{\mesc^4}{8\rat^{2}}\Of^2\dur^4.
 	\label{QFIscalingtime}
 \end{equation} 
Hence, the QFI increases like the quartic power of the protocol duration. By contrast, in the standard Ramsey protocol with a single spin, the QFI scales like $\dur^2$ \cite{pezze_quantum_2018}. Indeed, this scaling is achieved in any interferometric protocol based on time-independent Hamiltonian. By contrast, here, the adiabatic preparation is explicitly time-dependent. It belongs to the larger category of quantum control procedures, for which quartic time scalings of the QFI have been reported for other observables \cite{pang_optimal_2017,yang_quantum_2017}, as we discussed in Chapter 3.

This result shows that quantum critical protocols can achieve enhanced time-scaling of the sensitivity, despite the critical slowing down. Note, however, that the prefactor in \eqref{QFIscalingtime} is very small. Therefore, to achieve a precision comparable to the Ramsey limit $\dur^2$, it is necessary to consider very long protocol duration, \textit{i.e.}, to prepare the state very close to the critical point. In this regime, however, the quartic terms in \eqref{Hrabi4} become relevant. More precisely, we have $\pazocal{I}_\Oq\sim \dur^2$ when $\dur^2 > \frac{\rat^{2}}{\Of^2}$, which implies: 
$$1-\frac{\gind_t}{\gind_p}\ll\rat^{-2}<\rat^{-2/3},$$
meaning that the quartic terms become non-negligible before the precision becomes comparable with Ramsey. When the quartic terms are included, however, an analytical study is no longer feasible. Therefore, it is still unclear which precision may be reached in this region.

\section{Dissipative case}
\label{sec:Dissipative_case}

The above results have been obtained in the case of an isolated system. However, interaction with the environment generally affects the performances of metrological protocols. We will now study the effect of photon loss and spin decay. The dissipative dynamics of the system will be described by a master equation of the form:
$$\partial_t\hat{\rho}=-i[\hat{H},\hat{\rho}]+\kappa L[\aop]\hat{\rho}+ \Gamma L[\hat{\sigma}_-]\hat{\rho}.$$

Once more, we are considering a phenomenological master equation with undressed jump operators (as in \cite{hwang_dissipative_2018}), since we are interested in effective implementations of the model.

 Instead of the ground state of $\Hop$, we will now be interested in the steady-state of this equation. The idea goes as follows: instead of slowly tuning the coupling constant to its desired value, we make an instantaneous quench. Then we let the system settle to its steady-state, before performing a measurement. We expect that quenching the coupling constant closer and closer to $\gind_p$ will increase the achievable precision. However, the time needed to reach the steady-state will also increase in that case, due to critical slowing down. Therefore, we will again have a trade-off between precision and time in that case.

For convenience, we will assume $\Gamma=O(\Oq)$ and $\kappa=O(\Of)$, but our analysis can be also be extended to lower spin dissipation values (such as $\Gamma=O(\sqrt{\Oq\Of})$) with only minor adjustments. We apply the transformation $\Uop=e^{i\rat^{-1/2}\frac{\gind}{\gind_p}\sigy(\adag+\aop)}$, which yields:

\begin{align}
\label{lindblad_complet}
  	\dot{\rop}= & -i[\Uop^\dagger\Hop\Uop,\rop]+\kappa L[\aop](\rop)+\Gamma L[\sigm](\rop) \\ \nonumber
  	& + \frac{2\Gamma\gind}{\gind_p} \rat^{-1/2}\left((\aop+\adag)\sigz\rho\sigp +h.c.\right) + \frac{\Gamma\gind}{\gind_p} \rat^{-1/2}\left\{(\aop+\adag)\sigx,\rop\right\}\\ \nonumber
  	& -\frac{\Gamma\gind^2}{\gind_p^2}\rat^{-1}\left[(\aop+\adag)^2\sigx\rop\sigp+h.c.-2(\aop+\adag)\sigz\rop\sigz(\aop+\adag)\right] + \frac{\Gamma \gind^2 \rat^{-1}}{2\gind_p^2}\left\{(\aop+\adag)^2\sigz,\rop\right\}+O(\Of\rat^{-1/2}), 
\end{align}

First, we will focus on the spin dynamics. We decompose the state $\rop$ into its spin and bosonic components:
\begin{equation}
	\rop=\rop_{bd}\ket{\downarrow}\bra{\downarrow} + \rop_{bu}\ket{\uparrow}\bra{\uparrow} + \rop_{bc}\ket{\downarrow}\bra{\uparrow}+ \rop^{\dag}_{bc}\ket{\uparrow}\bra{\downarrow},
\end{equation}
where $\rop_{bd}$, $\rop_{bu}$, $\rop_{bc}$ and $\rop^{\dag}_{bc}$ are (unnormalized) matrices describing the state of the bosonic field (note that $\rop_{bc}$ may not be hermitian). We will be interested in the steady-state behavior of these operators.

In \eqref{lindblad_complet}, the term $\Gamma L[\sigm](\rop)$ will tend to bring to the $\ket{\downarrow}\bra{\downarrow}$ subspace. The other terms, which create non-zero value outside this subspace, are only of order $\Gamma\rat^{-1}$ and $\Gamma\rat^{-1/2}$, respectively. Therefore, we make the following ansatz for the four state components, which will be verified later on:
$$\rop_{bd}=O(1), \ \rop_{bc}=O(\rat^{-1/2}), \ \rop_{bu}=O(\rat^{-1}).$$

By projecting \eqref{lindblad_complet}, we obtain:

\begin{align}
	\label{compo_dissip}
	\dot{\rop}_{bu}= & -2\Gamma\rop_{bu}+\frac{\gind\Gamma}{2\gind_p}\rat^{-1/2}\left((\aop+\adag)\rop_{bc}+h.c.\right)+O(\Gamma\rat^{-2}), \\ \nonumber
	\dot{\rop}_{bc}= & (i\Oq-\Gamma)\rop_{bc} + \frac{\gind\Gamma}{2\gind_p}\rat^{-1/2}\rop_{bd}(\aop+\adag) +O(\Gamma\rat^{-3/2}), \\ \nonumber
	\dot{\rop}_{bd}= & -i[\Of\adag\aop-\Of\frac{\gind^2}{4\gind_p^2}(\aop+\adag)^2,\rop_{bd}] + 2\Gamma\rop_{bu} \\ \nonumber
	& -\frac{\gind\Gamma}{\gind_p}\rat^{-1/2}\left((\aop+\adag)\rop_{bc}+h.c.\right) + \frac{\gind\Gamma}{2\gind_p}\rat^{-1/2}\left((\aop+\adag)\rop^{\dag}_{bc}+h.c.\right) \\ \nonumber
	&  + \kappa L[\aop](\rop_{bd}) + \frac{\gind^2\Gamma}{2\gind_p^2}\rat^{-1}\left[(\aop+\adag)\rop_{bd}(\aop+\adag)\right] -\frac{\gind^2\Gamma}{4\gind_p^2}\rat^{-1}\left\{(\aop+\adag)^2\rop_{bd}\right\} + O(\Gamma\rat^{-2}).
\end{align}

$\rop_{bu}$ and $\rop_{bc}$ evolve quickly, at a rate $\Gamma$. Therefore, we suppose that they reach their steady-state very quickly, and suppress them using adiabatic elimination. We obtain:
\begin{align}
	\rop_{bc}=\frac{\gind\Gamma}{2\gind_p(\Gamma-i\Oq)}\rat^{-1/2} \rop_{bd} (\aop+\adag) + O(\rat^{-3/2}),\\ \nonumber
	\rop_{bu}=\frac{\gind^2}{4\gind_p^2}\frac{\Gamma^2}{\Gamma^2+\Oq^2}\rat^{-1} (\aop+\adag)\rop_{bd} (\aop+\adag) +O(\rat^{-2}).
\end{align}
These expressions give indeed $\rop_{bc}=O(\rat^{-1/2}), \ \rop_{bu}=O(\rat^{-1})$. We reinject these in Eq. \eqref{compo_dissip}, and obtain after straightforward calculations: 
\begin{align}
	\nonumber \dot{\rop}_{bd}= & -i[\Of\adag\aop-\Of \mathcal{P}\frac{\gind^2}{4\gind_p^2}(\aop+\adag)^2,\rop_{bd}] + \kappa L[\aop](\rop_{bd})\\ 
	& + \frac{\mathcal{P}\gind^2}{4\gind_p^2}\frac{\Of\Gamma}{\Oq} L[\aop+\adag](\rop_{bd})+O(\Of\rat^{-1}),
\end{align}
with $\mathcal{P}=\frac{\Oq^2}{\Gamma^2+\Oq^2}$. This equation describes the dynamics of the bosonic field inside the lower spin subspace. To solve it, we will now move to phase space \cite{ferraro_gaussian_2005}. The Lindblad equation is transformed into a Fokker-Planck equation for the Wigner function:
\begin{equation}
\frac{\partial W}{\partial t}(x,p)=-\Of p\frac{\partial W}{\partial x} - \Of\left(\mathcal{P}\frac{\gind^2}{\gind_p^2}-1\right)x\frac{\partial W}{\partial p}+\kappa\left(2W+\sum_{i=1}^2 x_i\partial_iW + \sum_{i,j=1}^2\partial_i\sigma^L_{ij}\partial_jW\right).\end{equation}

Here $x_1=x$, $x_2=p$, and $$\sigma^L =\frac{1}{2}\begin{bmatrix}1 & 0 \\ 0 & 1+\mathcal{P}\frac{\Gamma\Of\gind^2}{\Oq\kappa\gind_p^2}\end{bmatrix}.$$
Since this equation is quadratic in $x$ and $p$, it can be solved by a Gaussian ansatz $W=\frac{1}{\sqrt{\pi \text{det}(\sigma)}}\exp\left\{-\frac{1}{2}x_i(\sigma^{-1})_{ij}x_j\right\}$. The displacement (\textit{i.e.}, the first-order moment) decays at a rate $2\kappa$ and will quickly reach $0$. Thus, the Wigner function is entirely characterized by the covariance matrix, which is described by the following equation:

\begin{align}
\label{equationsigmabis}
\partial_t\sigma & = \mathcal{E}\sigma+\sigma \mathcal{E}^T- 2\kappa(\sigma - \sigma^L), \\ \nonumber
\mathcal{E} & =\begin{bmatrix}0 & \Of\\ \omega\left(\mathcal{P}\frac{\gind^2}{\gind_p^2}-1\right) & 0\end{bmatrix}.
\end{align}

The term $\mathcal{E}\sigma+\sigma \mathcal{E}^T$ in \eqref{equationsigmabis} originates from the Hamiltonian dynamics. We will define the eigenmatrices of this evolution: $\mathcal{E} M_i + M_i \mathcal{E}^T=\mu_iM_i$.
Let us emphasize that rigorously, we should distinguish between the cases $\frac{\gind^2}{\gind_p^2}\leq\frac{1}{\mathcal{P}}$ and $\frac{\gind^2}{\gind_p^2}\geq\frac{1}{\mathcal{P}}$. In the former case, the Hamiltonian is an ordinary squeezing Hamiltonian bounded from below, the $\mu_i$ are complex, and the $M_i$ correspond to oscillating solution. In the latter case, the Hamiltonian is no longer bounded from below, the $\mu_i$ are real, and the $M_i$ are diverging (or vanishing) in time. Here we will focus only on the case $\frac{\gind^2}{\gind_p^2}\geq\frac{1}{\mathcal{P}}$, however the formalism is very similar for $\frac{\gind^2}{\gind_p^2}\leq\frac{1}{\mathcal{P}}$. We find the following eigenmatrices and eigenvalues:

\begin{center}
$M_0=\begin{bmatrix} 0 & 1 \\ -1 & 0\end{bmatrix}$, $M_1=\begin{bmatrix} \frac{1}{\sqrt{\mathcal{P}(\gind/\gind_p)^2-1}} & 0 \\ 0 & -\sqrt{\mathcal{P}(\gind/\gind_p)^2-1}\end{bmatrix}$, $M_{\pm}=\begin{bmatrix} \frac{1}{\sqrt{\mathcal{P}(\gind/\gind_p)^2-1}} & \pm1 \\ \pm1 & \sqrt{\mathcal{P}(\gind/\gind_p)^2-1}\end{bmatrix}$

\vspace{10pt}

$\mu_0=\mu_1=0$, $\mu_{\pm}=\pm2\Of\sqrt{\frac{\mathcal{P}\gind^2}{\gind_p^2}-1}$.
\end{center}

The matrices $M_i$ form a complete basis, thus we may write $\sigma=\sum_i m_i M_i$ and $\sigma^L=\sum_i m^L_i M_i$: we find $m^L_0=0$, $m^L_1=\frac{\mathcal{P}(\gind/\gind_p)^2\left(1-\frac{\Of\Gamma}{\Oq\kappa}\right)-2}{4\sqrt{\mathcal{P}(\gind/\gind_p)^2-1}}$ and $m^L_\pm=\frac{\mathcal{P}(\gind/\gind_p)^2}{8\sqrt{\mathcal{P}(\gind/\gind_p)^2-1}}\left(1+\frac{\Of\Gamma}{\Oq\kappa}\right)$.
We can then rewrite \eqref{equationsigmabis} as an equation of evolution for the coefficients $m_i$:
$\partial_tm_i  = (\mu_i-2\kappa) m_i+2\kappa m_i^L = -\tilde{\mu}_i\left(m_i-\frac{2\kappa}{\tilde{\mu}_i}m_i^L\right)$ with $\tilde{\mu}_i=2\kappa-\mu_i$. We  obtain:

\begin{equation}
\label{citime}
m_i(t)=\left(m_i(t=0)-\frac{2\kappa}{\tilde{\mu}_i}m_i^L\right)e^{-\tilde{\mu}_it}+m_i^L\frac{2\kappa}{\tilde{\mu}_i} \,. 
\end{equation}

Finally, we can obtain the steady-state of the system. First we compute the coefficients $m_i$ when $t\rightarrow \infty$. We find $m_i\rightarrow m_i^L\frac{2\kappa}{\tilde{\mu}_i}$, and $\sigma=\sum_im_i^L\frac{2\kappa}{\tilde{\mu}_i}M_i$. Now, putting together the expressions of $m_i$, $m_i^L$ and $M_i$, we find the expression of the covariance matrix: 

\begin{equation}
	\sigma=\frac{1}{2}\begin{bmatrix}1 & 0 \\ 0 & 1-\frac{\gind^2\mathcal{P}}{2\gind_p^2}(1-\frac{\Of\Gamma}{\Oq\kappa})\end{bmatrix} + \frac{\gind^2}{4((\gind_p^{D})^2-\gind^2)}\left(1+\frac{\Of\Gamma}{\Oq\kappa}\right)\begin{bmatrix} 1 & \frac{\kappa}{\Of} \\ \frac{\kappa}{\Of} & \left(\frac{\kappa}{\Of}\right)^2
	\end{bmatrix},
\end{equation} 
with $\gind_p^{D}=\gind_p\sqrt{(1+\Gamma^2/\Oq^2)(1+\kappa^2/\Of^2)}$.
To summarize, the steady-state is a squeezed (undisplaced) thermal state. The system still exhibits critical behavior for $\gind\rightarrow\gind_p^D$; both the squeezing and thermal energies increase near this point. Since the state is Gaussian, the QFI may still be computed exactly, according to the formula we outlined in Chapter 3 (here dots denote the derivative with respect to $\Oq$):

$$
\pazocal{I}_\Oq=\frac{\nu^2}{2(\nu^2+1)}\left\{\text{Tr}\left [ (\sigma^{-1}\dot{\sigma})^2 \right ]\right\}\,,$$
with $\nu$ the symplectic eigenvalue of $\sigma$.\\

The time needed to reach the steady-state may also be estimated from the above analysis. From \eqref{citime}, we see that the $c_i-c_i(t\rightarrow\infty)$ decay at various rates, the smallest one being $ \tilde{\mu}_{+}=2\kappa-2\Of\sqrt{\mathcal{P}(\gind/\gind_p)^2-1}$, which near the transition is equal to $2\kappa\frac{\gind_p^D-\gind}{\gind_p^D}\left(1+\frac{\Of^2}{\kappa^2}\right)$. This vanishing decay rate will dominate the relaxation of the system near the critical point; hence, we can evaluate the duration of the protocol as $\dur\sim\frac{1}{2\kappa}\frac{\gind_p^D}{\gind_p^D-\gind}\frac{\kappa^2}{\kappa^2+\Of^2}$. Finally, we obtain the leading term of the QFI as a function of $\dur$:

\begin{align}\label{scalingdiss}
 \pazocal{I}_\Oq^{\text{diss}}\sim\left(\frac{\Gamma^2-\Oq^2}{\Gamma^2+\Oq^2}\right)^2\frac{\kappa^2}{2\Oq^2}\left(1+\frac{\Of^2}{\kappa^2} \right)^2\dur^2\,.
\end{align}

Hence, the presence of dissipation replaces the quartic scaling by a quadratic one. By contrast, the QFI of a Ramsey protocol in presence of spin decay is given by $\dur/\Gamma$, and so is linear in time. Therefore, in the critical protocol, a non-trivial scaling is achieved even in the presence of dissipation channels.

\section{Conclusion}

In this Chapter, we have presented a sensing protocol exploiting the critical behavior of the Rabi model. Many previous works have studied many-body phase transitions in the immediate vicinity of the critical point. In this case, the focus is on the scaling of precision with the number of components. Here, by contrast, we considered a finite-size system some distance away from the transition, where quadratic field terms dominate the transition and exact results can be obtained. The point then is to study the scaling of precision with time.

The system ground state is prepared through an adiabatic protocol, a time-dependent evolution that can be seen as a quantum control procedure. During this evolution, the bosonic field becomes squeezed. The amount of squeezing can be evaluated either through photon-number measurements or by studying the statistics of the homodyne signal. In turn, this allows us to reconstruct the value of the spin frequency. 

When the system is brought closer and closer to the critical point, the squeezing parameter becomes increasingly sensitive to the value of the spin frequency, improving the precision of the protocol. Due to the closure of the Hamiltonian gap, however, the preparation also becomes slower in this region. The QFI of the protocol scale like the fourth power of the protocol duration, in contrast with the quadratic scaling usually achieved in interferometric experiments with time-independent Hamiltonians. In the dissipative case, the scaling is quadratic, compared to the linear scaling of noisy interferometric protocols. However, this improved scaling comes with a small prefactor. It is possible to improve the performances by operating very close to the transition; however, when higher-order terms become non-negligible, exact results can no longer be obtained.

This work opens several interesting research directions. From an experimental perspective, one of the most promising platforms to implement this protocol would be atoms coupled to their vibrational degrees of freedom, in which the Rabi model in the ultrastrong-coupling regime has been achieved \cite{dareau_observation_2018}. These systems allow us to naturally realize the large-detuning regime in which the critical behavior could be observed. The use of optimal quantum control techniques could also be used to improve the preparation time.

From the theoretical side, more work needs to be conducted to understand the behavior of the system very close to the critical point. Although exact results can no longer be obtained, ansatz-based procedures have been used to successfully describe the critical exponents of the Rabi model in this regime \cite{hwang_quantum_2015}. Combined with numerical analysis, these techniques could be used to study also the metrological behavior of the system. Second, in the context of spin systems, it was shown that the time-scaling of the QFI could be obtained by studying the critical exponents of the model \cite{rams_at_2018}. By extending these studies to light-matter models, their sensing properties could be studied from a more general perspective. 

Finally, it was argued recently that exceptional points, like critical points, could be used for sensing applications. This claim, however, is still controversial. Exploiting tools from quantum metrology could help to understand which precision could be achieved with these protocols \cite{zhang_quantum_2019}.

\chapter{Towards a resource theory for quantum metrology}


\epigraph{\textit{Il n'est rien, dans ce monde ou dans l'autre, qui ne puisse être utile à un écrivain.}}{Gabriel Garc\'ia M\'arquez}
\epigraph{\textit{Je crois que c'est vrai aussi pour la physique.}}{Louis Garbe}

To design and improve quantum-enhanced protocols, it is useful to understand and characterize which properties are responsible for a quantum advantage. In particular, some works have pointed out the link between some measures of nonclassicality and the possibility to achieve sub-shot-noise metrology. In this Chapter, we will discuss these ideas in the context of interferometry protocols. We study the characterization of passive linear optical elements with two-mode Gaussian states, a large class of protocols which describes many relevant experimental situations. We introduce a notion of metrological advantage valid both for pure and mixed states, and show that this advantage can be unequivocally associated with squeezing of the probe state. This Chapter is divided into five parts. First, we introduce and motivate our analysis. Second, we discuss the protocols that we study. Third, we introduce a new criterion for metrological advantage, and show that squeezing is both necessary and sufficient to achieve an advantage in this sense. In the next section, we discuss our results in the context of resource theories, and point out a possible link with the notion of formal entanglement. Finally, we briefly conclude in the last section. The results presented here have been first published in \cite{garbe_metrological_2019}.

\section{Introduction}
Quantum mechanics imposes fundamental limitations to our ability to describe the world; however, non-classical correlations can also be used as tools to improve the performances of sensing or computation protocol. As we have discussed in Chapter 3 and 5, it is useful to study these problems by using the notion of resource. For a given amount of resource (for instance the number of particles, or the duration of the experiment), a quantum advantage is the possibility to achieve better performances by exploiting non-classical correlation. It is of course an important task to create new protocols, or improve existing ones. However, from a more fundamental perspective, it is also very interesting to understand which physical properties are responsible for the quantum advantage.  Metrology offers a good example: it is often implied that the improved performances of quantum metrology protocols come from entanglement. However, there are also several ways of achieving this advantage without using entanglement \cite{braun_quantum-enhanced_2018}. This begs the question: is there a property that is both necessary and sufficient to achieve metrological advantage? More generally, in addition to looking for which protocol will give the best performances, we can also try to characterize \textit{all} protocols which perform better than a subset of "classical" protocols. The goal here is to better define and characterize the notion of "non-classicality" in a given context.\\

These ideas can be formalized in the framework of resource theory \cite{chitambar_quantum_2019,coecke_mathematical_2016}.
This framework consists of two interconnected concepts. The first is the notion of \textit{free operations}, an ensemble of manipulations which, in a certain experimental context, are considered to be readily available, and can be applied without any restriction. The second is the notion of \textit{free states}, which similarly can be easily prepared. States and operations which do not fall in these categories are \textit{resources}. Which operations and states should be considered as free depends on the situation one is looking at. Then one can ask questions such as: can we quantify the amount of resource present in a given state? Given a certain amount of resources, which procedures can be done and which are impossible?

 Maybe the oldest example of resource theory is a famous geometrical problem: using only a compass and a straightedge, which figures can one produce? In this context, free operations are the drawing of strokes using the compass and straightedge. The free states are the figures so produced, which include, for instance, equilateral triangles with a given side length, or circles with a given radius. Some operations, however, cannot be performed using only these free operations, such as squaring the circle or trisecting an angle. Resources are figures which cannot be produced by compass-and-straightedge construction; one example is the Archimedean spiral. However, such a spiral is drawn on a piece of paper (by any means), it can then be combined with compass-and-straightedge construction to square the circle. In this way, the Archimedean spiral is indeed a resource: a difficult-to-produce entity which, when available, allows to realize a non-trivial task.

Another classical example is thermodynamics. Consider, for instance, a work engine. Free states can be defined, as objects thermalized at room temperature. By contrast, a source of heat, such as a furnace with a finite amount of coal, is a costly, resourceful, element. When used in combination with the work engine at ambient temperature $T_0$, this resource can be used to generate work. As work is extracted, the heat source gradually cools down until it reaches room temperature. The amount of resources contained in the heat source can be defined as the maximum amount of work that can be extracted from it. This quantity is given by the (adequately named) \textit{free energy} $E-T_0S$ of the heated object, where $E$ is the total energy of the object and $S$ its total entropy.

More recently, these ideas have been brought to the quantum domain, and many quantities have been studied from the perspective of resource theories. The best-known example is entanglement \cite{brandao_entanglement_2008}: here free states are separable states, and free operations are local operations and classical communications, which cannot create or increase entanglement. The amount of resource (\textit{i.e.}, of entanglement) in a given state can be meaningfully quantified by the distance in Hilbert space between the state considered and the set of separable states. 
 The formalism has also been extended to many other quantities, such as coherence \cite{baumgratz_quantifying_2014,streltsov_colloquium:_2017}, asymmetry \cite{ahmadi_wignerarakiyanase_2013}, non-Gaussianity and Wigner negativity \cite{takagi_convex_2018,albarelli_resource_2018,yadin_operational_2018} or  quantum thermodynamics \cite{brandao_second_2015,lostaglio_description_2015}.\\

Hence, we will be interested in the following questions: can we characterize all the states which achieve a quantum metrological advantage in a given context? Can the metrologically useful states be organized according to a resource theory? A similar question has been addressed in \cite{kwon_nonclassicality_2019} and \cite{yadin_operational_2018}. In these works, quantifiers of nonclassicality have been proposed, based on the capacity of a state to evaluate a \textit{translation} in phase space. 
However, in many experimental contexts, it is interesting to evaluate a \textit{rotation} in phase space. This is the case, for instance, when one wants to measure a phase shift in an optical circuit. More formally, previous results have studied the sensitivity of states with respect to displacements, \textit{i.e.}, operators of the form $e^{ix(\aop+\adag)+p(\adag-\aop)}$, which involve \textit{linear} field terms. By contrast, here, we will be interested in the evaluation of operators of the form $e^{i\phi\adag\aop}$, which involve \textit{quadratic} field terms. More precisely, in this Chapter, we will study the use of quantum states of light to study passive linear elements. We will focus on Gaussian states since they are largely available experimentally and allow to obtain exact results. Our goal then will be to characterize all the states which achieve a metrological advantage. We will then discuss how we tried, but only partially succeeded, to cast these results in the form of a resource theory. We will also discuss how this partial failure highlights some properties that are specific to phase estimation protocols.
 
\section{Setup}

\subsection{Protocol}
We consider a two-modes optical circuit composed of several passive linear elements (PLE), one of them being unknown. A probe state of light is sent into the circuit and measured; the measurement results are used to characterize the unknown optical elements. In general, we allow for all measurements, including homodyne measurements, which requires an additional, perfect, phase reference. The PLE composing the circuit include phase-shifters, beam splitters, and combinations of these elements. Let us define the following operators:

\begin{align}
	\Jpz=\frac{\adag_1\aop_1-\adag_2\aop_2}{2}, \hspace{5pt} \Jpx=\frac{\adag_1\aop_2+\aop_1\adag_2}{2}, \hspace{5pt} \Jpy=i\frac{\adag_1\aop_2-\aop_1\adag_2}{2}.
\end{align}
Any PLE can be described by a quantum channel $\Rop_\uvect(\phi)=e^{i\phi\uvect.\hat{\bm{J}}}$, with $\hat{\bm{J}}=(\Jpx,\Jpy,\Jpz)$, $\phi$ a real number and $\uvect$ any three-dimensional unitary vector. 
For instance, a phase difference of $\phi$ between the two modes of the circuit corresponds to an operator $\Rop_z(\phi)=e^{i\phi\Jpz}$. $\Rop_x(\phi)=e^{i\phi\Jpx}$ corresponds to the action of a beam-splitter with a transmissivity $\cos(\phi/2)$. Since these quantities describe passive, lossless optical elements, they conserve the total excitation number : $[\hat{J}_b,\adag_1\aop_1+\adag_2\aop_2]=0$ for $b=\{x,y,z\}$. If we restrict the full Hilbert space to a subspace where the total number of photons is fixed, these operators are formally equivalent to spin operators, which obey the SU(2) commutation relation $[\hat{J}_a,\hat{J}_b]=i\epsilon _{abc}\hat{J}_c$, with $\epsilon _{abc}$ the totally antisymmetric tensor. 

\begin{figure}
 	\includegraphics[angle=-90,width=\linewidth]{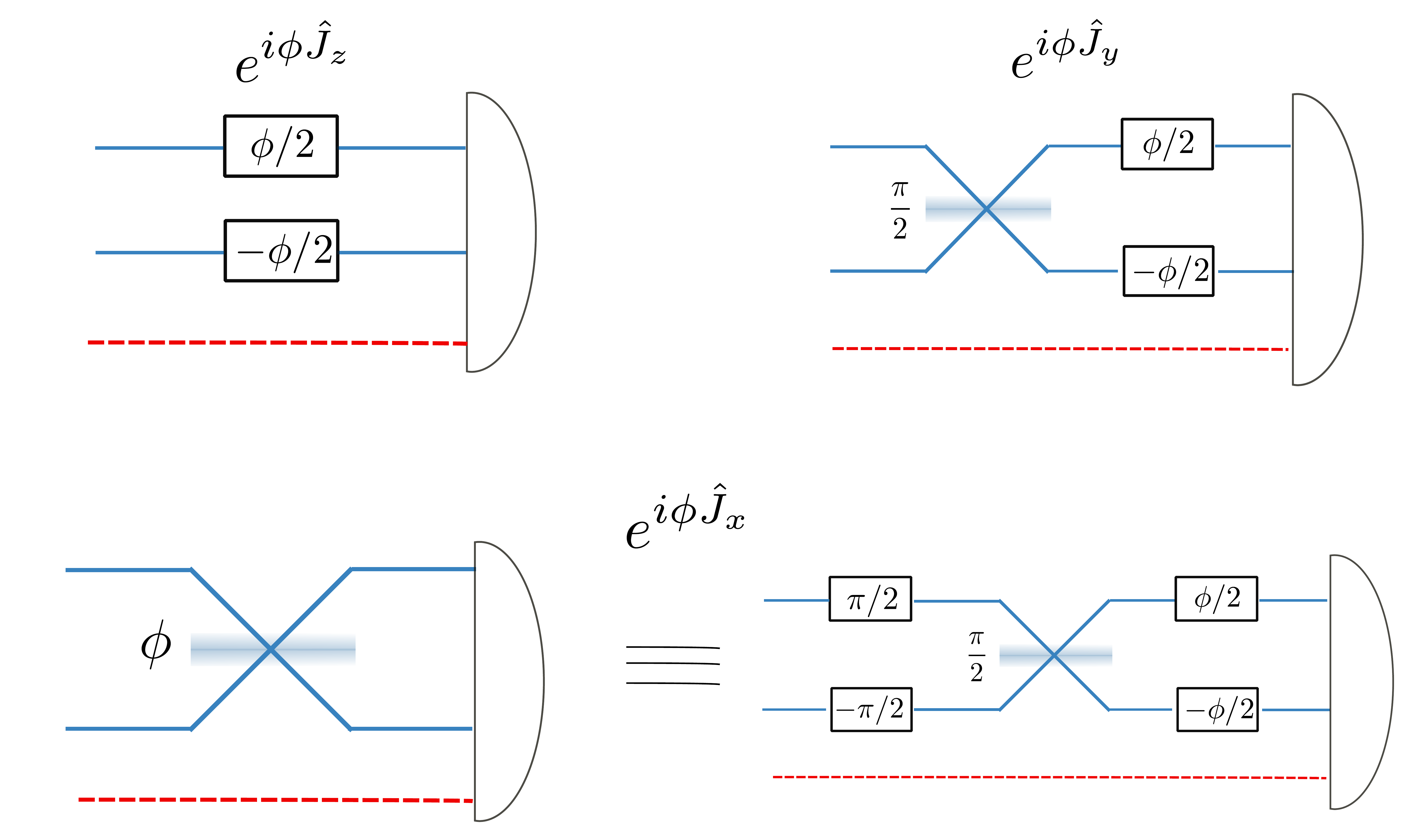}
 	\caption[Characterization of an optical element.]{Characterization of a given optical element. In each case, a two-mode probe state of light is sent into an optical circuit characterized by an unknown parameter $\phi$. The circuit can be modeled by a quantum channel $\Rop(\phi)$. Top left: evaluation of an unknown phase difference, which corresponds to a channel $\Rop=e^{i\phi\Jpz}$. Top right: channel $\Rop=e^{i\phi\Jpy}$. Bottom: estimation of the transmissivity of a beam-splitter, described by a channel $\Rop=e^{i\phi\Jpx}$. This problem and the phase-estimation problem can be mapped to each other by including more controlled PLE in the circuit (note that some elements have been absorbed in the measurement protocol). We allow for both photon-counting and homodyne measurements, which require an additional beam serving as a phase reference (dashed red line).}
 	\label{PLE}
 \end{figure}

In phase space, these transformations $\Rop_\uvect(\phi)$ can be described by rotations $\Rmat$ in 3-dimensional space. Importantly, since $\Rop_{\uvect1}(\theta_1)e^{i\phi\uvect.\bm{\hat{J}}}\Rop_{\uvect1}(\theta_1)^\dagger=e^{i\phi\Rmat_1(\uvect).\bm{\hat{J}}}$, different estimation problems can be mapped to each other. For instance, the problem of estimating the reflexivity of an unknown beam-splitter can be mapped to the problem of estimating a phase difference, simply by adding more beam-splitters and phase-shifters with controlled properties, and exploiting the relation $\Rop_x(\phi)=\Rop_y(\frac{\pi}{2})\Rop_z(\phi)\Rop_y(-\frac{\pi}{2})$. Different estimation procedures are depicted in Fig.\ref{PLE}. Note that one-mode Gaussian phase estimation also arises as a special case of this formalism: it corresponds to sending the vacuum in one arm and performing homodyne detection.\\

When an input probe state $\rop_0$ is sent into a circuit described by the channel $\Rop_\uvect(\phi)$,
and an optimized measurement is performed, the parameter $\phi$ can be evaluated with a precision bounded by the quantum CR bound:

\begin{equation}
 	\delta\phi\geq\frac{1}{\sqrt{\pazocal{I}_\phi^{\uvect}}},
 \end{equation} 
with $\pazocal{I}_\phi^{\uvect}$ the QFI associated with the channel $\Rop_\uvect(\phi)$ and the initial state $\rop_0$. As we have shown, various channels with different $\uvect$ can be mapped to each other just by adding more PLE. Since these elements are easily accessible experimentally, we allow, for a given input state, to optimize the channel in order to achieve the best possible precision. Hence, we will take the following quantity as a figure of merit:

\begin{equation}
	\pazocal{I}_\phi^{\text{opt}}=\text{Max}_{\{\uvect\}}\pazocal{I}_\phi^{\uvect},
\end{equation}
which we will call the \textit{metrological capacity}, and which depends only on the initial state $\rop_0$.\\

Alternatively, it is completely equivalent to say that the encoding channel is fixed, but we can optimize the input state by applying a PLE operation $\Rop_\uvect$ before sending it into the circuit. We then have an effective input state $\Rop_\uvect\rop_0\Rop_\uvect^\dagger$, and optimize the precision over the possible operations $\Rop_\uvect$. In the following, we will rather use this point of view. We adopt the following convention: we study the phase difference of a Mach-Zehnder interferometer, described by a quantum channel $\Rop_y(\phi)$, by using an optimized probe state $\rop^\uvect=\Rop_\uvect\rop_0\Rop_\uvect^\dagger$. We achieve a precision given by a QFI $\pazocal{I}_\phi^\uvect$.\\

Finally, although we will focus here on states of light, note that a similar formalism can be applied for atomic states. For instance, the estimation of a magnetic field using Ramsey protocol with a Bose-Einstein Condensate can be described by similar equations. The optimization step then corresponds to a spatial rotation of an atomic cloud, or equivalently to orienting the magnetic field of unknown amplitude in a different direction.

\subsection{Parameterization of input states}

In this Chapter, we will focus on Gaussian input states. The Gaussian properties of the state are preserved by the passive linear circuit. To properly describe the state, it is essential to have a good parameterization. This problem has been treated in some length in Chapter 3 and Appendix B, here we will only recall the key elements.\\ 

Any $q$-mode Gaussian state can be obtained by applying  Gaussian operations on a thermal state (see Chapter 3 and Ref.\cite{simon_quantum-noise_1994,safranek_gaussian_2016}):

\begin{equation}
\label{symplecform}
	\rop_0=\hat{G}\left(\bigotimes_i\rop_{\text{th}}^{(i)}(T_i)\right)\hat{G}^{\dagger},
\end{equation}
where $\hat{G}$ is a Gaussian (unitary) operation, and $\rop_{\text{th}}^{(i)}(T_i)=\frac{1}{Z_i}e^{-\frac{\hbar\omega_i \adag_i\aop_i}{k_BT_i}}$, with $k_B$ the Boltzmann constant, $T_i$ the temperature, $Z_i$ the partition function and $\omega_i$ the frequency of the $i$-th mode. Gaussian operations include PLE, displacement, squeezing, and composition thereof. 
The purity of the state is directly given by the temperatures $T_i$. We will restrict ourselves to \textit{isotropic} Gaussian states, which satisfy $\omega_1=\omega_2=\omega$ and $T_1=T_2=T$ (that is, the thermal noise is the same in each mode). 

 In phase space, all Gaussian states can be completely described by their covariance matrix $\sigma$ and their displacement vector $\dvect$. The action of a Gaussian operation $\Gop$ is described by a $2q\times2q$ matrix $\Gmat$ ($q$ being the number of modes): $\sigma\rightarrow\Gmat\sigma\Gmat^\dagger$, and $\dvect\rightarrow\Gmat\dvect$.\\

For isotropic states, the covariance matrix can be put in the form:

\begin{align}
\label{decomposition}
	\sigma=\nu\left[\Rmat_1(\phi_1)\Rmat_2(\phi_2)\Bmat(\theta)\Rmat_{as}(\Psi)\Smat_1(\xi_1)\Smat_2(\xi_2)\times h.c.\right],	
\end{align}
where $\nu=\text{cotanh}(\frac{\hbar\omega}{2k_BT})$ the \textit{symplectic eigenvalue} of the state. $\Rmat$, $\Smat$ and $\Bmat$ describe the effect of phase-shifting, squeezing, and mode-mixing, respectively. Their precise expressions can be found in Appendix B. The displacement $\dvect$ can always be parameterized as $(\bm{\dispmod},\bm{\dispmod}^*)$, with:

\begin{align}
	\bm{\dispmod}=\lvert\dispmod\rvert ( e^{-i\phi_{d1}} \cos{\dispan}, e^{-i\phi_{d2}} \sin{\dispan}),
\end{align}
with $\dispan$, $\dispmod$, $\phi_{d1}$, and $\phi_{d2}$ real parameters. 

Hence, all isotropic Gaussian states can be described by the following set of parameters:

 $(\nu,\lvert\dispmod\rvert,\dispan,\phi_{d1},\phi_{d2},\phi_1,\phi_2,\theta,\Psi,\xi_1,\xi_2)$. States with $\xi_1=\xi_2=0$ are \textit{displaced thermal states}; all the other states can be generically called \textit{squeezed states}. This definition of squeezing is not identical to the usual Wineland criterion \cite{wineland_spin_1992} based on the variance of the quadratures. Rather, squeezed states are defined here as those states that cannot be generated using only thermal states, displacement operations, and PLE. This definition of squeezing was first put forward in \cite{simon_quantum-noise_1994}.\\

By applying PLE, we can modify the various parameters describing the state. However, these parameters are not independent. For instance, applying a phase-shift $\chi_-$ on the first mode will change simultaneously $\phi_1\rightarrow\phi_1-\chi_-$ and $\phi_{d1}\rightarrow\phi_{d1}-\chi_-$. Properly understanding our level of control on the state parameters will be an important part of our analysis.
Finally, the following expression will be useful in the following:
\begin{equation}
\label{Nb}
	\langle\Nop\rangle=\sum_i\adag_i\aop_i = \frac{1}{4}Tr[\sigma]-\frac{q}{2}+\frac{\lvert \dvect\rvert^2}{2}.
\end{equation}

\section{Characterization of metrologically useful states}
\subsection{Definition of FTQL and metrological advantage}
We aim to characterize all states which lead to a metrological advantage. This requires to first define what a metrological advantage is. For this, we need to properly identify an ensemble of "classical" states. Quantum advantage can then be defined as reaching a precision that is unattainable by using those states. In this section only, we will consider general Gaussian states with $q$ modes. \\

For (isotropic) Gaussian states, displaced thermal states are good candidates to define such a reference. The goal then is to identify a notion of resource. To ensure a fair comparison, the state we want to study must be pitted against a "classical" state with the same amount of resources. The resource most commonly considered is the average total number of photons $\moy{\Nop}=\sum_q\moy{\adag_q\aop_q}$. To evaluate whether a given state $\rop_0$ is metrologically useful, we could imagine taking a \textit{pure} coherent state with the same $\moy{\Nop}$. The metrological capacity of such a state is given by the usual SQL $\pazocal{I}_\phi^{\text{ref}}=4\moy{\Nop}$. Then a state $\rop_0$ achieves metrological advantage if $\pazocal{I}_\phi(\rop_0)\geq 4\text{Tr}(\Nop\rop_0) $, independently of its purity. This definition would be relevant, for instance, in situations where the average number of photons is fixed by experimental constraints, but thermal noise is not. 
Here, we will follow a slightly different path. We will consider that both $\moy{\Nop}$ and the state purity (which is quantified by the symplectic eigenstate $\nu$) are resources and should be explicitly included in the definition of the "classical" reference. Specifically, for every state $\rop_0$, we associate a reference state which is an isotropic displaced thermal state with the same $\moy{\Nop}$ AND the same $\nu$ as $\rop_0$. Such a state has a metrological capacity $\pazocal{I}_\phi^{\text{ref}}=\frac{4\moy{\Nop}}{\nu}+2q\frac{1-\nu}{\nu}$ \cite{safranek_gaussian_2016}. We will call this bound the finite-temperature quantum limit, or FTQL. \footnote{Note that the reference state is not uniquely defined, since there are infinitely many displaced thermal states with the same {$\langle\hat{N}\rangle$} and {$\nu$}. However, all these states can be mapped to each other by PLE, and hence they all achieve the same metrological capacity. Thus, the FTQL is well-defined.} This precision increases with the number of photons and decreases with the temperature. Note that FTQL and SQL become equivalent in the limit of zero temperature, when the reference state becomes pure (\textit{i.e.}, for {$\nu=1$}).\\

To conclude, we consider an arbitrary isotropic two-mode Gaussian state $\rop_0$ with symplectic eigenvalue $\nu$ and an average number of photons $\moy{\Nop}$. If this state is used to characterize a quantum channel defined by a parameter $\phi$, we define the metrological advantage as:

\begin{eqnarray}
\label{metroadvFTQL}
	\pazocal{A}_G(\rop_0) & = &\text{Max}\Big( \pazocal{I}_\phi^{\text{opt}}-\pazocal{I}_\phi^{\text{ref}} , 0 \Big) \\ \nonumber
	 & = &\text{Max}\Big( \pazocal{I}_\phi^{\text{opt}}-\frac{4\moy{\Nop}}{\nu}-2q\frac{1-\nu}{\nu} , 0 \Big).
\end{eqnarray}

We aim to characterize the states which achieve $\pazocal{A}_G>0$. Note that a similar analysis has been carried out in \cite{hyllus_not_2010}, for pure states. By contrast here, we take explicitly the (im)purity of the state (that is, its thermal noise) into account, at the cost of restricting ourselves to Gaussian states.

\subsection{One-mode states}
\label{Onemode}
Let us first consider the specific case of one-mode interferometry. The covariance matrix and displacement vector of a one-mode Gaussian state can be parameterized like:

\begin{align}
	\sigma&=\nu\Rmat(\phi)\Smat(\xi)\Smat(\xi)^\dagger\Rmat(\phi)^\dagger\\ \nonumber,
	\dvect&=\lvert\dispmod\rvert(e^{i\phi_d},e^{-i\phi_d}).
\end{align}

For single-mode states, we can only act on the state through phase-shift. A phase-shift $\chi_-$ leads to $\phi\rightarrow\phi-\chi_-$, $\phi_d\rightarrow\phi-\chi_-$. This means that we can control the total phase $\phi+\phi_d$, but not the relative phase $\phi-\phi_d$, which describes the relative orientation of displacement and squeezing.

 A one-mode Gaussian state is sent through an unknown phase-shifter, and homodyne measurement is performed. Using equation (30) of Chapter 3, the QFI associated with phase estimation can be computed exactly. The result is \cite{safranek_gaussian_2016}:

\begin{equation}
	\pazocal{I}^{\text{opt}}_\phi=4\frac{\nu^2}{\nu^2+1}\sinh^2(2\xi) + \frac{4\lvert\dispmod\rvert^2}{\nu}\Big(e^{2\xi}\cos^2(\phi-\phi_d)+e^{-2\xi}\sin^2(\phi-\phi_d)\Big).
	\label{QFItwomode}
\end{equation}
Note that this expression is independent of the total phase $\phi+\phi_d$, and cannot be optimized any further.
The reference QFI is $\pazocal{I}^{\text{ref}}=\frac{4\moy{\Nop}}{\nu}+2\frac{1-\nu}{\nu}=4\sinh^2(\xi)+\frac{4\lvert\dispmod\rvert^2}{\nu}$.
Then we have: 

\begin{equation}
	\pazocal{I}^{\text{opt}}_\phi-\pazocal{I}^{\text{ref}}_\phi=4\left[\frac{\nu^2}{\nu^2+1}\sinh^2(2\xi)-\sinh^2(\xi)\right]+\frac{4\lvert\dispmod\rvert^2}{\nu}\Big[e^{2\xi}\cos^2(\phi-\phi_d)+e^{-2\xi}\sin^2(\phi-\phi_d)-1\Big].
	\label{Advonemode}
\end{equation}

The first term describes the effect of squeezing only, while the second results from the interplay between displacement and squeezing. When the squeezing vanishes, $\xi=0$, so does the advantage. Now we want to study when this quantity becomes strictly positive. We can distinguish two regimes of parameters: 

\begin{itemize}
	\item For $0\leq\phi-\phi_d\leq\frac{\pi}{4}$, we have $\pazocal{I}^{\text{opt}}-\pazocal{I}^{\text{ref}}>0$ if and only if $\xi\neq0$. This means that squeezed states always outperform displaced states. Physically, this regime describes states for which squeezing and displacement are aligned.
	\item  For $\frac{\pi}{4}\leq\phi-\phi_d\leq\frac{\pi}{2}$, squeezing and displacement are orthogonal, and the two terms in \eqref{Advonemode} have opposite signs. For small displacement $\dispmod$, squeezed state have still an advantage; however, it disappears for larger displacement. 
\end{itemize}
Hence, for one-mode states, the presence of squeezing does not guarantee a metrological advantage; the performances of a given state depend critically on the relative orientation of displacement and squeezing. These results are compatible with previous studies on metrology with one-mode Gaussian states \cite{pinel_quantum_2013}.

\subsection{Two-mode states}

For two-mode states, the inclusion of mode-mixing channels gives us an additional degree of control on the state, and lead to very different behavior. The main result of this section is the following theorem:

\begin{theorem}
	Let $\rop_0$ be an isotropic non-pure two-mode Gaussian state. Then $\pazocal{A}_G(\rop_0)=0$ if and only if $\rop_0$ is a displaced thermal state.
\end{theorem}
The "if" part of the Theorem holds by definition of $\pazocal{A}_G$. The converse is nontrivial: if a non-pure state is squeezed, then it has a nonzero metrological advantage. This does not mean that any squeezed state is useful for estimating, for instance, a phase difference in a Mach-Zehnder interferometer; rather, it means that any squeezed state can be transformed, by applying PLE, into a state which \textit{is} useful for such a task.

\begin{proof}
	We want to estimate the phase difference between the two arms of a Mach-Zehnder interferometer, by using a probe state $\rop^\uvect=\Rop_\uvect\rop_0\Rop_\uvect^\dagger$, with $\Rop_\uvect$ a PLE operation which can be freely optimized.
	The state $\rop_0$ is described by a set of parameters $(\nu,\lvert\dispmod\rvert,\dispan^0,\phi_{d1}^0,\phi_{d2}^0,\phi_1^0,\phi_2^0,\theta^0,\Psi^0,\xi_1,\xi_2)$. Without loss of generality, we can set $\xi_1\geq0$ and $\xi_2\geq0$ (any sign difference can be absorbed in the other parameters). 
	Let us apply the following operation: $\Rop_\uvect=\Rop_z(-2\Psi^0 + \frac{\pi}{2})\Rop_x(-2\theta^0)\Rop_z(\phi_2^0-\phi_1^0)$. After this operation, the covariance matrix is given by:
	\begin{eqnarray*}
\left[\Rmat_{as}\left(-\Psi^0 + \frac{\pi}{4})\right)\Bmat(-\theta^0)\Rmat_{as}\left(\frac{\phi_2^0-\phi_1^0}{2}\right) \Big(\Rmat_1(\phi_1^0)\Rmat_2(\phi_2^0)\Bmat(\theta^0)\Rmat_{as}(\Psi^0)\Big)\right]\times h.c.\\
 = \Rmat_1(\frac{\phi_1^0+\phi_2^0}{2})\Rmat_2(\frac{\phi_1^0+\phi_2^0}{2})\Rmat_{as}(\frac{\pi}{4})\times h.c.\\
  = \Rmat_1(\frac{\phi_1^0+\phi_2^0}{2} + \frac{\pi}{4})\Rmat_2(\frac{\phi_1^0+\phi_2^0}{2} - \frac{\pi}{4})\times h.c. 	
\end{eqnarray*}

(where we have used the notation $\Rmat_{as}(\phi)=\Rmat_1(\phi)\Rmat_2(-\phi)$, and the fact that $\Rmat_1(\phi)\Rmat_2(\phi)$ commutes with $\Bmat$). We have now a state $\rop^{\uvect_I}$, characterized by $\theta^I=\Psi^I=0$, $\phi_1^I=\frac{\phi_1^0+\phi_2^0}{2} + \frac{\pi}{4}$, $\phi_2^I=\frac{\phi_1^0+\phi_2^0}{2} - \frac{\pi}{4}$, and $\phi_1^I-\phi_2^I=\frac{\pi}{2}$.\\

Still using Eq. (30) of Chapter 3, the QFI associated with the evaluation of a Mach-Zehnder channel with a Gaussian state can be explicitly computed. This was done in \cite{safranek_gaussian_2016,safranek_optimal_2016}. With $\theta=\Psi=0$, the result can be simplified as:

\begin{eqnarray}
\label{eqpqfi}
	\pazocal{I}_\phi(\rop) = \frac{8\nu^2}{\nu^2+1} \Big(p^2\sinh(\xi_1-\xi_2)^2 +o^2\sinh(\xi_1+\xi_2)^2\Big) \nonumber +\frac{4\lvert \dispmod\rvert^2}{\nu}\Big(  e^{2\xi_1} \chi_+^2 + e^{-2\xi_1} \chi_-^2 + e^{2\xi_2} \upsilon_+^2 + e^{-2\xi_2} \upsilon_-^2\Big),
\end{eqnarray}
with:
\begin{align}
\label{defpara}
\nonumber
o&=\sin(\phi_1-\phi_2), \\ \nonumber
p&=\cos(\phi_1-\phi_2), \\ \nonumber
\chi_+&=\sin(\dispan)\cos(\phi_1-\phi_{d2}+\Psi), \\ \nonumber
\chi_-&=\sin(\dispan)\sin(\phi_1-\phi_{d2}+\Psi), \\ \nonumber
\upsilon_+&=\cos(\dispan)\cos(\phi_2-\phi_{d1}-\Psi), \\ 
\upsilon_-&=\cos(\dispan)\sin(\phi_2-\phi_{d1}-\Psi).
\end{align}
Note that $\chi_+^2+\chi_-^2+\upsilon_+^2+\upsilon_-^2=1$. Now, let us express the reference precision. Using \eqref{Nb}, we have:
\begin{equation*}
	\pazocal{I}_\phi^{\text{ref}}=4\frac{\langle \Nop\rangle +1}{\nu}-4=\frac{Tr[\sigma]}{\nu}+\frac{4\lvert\dispmod\rvert^2}{\nu} - 4 = 4(\sinh(\xi_1)^2+\sinh(\xi_2)^2)+\frac{4\lvert\dispmod\rvert^2}{\nu},	
\end{equation*}

and we obtain after a little rewriting:
\begin{align}
	\label{eqdep}
	\pazocal{I}_\phi^{\uvect_I}-\pazocal{I}_\phi^{\text{ref}} = & 2[o^2 \mathcal{X} +p^2\mathcal{Y}] + \frac{4\lvert\dispmod\rvert^2}{\nu}\Big(e^{2\xi_1}\chi_+^2 + e^{-2\xi_1}\chi_-^2 + e^{2\xi_2}\upsilon_+^2 + e^{-2\xi_2}\upsilon_-^2 -1 \Big)\\ \nonumber
 \mathcal{Y}= & 4 \frac{\nu^2}{\nu^2+1}\Big( \sinh^2(\xi_1-\xi_2) \Big) - 2\big(\sinh^2(\xi_1)+\sinh^2(\xi_2)\big)\\ \nonumber
 \mathcal{X}= & 4 \frac{\nu^2}{\nu^2+1}\Big( \sinh^2(\xi_1+\xi_2) \Big) - 2\big(\sinh^2(\xi_1)+\sinh^2(\xi_2)\big).	
 \end{align}

Importantly, we have $\mathcal{X}\geq\lvert \mathcal{Y}\rvert\geq0$. As for the one-mode estimation problem, we have two terms: one involves only the squeezing parameter, the second involves both squeezing and displacement. We are now equipped to complete the proof. We start from the state $\rop^{\uvect_I}$ with $\theta=\Psi=0$, $\phi_1-\phi_2=\frac{\pi}{2}$, $\dispan=\dispan^I$, $\phi_{d}=\phi_{d}^I$ (again, \textit{any} probe state can be put to this form). We call the associated parameters $o^I$, $p^I$, $\chi_+^I$, $\chi_-^I$, $\upsilon_+^I$ and $\upsilon_-^I$. Then we have $p^I=0$, and the expression above reduces to:

\begin{equation}
	\label{eqdep_2}
	\pazocal{I}_\phi^{\uvect_I}-\pazocal{I}_\phi^{\text{ref}} =2\mathcal{X}+\frac{4\lvert\dispmod\rvert^2}{\nu}V^I,
\end{equation}
with $V^I=\Big(e^{2\xi_1}(\chi_+^I)^2 + e^{-2\xi_1}(\chi_-^I)^2 + e^{2\xi_2}(\upsilon_+^I)^2 + e^{-2\xi_2}(\upsilon_-^I)^2 -1 \Big)$. Then three cases are possible:

\begin{itemize}
	\item  $\bm{V^I > 0}$: then (\ref{eqdep_2}) is strictly positive (\textit{i.e.}, we have an advantage) for all $\lvert\dispmod\rvert$ and all $\nu$. Note that this regime is only possible for $\xi_1\neq0$ or $\xi_2\neq0$. This corresponds to the case where both squeezing and displacement cooperate to increase the precision.\\

	\item $\bm{V^I < 0}$: this is the case where squeezing and displacement have competing effects. If we keep the same choice of parameters, \eqref{eqdep_2} will be positive for small $\lvert\dispmod\rvert$, but negative for high $\lvert\dispmod\rvert$: this is similar to what we obtained for one-mode metrology. However, we can solve this issue by applying an additional phase-shift when the value becomes negative. More precisely, let us define $\lvert\dispmod_s\rvert^2=\frac{\nu \mathcal{X}}{2\lvert V^I\rvert}$. For $\lvert\dispmod\rvert < \lvert\dispmod_s\rvert$ (which corresponds either to low displacement or high temperature), we still have $\pazocal{I}^{\uvect_I}_\phi>\pazocal{I}_\phi^{\text{ref}}$. When $\dispmod\geq\dispmod_s$, we apply an additional phase shift $\Rop_{z}(\frac{\pi}{2})$ (\textit{i.e.}, we apply an operation $\Gop^F=\Rop_z(-2\Psi^0 + \pi)\Rop_x(-2\theta^0)\Rop_z(\phi_2^0-\phi_1^0)$ on the initial state $\rop^0$). We obtain a new probe state $\rop^{\uvect^F}$, with parameters: $\phi_1^F=\phi^I_1 + \frac{\pi}{4}$, $\phi_2^F=\phi^I_2 - \frac{\pi}{4}$, $\phi_{d1}^F=\phi_{d1}^I + \frac{\pi}{4}$ and $\phi_{d2}^F=\phi_{d2}^I - \frac{\pi}{4}$, and still $\theta=\Psi=0$. This yields $\chi_+^F=-\chi_-^I$, $\chi_-^F=\chi_+^I$, $\upsilon_+^F=\upsilon_-^I$, $\upsilon_-^F=-\upsilon_+^I$. Of course, the parameter $p$ will not be equal to $0$ anymore (actually, it will even be equal to $1$). However, since $p\leq1$ and $\lvert \mathcal{Y} \rvert\leq \mathcal{X}$, we have:
	\begin{align*}
\nonumber
\pazocal{I}_\phi^{\uvect^F}-\pazocal{I}_\phi^{\text{ref}}  \geq -2\mathcal{X} + \frac{4\lvert\dispmod\rvert^2}{\nu}V^F \geq \frac{2\lvert\dispmod\rvert^2}{\nu}(V^F+V^I). 
\end{align*}
We have used $-\mathcal{X}\geq-\frac{2\lvert\dispmod\rvert^2 \lvert V^I\rvert}{\nu}$ (since $\lvert\dispmod\rvert\geq\lvert\dispmod_s\rvert$), and $V^I\leq0$. Finally, we note that:

\begin{align}
\nonumber
V^F + V^I & = ((\chi_+^I)^2+(\chi_-^I)^2)(e^{2\xi_1}+e^{-2\xi_1}) + ((\upsilon_+^I)^2+(\upsilon_-^I)^2)(e^{2\xi_2}+e^{-2\xi_2}) -2 \\ \nonumber
& > 2((\chi_+^I)^2+(\chi_-^I)^2) + 2((\upsilon_+^I)^2+(\upsilon_-^I)^2) -2=0.
\end{align}
The last inequality is strict as soon as $\xi_1\neq0$ or $\xi_2\neq0$.
	\item $\bm{V^I = 0}$: 
	We have $\pazocal{I}_\phi^{\uvect_I}-\pazocal{I}_\phi^{\text{ref}} = 2 \mathcal{X}$, which is strictly positive as soon as $\nu>1$ AND $\xi_1$ or $\xi_2$ is non-zero.

\end{itemize}
To summarize, for an arbitrary squeezed input state, we can always use PLE to set $\theta=\Psi=0$ (which corresponds to a separable two-mode squeezed state). 
Then by applying and additional phase-shift, we can set either $\phi_1-\phi_2=\frac{\pi}{2}$ or $\phi_1-\phi_2=\pi$. For $\nu>1$ and any value of the other parameters, one of these choices will always lead to $\pazocal{I}_\phi>\pazocal{I}_\phi^{\text{ref}}$, showing that a metrological advantage can be achieved with any non-pure squeezed Gaussian state, providing unlimited access to PLE is allowed. Conversely, this also means that a non-pure state which cannot beat the FTQL is necessarily non-squeezed, \textit{i.e.}, a displaced thermal state.\\
This proof works for non-pure states; however, the cases $V^I>0$ and $V^I<0$ remain the same when $\nu=1$.
\end{proof}

Note that the result above holds only for non-pure states, with $\nu>1$. It also holds for most, but not all, pure states. We have obtained the following Lemma (proof in Appendix B):

\begin{lemma}
	Pure two-mode squeezed states can always reach, but not always surpass, the FTQL.
\end{lemma}

\subsection{Examples}

\begin{figure}
	\begin{center}\includegraphics[angle=-90,width=\textwidth]{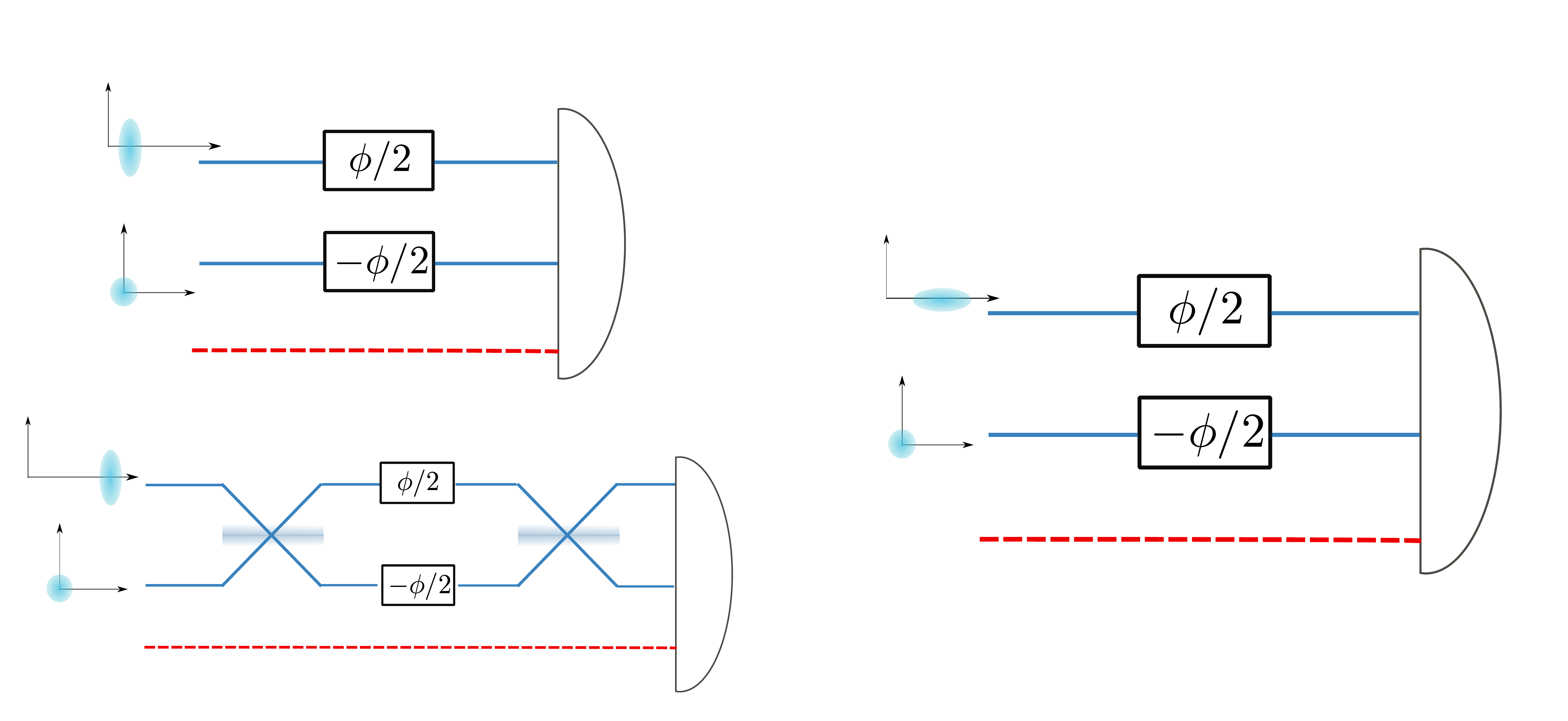}\end{center}
	\caption[Evaluation of PLE using a two-mode separable state.]{Evaluation of PLE using a two-mode separable state. Mode 1 is a displaced squeezed state, mode 2 is a thermal state. Left: displacement and squeezing orthogonal. For small displacement and large squeezing (top), the optimal precision consists in applying directly the unknown phase-shift, then perform a measurement. This is equivalent to a one-mode estimation protocol. For large displacement and small squeezing (bottom), by contrast, the optimal strategy is to use a Mach-Zehnder interferometer. Right: displacement and squeezing are parallel. Then one-mode protocol is always optimal.}
	\label{optimalstrat}
\end{figure}

Let us consider what happens if the first mode is in a displaced squeezed state, and the second is in a thermal state. For those states, we have found explicitly the optimal protocol. Two representative cases are illustrated in Fig.\ref{optimalstrat}. If the squeezing direction and the displacement are parallel (or more generally, if the relative angle between displacement and squeezing is between $0$ and $\pi/4$), the optimal precision is achieved by applying one-mode phase estimation. By contrast, if the squeezing and displacement are orthogonal (relative angle between $\pi/4$ and $\pi/2$), the optimal strategy depends on the displacement: for small displacement, it is still optimal to use one-mode protocol. By contrast, for large displacement, this strategy ceases to yield good results, as we discussed in Section \ref{Onemode}. In this limit, the best precision is achieved by sending the state in a Mach-Zehnder interferometer. This illustrates the differences between the one-mode and two-mode cases: in the two-mode case, even if the two modes are initially separable, we always have the possibility of mixing them before applying a phase-shift. It is this extra level of control which makes squeezing useful in all regimes of parameters. 

\begin{figure}
	\hspace{-18pt}\begin{tabular}{| p{.33\textwidth} | p{.33\textwidth}| p{.33\textwidth}|}
	\hline
	 \begin{center}\includegraphics[angle=-90,width=.33\textwidth]{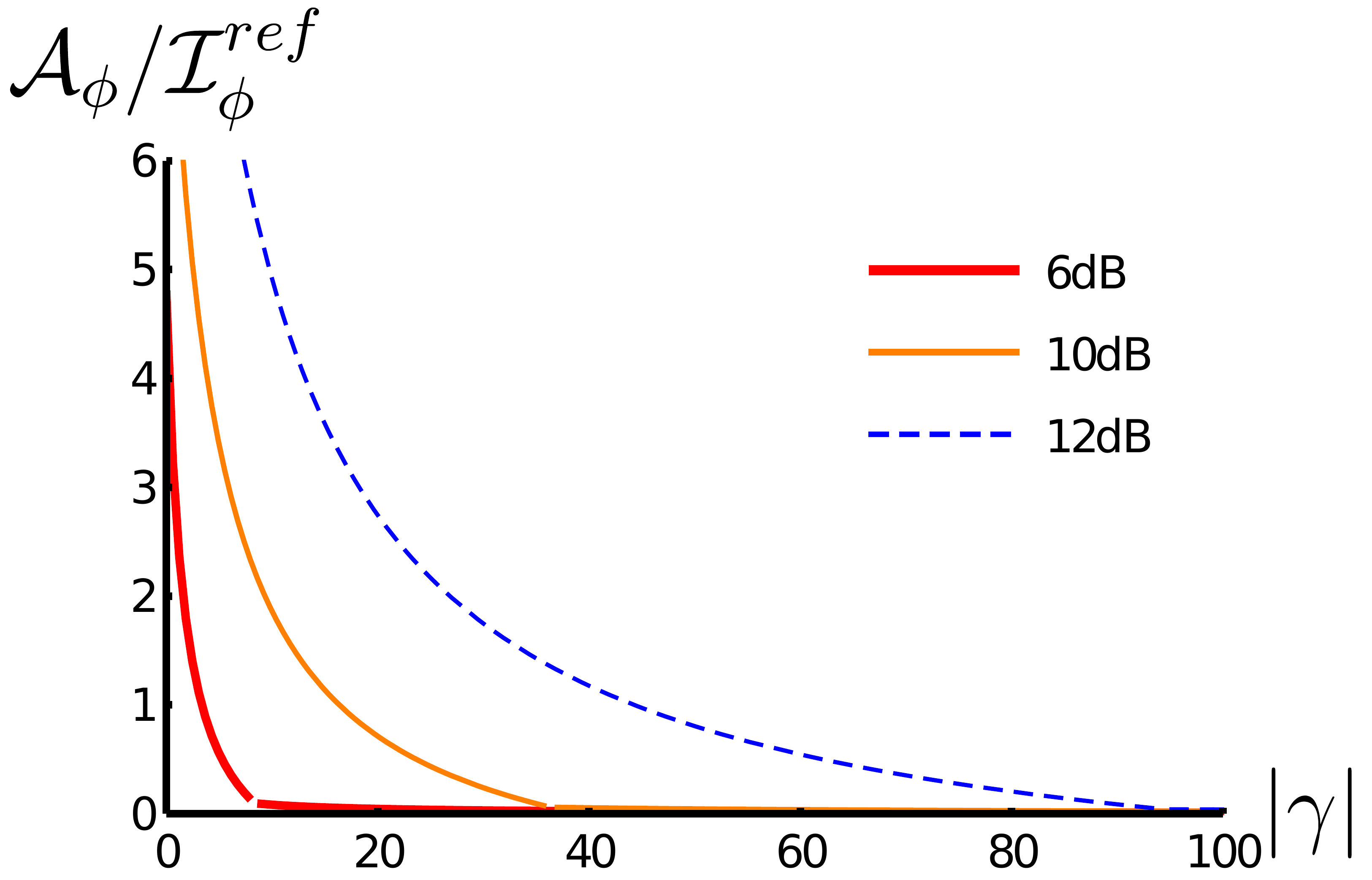}\end{center}  &  \begin{center}\includegraphics[angle=-90,width=.33\textwidth]{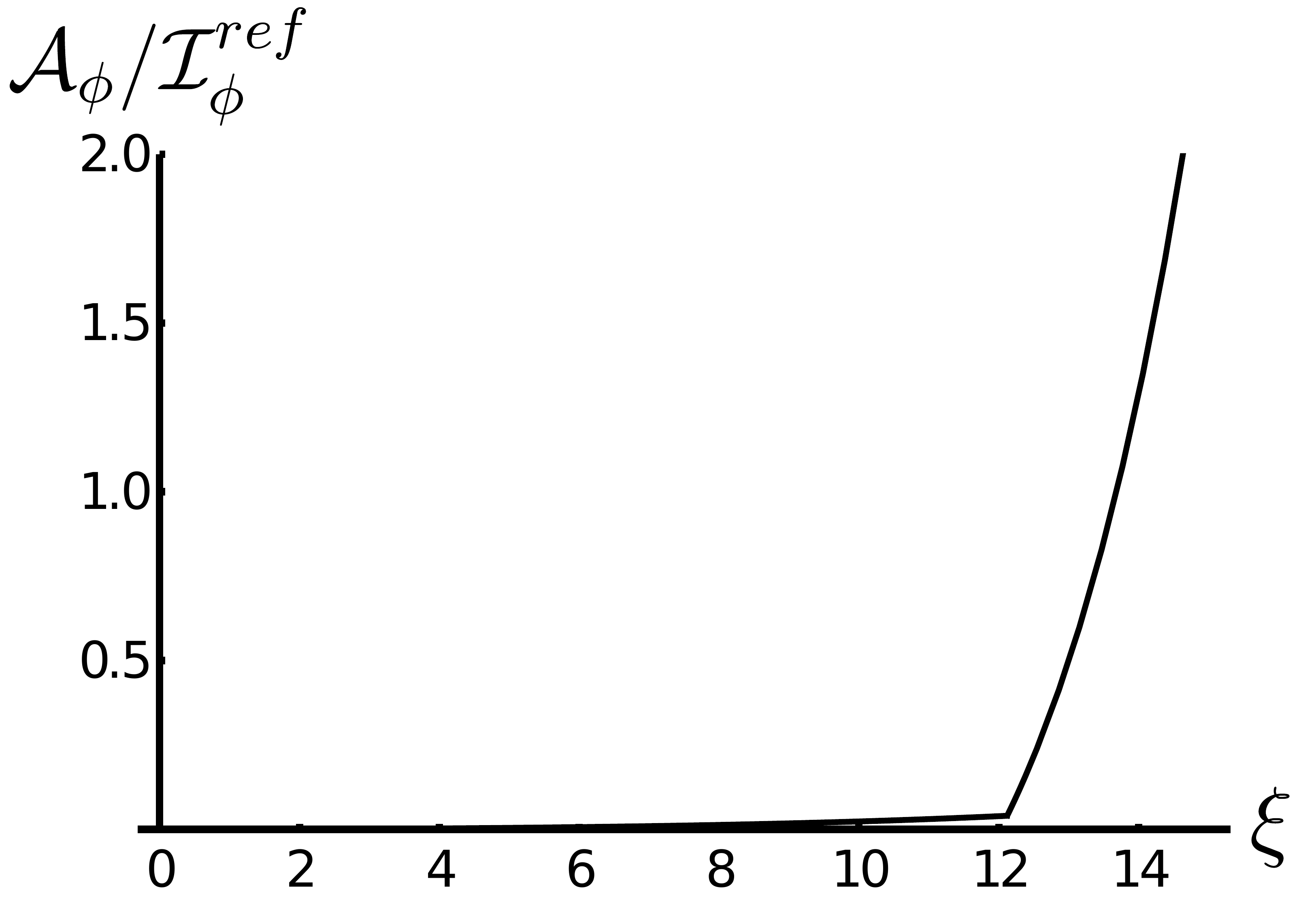}\end{center} & \begin{center}\includegraphics[angle=-90,width=.33\textwidth]{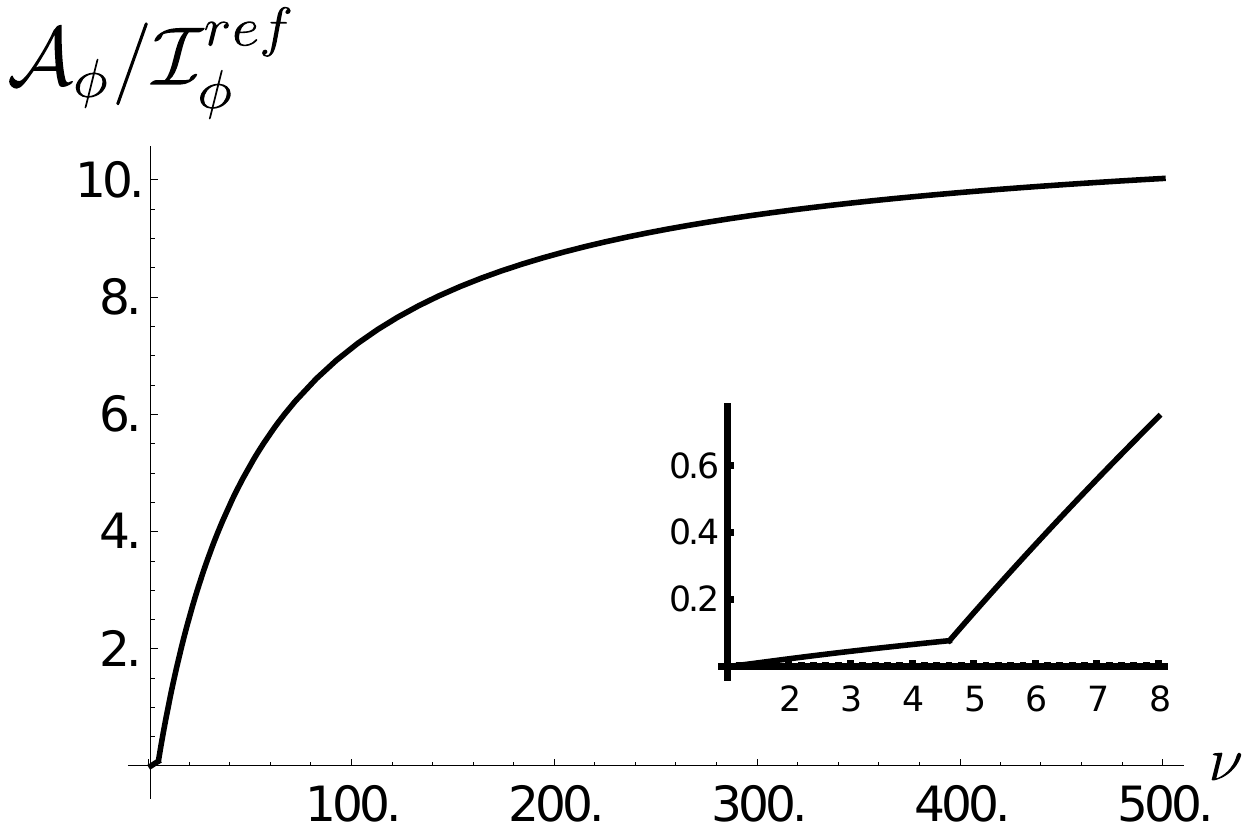}\end{center} \\
	   \begin{center}\includegraphics[angle=-90,width=.33\textwidth]{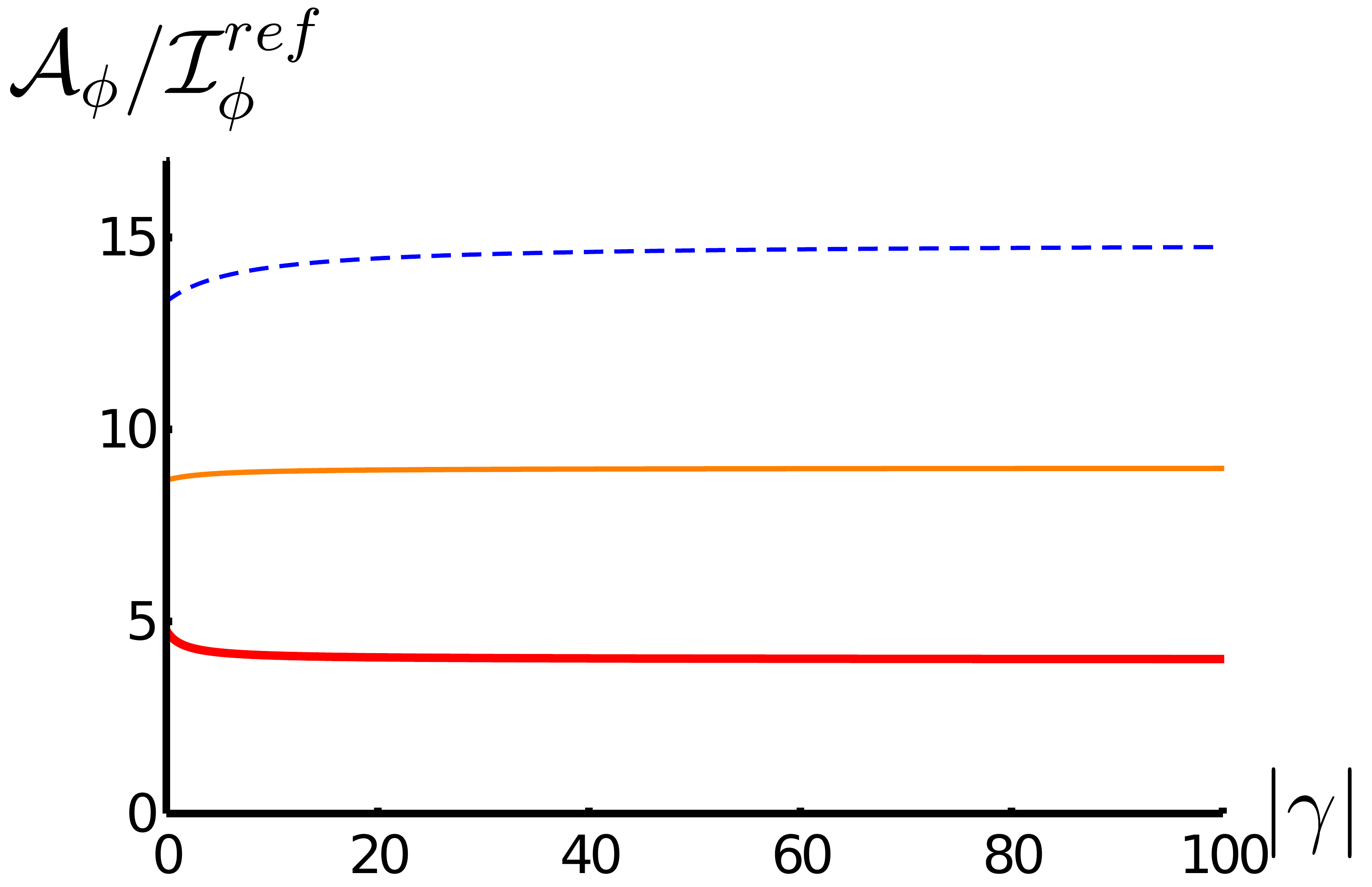}\end{center} & \begin{center}\includegraphics[angle=-90,width=.33\textwidth]{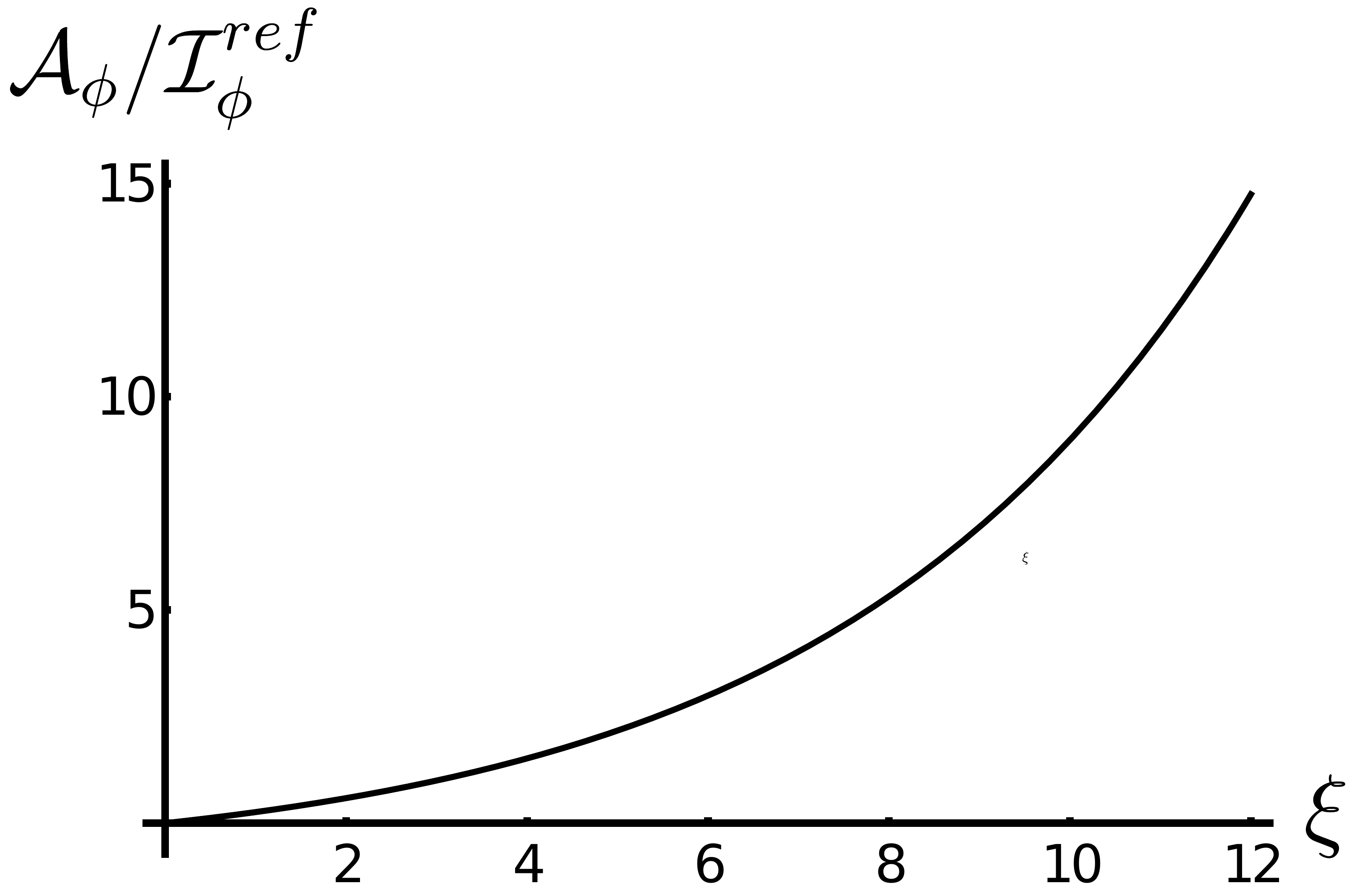}\end{center} &  \begin{center}\includegraphics[angle=-90,width=.33\textwidth]{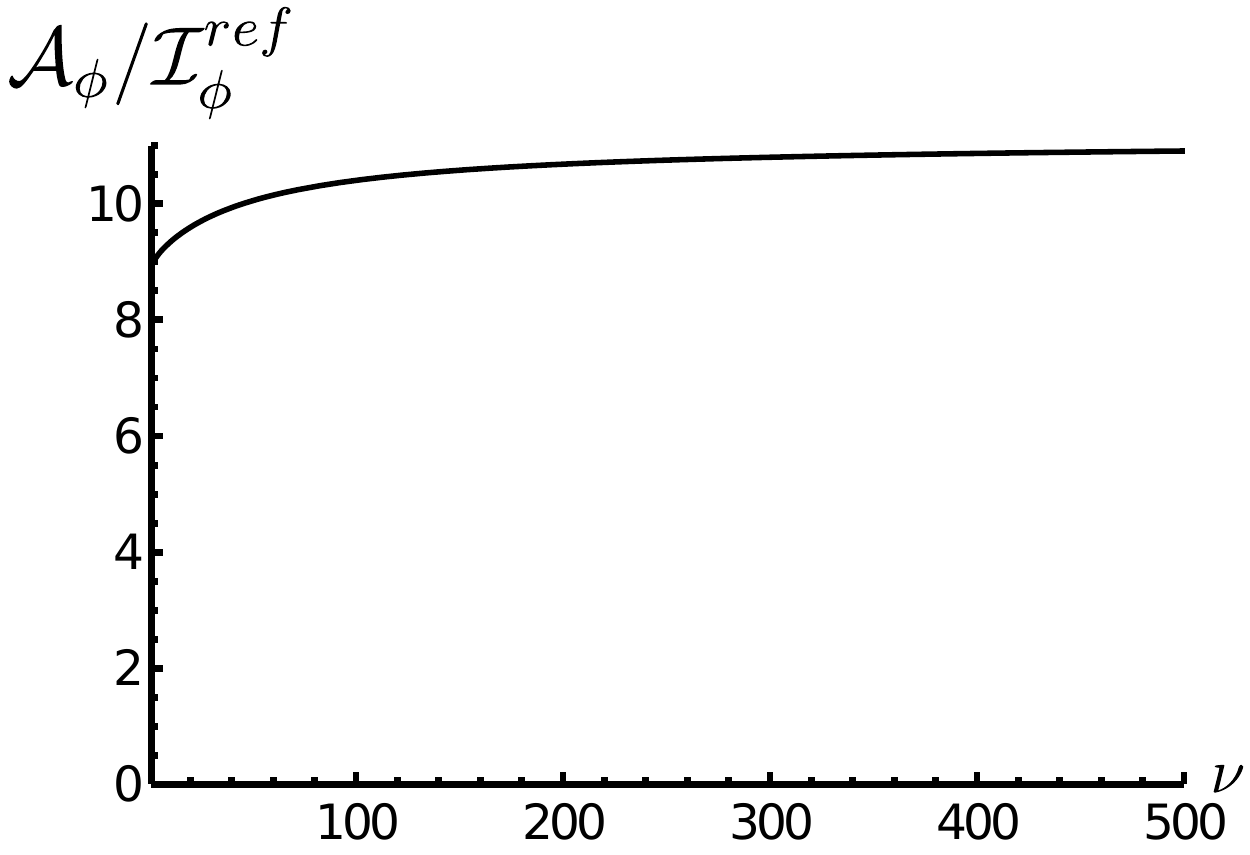}\end{center} \\
	 \hline
	\end{tabular}
	 \caption[Metrological advantage achievable with various input Gaussian states.]{Metrological advantage with squeezing and displacement orthogonal (upper row) or parallel (lower row). Left: metrological advantage, normalized by the FTQL, versus displacement, for $\nu=2$ and various squeezing. Center: same quantity versus squeezing (in dB), for $\lvert\dispmod\rvert=100$ and $\nu=2$. Right: same quantity versus symplectic eigenvalue $\nu$, for $\lvert\dispmod\rvert=100$ and $\xi=10$dB. In the top-right plot, the inset shows a zoom in the small $\nu$ region. On the upper row, we can clearly distinguish two regimes, which mark the point where the optimal strategy changes.}
	 \label{optimalprecision}
\end{figure}

The optimal precision, normalized by the FTQL, is plotted in Fig.\ref{optimalprecision}, in several regimes of parameters. Unsurprisingly, the relative advantage increases with the squeezing. It also decreases with the displacement, which shows that for very large displacement, a larger amount of squeezing is required to achieve a significant metrological advantage (said differently, a non-squeezed state with large displacement will already yield a very high precision; increasing significantly this precision requires a large amount of squeezing.) More intriguing is the fact that the relative advantage increases with $\nu$; that is, highly mixed states can actually yield a better relative advantage than pure states. In this case, this seemingly paradoxical situation mostly comes from the fact that we are renormalizing the advantage by the FTQL. Both $\pazocal{I}_\phi$ and $\pazocal{I}_\phi^\text{ref}$ decrease with $\nu$, but the reduction is faster for the latter. However, in certain cases (in particular, for zero displacement), the thermal noise can also increase the precision $\pazocal{I}_\phi$ \textit{in absolute value}, and not only relative to the reference $\pazocal{I}_\phi^\text{ref}$. \footnote{Take, for instance, Eq.\eqref{QFItwomode}. Increasing $\nu$ will reduce the term which combines displacement and squeezing, but increase the pure squeezing term. If we set the displacement $\dispmod$ equal to $0$, then when $\nu$ goes from $1$ to $\infty$, the QFI increases by a factor of $2$. The advantage \eqref{Advonemode}, and the advantage renormalized by the FTQL, also increase.} This is a counter-intuitive property of metrology with squeezed states, studied in detail in \cite{safranek_gaussian_2016}: for certain states, increasing the thermal noise is actually beneficial for the estimation, both in absolute value and compared with an unsqueezed reference. As we will see in the next section, this fact will be extremely important when we try to define a resource theory for phase estimation. However, this phenomenon only occurs for relatively small displacement. By contrast, most experiments in photonic metrology consider states with very large displacement and a relatively modest amount of squeezing; in these situations, the thermal noise tends to decrease the precision (although the ratio advantage/FTQL can still increase).

 These curves also clearly show the difference between the two strategies (which we will call the one-mode and Mach-Zehnder strategies, respectively). When the state is sent in a Mach-Zehnder, a nonzero, but very modest, advantage can be obtained. This advantage increases very slowly with the amount of squeezing. By contrast, when the one-mode protocol is optimal, much higher precision can be reached. The precision is also much less dependent on the displacement, and can actually increase or decrease with $\lvert\dispmod\rvert$.

\section{Discussion}
In this section, we will discuss some possible implications of our result. We will first present the possibility to create a resource theory for phase estimation. Then, we will discuss other possible sources for metrological advantage and their link with squeezing.
\subsection{Towards a resource theory of phase estimation}
Let us introduce in more detail the formalism of resource theories. The basic elements are a set of \textit{free states}, and a set of \textit{free operations}. Free operations cannot create or increase the amount of resources. This means in particular that the set of free state must be stable under the free operation; for every free state $\rop$ and free operation $\Lambda$, $\Lambda(\rop)$ is a free state. In general, free operations include some free unitary operations, adding ancilla systems in a free state, tracing out part of the system, or classical mixing. 

An important goal of the resource theory formalism is to define a \textit{resource monotone} $\pazocal{F}$, which quantifies the amount of resource in a given state. This monotone must satisfy a few important properties. Three properties, in particular, are common to almost all resource theories \cite{chitambar_quantum_2019}:

\begin{itemize}
	\item $\pazocal{F}(\rop)=0$ if and only if $\rop$ is a free state.
	\item $\pazocal{F}(\Lambda(\rop))\leq\pazocal{F}(\rop)$ for all free operation $\Lambda$ and for all $\rop$. That is, $\pazocal{F}$ can only decrease under free operations, hence the name resource monotone.
	\item $\pazocal{F}$ is convex: $\pazocal{F}(\sum_i q_i \rop_i)\leq\sum_iq_i\pazocal{F}(\rop_i)$, for all $\rop_i$ and all $\{q_i\}$ such that $\sum_i q_i=1$. That is, classically mixing a state cannot increase the resource.
\end{itemize}

There are several ways to define a valid resource quantifier: a common choice is to use a geometric distance between a given state and the set of free states \cite{brandao_entanglement_2008,baumgratz_quantifying_2014,brandao_second_2015,lostaglio_description_2015}.\\

Some works have already used metrological quantities as resource quantifiers. In particular, in \cite{kwon_nonclassicality_2019,yadin_operational_2018}, it was shown that the ability of a state to probe small quadrature displacements could be used to quantify the nonclassicality of the state. 
The question is, can the ability to probe a small phase-shift could similarly be used as a nonclassicality measure? We will now take a step in that direction by trying to interpret our metrological advantage $\pazocal{A}_G$ in a resource-theoretic framework.

The definition of quantum metrological advantage provided in Eq.(\ref{metroadvFTQL}) is only valid for Gaussian states. To make a resource-theoretic analysis, it is desirable to extend it to mixtures of Gaussian states. It would also make it possible to take into account preparation imperfections other than the thermal noise already considered. The difficulty here is that a mixture of Gaussian states is not Gaussian in general; therefore there is no longer a well-defined notion of temperature. It is unclear what the reference state should be if the probe is, \textit{e.g.}, a mixing of two Gaussian states with different temperatures. A possible solution is to perform convex roof minimization. Let us consider a state which can be decomposed into Gaussian components like $\rop=\sum_j p_j \rop_j$, where $\rop_j$ are isotropic Gaussian states. Then we define:
\begin{equation}
	\pazocal{A}(\rop) = \text{Min}_{\{p_j,\rop_j\} }\Big( \sum_j p_j \pazocal{A}_G(\rop_j) , 0 \Big).
\end{equation}
Here we are only performing minimization over Gaussian decomposition; that is, if a state can be written as a mixture of Gaussian states in several different ways, we optimize only over these expressions. Note that $\pazocal{A}$ can be extremely challenging to compute; to the best of our knowledge, it is not even known whether the decomposition of a state into Gaussian components is unique or not. Note also that not all states can be expressed as the sum of Gaussian states. Still, $\pazocal{A}$ yields interesting insight on phase estimation advantage as a resource theory, as we will discuss now.\\

We naturally define PLE as free operations, and displaced thermal states as free states. Then $\pazocal{A}$ has the following properties:

\begin{enumerate}
	\item $\pazocal{A}(\rop)=0$ if and only if $\rop$ is a (convex combination of) displaced thermal state,
	\item $\pazocal{A}$ is non-increasing (actually invariant) under PLE $\Rop$,
	\item $\pazocal{A}$ is convex: $\pazocal{A}$$\Big(\sum_i p_i \rop_i\Big)\leq \sum_i p_i \pazocal{A}(\rop_j)$.
\end{enumerate}
Clearly, $\pazocal{A}$ possesses several properties of a resource monotone. However, this simple resource theory framework cannot capture all the properties of phase-estimation protocol. In particular, we have found two operations that are usually considered free, but which can actually increase $\pazocal{A}$.

First, as we discussed in the previous section, $\pazocal{A}$ can increase with $\nu$. Contrary to our intuitive expectations, lowering the purity of a state can actually improve its metrological properties, and therefore cannot be considered a free operation in this sense.

Second, in most resource theories, adding free ancilla cannot create or increase the resource. This is not the case here. Consider the following protocol: we add a third mode containing a coherent state, we mix the modes with a BS, then we trace out the third mode. If the added coherent state is very large, this protocol results in a displacement of the probe state. This, in turn, enhance both $\pazocal{I}_\phi^{\text{opt}}$ and $\pazocal{I}_\phi^{\text{ref}}$. In some cases, this can lead to a net increase of $\pazocal{I}_\phi^{\text{opt}}-\pazocal{I}_\phi^{\text{ref}}$ (see for instance the lower-left panel of Fig.\ref{optimalprecision}: for high squeezing value, the advantage can increase with $\lvert\dispmod\rvert$, even when renormalized by the FTQL). Interestingly, this issue does not arise for the displacement estimation protocols studied in \cite{kwon_nonclassicality_2019} and \cite{yadin_operational_2018}, for the reasons explained below.

These two examples highlight the difficulty to define a proper resource theory for phase estimation. This shows an important difference between displacement and phase estimation: displacement estimation is not directly dependent on the number of photons in the probe state. For instance, both the vacuum and a coherent state yield the same precision for estimating a quadrature shift \cite{kwon_nonclassicality_2019}. By contrast, phase-estimation protocols are generally extensive in the number of photons. As a consequence, adding "classical" ancilla or increasing the thermal noise of the system are not generally detrimental to the protocol, because these operations can increase the total number of photons in the probe state. This, in turn, can increase the precision, both in absolute value AND relative to the FTQL. We have checked that even by considering the normalized quantity $\pazocal{A}/\pazocal{I}_\phi^{\text{ref}}$, the issue remains the same. This shows that operations increasing the total number of photons must be treated with care if one wants to develop a resource theory of phase estimation.

\subsection{Link between metrological advantage and nonclassical features}
So far, we have studied the link between metrological advantage and squeezing for Gaussian states. Squeezing is a property that is best understood and studied in phase space. To conclude this study, we will briefly consider how metrological advantage could instead be linked with quantities related to individual photons.\\

First, let us discuss formal entanglement. As we mentioned in Chapter 3 (and as our work also illustrates), metrological advantage cannot be univocally linked with entanglement between different modes. First, they are some mode-entangled states which are not useful for metrology (as shown explicitly, for instance, in \cite{hyllus_not_2010}). Second, certain states, which seemingly contain no mode-entanglement, can achieve a quantum advantage: for instance, one-mode displaced squeezed states, combined with homodyne detection. Finally, in a resource-theoretic perspective, mode-entanglement can be seemingly created for free using only PLE, and therefore it cannot be considered a resource. Several authors have proposed to focus instead on \textit{formal} entanglement between individual particles \cite{demkowicz-dobrzanski_chapter_2015}. \footnote{In this reference, entanglement between individual particles is called \textit{particle entanglement}; however, this name has been used in other works to refer to slightly different notions \cite{benatti_entanglement_2014,sciara_universality_2017,braun_quantum-enhanced_2018}, therefore we prefer to use the name \textit{formal entanglement} to avoid confusion.} This allows us to better understand the third issue discussed above: PLE do not create entanglement, they simply transform pre-existing formal entanglement into mode entanglement. However, considering formal entanglement does not solve the other two problems. For instance, some of the states studied in \cite{hyllus_not_2010} \textit{are} formally entangled, and still fail to achieve a metrological advantage.

However, we argue that in the specific context of Gaussian states, a stronger link might exist between formal entanglement and metrological advantage. We formulate the following conjecture:

\begin{conjecture}
If $\rop$ is a two-mode displaced thermal state, then $\rop$ is not formally entangled.
\end{conjecture}

In Chapter 3, we have already shown that Conjecture 1 is true for pure states (\textit{i.e.}, for coherent state). To the best of our knowledge, however, no one has extended this to mixed states. In Appendix B, we show that in the two-photon subspace, bimodal displaced thermal states are indeed separable. Proving the conjecture, however, would require to extend this result to all fixed-number subspaces.

If this conjecture is true, then using Theorem 1, we can show that for two-mode Gaussian states, formal entanglement implies squeezing, and therefore a metrological advantage. Note also that the condition of Gaussianity excludes the states considered in \cite{hyllus_not_2010}; thus this conjecture is consistent with previous results. 

Even if Conjecture 1 is true, however, we still need to account for the performances of single-mode displaced squeezed state, which of course can be neither mode- nor formally entangled. A possible origin for the metrological usefulness of these states is the superposition between different particle \textit{numbers}. (In other words, we have to consider not only entanglement within the subspaces corresponding to fixed $\moy{\Nop}$, but also coherences between those subspaces).

\begin{figure}
\begin{center}
\includegraphics[scale=0.35]{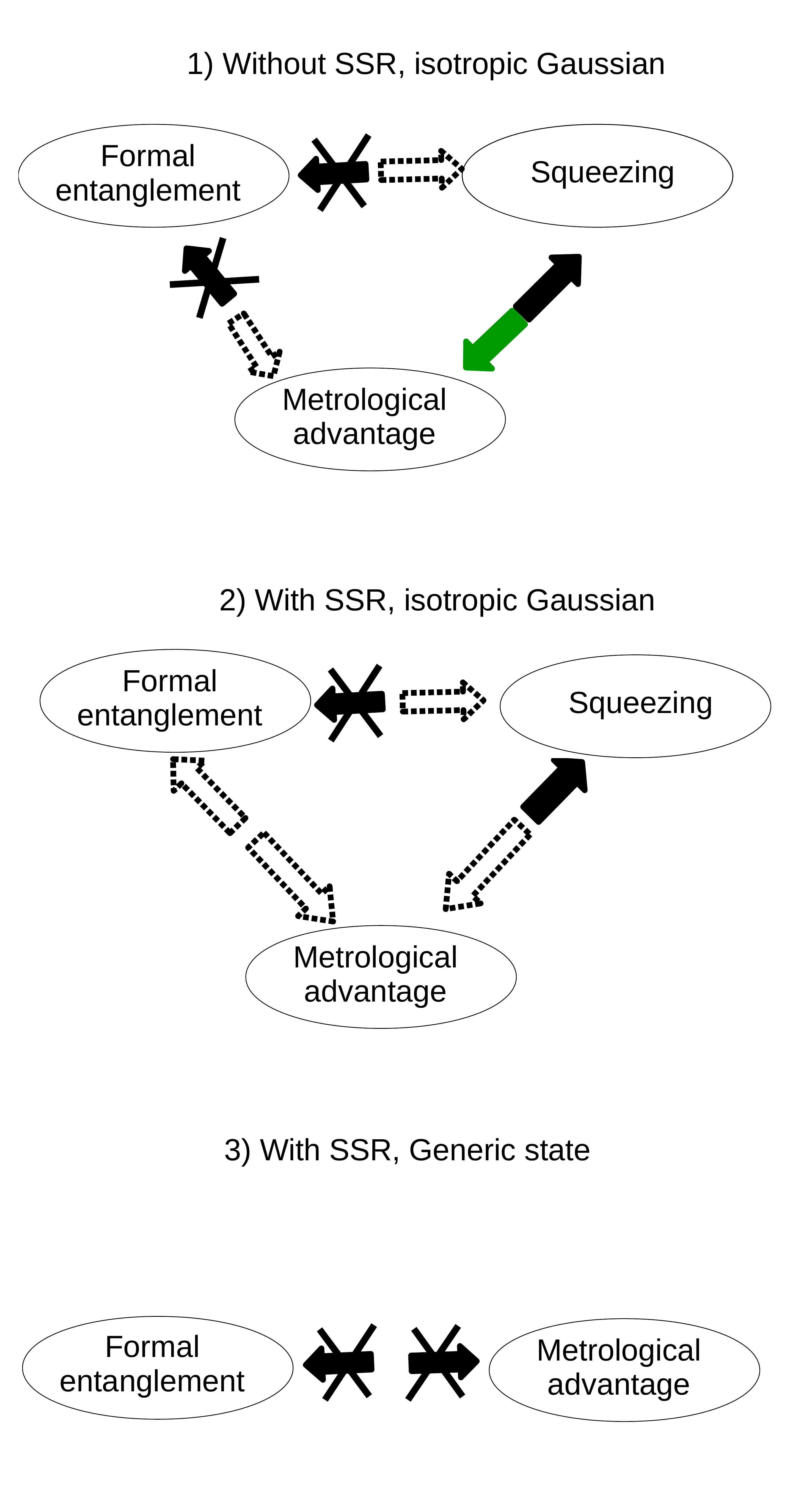}
\end{center}
\caption[Summary of the links between squeezing, formal entanglement, and metrological advantage.]{Summary of the links between squeezing, formal entanglement, and metrological advantage, with and without SSR, for Gaussian or arbitrary states. Full arrows are established results, dotted arrows are conjectures. Our results correspond to the green arrow in the first cartoon. The third cartoon corresponds to the results of \cite{hyllus_not_2010}. For non-Gaussian states without SSR, no general results are known to the best of our knowledge.}
\label{summary}	
\end{figure}

A simple way to study the role of number superposition is to see what happens when it is suppressed. For massive particles such as atoms, number superposition is naturally suppressed by superselection rules (SSR). For photons, an effective SSR can be achieved by applying restrictions at the measurement level. So far we have authorized the measure of all observables, including homodyne measurement. Restricting ourselves to observables that commute with the total number of photons (for instance, if we have access to photon-counter but no perfect phase reference) is equivalent to suppressing the number superposition in the input state (this problem can be expressed in the language of \textit{quantum reference frame} \cite{bartlett_reference_2007}). Metrological protocols with such limitations have already been studied in specific cases, in particular for pure input state \cite{jarzyna_quantum_2012,safranek_quantum_2015}. It would be quite interesting to see whether Theorem 1, or a modified version thereof, still holds in that case. In \ref{summary}, we have summarized the known results, and the conjectures we just formulated. We believe a further understanding of the link between formal entanglement, number superposition, squeezing, and metrological advantage will be important to define phase estimation-based nonclassicality quantifiers that can be formulated in a resource-theory framework.
 
\section{Conclusion}
\label{secconclusion}

To better understand the link between nonclassicality and metrological advantage, we have introduced a definition of metrological advantage which takes thermal noise into account. For bimodal non-pure isotropic Gaussian states, we have demonstrated that squeezing is a necessary and sufficient condition to achieve a metrological advantage in this sense. We have discussed the properties of our measure, and the challenges that remain to be overcome to achieve full-fledged resource theory of phase estimation. In particular, we have argued that "classical" operations which can increase the number of photons must be treated carefully to obtain a consistent resource theory, and to using phase estimation as a nonclassicality quantifier. Finally, we have discussed a possible link between our criterion and the notion of formal entanglement. \\

This work focused on the role of squeezing in phase estimation. To go further, it would be interesting to study in more detail the role of other quantities such as formal entanglement and number superposition. We have briefly reviewed some challenges and open questions concerning these issues. We may also generalize our study by considering multi-modes anisotropic states, or states that cannot be decomposed as a sum of Gaussian states. The key question then would be how to properly define a reference state, since in this case, we cannot match a displaced thermal state with each state. Finally, we may also try to include noise at the evolution level, in addition to the preparation level.


\chapter*{General Conclusions and Perspectives}

\addcontentsline{toc}{chapter}{General Conclusions and Perspectives}

In this Thesis, I have studied quantum optics systems from the perspectives of quantum metrology and quantum phase transitions. 
I have been particularly interested in the ultrastrong coupling, a regime in which several systems are so strongly coupled that they lose their individual properties and behave as a collective, hybridized system. An integral part of my research work was to understand and come to appreciate the exotic properties and conceptual problems that arise in this regime, in particular the notion of virtual excitations. An important question was the connection between ultrastrong coupling and the phenomenon of superradiance. The study of the literature has led me to defend the position that superradiant transitions are a generic property of ultrastrongly coupled systems. This is one of the main points which I have discussed in this Thesis, especially in Chapter I and II. 

Then, I have studied these effects from different perspectives. First, I have considered the effect of unusual, two-photon coupling, as discussed in Chapter IV. I have focused more specifically on the two-photon Dicke model, and shown that it could exhibit both a spectral collapse instability and a second-order, superradiant-like phase transition. I have striven to describe both transition and collapse in intuitive terms, in addition to determining formal properties such as the critical exponents. I have also emphasized that these two phenomena occur for different scalings of the coupling constant with the number of qubits, which means that these effects could be simultaneously observed only for a very particular regime of parameters. Then I have considered the effect of dissipation, which turned out to completely change the picture. First, the order of the transition changes from first-order to second-order. Second, the spectral collapse of the Hamiltonian, which is activated by the coupling strength, is replaced by an instability controlled by dissipation. An important consequence is the disappearance of the scaling behavior described above, which could make the transition easier to observe in practice.

Second, I have studied how quantum optics systems could be used to investigate connections between quantum metrology and quantum phase transitions. Indeed, in these platforms, it is possible to engineer exotic critical effects in finite-size systems. Precisely because of their simplicity, quantum correlations can be described and observed much more easily in these platforms than in many-body systems. Therefore, they provide ideal toy models to investigate quantum properties near critical points. In particular, these correlations are very sensitive to small perturbation, a fact which, beyond the fundamental interest, could be exploited to design new small-scale sensors.

 After an introduction to quantum metrology in Chapter III, I have focused on the Rabi model, which describes a single qubit coupled to a bosonic field. In Chapter V, I have studied how its critical behavior could be used to measure accurately either the bosonic or spin frequency. I have put particular emphasis on the duration of the protocol; indeed, the closure of the gap at the transition means that adiabatic control of the system will take longer as one approaches the critical point. I have analyzed the trade-off between improved sensing accuracy and increased protocol duration. For spin frequency estimation, I have shown that it was possible to achieve an inverse error scaling quadratically with time, instead of the usual linear scaling. This scaling emerges from the time-dependent behavior of the protocol. However, this improved scaling comes with a small prefactor, which eventually compromises the use of this protocol for practical purposes. This begs the question of whether such limitation is specific to this model, or is a fundamental phenomenon, as recent studies performed for spin systems tend to indicate. This last point represents a stimulating research direction to explore.

In the last Chapter of this Thesis, I have presented some work on a more fundamental issue. It is well-established that using quantum correlations can be beneficial for sensing protocols. However, the precise connection between nonclassicality and metrological advantage is still not completely understood. I have addressed this question from the perspective of resource theories. This formalism allows us to describe a system by its ability to perform certain tasks. Here, I have studied how the ability of a state to sense a small phase-shift could be used as a measure of nonclassicality. In the context of interferometry with Gaussian states, I have introduced a notion of metrological advantage taking into account the preparation imperfection of the probe state. I have discussed how the capacity to achieve this advantage could be used as a criterion to detect various nonclassical properties such as squeezing.\\

In summary, I have presented the rich phenomenology which arises in the ultrastrong coupling regime of light and matter, and showed that these systems are an excellent platform to study concepts coming from both quantum information and metrology, and quantum phase transitions. Some of the results concerning the metrological power of the Rabi model could find applications and contribute to the development of new critical sensors. I have also made several comments concerning the possible relationship between metrological advantage, squeezing, and the notion of formal entanglement, which hopefully will contribute to better understand the subtle connection between non-classicality and metrological advantage. 


The work presented in this dissertation can be the starting point of research conducted in several directions. Maybe the most promising is to understand better the connection between the critical and metrological properties of quantum optics models. In particular, it would be particularly interesting to derive general scaling properties for the Quantum Fisher Information, based on critical exponents. This would require extending previous studies, performed in the context of spin systems, to finite-size quantum optics transitions. Another direction concerns the preparation procedure; although our study focused on adiabatic preparation, the duration of the protocol could be reduced by considering shortcuts to adiabaticity. An important question, however, is that these protocols require previous knowledge of the Hamiltonian parameters one wants to estimate. Therefore, carefully accounting for the errors induced by imperfect previous knowledge is necessary to properly exploit these protocols. Finally, experimental implementations should be considered. The phase transition of the Rabi model has not been achieved yet. Cold atoms and trapped ions, in which the large frequency ratio can be naturally achieved, are promising platforms to observe this transition. Therefore, a first step would be to design a protocol to realize the transition in these systems.


Another possible research direction would be to extend the work on the two-photon Dicke model. In particular, it would be interesting to understand the full quantum behavior of the model in the presence of dissipation. The main challenge then is to properly deal with the local spin dissipation. This could be done numerically by exploiting permutational invariance, or by using quantum trajectories simulations. Also at the Hamiltonian level, more work is required to properly understand the behavior of the system near the spectral collapse.

Last, the connection between metrological advantage, non-classical correlations, and formal entanglement could be studied further. One interesting question to ask would be whether two-mode squeezed Gaussian states are necessarily formally entangled. The role of particle number superposition could also be investigated, notably by using the concept of superselection rules and quantum reference frames. Finally, other definitions of quantum metrological advantage could also be considered, in order to properly define a resource theory for quantum metrology.













.

 \begin{spacing}{0.9}

 \bibliographystyle{sapthesis}
 \cleardoublepage
 \bibliography{References/USC,References/Dicke,References/metrology,References/twophoton,References/metrocritic,References/Nonclassicality,References/Technicalmethods}

 \end{spacing}


 \begin{appendices}

\chapter{Technical methods}

\renewcommand{\theequation}{\thechapter.\arabic{equation}}

	In this Appendix, we give further technical details about the methods we have applied in this Thesis. First, we discuss the generalized rotating wave approximation, which allows us to describe the Rabi model in the deep-strong coupling regime. We present the method in detail and describe the mechanisms that make it a performant method in a wide regime of parameters. Second, we present a general formalism for Schrieffer-Wolff transformation that can be used to treat the Rabi, Dicke, two-photon Dicke, and other models in a unified way.  Finally, we detail the calculations of the eigenstates and eigenenergies for the two-photon Dicke model in its squeezed phase.

\section{Perturbative approach in the DSC regime: the gRWA}

In the DSC regime, one can apply the so-called generalized rotating wave approximation or gRWA. This is a perturbative approach which consists in keeping the field and interaction terms in the Rabi model as dominant and treat the bare qubit energy term as a perturbation. As we will now show, this approach is indeed justified in the limit of large coupling.
The eigenstates and eigenvalues of $\Hop_{DSC}=\Of\adag\aop+\gind(\adag+\aop)\sigx$ consist in a ladder of degenerate doublets: 
\begin{align}
	\ket{\phi_{N,\pm}} & =\frac{\ket{\leftarrow}\ket{N,\alpha_\gind} \pm \ket{\rightarrow}\ket{N,-\alpha_\gind}}{\sqrt{2}}, \\ \nonumber
	E_{N,+} & =E_{N,-}=N\Of-\alpha_\gind^2\Of,
\end{align}

where we recall $\alpha_\gind=\frac{\gind}{\omega}$. These properties will be useful in the following \cite{irish_dynamics_2005}:  
\begin{align*}
	\langle M,\alpha_\gind| N,-\alpha_\gind\rangle & =e^{-2\alpha_\gind^2}(2\alpha_\gind)^{M-N}\sqrt{\frac{N!}{M!}}L_N^{M-N}[4(\alpha_\gind)^2] \hspace{10pt} \text{if} \hspace{10pt} M\geq N,\\
	\langle M,-\alpha_\gind| N,\alpha_\gind\rangle & =(-1)^{N-M}	\langle N,-\alpha_\gind| M,\alpha_\gind\rangle,\\
	\langle M,\alpha_\gind| N,-\alpha_\gind\rangle & =\langle N,-\alpha_\gind| M,\alpha_\gind\rangle,
\end{align*}
where $L_i^j$ are associated Laguerre polynomials.
In the eigenbasis of $\Hop_{DSC}$, the perturbation $\sigz$ has the following action:

\begin{align}
	\sigz\ket{\phi_{N,+}}= \sum_{\lvert M-N\rvert \text{even}} \langle M,\alpha_\gind| N,-\alpha_\gind\rangle \ket{\phi_{M,+}} + \sum_{\lvert M-N\rvert \text{odd}} \langle M,\alpha_\gind| N,-\alpha_\gind\rangle \ket{\phi_{M,-}},\\ \nonumber
	\sigz\ket{\phi_{N,-}}= - \sum_{\lvert M-N\rvert \text{even}} \langle M,\alpha_\gind| N,-\alpha_\gind\rangle \ket{\phi_{M,-}} - \sum_{\lvert M-N\rvert \text{odd}} \langle M,\alpha_\gind| N,-\alpha_\gind\rangle \ket{\phi_{M,+}}.
\end{align}

Hence, the qubit energy term $\frac{\Oq}{2}\sigz$ can be decomposed as follows: 
\begin{align*}
\frac{\Oq}{2}\sigz & =V_D+V_O,\\
\frac{2}{\Oq}V_D & =\sum_N\langle N,\alpha_\gind| N,-\alpha_\gind\rangle \Big(\ket{\phi_{N,+}}\bra{\phi_{N,+}}-\ket{\phi_{N,-}}\bra{\phi_{N,-}}\Big),\\
\frac{2}{\Oq}V_O & =\sum_{\lvert M-N\rvert \text{even},\neq0} \langle M,\alpha_\gind| N,-\alpha_\gind\rangle \Big(\ket{\phi_{M,+}}\bra{\phi_{N,+}}-\ket{\phi_{M,-}}\bra{\phi_{N,-}}\Big)\\
& + \sum_{\lvert M-N\rvert \text{odd}} \langle M,\alpha_\gind| N,-\alpha_\gind\rangle \Big(\ket{\phi_{M,-}}\bra{\phi_{N,+}}-\ket{\phi_{M,+}}\bra{\phi_{N,-}}\Big).
\end{align*}

$V_D$ acts within each doublet, while $V_O$ connects different doublets. Both terms have a typical energy scale $\Oq$. This structure can also be connected with the usual RWA. For $g=0$ and $\Oq=\Of$, the eigenstates of the Hamiltonian also form degenerate doublets $\frac{\ket{\lspin N+1}\pm\ket{\hspin N}}{\sqrt{2}}$. The corotating terms act within each doublet, and lift the degeneracy between the symmetric and antisymmetric states. The counter-rotating terms connect different doublets and are neglected. The perturbative approach consists in keeping $V_D$, and neglecting $V_O$ (or, more generally, treat it as a perturbation). This leaves the eigenstates unchanged, but lift the degeneracy within each doublet. The new eigenvalues read:
\begin{equation}
	E_{N,\pm}=(N-\alpha_\gind^2)\Of\pm\frac{\Oq}{2}e^{-2\alpha_\gind^2}L_N(4\alpha_\gind^2).
\end{equation}

This approximation becomes increasingly accurate when $\gind$ increases. Why this should be the case is not immediately apparent, and require to study the matrix elements of $V_D$ in some details. We will focus here on the ground state $\ket{\phi_{0,-}}$. $V_D$ connects it with other eigenstates through matrix elements of the form:

\begin{equation}
	\bra{\phi_{0,-}}V_D\ket{\phi_{M,\pm}}\propto\frac{\Oq}{2}\langle 0,\alpha_\gind|M,-\alpha_\gind\rangle\propto\Oq e^{-2\alpha_\gind^2}\frac{(2\alpha_\gind)^M}{\sqrt{M!}}.
\end{equation} 

To evaluate the validity of perturbation theory, we need to  compare this matrix element with the difference in energy between the eigenstates:
\begin{equation}
	\frac{\bra{\phi_{0,-}}V_D\ket{\phi_{M,\pm}}}{E_{M,\pm}-E_{0,-}}\propto\frac{\Oq}{M\Of+\Oq e^{-2\alpha_\gind^2}(L_0(4\alpha_\gind^2)\pm L_M(4\alpha_\gind^2))} e^{-2\alpha_\gind^2}\frac{(2\alpha_\gind)^M}{\sqrt{M!}}.
\end{equation}
A perturbative treatment will be valid as long as these terms are small. This can be achieved in two ways. The first possibility is to set $\Oq\ll\Of$; then, intuitively, the spin term acts as a perturbation simply because the spin frequency is small. The second regime is the limit $\gind\rightarrow\infty$, which means $\alpha_\gind\rightarrow\infty$. The term $e^{-2\alpha_\gind^2}\frac{(2\alpha_\gind)^M}{\sqrt{M!}}$ is always smaller than $1$, and will vanish for small $M$. It will be maximal for $M_o\sim\alpha_g\gg1$. The associated matrix-element-to-energy ratio is $\sim\frac{\Oq}{M_o\Of}\sim\frac{\Oq}{\alpha_g\Of}$, which is small \textit{even} if $\Oq=\Of$. In more physical terms, in the DSC regime, the coupling between neighboring doublets is suppressed. Instead, the spin term connects doublets that are well separated in energies.\\

To summarize, the bare qubit energy term can be treated as a perturbation in the limit of small qubit frequency or large coupling. In the limit $\Oq\ll\Of$, the qubit energy term may connect neighboring doublets, but they are separated by an energy which is large compared with the qubit energy scale. In the limit $\gind\rightarrow\infty$, the qubit energy term connects only doublets far apart in the spectrum, and hence act perturbatively even if its energy scale is comparable with the energy separation between \textit{neighboring} doublets.

 \section{General formalism for SW transformation}

 In this section, we will adopt a more general perspective on the SW transformation, and apply it to several Dicke-like models. Let us consider a Hamiltonian which can be decomposed in a dominant term and perturbations at various order:

 \begin{equation}
 	\Hop=\Aop +\epsilon \Bop +\epsilon^2 \Cop + ...
 	\label{baseSW}
 \end{equation}
Where $\Aop$, $\Bop$, and $\Cop$ are arbitrary operators, and $\epsilon\ll1$. Let us define the eigenbasis $\ket{n}$ of $\Aop$, such that $\Aop\ket{n}=E_n\ket{n}$. 
We will assume here that the operator is non-degenerate. We can order the eigenstates such that $E_{n+1}\geq E_n$. Then any operator can be decomposed in terms of diagonal and off-diagonal terms in this eigenbasis. In general, we will have $\hat{O}=\sum_{i=0}^d\hat{O}^{(i)}$, with $\hat{O}^{(0)}=\sum_{n=0}^d c_n^{(0)}\ket{n}\bra{n}$, $\hat{O}^{(1)}=\sum_{n=0}^dc_n^{(1)}\ket{n}\bra{n+1}+h.c.$, $\hat{O}^{(2)}=\sum_{n=0}^d c_n^{(2)}\ket{n}\bra{n+2}+h.c.$ and so on; here $d$ is the dimension of the Hilbert space. By definition, we have $\Aop=\Aop^{(0)}$. In general, we will call $\mathbb{M}_A^i$ the ensemble of all operators which can be put in the form $\sum_{n=0}^d c_n^{(i)}\ket{n}\bra{n+i}+h.c.$. For every operators $\hat{O}$ and $\hat{P}$, and for all integer $i,j$, we can define an operator $\hat{Q}$ such that $$[\hat{O}^{(i)},\hat{P}^{(j)}]=\hat{Q}^{\lvert (i-j)\rvert}+\hat{Q}^{(i+j)},$$

which we can write down symbolically as 
  
\begin{equation}
	[\mathbb{M}_A^i,\mathbb{M}_A^j]\subset\mathbb{M}_A^{\lvert i-j\rvert}\oplus\mathbb{M}_A^{i+j}.
\end{equation}

In particular, we have $[\Aop,\mathbb{M}_A^i]\subset\mathbb{M}_A^i$, since $\Aop$ is diagonal.\\

We want to diagonalize the Hamiltonian in the $\ket{n}$ eigenbasis, at first order in $\epsilon$. We apply a transformation $e^{\epsilon\Sop_1}$ with $\Sop_1=\sum_i\Sop_1^{(i)}$. Under this transformation, the Hamiltonian now reads $\Hop=\Aop+\epsilon([\Sop_1,\Aop]+\Bop)$ plus higher-order terms. The goal is to ensure that all off-diagonal terms in this expression are canceled. For this, we need to choose an operator $\Sop_1$ which satisfies the following property:

\begin{equation}
	[\Sop_1^{(i)},\Aop]=-\Bop^{(i)}
\end{equation}
for all $i>1$; or, equivalently, 
$$[\Sop_1,\Aop]=\Bop^{(0)}-\Bop.$$
 By systematically choosing each term $\Sop_1^{i}$ one after the other, we can bring the Hamiltonian to the form $\Sop_1=\Aop+\epsilon \Bop^{(0)}$ (plus higher-order terms), which is indeed diagonal at first order in $\epsilon$. In general, such a procedure can require to compute a large (or even infinite) number of terms. However, it can be considerably simplified when the perturbation term only involves a low degree of off-diagonality, and when the operators involved obey simple algebraic commutation relations. We will give several examples later on.\\

 Let us now consider higher-order terms in the development in terms of $\epsilon$. In addition to the operators already present at the beginning in the Hamiltonian \eqref{baseSW}, the transformation $\Sop_1$ will bring additional corrections. At order $\epsilon^2$, we have a term 
 $$\tilde{C}=\Cop+\frac{1}{2}[\Sop_1,[\Sop_1,\Aop]]+[\Sop_1,\Bop]=\Cop+\frac{1}{2}[\Sop_1,\Bop+\Bop^{(0)}].$$
To eliminate off-diagonal terms at order $\epsilon^2$, we need to apply the operator $e^{\epsilon \Sop_1+\epsilon^2\Sop_2}$, with $$[\Sop_2^{(i)},\Aop]=-\tilde{C}^{(i)}.$$ 
Hence, by including more and more terms in the SW transformation, we can eliminate the off-diagonal terms at  a higher and higher order.\\

To summarize, the SW transformation consists of defining a transformation $e^{\sum_{n=1}^{n_0}\sum_i^d\epsilon^n \Sop_n^{(i)}}$ to eliminate all the off-diagonal terms of the Hamiltonian at a given order $\epsilon^{n_0}$. This operator will also create higher-order corrections. Once these have been computed and added to the initial Hamiltonian, we can add another series of terms $\sum_i^d\epsilon^{n_0+1}\Sop_{n_0}^{(i)}$ to eliminate the off-diagonal terms also at order $\epsilon^{n_0+1}$. \footnote{Instead of applying the transformation as the exponential of a single unitary operator, we can also apply a series of transformation $e^{\epsilon^N\tilde{S}_N}...e^{\epsilon^3\tilde{S}_3}e^{\epsilon^2\tilde{S}_2}e^{\epsilon\tilde{S}_1}$. Note that in general $\tilde{S}\neq S$.}\\

Let us now show how this general method applies to several models. Perhaps the simplest example is a spin model of the form:

\begin{equation}
	\Hop=\Jz+\epsilon\Jx,
\end{equation}
with the spin operators $\Jx$ and $\Jz$ obeying the usual SU(2) commutation relations. The perturbation $\Jx=\frac{\Jp+\Jm}{2}$ belongs to $\mathbb{M}_{\Jz}^{(1)}$. To eliminate the off-diagonal term at first order in $\epsilon$, we need to apply an operator $e^{\epsilon \Sop_1}$, with $[\Sop_1,\Jz]=\Jx^{(0)}-\Jx=-\Jx$. Since $\Jx\in\mathbb{M}_{\Jz}^{(1)}$, it is sufficient to choose $\Sop_1\in\mathbb{M}_{\Jz}^{(1)}$. This suggests using a linear combination of $\Jp$ and $\Jm$. Indeed, the combination $\Sop_1=\Jp-\Jm=i\Jy$ achieves the desired result. The term of order $1$ is now diagonalized (actually suppressed). At order $2$, the transformation creates a correction $\epsilon^2\frac{1}{2}[i\Jy,[i\Jy,\Jz]]=-\frac{1}{2}\Jz$, which is already diagonal: no further term is needed. At order $3$, we have a term $\epsilon^3\frac{1}{3!}[i\Jy,[i\Jy,[i\Jy,\Jz]]]=\frac{\epsilon^3}{6}\Jx$, which is in $\mathbb{M}_{\Jz}^{(1)}$, and can be eliminated by adding a further term $\epsilon^3\Sop_3$ in the transformation, with $\Sop_3=\frac{-i}{6}\Jy$. 
Of course, this is only a very simple example; the SW transformation in this case converges towards a simple rotation around a single spin axis. 

Let us now consider a slightly more complicated Hamiltonian, $\Hop=\Jz+\epsilon\Jx^2$ (this is simply the Lipkin-Meshkov-Glick Hamiltonian with a small spin squeezing term). This time the perturbation has a non-zero diagonal part $(\Jx^2)^{(0)}=(\Jp\Jm+\Jm\Jp)$, and an off-diagonal part $(\Jx^2)^{(2)}=(\Jp^2+\Jm^2)\in\mathbb{M}_{\Jz}^{(2)}$. To eliminate it, we need an operator $\Sop_1$ which belongs to $\mathbb{M}_{\Jz}^{(2)}$. Given the SU(2) algebraic structure, it is easy to find that $\Sop_2=\frac{1}{2}(\Jp^2-\Jm^2)$ is a suitable choice. We obtain a diagonal Hamiltonian $\Jz+\epsilon(\Jp\Jm+\Jm\Jp)$ plus higher-order terms, which can be treated as earlier.\\

So far, we have considered only one degree of freedom. In that case, the SW transformation is simply an alternative formulation of standard perturbation theory. SW transformation, however, is most useful when we have several degrees of freedom, with vastly different timescales and weak coupling. Then SW allows to adiabatically eliminate one subsystem while keeping all quantum correlations. This is the principle that has been applied in our study of the various versions of the Dicke model. All of these models can be put in the following general form:

\begin{equation}
	\Hop=\Pop_z+ \epsilon\lambda \Pop_x\Qop_x +\epsilon^2 \Qop_z=\Aop+\epsilon \Bop +\epsilon^2 \Cop.
\end{equation}
We have $\hat{O}_x=(\hat{O}_++\hat{O}_-)/2$ and $\hat{O}_y=i(\hat{O}_--\hat{O}_+)/2$, with $\hat{O}_+=\hat{O}_-^\dagger$  (here $\hat{O}=\Pop$ or $\Qop$). The operators $\hat{O}_+$ and $\hat{O}_-$ are off-diagonal in the eigenbasis of $\hat{O}_z$; more specifically, we have $\hat{O}_\pm\in \mathbb{M}_{\hat{O}_z}^{(1)}$.

 Various versions of the Dicke model are described by various operators $\Pop$ and $\Qop$, which obey specific commutations relations. The following relations are particularly relevant:

\begin{itemize}
	\item SU(2) commutation relations: $[\hat{O}_z,\hat{O}_\pm]=\pm \hat{O}_\pm$ and $[\hat{O}_+,\hat{O}_-]= 2\hat{O}_z$,
	\item SU(1,1) relation: $[\hat{O}_z,O_\pm]=\pm \hat{O}_\pm$ and $[\hat{O}_+,\hat{O}_-]=- 2\hat{O}_z$,
	\item Bosonic commutation relations: $[\hat{O}_z,\hat{O}_\pm]=\pm \hat{O}_\pm$ and $[\hat{O}_+,\hat{O}_-]=-1$.
\end{itemize}

The various models we have studied can then be categorized according to the commutation relations of $\Pop$ and $\Qop$. We can define the following classes of models:

\begin{itemize}
	\item \textit{Rabi-like models:} $\Pop$ and $\Qop$ satisfy $SU(2)$ and bosonic commutation relations, respectively. This is indeed the case in the Rabi model, where $\Pop_i=\hat{\sigma}_i$, $\Qop_z=\adag\aop$, and $\Qop_+=\adag$. However, this class also includes the multi-qubit version of the Rabi model, where $\Pop_i=\hat{J}_i$ describes the collective spin. Let us see how the procedure described above can be applied in this case. To diagonalize the Hamiltonian at first order in $\epsilon$, we apply the operator $e^{\epsilon\Sop_1}$, with $[\Sop_1,\Pop_z]=-(\Bop+\Bop^{(0)})=-\Bop=-\lambda \Pop_x\Qop_x$. In the eigenbasis of $\Pop_z$, $\Bop$ is an off-diagonal operator; more precisely, we have $\Bop\in\mathbb{M}_{\Pop_z}^{(1)}$. Therefore, $\Sop_1$ must also belong to $\mathbb{M}_{\Pop_z}^{(1)}$. Given the $SU(2)$ commutations relations obeyed by $\Pop$, it is straightforward to show that $\Sop_1=i\lambda \Pop_y\Qop_x$ achieves the desired result.

	 At order $2$, we have now a term $\tilde{C}=\Cop+\frac{1}{2}[\Sop_1,\Bop]=\Qop_z+\frac{\lambda^2}{2}\Qop_x^2\Pop_z$. This term is diagonal in $\Pop_z$, which means that the fast degree of freedom $\Pop$ can be readily eliminated, leaving an effective Hamiltonian for $\Qop$ only.

	At order $3$, we have now a correction $\tilde{D}=\frac{1}{3!}[\Sop_1,[\Sop_1,[\Sop_1,\Aop]]]+\frac{1}{2}[\Sop_1,[\Sop_1,\Bop]]+[\Sop_1,\Cop]=\frac{1}{3}[\Sop_1,[\Sop_1,\Bop]]+[\Sop_1,\Cop]=-\frac{\lambda^3}{3}\Qop_x^3\Pop_y+\lambda \Pop_y\Qop_y\in \mathbb{M}_{\Pop_z}^{(1)}$. To remove this correction, we need an operator $\epsilon^3\Sop_3$ such that $[\Sop_3,\Aop]=-\tilde{D}$; and once more, the operator $\Sop_3$ must belong to $\mathbb{M}_{\Pop_z}^{(1)}$. Once again, the commutation relations of $\Pop$ allow to easily find the following expression: $\Sop_3=-i\frac{\lambda^3}{3}\Qop_x^3\Pop_y+i\lambda \Pop_x\Qop_y$. The transformation $e^{\epsilon\Sop_1+\epsilon^3\Sop_3}$ allows to completely suppress the term of order $3$.

	Finally, at order $4$, we obtain the following correction: $$\tilde{E}=\frac{1}{4!}[\Sop_1,\Aop]_4+\frac{1}{2}([\Sop_1,[\Sop_3,\Aop]]+[\Sop_3,[\Sop_1,\Aop]])+\frac{1}{3!}[\Sop_1,\Bop]_3+[\Sop_3,\Bop]+\frac{1}{2}[\Sop_1,\Cop]_2=\frac{1}{4!}[\Sop_1,\Aop]_4+\frac{1}{2}[\Sop_3,\Bop],$$
	where nested commutators $[]_i$ are defined by the recurrence relation $[\Sop_1,\Xop]_{i+1}=[\Sop_1,[\Sop_1,\Xop]_i]$. Developing the last expression yields: $\tilde{E}=-\frac{\lambda^4}{8}\Qop_x^4\Pop_z+\frac{\lambda^2}{2}\Pop_x^2$, which has both a diagonal and off-diagonal part: $\tilde{E}^{(0)}=-\frac{\lambda^4}{8}\Qop_x^4\Pop_z+\frac{\lambda^2}{8}(\Pop_+\Pop_-+\Pop_-\Pop_+)$, $\tilde{E}^{(2)}=\frac{\lambda^2}{8}(\Pop_+^2+\Pop_-^2)$. This last term must be suppressed by an operator $\Sop_4\in\mathbb{M}_{\Pop_z}^{(2)}$, which satisfies $[\Sop_4,\Aop]=-\tilde{E}^{(2)}$. The appropriate combination is $\Sop_4=\frac{\lambda^2}{16}(\Pop_+^2-\Pop_-^2)$. In summary, by applying the operator $\Uop=e^{\epsilon\Sop_1+\epsilon^3\Sop_3+\epsilon^4\Sop_4}$, we obtain the following Hamiltonian:
	\begin{equation}
		\Uop\Hop\Uop^\dagger=\Pop_z+\epsilon^2\left(\Qop_z+\frac{\lambda^2}{2}\Qop_x^2\Pop_z\right)+\epsilon^4\left(\frac{\lambda^2}{8}(\Pop_+\Pop_-+\Pop_-\Pop_+)-\frac{\lambda^4}{8}\Qop_x^4\Pop_z\right).
		\label{order4Rabilike}
	\end{equation}
	Since we have $\Pop_+\in \mathbb{M}_{\Pop_z}^{(1)}$, $\Pop_-=\Pop_+^\dagger$, and we assume no degeneracies, we have $\Pop_+\Pop_-\in \mathbb{M}_{\Pop_z}^{(0)}$. That is, this Hamiltonian commutes with $\Pop_z$, which allows us to eliminate the fast degree of freedom and obtain an effective Hamiltonian involving only $\Qop$.

	In the case of the Rabi model, we have followed exactly this procedure to obtain the quartic boson Hamiltonian. The only difference is that for a single spin, $\Jp^2=\Jm^2=0$, which means we did not need to apply the last operator $\Sop_4$.
\item \textit{Two-photon Dicke model:} P and Q obey $SU(1,1)$ and bosonic commutation relations respectively. This is the case in the two-photon Dicke model with many qubits, where $\Pop_z=\Ko$ and $\Qop_x=\frac{\Jx}{\sqrt{N}}\sim(\frac{\bop+\bdag}{2}-\frac{1}{4N}({\bop}^{\dagger2}\bop+\bdag\bop^2))$. The procedure is the same as for the Rabi-like model. We find the following operators: $\Sop_1=i\lambda \Qop_x\Pop_y$, $\Sop_2=0$, $\Sop_3=-i\lambda \Qop_y\Pop_x+i\frac{\lambda^3}{3}\Pop_y\Qop_x^3$ and $\Sop_4=\frac{\lambda^2}{32}(\Pop_-^2-\Pop_+^2)=i\frac{\lambda^2}{16}(\Pop_x\Pop_y+\Pop_y\Pop_x)$. The transformed Hamiltonian reads: 
\begin{equation}
	\Uop\Hop\Uop^\dagger=\Pop_z+\epsilon^2\left(\Qop_z-\frac{\lambda^2}{2}\Qop_x^2\Pop_z\right)+\epsilon^4\left(\frac{\lambda^2}{8}(\Pop_+\Pop_-+\Pop_-\Pop_+)-\frac{\lambda^4}{8}\Pop_z\Qop_x^4\right).
	\label{order4twophotlike}
\end{equation}
By replacing $\epsilon=\frac{\ratbis}{N}$, $\Qop_z=-\frac{N}{2}+\bdag\bop$ and $\Qop_x=(\frac{\bop+\bdag}{2}-\frac{1}{4N}({\bop}^{\dagger2}\bop+\bdag\bop^2))$, we obtain the expression given in the main text.\\

\item \textit{Two-photon Rabi model:} $\Pop$ and $\Qop$ obey respectively $SU(1,1)$ and $SU(2)$ commutations relations; this is the case with the two-photon Rabi model, provided we have still $\Omega\ll\omega$. The transformation operators are $\Sop_1=i\lambda \Qop_x\Pop_y$, $\Sop_2=0$, $\Sop_3=-i\lambda \Qop_y\Pop_x+i\frac{\lambda^3}{3}\Pop_y\Qop_x^3$ and $\Sop_4=-\frac{\lambda^2}{16}(\Pop_-^2-\Pop_+^2)\Qop_z=-i\frac{\lambda^2}{8}(\Pop_x\Pop_y+\Pop_y\Pop_x)\Qop_z$, and the transformed Hamiltonian is:
\begin{equation}
 \Uop\Hop\Uop^\dagger=\Pop_z+\epsilon^2\left(\Qop_z-\frac{\lambda^2}{2}\Qop_x^2\Pop_z\right)+\epsilon^4\left(-\frac{\lambda^2}{8}(\Pop_+\Pop_-+\Pop_-\Pop_+)\Qop_z-\frac{\lambda^4}{8}\Qop_x^4\Pop_z\right).
 \label{order4twophotonRabi}
\end{equation}
Near the transition, the effective potential is $\epsilon^2(\Qop_z-\frac{\lambda^2}{2}\Qop_x^2)-\epsilon^4\frac{\lambda^4}{8}\Qop_x^4-\epsilon^4\frac{\lambda^2}{4}\Qop_z$, which is still unstable.

\item \textit{Boson-boson model:} Both $\Pop$ and $\Qop$ obey bosonic commutation relations. This is the case when we consider two coupled bosonic fields with vastly different frequencies. In this case, we find the following operators: $\Sop_1=i\lambda \Qop_x\Pop_y$, $\Sop_2=0$, $\Sop_3=-i\lambda \Qop_y\Pop_x$ and $\Sop_4=\frac{\lambda^2}{32}(\Pop_+^2-\Pop_-^2)$. The transformed Hamiltonian reads: 
\begin{equation}
	\Uop\Hop\Uop^\dagger=\Pop_z+\epsilon^2\left(\Qop_z-\frac{\lambda^2}{4}\Qop_x^2\right)+\epsilon^4\left(\frac{\lambda^2}{8}(\Pop_+\Pop_-+\Pop_-\Pop_+)\right).
	\label{order4twophotlike}
\end{equation}

Importantly, the quartic term does not depend on $\Qop_x$. As a consequence, the quadratic term is not compensated, and the Hamiltonian becomes unstable when $\lambda$ increases. This is not unexpected, since a system of two coupled harmonic oscillators can generally become unstable (the same conclusion can be reached by diagonalizing the system directly).
\end{itemize}

\section{Two-photon Dicke model: squeezed phase}

 The analysis for the two-photon Dicke model in the squeezed phase is similar to the one in the normal phase. We start by displacing the $\bop$ field according to the mean-field prediction: $\bop=\sqrt{N}\beta+\dop$ with $\beta=\pm\beta_0=O(1)$. Because the field is centered around a large value, some of the nonlinear terms that could be neglected in the normal phase will start to play a role. To take this into account, we need to develop the spin term $\Jp=(\sqrt{N}\beta+\dop)\sqrt{1-\beta^2-\frac{\beta(\dop+\opddag)}{\sqrt{N}}-\frac{\opddag\dop}{N}}$ in powers of $N$. We obtain: 
 $$2\Jx=2\tilde{\chi} N\beta + \sqrt{N}\tilde{\chi}\left(1-\frac{\beta^2}{\tilde{\chi}^2}\right)(\dop+\opddag) - \mathcal{W}(\dop) + O\left(\frac{1}{\sqrt{N}}\right),$$ where we have defined $\tilde{\chi}=\sqrt{1-\beta^2}=O(1)$ and 
  $$\mathcal{W}(\dop)=\frac{\beta}{\tilde{\chi}}\opddag\dop+\frac{\beta^3}{4\tilde{\chi}^3}(\dop+\opddag)^2+\frac{\beta}{2\tilde{\chi}}(\dop+\opddag)^2-\frac{\beta}{2\tilde{\chi}}.$$

 The two-photon Hamiltonian can then be developed as:

 \begin{equation}
 	\Hop=(2\omega\Ko+8\gtwo\tilde{\chi}\beta\Kx)+\frac{4\gtwo\tilde{\chi}}{\sqrt{N}}\left(1-\frac{\beta^2}{\tilde{\chi}^2}\right)\Kx(\dop+\opddag) -\frac{4\gtwo}{N}\mathcal{W}(\dop)\Kx+\Omega\sqrt{N}\beta(\dop+\opddag)+ \Omega\opddag\dop + O\left(\frac{\omega}{N\sqrt{N}}\right).
 \end{equation}

 We can rewrite the dominant term using the Bogoliubov transformation on the $\hat{K}$ operator that we defined during the mean-field analysis. 
 Eliminating additive constants, we obtain:

 \begin{align}
 \nonumber
  \frac{\Hop}{2\omega} = & c_0 \Kop + \frac{c_1}{\sqrt{N}}(\dop+\opddag) + \frac{c_2}{\sqrt{N}}(d+\opddag)\Kxp + \frac{c_3}{\sqrt{N}}(\dop+\opddag)\Kop + \frac{\ratbis}{2N}\opddag d \\
 & - \frac{2\gtwo}{N\omega}\cosh(2\xi_b)\Kxp \mathcal{W}(\dop) + \frac{2\gtwo}{N\omega}\sinh(2\xi_b)\Kop \mathcal{W}(\dop) ,
 \end{align}
 plus some corrections of order $\frac{1}{N\sqrt{N}}$. We have defined the following quantities:
 \begin{equation*}
 c_0=\sqrt{1-\frac{4\gprime^2}{\omega^2}},\hspace{5pt} c_1=\frac{\Omega N\beta}{2\omega}, \hspace{5pt} c_2 = 2\cosh(2\xi_b)\frac{\gtwo\tilde{\chi}}{\omega}\left(1-\frac{\beta^2}{\tilde{\chi}^2}\right), \hspace{5pt} c_3=-2\sinh(2\xi_b)\frac{\gtwo\tilde{\chi}}{\omega}\left(1-\frac{\beta^2}{\tilde{\chi}^2}\right),
 \end{equation*}

 which are all of order $O(1)$. Next, we apply the following transformation upon this Hamiltonian: $e^{i\Sop}$ with $\Sop=\frac{1}{\sqrt{N}}\frac{c_2}{c_0}(\dop+\opddag)\Kyp - \frac{1}{N}\Kyp\left(\frac{c_3c_2}{c_0^2}(\dop+\opddag)^2+ \frac{2\gtwo}{\omega c_0}\cosh(2\xi_b)\mathcal{W}(\dop) \right)$. This yields a Hamiltonian commuting with $\Kop$:

 \begin{equation}
 e^{i\Sop} \frac{\Hop}{2\omega} e^{-i\Sop} = c_0\Kop + \frac{c_3}{\sqrt{N}}(\dop+\opddag)\Kop + \frac{c_1}{\sqrt{N}}(\dop+\opddag) + \frac{\ratbis}{2N}\opddag \dop + \frac{2\gtwo}{N\omega}\sinh(2\xi_b)\Kop \mathcal{W}(\dop) - \frac{c_2^2}{2Nc_0}(\dop+\opddag)^2\Kop,
 \end{equation}

 Plus some terms of order $\frac{1}{N\sqrt{N}}$. The final step is a projection into the lowest eigenspace of $\Kop$, which yields an effective Hamiltonian for $\dop$. This Hamiltonian contains both quadratic and linear terms. The linear terms can be absorbed by a further displacement of the $\bop$ field, which corresponds to adding a small correction to $\beta$. The remaining quadratic terms may then be treated through Bogoliubov transformation. Once more, we obtain a ladder of squeezed Fock states for the $\dop$ field. The squeezing parameter and eigenvalues are, respectively: 

 $$\xi_b^{(S)}=\frac{1}{4} \text{ln}\left(\frac{1}{(1-\beta^2)^2}\left(1-\frac{\gtwo_c^2(1-2\beta^2)^2}{\gtwo_c^2-\gprime^2}\right)\right)=-\frac{1}{4}\text{ln}\left(\frac{\lambda^4}{4(\lambda^4-1)}\left(1+\sqrt{\frac{1-\frac{\gtwo^2}{\gtwo_c^2}}{\lambda^4-\frac{\gtwo^2}{\gtwo_c^2}}}\right)^2\right),$$ 
 $$E^{(S)}_m=m\Omega\sqrt{\left(\frac{1}{(1-2\beta^2)^2}-\frac{\gtwo_c^2}{\gtwo_c^2-\gprime^2}\right)}=m\Omega\sqrt{\frac{\left(\lambda^4-\frac{\gtwo^2}{\gtwo_c^2}\right)\left(1-\frac{1}{\lambda^4}\right)}{\left(1-\frac{\gtwo^2}{\gtwo_c^2}\right)}}.$$

\chapter{Metrology with Gaussian states}

\renewcommand{\theequation}{\thechapter.\arabic{equation}}

	In this Appendix, we provide some useful results for quantum metrology with Gaussian state. In the first section, we discuss the expression of Gaussian states and PLE in phase space. In the second one, we consider the evaluation of a quantum channel by a Gaussian state, and give a general expression for the QFI. Third, we prove Lemma 1 of Chapter 6. Finally, we show that a two-mode displaced thermal state, when reduced to a two-photon subspace, is not formally entangled.

	\section{Gaussian states in phase space}
 \label{appendixB}

In the following, it will be very useful to work with a phase-space representation of our states, instead of the usual Hilbert space representation. We will make extensive use of the parameterization and expression developed in \cite{safranek_gaussian_2016} and \cite{safranek_optimal_2016}.
 For a $q$-mode state, we can collect the creation and annihilation operators in a vector $\mathbf{A}=(\hat{\textbf{a}},\hat{\textbf{a}}^{\dagger})^T$, with $\hat{\textbf{a}}=(\hat{a}_1,...,\hat{a}_q$). We will study the symmetric characteristic function $tr[\rop e^{\mathbf{A}^{\dagger} \Kmat \xivect}]$, where $\xivect$ is in $\mathbb{C}^{2q}$ and can be written like $\xivect=\mathbf{x}\oplus\mathbf{x}^*$, and $\Kmat$ is the so-called symplectic form, which here reads as $\Kmat=\left[ {\begin{array}{cc}
\mathds{1}_q & 0\\
0 & -\mathds{1}_q\\	
\end{array}} \right]$
where $\mathds{1}_q$ is the $q\times q$ identity matrix. For Gaussian states, these functions take a Gaussian form, \textit{i.e.}
\begin{equation}
\text{Tr}[\rop e^{\mathbf{A}^{\dagger} \Kmat \xivect}]=\exp\Big(-\frac{\xivect^{\dagger}\sigma\xivect}{4}-i\ddagvect \Kmat\xivect\Big).
\end{equation}

Here $\sigma$ is the covariance matrix and $\dvect$ the displacement vector, defined as:
\begin{subequations}

\begin{equation}
	\dvect^i=\text{Tr}[\rop\mathbf{A}^i],
\end{equation}
\begin{equation}
	\sigma^{ij}=\text{Tr}[\rop\{\mathbf{A}^i-\dvect^i,\mathbf{A}^{j\dagger}-\dvect^{j\dagger}\}].	
\end{equation}
\end{subequations}

For a $q$-mode thermal state, one has $\dvect=0$ and $\sigma=\text{diag}(\nu_1,...\nu_q,\nu_1,...,\nu_q)$, where $\nu_i=\text{cotanh}(\frac{\hbar\omega_i}{2k_BT_i})$.

In phase space, phase-shifting, beam-splitter, and squeezing unitaries act as symplectic operations $\Gmat$, \textit{i.e.}, $q\times q$ matrices that verify $\Gmat \Kmat \Gmat^{\dagger}=\Kmat$. These matrices define a complex representation of the \textit{real symplectic group} Sp(2q,$\mathbb{R}$). Under such unitaries, first and second moments transform as: $\dvect\rightarrow \Gmat \dvect$ and $\sigma\rightarrow \Gmat \sigma \Gmat^{\dagger}$. Displacement operators $\Dop(\bm{\alpha})$ act as $\dvect\rightarrow\dvect+(\bm{\alpha},\bm{\alpha}^*)^T$ and leave $\sigma$ invariant. \\

The covariance matrix of a Gaussian state can be parameterized according to the so-called Williamson decomposition. We have the following:
\begin{subequations}
\label{dsigma}
\begin{equation}
	\dvect=(\bm{\dispmod},\bm{\dispmod}^*)^T,
\end{equation}
\begin{eqnarray}
	 \nonumber \sigma = \mathpzc{G}_\sigma^{\dagger} \hspace{5pt} \text{diag}(\nu_1,\nu_2,\nu_1,\nu_2) \hspace{5pt} \mathpzc{G}_\sigma,
\end{eqnarray}
\end{subequations}
where $\text{diag}$ is the diagonal matrix, and $\Gmat_\sigma$ a symplectic matrix, which can itself be decomposed like: 
$$\Gmat_\sigma=\Rmat_1(\phi_1)\Rmat_2(\phi_2)\Bmat(\theta_2)\Rmat_{as}(\Psi_2)\Smat_1(\xi_1)\Smat_2(\xi_2)\Rmat_{as}(\Psi_1) \Bmat(\theta_1).$$ 
 $\Rmat$, $\Bmat$ and $\Smat$ represent the action of PLE in phase space, and read:\\

$$\Rmat_1(\phi_1)=\begin{bmatrix}
e^{-i\phi_1} & 0 & 0 & 0 \\
0 & 1 & 0 & 0 \\
0 & 0 & e^{i\phi_1} & 0 \\
0 & 0 & 0 & 1\\
\end{bmatrix}, \hspace{8pt} \Rmat_2(\phi_2)=\begin{bmatrix}
1 & 0 & 0 & 0 \\
0 & e^{-i\phi_2} & 0 & 0 \\
0 & 0 & 1 & 0 \\
0 & 0 & 0 & e^{i\phi_2} \\
\end{bmatrix}, \hspace{8pt} \Bmat(\theta)=\begin{bmatrix}
\cos(\theta) & \sin(\theta) & 0 & 0 \\
-\sin(\theta) & \cos(\theta) & 0 & 0 \\
0 & 0 & \cos(\theta) & \sin(\theta) \\
0 & 0 & -\sin(\theta) & \cos(\theta) \\
\end{bmatrix},$$\vspace{8pt}

$$\Smat_1(\xi_1)= \begin{bmatrix}
\cosh(\xi_1) & 0 & -\sinh(\xi_1) & 0 \\
0 & 0 & 0 & 0 \\
-\sinh(\xi_1) & 0 & \cosh(\xi_1) & 0 \\
0 & 0 & 0 & 0 \\
\end{bmatrix}, \hspace{8pt} \Smat_2(\xi_2)= \begin{bmatrix}
0 & 0 & 0 & 0 \\
0 & \cosh(\xi_2) & 0 & -\sinh(\xi_2) \\
0 & 0 & 0 & 0 \\
0 & -\sinh(\xi_2) & 0 & \cosh(\xi_2) \\
\end{bmatrix}, \hspace{8pt} \Rmat_{as}(\phi)=\Rmat_1(\phi)\Rmat_2(-\phi).$$ \\

Physically, $\Rmat_i$ represent the action in phase space of a phase-shift applied in mode $i$, $\Bmat$ represents the action of mode-mixing, and $\Smat_i$ is associated with the squeezing of mode $i$.\\

As of the symplectic eigenvalues $\nu_i$, they can be obtained by solving the usual eigenvalue problem for the matrix $\Mmat=\Kmat\sigma$. Since $\Kmat$ is invertible, the $\nu_i$ are uniquely defined for a given Gaussian state. From a physical perspective, these symplectic eigenvalues quantify the purity of the state: we have $\nu_i=1$ for a pure state and $\nu\rightarrow\infty$ for a completely mixed state. \\

Note that $\Bmat^{\dagger}=\Bmat^{-1}$ and $\Rmat_i^{\dagger}=\Rmat_i^{-1}$; thus, for isotropic states $\nu_1=\nu_2=\nu$, we can simplify Eq.(\ref{dsigma}) and obtain:
\begin{equation}
\label{decomposition}
	\sigma=\nu\Rmat_1(\phi_1)\Rmat_2(\phi_2)\Bmat(\theta)\Rmat_{as}(\Psi)\Smat_1(\xi_1)\Smat_2(\xi_2)(.)^{\dagger}.
\end{equation}

Finally, let us give the complete expression for the mean number of photons, still for a two-mode state:

\begin{equation}
\label{Nb}
	\langle\Nop\rangle=\adag\aop+\bdag\bop = \frac{1}{4}\text{Tr}[\sigma]-1+\lvert \bm{\dispmod}\rvert^2=\frac{1}{4}\text{Tr}[\sigma]-1+\frac{\lvert \bm{d}\rvert^2}{2}.
\end{equation}
For a $q$-mode state, we have:

\begin{equation}
	\langle\Nop\rangle = \frac{1}{4}Tr[\sigma]-\frac{q}{2}+\frac{\lvert \bm{d}\rvert^2}{2}.
\end{equation}

Due to the unitarity of $\Rmat_1$, $\Rmat_2$ and $\Bmat$, the phase-shifting and mode-mixing operation preserve the trace of the covariance matrix. Therefore, we have:

\begin{eqnarray*}
\text{Tr}[\sigma]& = & \text{Tr}[\nu\Smat_1(\xi_1)\Smat_2(\xi_2)\Smat_2(\xi_2)^{\dagger}\Smat_1(\xi_1)^{\dagger}]\\
 & = & \text{Tr}[\nu\Smat_1(2\xi_1)\Smat_2(2\xi_2)],
\end{eqnarray*}

which yields:
\begin{eqnarray}
	\text{Tr}[\sigma] & = & 2\nu(\cosh{2\xi_1}+\cosh{2\xi_2})\\
	 & = & 4\nu+4\nu(\sinh(\xi_1)^2+\sinh(\xi_2)^2). \nonumber
\end{eqnarray}

\section{Metrology with Gaussian state}

We consider a Gaussian probe used to evaluate a quantum channel $\Lambda_x$. During the evolution, the covariance matrix and the displacement vector evolve according to $\dvect_0\rightarrow\dvect_x$ and $\sigma_0\rightarrow\sigma_x$. We will call $\dot{\dvect_x}$ and $\dot{\sigma_x}$ the derivatives with respect to $x$. For now, we will consider a general, multi-mode state.
Then the QFI for the estimation procedure is given by the following, implicit equation \cite{monras_phase_2013,banchi_quantum_2015,safranek_quantum_2015-1}:

\begin{equation}
	\pazocal{I}_x=\frac{1}{2}Tr[\dot{\sigma}_x\mathpzc{Y}] + 2\dot{\dvect_x}^\dagger\sigma_x^{-1}\dot{\dvect_x},
\end{equation}

where $\mathpzc{Y}$ is a solution of the Stein equation: $\dot{\sigma}_x=\sigma_x \mathpzc{Y}\sigma_x -\Kmat \mathpzc{Y}\Kmat$. If all symplectic eigenvalues of $\sigma_x$ are larger than one, we can obtain the QFI in an \textit{explicit} form \cite{safranek_quantum_2015-1}:

\begin{equation}
	\pazocal{I}_x=\frac{1}{2}\sum_{n=1}^\infty \text{Tr}[(\Mmat_x^{-n}\dot{\Mmat}_x)^2] + 2\dot{\dvect}_x^\dagger\sigma_x^{-1}\dot{\dvect}_x,
	\label{sumQFIinfinite}
\end{equation}

with $\Mmat=K\sigma_x$. This expression, however, still involves an infinite number of terms. To simplify it, we will make a few additional assumptions. First, we have the following properties, which hold in general: 

\begin{itemize}
	\item $\sigma_x=\Sigmatd_x \Dmat_x\Sigmatn_x$,
	\item $\Dmat_x=\text{diag}(\nu_1,\nu_2,\nu_1,\nu_2)$ and $\Kmat \Dmat_x=\text{diag}(\nu_1,\nu_2,-\nu_1,-\nu_2)$,
	\item $\Sigmatn_x \Kmat \Sigmatd_x=\Kmat$,
	\item $\Sigmati_x=\Kmat\Sigmatd_x \Kmat$,
	\item $\Mmat_x=\Kmat\sigma_x=\Sigmati_x \Kmat \Dmat_x\Sigmatn_x$,
	\item $\Mmat_x^{-1}=\Kmat\Sigmatd_x \Dmat_x^{-1} \Sigmatn_x=\Sigmati_x (\Kmat \Dmat_x)^{-1} \Sigmatn_x$,
	\item $\Mmat_x^2=\Sigmati_x (\Kmat \Dmat_x)^2 \Sigmatn_x=\Sigmati_x \Dmat_x^2 \Sigmatn_x$, 
	\item $\Mmat_x^{-1}\dot{\Mmat_x}=\sigma_x^{-1}\dot{\sigma}_x$,
	\item $\lvert \Mmat_x \rvert=\Pi_{i=1}^q\nu_i^2$.
\end{itemize}

Now, we will assume that the channel $\Lambda$ is a PLE. Then the symplectic eigenvalues $\nu_i$ are independent of $x$, and we have $\Dmat_x=\Dmat_0$. Furthermore, we will consider isotropic state, for which $\nu_i=\nu$ for all $i$. Then we have:

\begin{itemize}
	\item $\sigma_x=\nu\Sigmatd_x\Sigmatn_x$,
	\item $\Mmat_x^2=\nu^2\mathbb{1}$,
	\item $\dot{\Mmat}_x\Mmat_x+\Mmat_x\dot{\Mmat}_x=0$,
	\item $\Kmat\sigma_x^{-1}\Kmat=\frac{1}{\nu^2}\sigma_x$,
	\item $\Mmat_x^{-1}=\frac{1}{\nu^2}\Mmat_x$,
	\item $\dot{\Mmat}_x=-\nu^2\Mmat_x^{-1}\dot{\Mmat}_x\Mmat_x^{-1}$, and therefore $(\Mmat_x^{-1}\dot{\Mmat}_x)^2=(\sigma_x^{-1}\dot{\sigma}_x)^2=-\frac{1}{\nu^2}\dot{\Mmat}_x^2$.
\end{itemize}

Using the properties above, we can deduce that $\Mmat^{-2p}=\frac{1}{\nu^{2p}}$ and $\Mmat^{-(2p+1)}\dot{\Mmat}=\frac{1}{\nu^{2p}}\sigma^{-1}\dot{\sigma}$. Then we can develop the sum \eqref{sumQFIinfinite}, and we find that when an \textbf{isotropic} Gaussian state is used to measure a PLE channel, the QFI is given by:

\begin{align}
	\pazocal{I}_x & =\frac{1}{2(\nu^4-1)}\left[\nu^4Tr[(\sigma_x^{-1}\dot{\sigma}_x)^2]+Tr[(\Kmat\dot{\sigma}_x)^2]\right]+ 2\dot{\dvect}_x^\dagger\sigma_x^{-1}\dot{\dvect}_x\\ \nonumber
	&= -\frac{Tr[(\Kmat\dot{\sigma}_x)^2]}{2(1+\nu^2)}+2\dot{\dvect}_x^\dagger\sigma_x^{-1}\dot{\dvect}_x\\ \nonumber
	&=\frac{\nu^2Tr[(\sigma_x^{-1}\dot{\sigma}_x)^2]}{2(1+\nu^2)}+2\dot{\dvect}_x^\dagger\sigma_x^{-1}\dot{\dvect}_x.
\end{align}

For a non-isotropic state and an arbitrary Gaussian channel (including also channels which can increase $\nu$), no general expression is known to the best of our knowledge. However, if we restrict ourselves to \textbf{two-mode} states, then the exact expression for the QFI was found in \cite{safranek_quantum_2015-1}, and reads:

\begin{align}
	\nonumber
	\pazocal{I}_x & =\frac{1}{2(\lvert \Mmat_x\rvert-1)}\Big[\lvert \Mmat_x\rvert Tr[(\Mmat_x^{-1}\dot{\Mmat}_x)^2]+\sqrt{\lvert\mathbb{1}+\Mmat_x^2\rvert}Tr\Big[((\mathbb{1}+\Mmat_x^2)^{-1}\dot{\Mmat}_x)^2\Big] \\
& + 4(\nu_1^2-\nu_2^2)\big(\frac{\dot{\nu_2}^2}{\nu_2^4-1}-\frac{\dot{\nu_1}^2}{\nu_1^4-1}\big)  \Big] + 2\dot{\dvect}_x^\dagger\sigma_x^{-1}\dot{\dvect_x}.
\end{align}

\section{Proof of Lemma 1, and specific examples}
We proceed as for Theorem 1. The cases $V_I>0$ and $V_I<0$ remain the same when $\nu=1$; in this case, the squeezed state can be used to strictly surpass the FTQL. Therefore, we will only consider the case $V_I=0$. 

As we said, this case corresponds to a displaced squeezed state in one mode and a thermal state in the other. Let us assume, without loss of generality, that the thermal state is in the second mode. We start from a state with $\sigma=\Rmat_1(\phi_1)\Rmat_2(\phi_2)\Smat_1(\xi_1)\times h.c.$ and $\dvect=\lvert\dispmod\rvert(e^{i\phi_d},0,e^{-i\phi_d},0)$. In this case, we can find explicitly the protocol which gives the optimal precision. We apply a general PLE, which can be decomposed using Euler's angles: $\Gop=\Rop_z(a)\Rop_x(b)\Rop_z(c)$. The resulting state $\rop^\uvect=\Gop\rop\Gop^\dagger$ is described by the following set of parameters: $\phi_1^\uvect=a+\frac{\phi_1+\phi_2}{2}$, $\phi_2^\uvect=-a+\frac{\phi_1+\phi_2}{2}$, $\theta^\uvect=b$, $\Psi^\uvect=\frac{\phi_1-\phi_2}{2}+c$, $\phi_{d1}^\uvect=\phi_d-c-a$, $\phi_{d2}^\uvect=\phi_d-c+a$, and $l^\uvect=-b$. This yields:

\begin{align}
\label{specificase}
	\pazocal{I}_\phi^\uvect= & \frac{8\nu^2}{\nu^2+1}\sinh{\xi_1}^2 + \frac{4\lvert\dispmod\rvert^2}{\nu} + \sin^2(2b)\sin^2(2a)\Big[\frac{4\nu^2}{\nu^2+1}\big(\sinh^2(2\xi_1)-2\sinh^2(\xi_1)\big) \\
& + \frac{4\lvert\dispmod\rvert^2}{\nu}(e^{2\xi_1}\sin^2\big(\phi_d-\phi_1)+e^{-2\xi_1}\cos^2(\phi_d-\phi_1) - 1\big)\Big].
\end{align}
We can now optimize \eqref{specificase} with respect to $a$, $b$ and $c$. Three cases are possible:
\begin{itemize}
	\item $V=e^{2\xi_1}\sin^2\big(\phi_d-\phi_1)+e^{-2\xi_1}\cos^2(\phi_d-\phi_1) - 1\geq 0$. Then the optimal protocol is always to set $b=a=\frac{\pi}{4}$. The total protocol is then equivalent to a simple one-mode phase estimation protocol. 
	We have then: 
	\begin{equation}
	\pazocal{I}_\phi^{\text{opt}}= \frac{4\nu^2}{\nu^2+1}\sinh^2(2\xi_1) + \frac{4\lvert\dispmod\rvert^2}{\nu}(V+1).
	\end{equation}

	\item $V<0$, and $-4\frac{\lvert\dispmod\rvert^2 V}{\nu} < \frac{4\nu^2}{\nu^2+1}\big(\sinh^2(2\xi_1)-2\sinh^2(\xi_1)\big)$. Then the optimal strategy and $\pazocal{I}_\phi^{\text{opt}}$ remain the same.
	\item $V<0$, and $-4\frac{\lvert\dispmod\rvert^2}{\nu}V \geq \frac{4\nu^2}{\nu^2+1}\big(\sinh^2(2\xi_1)-2\sinh^2(\xi_1)\big)$. Then the optimal precision is achieved by setting $a=b=0$, that is, sending directly the state in the Mach-Zehnder interferometer. In this case, we have
\begin{equation}
	\pazocal{I}_\phi^{\text{opt}}=\frac{8\nu^2}{\nu^2+1}\sinh{\xi_1}^2 + \frac{4\lvert\dispmod\rvert^2}{\nu}.
\end{equation}
	
\end{itemize}

The FTQL is $\pazocal{I}_\phi^{\text{ref}}=4\frac{\lvert\dispmod\rvert^2}{\nu} + 4\sinh{\xi_1}^2$. In the first two cases, $\pazocal{I}_\phi^{\text{opt}}>\pazocal{I}_\phi^{\text{ref}}$ for all $\nu$. In the third case, $\pazocal{I}_\phi^{\text{opt}}\geq\pazocal{I}_\phi^{\text{ref}}$, with equality if and only if $\nu=1$. Hence, if the state is pure, the FTQL can be attained, but not surpassed. This shows that, although pure squeezed states can always reach the FTQL, there are some specific cases in which they cannot beat this bound.

\section{Separability of displaced thermal states}
\label{separability displtherm}

In this section, we will prove that a two-mode displaced thermal state, when restricted in the subspace with $2$ photons, is not formally entangled.
We will start by considering a general two-mode displaced Fock state 
$\ket{p,\beta_1}\ket{m,\beta_2}=\Dop(\beta_1)\ket{p}\Dop(\beta_2)\ket{q}=\sum_{N}\ket{p,\beta_1}\ket{m,\beta_2}^{(N)}$.
Here $\ket{p, \beta_1}\ket{m, \beta_2}^{(N)}$ is the component of $\ket{p, \beta_1}\ket{m, \beta_2}$ with a total average number of particle equal to $N$.
We will focus on the $N=2$ subspace. By direct computation, the state in this subspace reads:

\begin{align}
\label{Dispfock}
	\ket{p, \beta_1}\ket{m, \beta_2}^{(2)}= & \frac{(\beta^*_1)^p(\beta^*_2)^m}{\sqrt{p!m!}}e^{-\frac{\lvert\beta_1\rvert^2+\lvert\beta_2\rvert^2}{2}} \Big[ \frac{(\beta_1)^2}{\sqrt{2}} \Big(1+\frac{2p}{\lvert\beta_1\rvert^2}+\frac{p(p-1)}{\lvert\beta_1\rvert^4}\Big) \ket{20} \\ \nonumber
	& + \frac{(\beta_2)^2}{\sqrt{2}} \Big(1+\frac{2m}{\lvert\beta_2\rvert^2}+\frac{m(m-1)}{\lvert\beta_2\rvert^4}\Big)\ket{02} + \beta_1\beta_2\Big(1+\frac{p}{\lvert\beta_1\rvert^2}\Big)\Big(1+\frac{m}{\lvert\beta_2\rvert^2}\Big)\ket{11} \Big],
\end{align}

where $\ket{m p}$ means m(p) photons in the first(second) mode. We have omitted a global normalization factor to avoid cluttering. 

We are now ready to study two-mode displaced thermal states. These states can be decomposed using displaced Fock state, according to: $$\rop_{th}(\beta_1,\beta_2)=\sum_{p,m}\frac{\Theta_1^p\Theta_2^m}{Z_1Z_2}\ket{p, \beta_1}\ket{m, \beta_2}\bra{p, \beta_1}\bra{m, \beta_2},$$

where $\Theta_i=e^{-\frac{\hbar\omega_i}{k_BT_i}}$, and the $Z_i$ are partition functions. We can now find the expression of this state in the $N=2$ subspace, by using \eqref{Dispfock}. Tedious but straightforward computations yield (up to normalization):
\begin{align}
\label{expression_discoh} \nonumber
	\rop_{th}(\beta_1,\beta_2)^{(2)}&=C\Big[\Phi_1 \ket{20}\bra{20} + \Phi_2 \ket{02}\bra{02} + \aleph \ket{11}\bra{11} + (\Upsilon_1 \ket{11}\bra{20} + \Upsilon_2 \ket{11}\bra{02} + \Xi\ket{20}\bra{02} + c.c)\Big],\\
	C&=\frac{1}{Z_1Z_2}e^{\lvert\beta_1\rvert^2(\Theta_1-1)+\lvert\beta_2\rvert^2(\Theta_2-1)},\\ \nonumber
	\Phi_i&=\frac{1}{2}\Big(\lvert\beta_i\rvert^4(1+\Theta_i)^4+4\lvert\beta_i\rvert^2\Theta_i(1+\Theta_i)^2+2\Theta_i^2),\\ \nonumber
	\Upsilon_1&=\frac{(\beta^*_1\beta_2}{\sqrt{2}}(1+\Theta_1)(1+\Theta_2)\Big(\lvert\beta_1\rvert^2(1+\Theta_1)^2 + 2\Theta_1 \Big),\\ \nonumber
	\Xi&=\beta_1^2(\beta^*_2)^2\frac{(1+\Theta_1)^2(1+\Theta_2)^2}{2},\\ \nonumber
	\aleph&=\Big(\lvert\beta_1\rvert^2(1+\Theta_1)^2+\Theta_1\Big)\Big(\lvert\beta_2\rvert^2(1+\Theta_2)^2+\Theta_2\Big).
\end{align}

 In the single particle basis ($\ket{\psi_1^a\psi_1^b}$,$\ket{\psi_1^a\psi_2^b}$, $\ket{\psi_2^a\psi_1^b}$, $\ket{\psi_2^a\psi_2^b}$), we can rewrite the density matrix as:

\begin{equation}
 	\begin{bmatrix}
 	\Phi_1 & \frac{\Upsilon_1}{\sqrt{2}} & \frac{\Upsilon_1}{\sqrt{2}} & \Xi\\
 	\frac{\Upsilon^*_1}{\sqrt{2}} & \frac{\aleph}{2} & \frac{\aleph}{2} & \frac{\Upsilon_2}{\sqrt{2}}\\ 
 	\frac{\Upsilon^*_1}{\sqrt{2}} & \frac{\aleph}{2} & \frac{\aleph}{2} & \frac{\Upsilon_2}{\sqrt{2}}\\ 
 	\Xi^* & \frac{\Upsilon^*_2}{\sqrt{2}} & \frac{\Upsilon^*_2}{\sqrt{2}} & \Phi_2 
 	\end{bmatrix}.
 \end{equation} 
To study the separability of this state, we used Peres-Horodecki criterion \cite{peres_separability_1996,horodecki_separability_1996}. We compute the partial transpose of the matrix, then rewrite it in the $\ket{\psi_1^a\psi_1^b},\frac{\ket{\psi_1^a\psi_2^b} + \ket{\psi_2^a\psi_1^b}}{2}, \ket{\psi_2^a\psi_2^b}, \frac{\ket{\psi_1^a\psi_2^b} - \ket{\psi_2^a\psi_1^b}}{2}$ basis. We obtain:
\begin{equation}
	\begin{bmatrix}
 	\Phi_1 & \frac{\Upsilon_1+\Upsilon^*_1}{2} & \frac{\aleph}{2} & \frac{\Upsilon^*_1-\Upsilon_1}{2}\\ 
 	\frac{\Upsilon_1+\Upsilon^*_1}{2} & \frac{\aleph}{2} + \frac{\Xi+\Xi^*}{2} & \frac{\Upsilon_2+\Upsilon^*_2}{2} & 0\\ 
 	\frac{\aleph}{2} & \frac{\Upsilon_2+\Upsilon^*_2}{2} & \Phi_2 & \frac{\Upsilon^*_2-\Upsilon_2}{2}\\ 
 	\frac{\Upsilon_1-\Upsilon^*_1}{2} & 0 & \frac{\Upsilon_2-\Upsilon^*_2}{2} & \frac{\aleph}{2} - \frac{\Xi+\Xi^*}{2}
 	\end{bmatrix}.
\end{equation}
$\rop_{th}(\beta_1,\beta_2)^{(2)}$ is separable if and only if this matrix is positive.
In what follows, we will consider only the symmetric case $\Theta_1=\Theta_2=\Theta$ and $\beta_1=\beta_2=\beta$. This means $\Phi_1=\Phi_2=\Phi$, $\Upsilon_1=\Upsilon_2=\Upsilon=\Upsilon^*$, $\Xi=\Xi^*$.

The matrix reduces to :
\begin{equation}
	\begin{bmatrix}
 	\Phi & \Upsilon & \frac{\aleph}{2} & 0\\ 
 	\Upsilon & \frac{\aleph}{2} + \Xi & \Upsilon & 0\\ 
 	\frac{\aleph}{2} & \Upsilon & \Phi & 0\\ 
 	0 & 0 & 0 & \frac{\aleph}{2} - \Xi
 	\end{bmatrix}.
\end{equation}
We can immediately isolate the eigenvalue $\frac{\aleph}{2}-\Xi$. Given the expressions \eqref{expression_discoh}, this quantity is positive. We note $y_1$, $y_2$, $y_3$ the other eigenvalues. Using the derivatives of the characteristic polynomial, we obtain: 
\begin{align}
	y_1+y_2+y_3 & =2\Phi+\Xi+\frac{\aleph}{2},\\ \nonumber
	y_1y_2+y_3y_2+y_1y_3 & = \Phi^2-\frac{\aleph^2}{4} + 2\Big(\Xi+\frac{\aleph}{2}\Big)\Phi - 2\Phi^2,\\\nonumber
	y_1y_2y_3 & =\Phi^2(\aleph-2\Phi)+\Big(\Xi+\frac{\aleph}{2}\Big)\Big(\Phi^2-\frac{\aleph^2}{4}\Big)\\\nonumber
	 & =\Big(\Phi-\frac{\aleph}{2}\Big)\Big[\Big(\Xi+\frac{\aleph}{2}\Big)\Big(\Phi+\frac{\aleph}{2}\Big)-2\Upsilon^2\Big].
\end{align}

Using \eqref{expression_discoh}, it is straightforward to show that these three expressions are positive. We can directly deduce that $y_1$, $y_2$ and $y_3$ are positive, which means the $N=2$ substate of a symmetric displaced thermal state is not formally entangled.

\chapter{Résumé substantiel en français}

\renewcommand{\theequation}{\thechapter.\arabic{equation}}

Cette thèse porte sur l’étude théorique de systèmes lumière-matière en régime de couplage ultrafort. En particulier, nous avons étudié l’interaction entre des systèmes à deux niveaux (qubits) et un champ bosonique, qui peut être décrite par le modèle de Rabi :

\begin{equation}
	\Hop=\Of\adag\aop+\frac{\Oq}{2}\sigz+\gind(\adag+\aop)\sigx,
\end{equation}

ou, dans le cas de plusieurs qubits, par celui de Dicke :

\begin{equation}
	\Hop=\Of\adag\aop+\frac{\Oq}{2}\sum_n\sigz_n+\frac{\gcoll}{\sqrt{N}}(\adag+\aop)\sum_n\sigx_n.
\end{equation}

Ces deux modèles décrivent le comportement de systèmes très différents, tels que des circuits supraconducteurs, des atomes en cavité, des ions piégés, ou des excitations électroniques dans des systèmes à l’état solide. Bien que le champ bosonique corresponde généralement à un champ électromagnétique, le même comportement peut s’observer avec un mode mécanique, par exemple lorsque l’état interne d’un ion piégé est couplé à son mouvement de vibration.

Lorsque la constante de couplage $\gcoll$ est faible devant les fréquences propres $\Of$ et $\Oq$ des deux sous-systèmes isolés, les composantes de lumière et de matière conservent leurs propriétés propres. Il est alors justifié de négliger les termes contra-rotatifs $\aop\sum_n\sigm_n+\adag\sum_n\sigp_n$ et de décrire l’interaction par un simple échange de quanta d’énergie entre deux sous-systèmes bien définis. En revanche, lorsque $\gcoll$ devient une fraction significative des fréquences propres (typiquement pour $\gcoll/\Of>0.1$), nous entrons alors dans le régime de couplage ultrafort. Dans ce régime, les termes contra-rotatifs ne peuvent plus être négligés. En termes physiques, lumière et matière sont désormais si fortement couplés qu’ils ne peuvent plus être définis comme des entités séparées; il convient alors de décrire le système par des excitations hybrides, les polaritons. L’état fondamental correspond alors à un vide de polaritons, qui contient un nombre non nul de photons. 

Ainsi, l’augmentation du couplage mène à une modification qualitative des états propres du système. Pour des systèmes résonants (c’est-à-dire tels que $\Of=\Oq$), et avec un nombre de qubits peu élevé, la transition s’effectue de façon continue. En revanche, lorsque le nombre de qubit $N$ tend vers l’infini, on assiste alors à une transition de phase super-radiante. Pour $\gcoll<\gcoll_p=\sqrt{\frac{\Of\Oq}{4}}$, le système est dans la phase normale, pour laquelle l’état fondamental correspond au vide de photons. En revanche, pour $\gcoll>\gcoll_p$, le système entre dans une phase dite super-radiante, dans laquelle le champ bosonique est peuplé par un nombre macroscopique d’excitations. Cette transition a été observé expérimentalement à plusieurs reprises. Bien qu’elle ait été souvent présentée comme un effet collectif dû à la présence d’émetteurs multiples, plusieurs études récentes tendent à montrer qu’il s’agit en fait d’un effet générique lié au couplage ultrafort. En particulier, la même transition peut s’observer dans le modèle de Rabi, lorsque la fréquence propre du qubit est très large devant celle du boson ($\Oq\gg\Of$), ou avec deux champs bosoniques présentant une très faible non-linéarité. 

L’un des travaux de cette thèse a porté sur l’étude d’un mécanisme de couplage à deux photons, qui décrit l’émission ou l’absorption de photons par paires. Un tel couplage peut être réalisé dans des circuits supraconducteurs ou par des techniques de simulation quantique. Un ensemble de qubits couplé à un champ bosonique par ce mécanisme peut être décrit par le modèle de Dicke à deux photons : 
\begin{eqnarray}
\hat{H} = \Of \adag \aop + \frac{\Oq}{2}\sum_n\sigz_n + \frac{\gtwo}{N}\sum_n\sigx_n(\aop^2+\adagsq).
\end{eqnarray}

Pour $\gtwo=\gtwo_c=\Of/2$, les niveaux d’énergie de ce modèle fusionnent en une bande d’énergie continue, un phénomène appelé effondrement spectral. Dans une première étude, nous avons étudié la présence de transitions de phase au sein de ce modèle. Une étude de champ moyen prédit qu’une telle transition peut se produire pour $\gtwo_p=\sqrt{\frac{N\Of\Oq}{4}}$. Pour $\gtwo<\gtwo_p$, l’état fondamental correspond à un vide de photons. Pour $\gtwo>\gtwo_p$, le champ bosonique entre dans un état squeezé, lors d’une transition du second ordre.

 Néanmoins, pour que cette transition puisse être observable, elle doit avoir lieu avant l’effondrement spectral ; ce qui ne peut se produire que pour $\Oq<\frac{\Of}{N}$, donc pour une fréquence de qubit très \textit{faible} devant celle du boson. Dans ce régime de paramètres, il est possible de faire une élimination adiabatique du boson par une transformation de Schrieffer-Wolff, ce qui permet d’étudier les fluctuations quantiques du spin, et d’obtenir les exposants critiques. Ceci a permis de conclure que la transition du modèle à deux photons était dans la même classe d’universalité que la transition super-radiante du modèle à un photon.

 Nous discuterons ensuite du comportement du modèle à deux photons en présence d’une interaction avec l’environnement. Dans ce cas, le système doit être décrit par une équation maîtresse de Lindblad :
 \begin{align}
 	\frac{\partial \rop}{\partial t}=-i[\Hop,\rop]+\sum_j \kappa_j \left( \hat{c}_j\rop \hat{c}_j^\dagger - \frac{1}{2}(\hat{c}_j^\dagger\hat{c}_j\rop+\rop\hat{c}_j^\dagger\hat{c}_j) \right), 
 \end{align}

 Trois canaux de dissipation ont été considéré : la perte de photon( $\hat{c}_j=\aop$), le déphasage et la désexcitation des qubits  ($\hat{c}_j=\sigz$ et $\hat{c}_j=\sigm$, respectivement). \'Etant donné que les qubits sont affectés par les pertes de manière individuelle, il n’est plus possible de les modéliser par des modes collectifs. Néanmoins, il est toujours possible d’étudier le comportement de plusieurs observables par une méthode de type champ moyen. Cette étude a montré que, en présence de dissipation, l’effondrement spectral semble disparaître au profit d’une instabilité contrôlée par le taux de dissipation. La transition de phase du second ordre devient du premier ordre, et présente notamment un phénomène de bistabilité au niveau champ moyen. En ajustant la dissipation, il est possible de faire apparaître ou disparaître l’instabilité ; le diagramme de phase comprend donc des phases stable, bistables, et instables.

 Le second axe de recherche présenté dans ce manuscrit concerne l’utilisation de ces phénomènes critiques pour développer des protocoles de métrologie quantique. Tout protocole de mesure est affecté par une incertitude fondamentale, imposée par la mécanique quantique. Ce bruit d’origine quantique est déjà un facteur limitant dans plusieurs systèmes métrologiques tels que des horloges atomiques. Lorsque tout le bruit d’origine technique a été supprimé, l’incertitude est alors donnée par la limite de Cramer-Rao quantique :
 \begin{equation}
 	\delta x\geq\frac{1}{\sqrt{\nbrep \pazocal{I}_x}}
 \end{equation}
 où $\pazocal{I}_x$ est \textit{l’information de Fisher quantique}. L’utilisation d’états quantiques convenablement choisis (par exemple des états de spin comprimés dans des horloges atomiques) permet d’augmenter $\pazocal{I}_x$ et peut ainsi contribuer à améliorer les protocoles métrologiques existants.

 \`A proximité d’un point critique, un système physique devient extrêmement sensible à des perturbations extérieures. Ceci suggère qu’une transition de phase quantique pourrait être utilisée pour réaliser un protocole métrologique. Nous avons appliqué cette idée au modèle de Rabi. Par un protocole de préparation adiabatique, il est possible d’amener le système près du point critique tout en le maintenant dans son état fondamental. En mesurant le champ bosonique par une mesure homodyne ou un comptage de photons, il est possible de déterminer la fréquence propre du qubit ou du champ. Plus le système est proche du point critique, plus la précision de ce protocole, quantifiée par la QFI, devient grande. Néanmoins, le temps nécessaire pour préparer le système diverge aussi, à cause de la fermeture du gap, qui nécessite une évolution adiabatique de plus en plus lente pour maintenir le système dans son état fondamental. Nous avons étudié la précision atteignable en fonction de la durée du protocole $\dur$ ; pour l’estimation de la fréquence du spin, nous avons mis en évidence une loi  d’échelle $\pazocal{I}_\Oq\propto\dur^4$, différente de la loi $\dur^2$ obtenue dans la plupart des dispositifs métrologiques. Ce protocole pourrait par exemple être réalisé en utilisant des atomes froids, en utilisant une transition entre deux niveaux ultrafins pour réaliser le système à deux niveaux. Puisque la fréquence propre de la transition est alors modulée par la présence d’un champ magnétique externe, un tel système pourrait être utilisé pour réaliser des mesures de magnétométrie  à une échelle microscopique.

 Le dernier projet de recherche, plus fondamental, concerne le développement d’une théorie de ressource pour la métrologie quantique. Les corrélations quantiques, comme l’intrication, peuvent être mises à profit pour améliorer des protocoles de calcul ou de métrologie. Inversement, la capacité d’un système à réaliser de telles tâches pourrait être utilisée pour définir et quantifier son caractère non-classique. Cette idée peut être exprimée par le formalisme des théories de ressources. Une théorie de ressources distingue des états ou opérations « libres », qui sont aisément accessibles, et des ressources, considérées comme étant en nombre limité (ainsi, dans le cas de l’intrication, les états intriqués sont des ressources, alors que les états séparables, les opérations locales et la communication classique sont considérées comme facilement accessibles et en quantité illimitée). L’objectif est alors de déterminer quelle quantité de ressources est nécessaire pour réaliser une tâche donnée. Nous avons considéré l’utilisation d’états Gaussiens pour caractériser des éléments optiques passifs, tels que des séparateurs de faisceau ou des retards de phase. Nous avons alors défini un critère de ressource, fondé sur la capacité d’un état donné à réaliser une estimation précise. Nous avons montré comment ce critère reproduisait certaines propriétés des théories de ressources, tout en mettant en évidence certaines différences.

 \end{appendices}


\end{document}